\documentclass[preprint2]{aastex}
\usepackage{epsfig}
\usepackage{graphicx} 
\usepackage{subfigure}
\usepackage{lscape}
\newcommand{\mo}{\ifmmode ^{-1}\else $^{-1}$\fi}
\newcommand{\mt}{\ifmmode ^{-2}\else $^{-2}$\fi}
\newcommand{\microns}{\ifmmode \,\mu$m$\else \,$\mu$m\fi}
\newcommand{\WHzsr}{\ifmmode $\,W\,Hz\mo\,sr\mo$\else \,W\,Hz\mo\,sr\mo\fi}

\def\ltsim{\ifmmode\stackrel{<}{_{\sim}}\else$\stackrel{<}{_{\sim}}$\fi}
\def\gtsim{\ifmmode\stackrel{>}{_{\sim}}\else$\stackrel{>}{_{\sim}}$\fi}

\shorttitle{Spitzer Observations of 3C Quasars and Radio Galaxies}
\shortauthors{}

\begin{document}


\title{Spitzer Observations of 3C Quasars and Radio Galaxies: Mid-Infrared Properties of Powerful Radio Sources}


\author{K.\ Cleary\altaffilmark{1}, C.\ R.\ Lawrence\altaffilmark{1}, J.\ A.\ Marshall\altaffilmark{2}, L.\ Hao\altaffilmark{2} and D.\ Meier\altaffilmark{1}}
\altaffiltext{1}{Jet Propulsion Laboratory, California Institute of Technology; Kieran.A.Cleary@jpl.nasa.gov}
\altaffiltext{2}{Astronomy department, Cornell University, Ithaca, NY 14853}

\begin{abstract}
We have measured mid-infrared radiation from an orientation-unbiased
sample of 3CRR galaxies and quasars at redshifts $0.4\leq z\leq 1.2$ with the
IRS and MIPS instruments on the {\sl Spitzer Space Telescope\/}.
Powerful emission ($L_{24\mu{\rm m}} > 10^{22.4}$\,\WHzsr ) was
detected from all but one of the sources.  We fit the {\sl Spitzer\/} data as well as
other measurements from the literature with synchrotron and dust components.
The IRS data provide powerful constraints on the fits.  At 15\microns, quasars
are typically four times brighter than radio galaxies with the same isotropic
radio power.  Based on our fits, half of this difference can be attributed to the presence
of non-thermal emission in the quasars but not the radio galaxies.  The other half 
is consistent with dust absorption in the radio galaxies but not the quasars.  Fitted
optical depths are anti-correlated with core dominance, from which we infer an
equatorial distribution of dust around the central engine.  The median optical
depth at 9.7\microns\ for objects with core-dominance factor ${\rm R} > 10^{-2}$ is
$\approx 0.4$; for objects with ${\rm R} \leq 10^{-2}$, it is $\approx 1.1$.
We have thus addressed a long-standing question in the unification of FR\,II
quasars and galaxies: quasars are more luminous in the mid-infrared than galaxies
because of a combination of Doppler-boosted synchrotron emission in quasars and extinction in
galaxies, both orientation-dependent effects.
\end{abstract}



\section{INTRODUCTION}

From an observational perspective, active galactic nuclei are diverse. The existence of
relativistic jets in some AGN, with associated Doppler boosting \citep{blandford_rees_78}, guarantees that some of this diversity
of appearance must be due to orientation.  It is equally clear that not all
differences can be due to orientation, and that there are real physical
differences between \hbox{AGN}.   Much observational and theoretical work over
the last three decades has been devoted to sorting out orientational and
physical differences.  The greatest successes have been achieved in the
``unified model'' of Fanaroff-Riley Class II galaxies and quasars \citep{barthel_89}, and in the unification of Seyfert~1 and Seyfert~2
galaxies \citep[see e.g.,][]{antonucci_miller_85}.  In both cases a dusty torus or disk-like
structure obscures optical and ultraviolet emission from the accretion disk and
environs along some lines-of-sight, accounting for apparent differences between these classes of AGN \citep[see e.g.,][]{urry_padovani_95}.

The FR\,II/quasar  case is particularly important for the following reason. These objects are selected to have low frequency (178\,MHz) radio luminosity greater than about
$10^{25}$\,\WHzsr\ \citep{fanaroff_riley_74}.  Low frequency emission is synchrotron emission from giant,
optically thin clouds that therefore emit {\em isotropically}.  The low
frequency emission of dense and possibly Doppler-boosted regions of these
sources is suppressed by self-absorption, which produces a $\nu^{2.5}$
spectrum.  As a result, bolometrically insignificant but highly anisotropic
emission, which may dominate the source at higher frequencies, is almost always
insignificant at a few hundred megahertz or less.  Moreover, high spatial
resolution observations of these sources with interferometers allow one to
measure and subtract the small contribution of potentially anisotropic emitting
regions at low frequency.  Of all the identifying signatures of nuclear
activity, only low frequency radio emission originates far from the nucleus. 
Only low frequency radio emission comes from outside the regions where
anisotropic emission and obscuration are potentially important.  {\it Low
frequency radio emission, therefore, provides a unique way to obtain an
orientation-unbiased sample of \hbox{AGN}}.  

Only a small fraction of AGN produces the prodigious radio power of FR\,II galaxies
and quasars.   Nevertheless, the fact that FR\,IIs can be
studied in an orientation-unbiased sample, selected on the basis of low
frequency radio emission, makes them uniquely valuable in separating the
effects of orientation from physical differences.  Visible-light and X-ray
selected samples do not have this property.

The mid- and far-infrared properties of these powerful radio sources are largely
unknown.  Their space density is so low that only a few (e.g., 3C\,405 = Cygnus
A) are at low redshifts. The Spitzer Space Telescope \citep{werner_etal_04} promised a major advance
in sensitivity over previous telescopes, and the capability to measure these
objects. We therefore undertook observations with {\em Spitzer\/} of powerful radio
sources.  The overall goal was straightforward: to measure for the first time the mid-
and far-infrared emission from an orientation-unbiased sample of the powerful
radio sources.  

More specific goals can also be considered.  For example, the central
proposition of FR\,II galaxy/quasar unification is that FR\,II galaxies are just
quasars seen from an angle in which the optical-to-uv emission of the
central region is intercepted by the opaque, dusty torus, which re-emits this
radiation at longer wavelengths.  Conversely, quasars are just FR\,II
galaxies seen closer to the jet axis.  Models suggest that the dusty
torus becomes optically thin at wavelengths in the far-infrared \citep[see e.g.,][]{pier_krolik_92,granato_danese_94,granato_etal_97,nenkova_etal_02, schartmann_etal_05, fritz_etal_06}, suggesting a
direct test of FR\,II galaxy/quasar orientation-based unification. If radio-lobe luminosity is an indicator (albeit
time-averaged) of central engine energy, then FR\,II galaxies and quasars with
comparable isotropic radio-lobe luminosity should exhibit ``comparable''
isotropic far-infrared luminosity.  Comparable does not necessarily mean equal, but there should be no FR\,IIs
with low nuclear infrared luminosity at wavelengths where the emission is optically thin.

This test has been performed by other investigators using infrared measurements from the Infrared Astronomical Satellite ({\it IRAS\/}) and the Infrared Space Observatory \citep[{\it ISO\/};][]{kessler_etal_96}. \cite{heckman_etal_92} compared {\it IRAS\/} measurements of 3CR quasars and radio galaxies ($z > 0.3$) selected on the basis of their 178 MHz radio flux density. They found that the rest-frame 6--50\microns\ emission is on average four times stronger in the quasars than in the radio galaxies, suggesting that either the quasars are intrinsically stronger sources of mid-far infrared emission than quasars, or that this emission is anisotropic. The latter requires large optical depths in the far-infrared, possibly requiring implausibly large dust masses. The former possibility is problematic for pure orientation-based unification of quasars and radio galaxies, but may be due to a significant non-thermal contribution in this wavelength range from the radio jet.  \cite{hes_etal_95} re-analyzed the {\it IRAS\/} 60\microns\ data for 3CR quasars and radio galaxies in the redshift range $0.3 < z < 0.8$. They confirm the finding of \cite{heckman_etal_92} that 3CR quasars are systematically brighter than radio galaxies in the mid-far infrared and conclude that a beamed 60\microns\ component may account for this difference. In subsequent work, \cite{hoekstra_etal_97} find that the observed {\it IRAS\/} 60\microns\ infrared-to-radio flux-density ratios are consistent with an orientation-based model incorporating beamed emission, but conclude that other processes such as torus optical depth also contribute.

The {\it ISO\/} photometry for four pairs of 3CR quasars and radio galaxies matched in redshift and radio power was compared by \cite{vanbemmel_etal_00}. Although the non-thermal contribution was estimated at less than 2\%, a systematic infrared excess was found for the quasars with respect to the galaxies, suggesting the presence of optically thick dust. However the small sample size did now allow any firm conclusions to be drawn. \cite{meisenheimer_etal_01} compared {\it ISO\/} photometry for ten pairs of 3C quasars and radio galaxies, and found no difference in the distribution of their (radio-lobe normalized) far-infrared luminosities. \cite{andreani_etal_02} obtained millimeter data and {\it ISO\/} photometry for a sample of 3C radio galaxies and quasars (selected on the basis of visibility to {\it ISO\/}) and found that the quasar composite spectrum is elevated by over a factor of three compared with the radio galaxy composite spectrum in the millimeter region, which the authors attribute to a beamed non-thermal component. \cite{siebenmorgen_etal_04} apply radiative transfer models to all 68 3CR sources imaged by {\it ISO\/}. The model spectra of Type~1 objects (i.e.\ viewed at inclinations close to the radio axis) peak around 40\microns\ while those of Type~2 (i.e.\ viewed at inclinations away from the radio axis) peak at wavelengths over 100\microns . This is interpreted as being in support of orientation-based unification. \cite{haas_etal_04} examined the spectral energy distributions of 3CR quasars and radio galaxies based on {\em ISO\/} photometry, millimeter observations using MAMBO and supplementary data from the Nasa/IPAC Extragalactic Database (NED). The sample was selected on the basis of good visibility to {\it ISO\/}. At rest-frame 70\microns , the authors found that the (radio-lobe normalized) thermal dust luminosity of quasars and galaxies had similar distributions, in line with orientation-based unification.

It is clear from previous work that beamed synchrotron emission and dust extinction modulate the mid-infrared emission of quasars and radio galaxies to some degree. However, it has not been possible to assess quantitatively whether a combination of these two effects is sufficient to unify an orientation-unbiased sample of quasars and radio galaxies. The increased photometric sensitivity and spectroscopic data provided by {\em Spitzer\/} provide additional constraints in the infrared, allowing us to perform a spectral decomposition of the thermal and non-thermal components of emission for the most powerful radio sources and also to estimate the silicate optical depth.

In this paper, we describe the selection methodology (\S\,\ref{sample}), {\em Spitzer\/} observations (\S\,\ref{obs}) and data reduction (\S\,\ref{reduction}) for our sample of powerful 3C radio sources. The photometric and spectroscopic results are presented (\S\,\ref{mips_irs_mmts}) and the spectral energy distributions (SED) are fitted with a combination of synchrotron and dust components (\S\,\ref{sed_section}). We then explore what these results can tell us about the physical conditions (\S\,\ref{phys_cond}) and the origin of the infrared emission (\S\,\ref{starformation}, \S\,\ref{nontherm}) in these objects. Finally, the comparative infrared luminosity of quasars and galaxies is discussed (\S\,\ref{comparative}).

\section{THE SAMPLE \label{sample}}

We require a sample selected at low frequency, with  $L_{\rm 178\,MHz} > 10^{26}$\,\WHzsr\ and with a reasonable balance between FR\,II radio galaxies and quasars\footnote{Since the low frequency radio emission is isotropic, one would be justified in multiplying by $4\pi$ and writing the full luminosity per hertz;  but since luminosities at other frequencies are {\it not\/} necessarily isotropic, we will generally use units of W\,Hz\mo\,sr\mo .  We express even the isotropic low frequency radio value in the same way for easy comparison.}. 

The sample of \cite{barthel_89} provides the ideal starting point.  It contains the 50
sources in the complete low frequency 3CRR catalogue of \cite{laing_etal_83} with
$0.5 \leq z \leq 1.0$.  All have emitted radio luminosity $L_{\rm 178\,MHz} > 10^{26}$\,\WHzsr.  Of the
50, 33 are identified with galaxies and 17 with quasars, with classifications based on
the optical type designated by \cite{laing_etal_83}.  On the basis of radio
morphology, all sources in the sample are FR\,II, as expected from the high
radio luminosity.   

To reduce the observing time required, we reduced the sample to
33 objects based on ecliptic latitude.  Sources at high ecliptic latitude are
easier to schedule with {\em Spitzer}.   We also added one source to the sample, 3C\,200, at
$z=0.458$, because we already had a 16\,ks {\it Chandra\/} observation of it.  We emphasize
that neither of these changes introduces a bias in orientation in the sample.

\begin{figure}[!t]
\epsscale{1.0}
\plotone{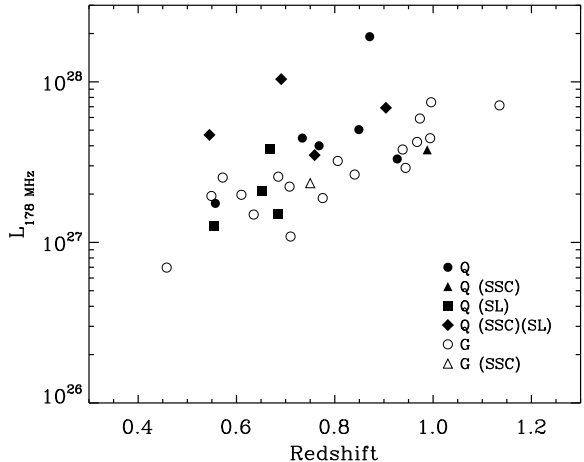}
\caption{Luminosity (W\,Hz\mo\,sr\mo) at rest-frame 178\,MHz vs.\ redshift. Objects from our sample are labeled as follows; closed circles: quasars --- closed triangles:
steep-spectrum compact (SSC) quasars --- closed squares: superluminal quasars --- closed diamonds: SSC superluminal quasars --- open circles: galaxies --- open triangles: SSC galaxies.\label{L178vsz}}
\end{figure}

The {\em Spitzer\/} sample (hereafter, the ``primary'' sample) then consists of 16~quasars and 18~galaxies.  The general
properties of the sample are summarised in Table~\ref{sample_props}. Figure~\ref{L178vsz}
shows the radio lobe luminosity of the sample at an emitted frequency of 178\,MHz as a
function of redshift. The galaxies and quasars span similar ranges in redshift and
radio-lobe luminosity. The mean (median) redshift for quasars in the primary sample is 0.74 (0.73) and for galaxies is 0.79 (0.78). The mean (median) radio luminosity, $L_{\rm 178\,MHz}$, is $5.0 (3.8)\times 10^{27}$\,\WHzsr\ for quasars and is $3.2 (2.6)\times 10^{27}$\,\WHzsr\ for galaxies. The fit to the synchrotron SED used to interpolate the observed low-frequency flux density to the emitted frame is
described in \S\,\ref{sed_section}. The sample includes five steep-spectrum compact (SSC) quasars
and one SSC galaxy. These objects are characterised by small projected linear sizes
($<15$~kpc) and convex radio spectra peaking around 100\,MHz, steepening to
$\alpha < -0.5$ ($S_{\nu} \propto \nu^{\alpha}$) at higher frequencies \citep{odea_98}. The sample
also includes seven quasars with reported superluminal motion, 3C\,138, 3C\,147, 3C\,216, 3C\,263, 3C\,334, 3C\,336 and 3C\,380.

For comparison, we will show results from {\it ISO\/} observations of six~3CRR objects from the compilation of \cite{haas_etal_04}, {\em Spitzer\/}
observations of thirteen~3CRR objects from \cite{shi_etal_05}, {\em Spitzer\/} observations
of ten~3CRR radio galaxies from Birkinshaw et al.\ (2006, in preparation) and {\em Spitzer\/} observations of nineteen~3CRR galaxies from \cite{ogle_etal_06}.
Table~\ref{supplementary_props.tbl} summarises the properties of these 48~additional objects
which comprise the supplementary sample. Together, the primary and supplementary samples comprise 82 3CRR objects.

\begin{deluxetable}{llccc}
\tabletypesize{\scriptsize}
\tablecaption{Summary of primary 3CRR sample properties.
\label{sample_props}}
\tablewidth{0pt}
\tablenum{1}
\tablehead{
\colhead{} & \colhead{} & \colhead{} & \colhead{$\log L^{\rm e}_{\rm 178\,MHz}$\tablenotemark{b}} & \colhead{}\\
\colhead{Source} & \colhead{ID\tablenotemark{a}} & \colhead{z}  & \colhead{[W\,Hz$^{-1}$\,sr$^{-1}$]} & \colhead{$\beta_{app}$\tablenotemark{c}}
}
\startdata
3C\,138	  & Q(SSC)(SL) 	& 0.759\phn      & 27.43	& $2.1 h^{-1}$\\
3C\,147	  & Q(SSC)(SL) 	& 0.545\phn      & 27.48	& $1.3 h^{-1}$\\
3C\,175	  & Q      	& 0.768\phn      & 27.60	& \ldots\\
3C\,196	  & Q      	& 0.871\phn      & 28.25	& \ldots\\
3C\,207	  & Q      	& 0.684\phn      & 27.36	& \ldots\\
3C\,216	  & Q(SL)  	& 0.67\phn\phn   & 27.48	& $2.6 h^{-1}$\\
3C\,254	  & Q      	& 0.734\phn      & 27.61	& \ldots\\
3C\,263	  & Q(SL)  	& 0.646\phn      & 27.32	& $1.3 h^{-1}$\\
3C\,275.1 & Q      	& 0.557\phn      & 27.14	& \ldots\\
3C\,286	  & Q      	& 0.849\phn      & 27.67	& \ldots\\
3C\,309.1 & Q(SSC) 	& 0.904\phn      & 27.76	& \ldots\\
3C\,334	  & Q(SL)  	& 0.555\phn      & 27.03	& $2.0 h^{-1}$\\
3C\,336	  & Q(SL)  	& 0.927\phn      & 27.64	& $2.6 h^{-1}$\\
3C\,343	  & Q(SSC) 	& 0.988\phn      & 27.40	& \ldots\\
3C\,380	  & Q(SSC)(SL) 	& 0.691\phn      & 27.99	& $8.0 h^{-1}$\\
3C\,6.1	  & G		& 0.84\phn\phn   & 27.54       & \dots\\
3C\,22	  & G      	& 0.937\phn      & 27.63	& \ldots\\
3C\,184	  & G      	& 0.994\phn      & 27.65	& \ldots\\
3C\,200	  & G      	& 0.458\phn      & 27.86	& \ldots\\
3C\,220.1 & G      	& 0.62\phn\phn   & 27.26	& \ldots\\
3C\,220.3 & G      	& 0.685\phn      & 27.39       & \ldots\\
3C\,268.1 & G      	& 0.9737         & 27.80	& \ldots\\
3C\,272	  & G      	& 0.944\phn      & 27.46	& \ldots\\
3C\,280	  & G      	& 0.9975         & 28.94	& \ldots\\
3C\,289	  & G      	& 0.9674         & 27.65	& \ldots\\
3C\,292	  & G      	& 0.713\phn      & 27.16	& \ldots\\
3C\,325	  & G      	& 1.135\phn\tablenotemark{d}      & 27.88	& \ldots\\
3C\,330	  & G      	& 0.55\phn\phn   & 27.41	& \ldots\\
3C\,337	  & G      	& 0.635\phn      & 27.19	& \ldots\\
3C\,340	  & G      	& 0.7754         & 27.30	& \ldots\\
3C\,343.1 & G(SSC) 	& 0.75\phn\phn   & 26.13	& \ldots\\
3C\,352	  & G      	& 0.8057         & 27.50	& \ldots\\
3C\,427.1 & G      	& 0.572\phn      & 27.48	& \ldots\\
3C\,441	  & G      	& 0.707\phn      & 27.35	& \ldots\\\enddata
\tablenotetext{a}{G = galaxy, Q = quasar, SSC = steep spectrum compact and SL = superluminal source.}\tablenotetext{b}{See \S\,\ref{sed_section}}
\tablenotetext{c}{Data on superluminal motion taken from \cite{zensus_etal_87},  \cite{venturi_etal_93}, \cite{mantovani_etal_98}, \cite{shen_etal_01}, and \cite{hough_etal_02}.}
\tablenotetext{d}{We use the updated redshift of $z=1.135$ for 3C\,325 from \cite{grimes_etal_05}.}
\end{deluxetable}

\begin{deluxetable}{llccc}
\tabletypesize{\tiny}
\tablecaption{Summary of supplementary 3CRR sample properties.
\label{supplementary_props.tbl}}
\tablewidth{0pt}
\tablenum{2}
\tablehead{
\colhead{} & \colhead{} & \colhead{} & \colhead{$\log L_{\rm e}({\rm 178\,MHz})$} & \colhead{} \\
\colhead{Source} & \colhead{ID} & \colhead{z}  & \colhead{[W\,Hz$^{-1}$\,sr$^{-1}$]} & \colhead{Reference}
}
\startdata
     3C\,47 &     Q & 0.4250 & 27.17 & 1 \\
     3C\,48 &     Q & 0.3670 & 27.29 & 2 \\
    3C\,109 &     Q & 0.3056 & 26.73 & 1 \\
  3C\,249.1 &     Q & 0.3110 & 26.44 & 2 \\
    3C\,351 &     Q & 0.3710 & 26.71 & 2 \\
  3C\,454.3 & Q(SL) & 0.8590 & 27.35 & 1 \\
     3C\,28 &     G & 0.1952 & 26.19 & 4 \\
     3C\,33 &     G & 0.0595 & 25.60 & 4 \\
   3C\,33.1 &     G & 0.1810 & 25.99 & 1 \\
     3C\,55 &     G & 0.7350 & 27.67 & 4 \\
   3C\,61.1 &     G & 0.1860 & 26.41 & 2 \\
    3C\,66B &     G & 0.0215 & 24.34 & 3 \\
  3C\,83.1B &     G & 0.0255 & 24.53 & 3 \\
    3C\,123 &     G & 0.2177 & 27.33 & 4 \\
    3C\,153 &     G & 0.2769 & 26.47 & 4 \\
    3C\,171 &     G & 0.2384 & 26.45 & 2 \\
    3C\,172 &     G & 0.5191 & 27.11 & 4 \\
  3C\,173.1 &     G & 0.2920 & 26.55 & 2 \\
    3C\,192 &     G & 0.0598 & 25.19 & 3 \\
    3C\,219 &     G & 0.1744 & 26.47 & 2 \\
    3C\,234 &     G & 0.1848 & 26.41 & 1 \\
    3C\,236 &     G & 0.0989 & 25.47 & 2 \\
  3C\,244.1 &     G & 0.4280 & 27.04 & 4 \\
  3C\,263.1 &     G & 0.8240 & 27.67 & 4 \\
    3C\,265 &     G & 0.8108 & 27.71 & 4 \\
  3C\,274.1 &     G & 0.4220 & 26.94 & 4 \\
    3C\,284 &     G & 0.2394 & 26.22 & 2 \\
    3C\,285 &     G & 0.0794 & 25.18 & 3 \\
    3C\,288 &     G & 0.2460 & 26.46 & 4 \\
    3C\,300 &     G & 0.2720 & 26.53 & 4 \\
    3C\,310 &     G & 0.0540 & 25.52 & 4 \\
    3C\,315 &     G & 0.1083 & 25.65 & 2 \\
    3C\,319 &     G & 0.1920 & 26.14 & 4 \\
    3C\,321 &     G & 0.0960 & 25.42 & 3 \\
    3C\,326 &     G & 0.0895 & 25.54 & 4 \\
    3C\,349 &     G & 0.2050 & 26.13 & 2 \\
    3C\,381 &     G & 0.1605 & 25.99 & 2 \\
    3C\,382 &     G & 0.0578 & 25.13 & 3 \\
    3C\,386 &     G & 0.0177 & 24.16 & 3 \\
    3C\,388 &     G & 0.0908 & 25.63 & 2 \\
  3C\,390.3 &     G & 0.0569 & 25.50 & 1 \\
    3C\,401 &     G & 0.2010 & 26.30 & 4 \\
    3C\,433 &     G & 0.1016 & 26.10 & 4 \\
    3C\,436 &     G & 0.2145 & 26.31 & 4 \\
    3C\,438 &     G & 0.2900 & 27.00 & 4 \\
   3C\,442A &     G & 0.0270 & 24.37 & 3 \\
    3C\,452 &     G & 0.0811 & 25.88 & 3 \\
    3C\,465 &     G & 0.0293 & 24.81 & 3 \\
\enddata
\tablerefs{(1) \cite{haas_etal_04}, (2) \cite{shi_etal_05}, (3) Birkinshaw et al.\ (2006, in preparation), (4) \cite{ogle_etal_06}}
\end{deluxetable}

\section{OBSERVATIONS\label{obs}}

All sources in the primary sample were scheduled for observation with the Long-Low module of the Spitzer Infrared Spectrograph \citep[IRS;][]{houck_etal_04} resulting in spectra in the range 15--37\microns , as well as with the Multiband Imaging Photometer \citep[MIPS;][]{rieke_etal_04} in photometry mode at 24, 70, and 160\microns . Of the 34 observations with IRS, four (3C\,200, 3C\,220.3, 3C\,272 and 3C\,289) peaked-up incorrectly, either on transient pixels or on a brighter object in the field. The galaxy 3C\,427.1 was not detected with IRS and 3C\,6.1 was detected in LL1 only. All objects in the sample are unresolved by the {\em Spitzer\/} instruments.

\section{DATA REDUCTION\label{reduction}}

The IRS spectra were extracted from the basic calibrated data (BCD) images provided by the Spitzer Science Center, using
the Spectroscopy Modeling and Analysis Tool \citep[\textsc{Smart};][]{higdon_etal_04}.
The images were background-subtracted by forming difference images from the BCD data at the two nodding positions. Spectra were
then extracted using narrow extraction apertures to maximise the signal-to-noise ratio, and calibrated using
a standard IRS Long-Low stellar calibrator, HD$\:$173511, correcting the flux calibration for the width of the narrow extraction aperture.  The spectra were rescaled
to the MIPS 24\microns\ point in order to provide a consistent flux calibration
between MIPS and IRS measurements. The
MIPS and IRS observations were not contemporaneous; if the source
is variable, this cross-calibration introduces an error.

Mosaics of the MIPS 24 and 70\microns\ BCD images were produced using the Mosaicking and Point
Source Extraction tool, \textsc{Mopex} \citep{makovoz_khan_05}, while for 160\microns , the
post-BCD mosaics provided by the SSC were used.  The radius of the photometric aperture was
set to $15\arcsec$, $15\arcsec$, and $48\arcsec$ at 24, 70, and 160\microns , respectively. Variations in the background on the scale of
the aperture were used to estimate the uncertainty in the extracted flux densities. 

\section{RESULTS}

This section presents the infrared measurements for the primary and supplementary
samples. The infrared luminosities\footnote{We used H$_{0}=70$\,km\,s\mo \,Mpc\mo, $\Omega_{\rm M}=0.3$ and
$\Omega_{\Lambda}=0.7$.} of the sample are characterised at rest-frame wavelengths of 15 and 30\microns . 

A rest wavelength of 15\microns\ was chosen because it lies between the broad silicate features around 9.7 and 18\microns\ \citep[see e.g.,][]{siebenmorgen_etal_05, hao_etal_05} and avoids the [Ne\,{\sc v}]$\lambda14.3\,\mu{\rm m}$ and [Ne\,{\sc III}]$\lambda15.5\,\mu{\rm m}$ lines. All objects in the primary sample are at redshifts such that rest-frame 15\microns\ lies within IRS Long-Low. For four objects without IRS spectra (3C\,200, 3C\,220.3, 3C\,272 and 3C\,289), the rest-frame 15\microns\ flux density was estimated by assuming a power-law SED between the MIPS 24 and 70\microns\ measurements. The galaxy 3C\,200 was detected in the MIPS 24\microns\ channel but not at 70\microns , so the derived 15 and 30\microns\ flux densities are considered lower and upper limits respectively. The galaxy 3C\,427.1 was not detected using MIPS or IRS. However, since rest-frame 15\microns\ corresponds to an observed wavelength of $\approx 24$\microns\ for this object, we use the observed MIPS 24\microns\ measurement to place an upper limit on the rest-frame 15\microns\ flux density. This galaxy is of interest since it is the faintest 15\microns\ object in the primary sample. The rest-frame 30\microns\ flux density was determined by assuming a power-law SED to interpolate between the observed MIPS 24 and 70\microns\ flux densities. Where the 70\microns\ measurement resulted in an upper limit, the interpolated 30\microns\ flux density is also considered an upper limit. The 178\,MHz flux densities are corrected to the rest-frame using a fit to the radio synchrotron spectra as described in \S\,\ref{sed_section}. Table~\ref{char_lums} in Appendix~A gives the characteristic 15 and 30\microns\ flux density and luminosity for objects in the primary sample as well as the rest-frame 178\,MHz luminosity.

For the supplementary sample, power-law SEDs are also assumed in order to interpolate the available photometry (e.g., {\it ISO}, MIPS, or IRAC) to rest-frame wavelengths of 15 and 30\microns . Only 15\microns\ data were available for the objects in the supplementary sample from \cite{ogle_etal_06}. Extrapolation to rest-frame 178\,MHz was achieved using the observed 178\,MHz flux density and the 178--750\,MHz spectral index.

\subsection{MIPS and IRS measurements\label{mips_irs_mmts}}

Table~3 gives the MIPS 24, 70 and 160\microns\ flux density measurements for the primary sample. All objects in the primary sample were detected at 24\microns\  except for the galaxy 3C\,427.1. At 70\microns , 6 out of 19 galaxies and all the quasars were detected. At 160\microns , there was just a single detection, the galaxy 3C\,220.3.

\begin{figure*}
\epsscale{1.9}
\plotone{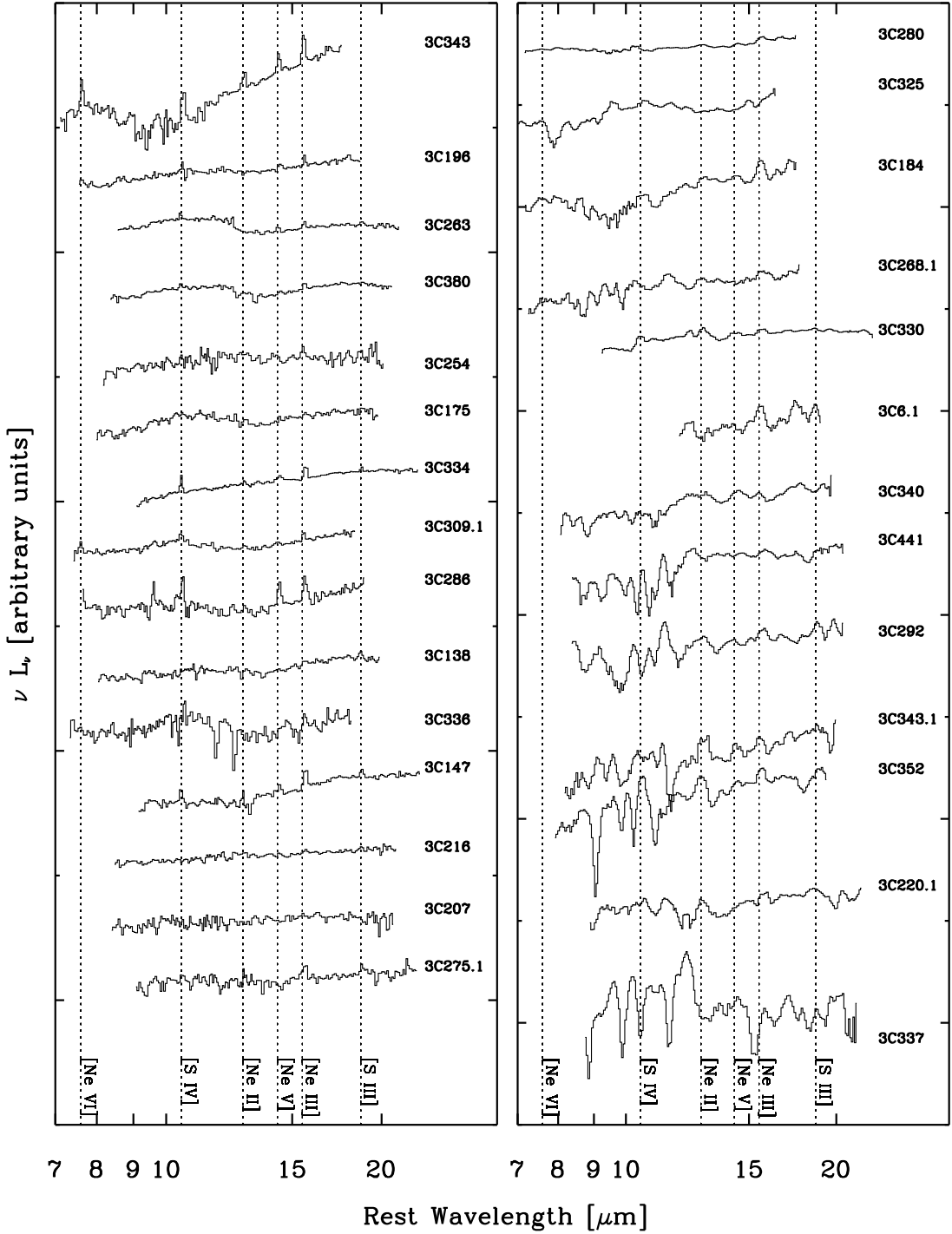}
\caption{This plot shows the spectra for all the quasars (left) and galaxies (right)  in the sample, so that the general morphology and spectral features may be compared. The spectra are arranged in order of decreasing total dust luminosity (see Table~7). The data for the galaxies have been smoothed over four spectral resolution elements. LL2 was not detected for the galaxy 3C\,6.1. For clarity, the errors on the spectra are not shown (see Figure~\ref{fits} instead). \label{Spectra}}
\end{figure*}

Figure~\ref{Spectra} shows the IRS Long-Low spectra for the quasars (left panel) and
galaxies (right panel) plotted in (arbitrarily scaled) units of luminosity, $\nu L_{\nu}$,  at the emitted
wavelength, arranged in order of decreasing total dust luminosity (see \S\,\ref{sed_section}) from top to bottom. This figure allows the general morphology and features to be compared (the spectra are presented in greater detail with error estimates in Figure~\ref{fits}). Table~4 gives the emission line flux densities for the objects in the primary sample. 

The [Ne\,{\sc iii}]$\lambda15.5\,\mu{\rm m}$ line is present at some level in most of the galaxies and [Ne\,{\sc v}]$\lambda14.3\,\mu{\rm m}$ is detected in two galaxies, 3C\,184 and 3C\,330. The [Ne\,{\sc iii}]$\lambda15.5\,\mu{\rm m}$ line can also be seen in many of the quasars in the sample, as well as [Ne\,{\sc ii}]$\lambda12.8\,\mu{\rm m}$, [Ne\,{\sc v}]$\lambda14.3\,\mu{\rm m}$, [S\,{\sc iii}]$\lambda18.7\,\mu{\rm m}$, and [S\,{\sc iv}]$\lambda10.5\,\mu{\rm m}$. [Ne\,{\sc vi}]$\lambda7.6\,\mu{\rm m}$ is detected in two quasars, 3C\,309.1 and 3C\,343.

The galaxy continua are generally fainter in the mid-infrared than those of the quasars and so the signal-to-noise ratio is typically lower for the galaxies. The presence of silicates in emission is evident in some quasars (e.g., 3C\,263, 3C\,175, 3C\,380). Silicate absorption is not obvious from a visual inspection of the spectra, except for the quasar 3C\,343 and the galaxy 3C\,184.

\begin{deluxetable}{lcccccc}
\tabletypesize{\scriptsize}
\tablecaption{MIPS photometry for the primary sample.}
\label{mips_obs.tbl}
\tablewidth{0pt}
\tablenum{3}
\tablehead{
\colhead{} & \colhead{$S_{24\mu{\rm m}}$} & \colhead{$\sigma_{24\mu{\rm m}}$} & \colhead{$S_{70\mu{\rm m}}$} & \colhead{$\sigma_{70\mu{\rm m}}$} & \colhead{$S_{160\mu{\rm m}}$} & \colhead{$\sigma_{160\mu{\rm m}}$}\\
\colhead{Name} & \colhead{[mJy]} & \colhead{[mJy]} & \colhead{[mJy]} & \colhead{[mJy]} & \colhead{[mJy]} & \colhead{[mJy]}\\
}
\startdata
3C\,6.1   &	\phm{$<$}\phn{\bf2.8}  &	0.3  &  $<$13.8		      &	\phn4.6   &$<$45.6 & 15.2 \\
3C\,22    &	\phm{$<$}{\bf13.4}     &  	1.3  &  \phm{$<$}{\bf22.8}    &	\phn5.6   &$<$45.6 & 15.2 \\
3C\,138   &	\phm{$<$}{\bf15.9}		       & 1.6	     &	\phm{$<$}{\bf24.3}		      &	\phn6.5	  & $<$45.6     &15.2 \\
3C\,147   &	\phm{$<$}{\bf22.4}     & 	2.2  &	\phm{$<$}{\bf57.2}    &	12.4      &$<$88.8 & 29.6 \\
3C\,175   &	\phm{$<$}{\bf15.5}     & 	1.6  &	\phm{$<$}{\bf25.1}    &	\phn6.4   &$<$60.0 & 20.0 \\
3C\,184   &	\phm{$<$}\phn{\bf2.7}  &	0.3  &	\phm{$<$}{\bf10.7}    &	\phn2.9   &$<$45.6 & 15.2 \\
3C\,196   &	\phm{$<$}{\bf18.1}     & 	1.8  &	\phm{$<$}{\bf12.2}    &	\phn3.3   &$<$95.4 & 31.8 \\
3C\,200   &	\phm{$<$}\phn{\bf1.3}  &	0.1  & $<$10.8                & \phn3.6   &$<$45.6 & 15.2 \\
3C\,207   &	\phm{$<$}{\bf11.4}     & 	1.1  &	\phm{$<$}{\bf21.5}    &	\phn6.9   &$<$77.1 & 25.7 \\
3C\,216   &	\phm{$<$}{\bf20.9}     & 	2.1  &	\phm{$<$}{\bf69.5}    &	14.6      &$<$45.6 & 15.2 \\
3C\,220.1 &	\phm{$<$}\phn{\bf2.2}  &	0.2  & \phn$<$7.8             & \phn2.6   &$<$83.4 & 27.8 \\
3C\,220.3 &	\phm{$<$}\phn{\bf4.8}  &	0.5  &	\phm{$<$}{\bf20.7}    &	\phn4.8   &{\bf136.3}   & 31.1 \\
3C\,254   &	\phm{$<$}{\bf12.4}     & 	1.2  &	\phm{$<$}{\bf11.7}    &	\phn2.5   &$<$84.9 & 28.3 \\
3C\,263   &	\phm{$<$}{\bf27.0}     & 	2.7  &	\phm{$<$}{\bf31.1}    &	\phn6.5   &$<$46.5 & 15.5 \\
3C\,268.1 &	\phm{$<$}\phn{\bf2.4}  &	0.2  & \phn$<$6.9             & \phn2.3   &$<$50.7 & 16.9 \\
3C\,272   &	\phm{$<$}\phn{\bf2.1}  &	0.2  &	\phm{$<$}\phn{\bf9.8} & \phn2.2   &$<$85.8 & 28.6 \\
3C\,275.1 &	\phm{$<$}\phn{\bf8.1}  &	0.8  &	\phm{$<$}{\bf23.6}    & \phn7.8   &$<$45.6 &15.2      \\
3C\,280   &	\phm{$<$}\phn{\bf9.3}  &	0.9  &	\phm{$<$}{\bf18.2}    &	\phn4.0   &$<$61.2 & 20.4 \\
3C\,286   &	\phm{$<$}\phn{\bf7.7}  &	0.8  &	 \phm{$<$}{\bf29.3}   & \phn6.6   &$<$45.6  & 15.2     \\
3C\,289   &	\phm{$<$}\phn{\bf3.7}  &	0.4  &	\phm{$<$}{\bf17.4}    &	\phn4.0   &$<$81.3 & 27.1 \\
3C\,292   &	\phm{$<$}\phn{\bf1.9}  &	0.2  & \phn$<$8.7             & \phn2.9   &$<$83.4 & 27.8 \\
3C\,309.1 &	\phm{$<$}{\bf17.8}     & 	1.8  &	\phm{$<$}{\bf36.1}    &	\phn8.0   &$<$45.6 & 15.2 \\
3C\,325   &	\phm{$<$}\phn{\bf3.4}  &	0.3  & \phn$<$7.2             & \phn2.4   &$<$45.6 & 15.2 \\
3C\,330   &	\phm{$<$}\phn{\bf7.5}  &	0.7  & $<$13.5                & \phn4.5   &$<$45.6 & 15.2 \\
3C\,334   &	\phm{$<$}{\bf35.0}     & 	3.5  &	\phm{$<$}{\bf65.3}    &	13.6      &$<$45.6 & 15.2 \\
3C\,336   &	\phm{$<$}\phn{\bf4.7}  &	0.5  &	\phm{$<$}{\bf12.0}    &	\phn3.8   &$<$46.5 & 15.5 \\
3C\,337   &	\phm{$<$}\phn{\bf1.0}  &	0.1  & \phn$<$7.8             & \phn2.6   &$<$47.7 & 15.9 \\
3C\,340   &	\phm{$<$}\phn{\bf2.4}  &	0.2  & \phn$<$9.3             & \phn3.1   &$<$68.1 & 22.7 \\
3C\,343   &	\phm{$<$}\phn{\bf7.8}  &	0.8  &	\phm{$<$}{\bf41.2}    &	10.8      &$<$70.8 & 23.6 \\
3C\,343.1 &	\phm{$<$}\phn{\bf1.3}  &	0.1  & \phn$<$9.0             & \phn3.0   &$<$86.4 & 28.8 \\
3C\,352   &	\phm{$<$}\phn{\bf1.2}  &	0.1  & \phn$<$6.3             & \phn2.1   &$<$47.4 & 15.8 \\
3C\,380   &	\phm{$<$}{\bf28.1}     & 	2.8  &	\phm{$<$}{\bf34.6}    &	\phn8.4   &$<$60.0 & 20.0 \\
3C\,427.1 &      \phn$<$0.4	       &	0.1  &	\phn$<$7.8            &	\phn2.6   &$<$45.6 & 15.2 \\
3C\,441   &     \phm{$<$}\phn{\bf2.4}  &	0.2  &	\phn$<$4.8            &	\phn1.6   &$<$60.9 & 20.3 \\
\enddata
\tablecomments{The uncertainties include an absolute flux calibration uncertainty of 10\% at 24\microns\ and 20\% at 70 and 160\microns . The upper limits are $3\sigma$; detections are in bold font.}
\end{deluxetable}

\begin{deluxetable}{lcccccc}
\tabletypesize{\scriptsize}
\tablecaption{Emission line flux densities for the primary sample.}
\label{line_lums.tbl}
\tablewidth{0pt}
\tablenum{4}
\tablehead{
\colhead{} & \colhead{[Ne\,{\sc ii}]} & \colhead{[Ne\,{\sc iii}]} & \colhead{[Ne\,{\sc v}]} & \colhead{[Ne\,{\sc vi}]} & \colhead{[S\,{\sc iii}]} & \colhead{[S\,{\sc iv}]}\\
\colhead{} & \colhead{$12.8\,\mu{\rm m}$} & \colhead{$15.5\,\mu{\rm m}$} & \colhead{$14.3\,\mu{\rm m}$} & \colhead{$7.6\,\mu{\rm m}$} & \colhead{$18.7\,\mu{\rm m}$} & \colhead{$10.5\,\mu{\rm m}$}\\
\colhead{Name} & \colhead{21.6\,eV} & \colhead{41.0\,eV} & \colhead{97.1\,eV} & \colhead{126.2\,eV} & \colhead{23.3\,eV} & \colhead{34.8\,eV}
}
\startdata
3C\,6.1	    &	$<\phn8.6\phm{\bf \pm 0.0}$		&	\phm{$<$}$\phn\phn{\bf 6.3\pm1.3}$   &	$<\phn9.2\phm{\bf \pm 0.0}$        &	$\ldots$		          & $<15.5\phm{\bf \pm 0.0}$        &	$\ldots$     \\
3C\,22	    &	$<\phn4.2\phm{\bf \pm 0.0}$   	 	&	\phm{$<$}$\phn\phn{\bf 3.4\pm0.8}$   &	$<\phn5.1\phm{\bf \pm 0.0}$  	&	$<\phn5.9\phm{\bf \pm 0.0}$   	  & $\ldots$	 	  	    &	\phm{$<$}$\phn\phn{\bf 5.4\pm1.5}$     \\
3C\,138	    &	$<\phn9.7\phm{\bf \pm 0.0}$   	 	&	$<\phn5.5\phm{\bf \pm 0.0}$   	 &	$<\phn8.0\phm{\bf \pm 0.0}$  	&	$\ldots$	 	   	  & \phm{$<$}$\phn\phn{\bf 5.3\pm0.8}$  &	$<\phn8.8\phm{\bf \pm 0.0}$     \\
3C\,147	    &	$<10.5\phm{\bf \pm 0.0}$   	 	&	\phm{$<$}$\phn\phn{\bf 8.3\pm0.7}$   &	$<\phn6.3\phm{\bf \pm 0.0}$  	&	$\ldots$	 	   	  & \phm{$<$}$\phn\phn{\bf 5.3\pm0.5}$ 	    &	\phm{$<$}$\phn\phn{\bf 6.6\pm0.3}$     \\
3C\,184	    &	\phm{$<$}$\phn\phn{\bf 1.9\pm0.4}$   	&	\phm{$<$}$\phn\phn{\bf 6.5\pm1.1}$   &	\phm{$<$}$\phn\phn{\bf 1.7\pm0.5}$  &	$<\phn4.4\phm{\bf \pm 0.0}$   	  & $\ldots$		  	    &	$<\phn5.7\phm{\bf \pm 0.0}$     \\
3C\,196	    &	$<\phn6.4\phm{\bf \pm 0.0}$   	 	&	\phm{$<$}$\phn\phn{\bf 4.6\pm0.4}$   &	\phm{$<$}$\phn\phn{\bf 3.0\pm0.4}$  &	$\ldots$		   	  & $\ldots$		  	    &	$<\phn9.8\phm{\bf \pm 0.0}$     \\
3C\,216	    &	$<\phn6.7\phm{\bf \pm 0.0}$   	 	&	$<\phn3.8\phm{\bf \pm 0.0}$   	 &	$<\phn6.1\phm{\bf \pm 0.0}$  	&	$\ldots$		   	  & $<\phn5.2\phm{\bf \pm 0.0}$ 	    &	$<\phn3.8\phm{\bf \pm 0.0}$     \\
3C\,220.1   &	$<\phn3.8\phm{\bf \pm 0.0}$   	 	&	$<\phn2.1\phm{\bf \pm 0.0}$   	 &	$<\phn2.1\phm{\bf \pm 0.0}$  	&	$\ldots$		   	  & $<\phn2.1\phm{\bf \pm 0.0}$ 	    &	$<\phn4.1\phm{\bf \pm 0.0}$     \\
3C\,254	    &	$<\phn8.2\phm{\bf \pm 0.0}$   	 	&	\phm{$<$}$\phn\phn{\bf 5.7\pm1.3}$   &	$<11.9\phm{\bf \pm 0.0}$  	&	$\ldots$		   	  & $<18.0\phm{\bf \pm 0.0}$ 	    &	$<10.3\phm{\bf \pm 0.0}$     \\
3C\,263	    &	$<17.9\phm{\bf \pm 0.0}$   	 	&	\phm{$<$}$\phn\phn{\bf 4.7\pm1.0}$   &	$<\phn9.9\phm{\bf \pm 0.0}$  	&	$\ldots$	 	   	  & \phm{$<$}$\phn\phn{\bf 3.2\pm0.3}$ 	    &	\phm{$<$}$\phn\phn{\bf 5.0\pm0.9}$     \\
3C\,268.1   &	$<\phn0.5\phm{\bf \pm 0.0}$   	 	&	$<\phn0.9\phm{\bf \pm 0.0}$   	 &	$<\phn0.5\phm{\bf \pm 0.0}$  	&	$<\phn0.6\phm{\bf \pm 0.0}$   	  & $\ldots$		  	    &	$<\phn0.9\phm{\bf \pm 0.0}$     \\
3C\,275.1   &	$<\phn5.1\phm{\bf \pm 0.0}$   	 	&	\phm{$<$}$\phn\phn{\bf 6.7\pm0.7}$   &	$<\phn5.4\phm{\bf \pm 0.0}$  	&	$\ldots$		   	  & $<\phn3.1\phm{\bf \pm 0.0}$ 	    &	$<\phn5.6\phm{\bf \pm 0.0}$     \\
3C\,280	    &	$<\phn3.6\phm{\bf \pm 0.0}$   	 	&	\phm{$<$}$\phn\phn{\bf 4.2\pm0.5}$   &	$<\phn6.7\phm{\bf \pm 0.0}$  	&	$<\phn5.2\phm{\bf \pm 0.0}$   	  & $\ldots$		  	    &	$<13.7\phm{\bf \pm 0.0}$     \\
3C\,286	    &	$<\phn9.7\phm{\bf \pm 0.0}$   	 	&	\phm{$<$}$\phn\phn{\bf 6.9\pm0.6}$   &	\phm{$<$}$\phn\phn{\bf 7.5\pm0.4}$  &	$\ldots$		   	  & $<10.4\phm{\bf \pm 0.0}$ 	    &	\phm{$<$}$\phn\phn{\bf 7.4\pm0.9}$     \\
3C\,292	    &	$<\phn7.2\phm{\bf \pm 0.0}$   	 	&	$<\phn4.6\phm{\bf \pm 0.0}$   	 &	$<\phn7.7\phm{\bf \pm 0.0}$  	&	$\ldots$		   	  & $<\phn6.1\phm{\bf \pm 0.0}$ 	    &	$<10.4\phm{\bf \pm 0.0}$     \\
3C\,309.1   &	$<\phn5.5\phm{\bf \pm 0.0}$   	 	&	\phm{$<$}$\phn\phn{\bf 4.3\pm0.6}$   &	\phm{$<$}$\phn\phn{\bf 3.3\pm0.4}$  &	\phm{$<$}$\phn\phn{\bf 7.2\pm1.5}$    & $\ldots$		  	    &	$<10.6\phm{\bf \pm 0.0}$     \\
3C\,325	    &	$<\phn3.7\phm{\bf \pm 0.0}$   	 	&	$<\phn4.3\phm{\bf \pm 0.0}$   	 &	$<\phn3.3\phm{\bf \pm 0.0}$  	&	$<\phn4.6\phm{\bf \pm 0.0}$   	  & $\ldots$		  	    &	$<\phn9.7\phm{\bf \pm 0.0}$     \\
3C\,330	    &	$<\phn6.1\phm{\bf \pm 0.0}$   	 	&	\phm{$<$}$\phn\phn{\bf 4.0\pm0.2}$   &	\phm{$<$}$\phn\phn{\bf 4.2\pm0.6}$  &	$\ldots$		   	  & $<\phn2.2\phm{\bf \pm 0.0}$ 	    &	\phm{$<$}$\phn\phn{\bf 3.5\pm0.7}$     \\
3C\,334	    &	\phm{$<$}$\phn\phn{\bf 6.6\pm0.4}$   	&	\phm{$<$}$\phn\phn{\bf 16.2\pm1.3}$   &	\phm{$<$}$\phn\phn{\bf 11.8\pm1.7}$  &	$\ldots$	 	   	  & \phm{$<$}$\phn\phn{\bf 3.1\pm0.6}$ 	    &	\phm{$<$}$\phn\phn{\bf 11.8\pm0.6}$     \\
3C\,336	    &	$<\phn4.9\phm{\bf \pm 0.0}$   	 	&	\phm{$<$}$\phn\phn{\bf 1.8\pm0.1}$   &	$<\phn5.4\phm{\bf \pm 0.0}$  	&	$<\phn5.0\phm{\bf \pm 0.0}$   	  & $\ldots$		  	    &	$<\phn8.9\phm{\bf \pm 0.0}$     \\
3C\,340	    &	$<10.9\phm{\bf \pm 0.0}$   	 	&	$<\phn3.2\phm{\bf \pm 0.0}$   	 &	$<\phn5.6\phm{\bf \pm 0.0}$  	&	$\ldots$	 	   	  & $<\phn6.3\phm{\bf \pm 0.0}$ 	    &	$<\phn9.5\phm{\bf \pm 0.0}$     \\
3C\,337	    &	$<\phn3.2\phm{\bf \pm 0.0}$   	 	&	$<\phn3.1\phm{\bf \pm 0.0}$   	 &	$<\phn3.4\phm{\bf \pm 0.0}$  	&	$\ldots$	 	   	  & $<\phn2.4\phm{\bf \pm 0.0}$ 	    &	$\ldots$		     \\
3C\,343	    &	\phm{$<$}$\phn\phn{\bf 7.0\pm0.9}$   	&	\phm{$<$}$\phn\phn{\bf 13.9\pm1.3}$   &	\phm{$<$}$\phn\phn{\bf 8.3\pm1.1}$  &	\phm{$<$}$\phn\phn{\bf 6.2\pm0.9}$    & $\ldots$	 	    &	\phm{$<$}$\phn\phn{\bf 7.9\pm0.6}$     \\
3C\,343.1   &	\phm{$<$}$\phn\phn{\bf 3.9\pm0.5}$   	&	\phm{$<$}$\phn\phn{\bf 2.2\pm0.3}$   &	$<\phn3.8\phm{\bf \pm 0.0}$  	&	$\ldots$		   	  & $<\phn5.8\phm{\bf \pm 0.0}$ 	    &	$<\phn5.2\phm{\bf \pm 0.0}$     \\
3C\,352	    &	\phm{$<$}$\phn\phn{\bf 2.9\pm0.3}$   	&	\phm{$<$}$\phn\phn{\bf 1.4\pm0.4}$   &	$<\phn2.7\phm{\bf \pm 0.0}$  	&	$\ldots$		   	  & $<\phn4.0\phm{\bf \pm 0.0}$ 	    &	$<\phn6.4\phm{\bf \pm 0.0}$     \\
3C\,380	    &	$<17.1\phm{\bf \pm 0.0}$   	 	&	\phm{$<$}$\phn\phn{\bf 4.4\pm1.1}$   &	$<\phn5.3\phm{\bf \pm 0.0}$  	&	$\ldots$		   	  & $<\phn4.3\phm{\bf \pm 0.0}$ 	    &	\phm{$<$}$\phn\phn{\bf 4.3\pm0.6}$     \\
3C\,441	    &	$<\phn4.9\phm{\bf \pm 0.0}$   	 	&	$<\phn5.4\phm{\bf \pm 0.0}$   	 &	$<\phn6.2\phm{\bf \pm 0.0}$  	&	$\ldots$		   	  & $<\phn5.3\phm{\bf \pm 0.0}$ 	    &	$<\phn9.7\phm{\bf \pm 0.0}$     \\
\enddata
\tablecomments{The emission line flux densities are given in units of $10^{-22}$~W\,cm\mt . The upper limits are $3\sigma$; detections are in bold font. The ionisation potential shown above each column is that needed to create the species from the next lower ionisation stage.}
\end{deluxetable}

\begin{figure*}
\epsscale{1.2}
\plotone{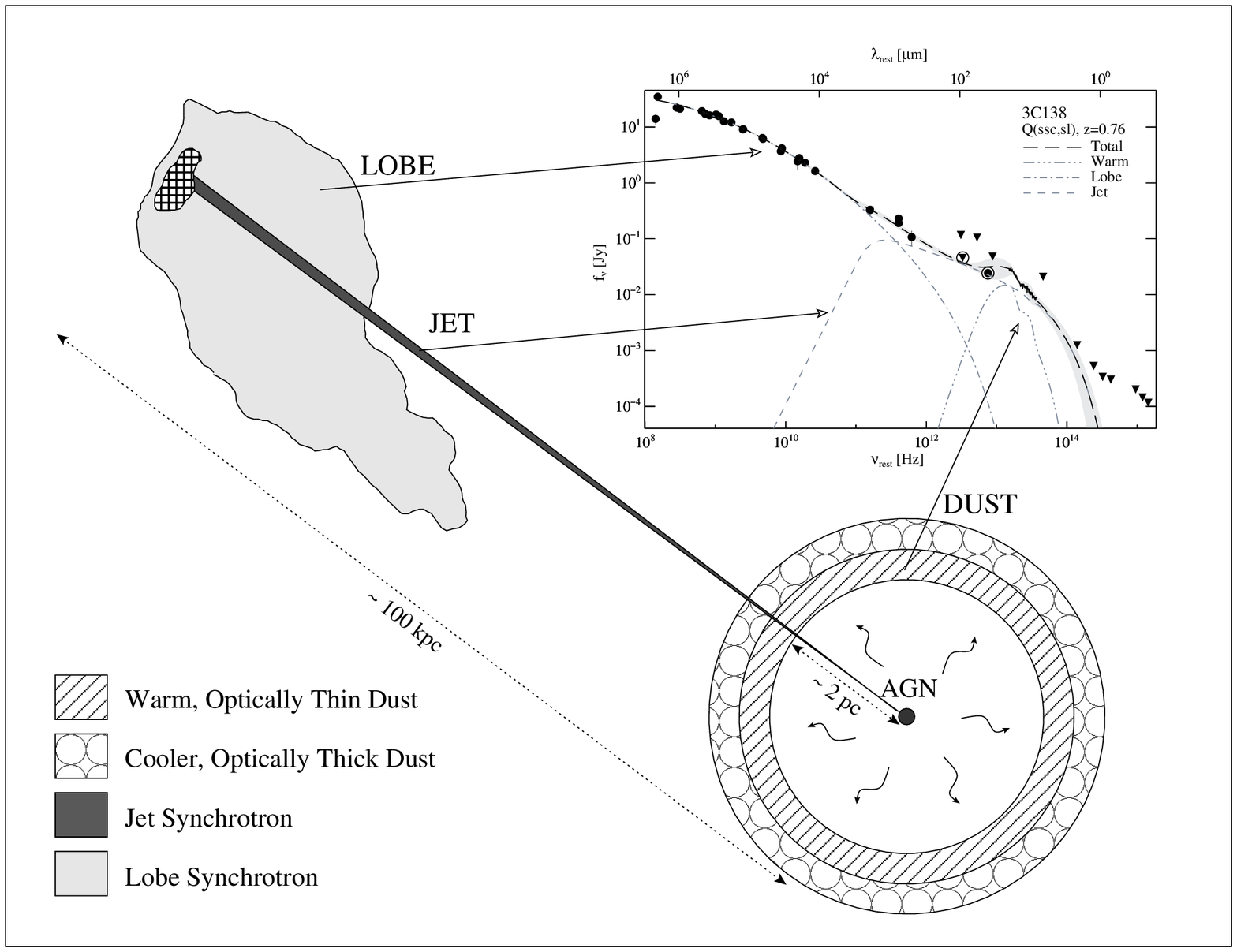}
\caption{This cartoon indicates the physical origin of the broadband spectral components used to fit the spectral energy distributions, using the SED of the quasar 3C\,138 as an example (see Fig.~\ref{fits}). The drawing is not to scale --- the dust shell has been drawn $\approx 10^{4}$ times its correct size relative to the extent of the jet. \label{cartoon}}
\end{figure*}

\subsection{Spectral Fitting\label{sed_section}}

In this section we fit the IRS and MIPS data, as well as additional photometry from the NED, with models for the thermal and non-thermal emission, in order to separate their respective contributions to the mid-infrared. Figure~\ref{cartoon} shows a cartoon of the physical origin of the broadband spectral components used in the fits: (i)~synchrotron emission from the radio lobes, (ii)~synchrotron emission from the radio jet, and (iii)~thermal emission from dust. Four combinations of components are fitted to all objects, and the combination with the best reduced $\chi^{2}$ is selected as the best fit. We now describe the specific functional forms used for each component.

In order to characterise the synchrotron emission from the radio lobes, $F^{\rm lobe}_{\nu}$,  the following parabolic function \citep[see e.g.,][]{polletta_etal_00,andreani_etal_02} is fit to the data,
\begin{equation}
\log{F^{\rm lobe}_{\nu}(\nu)} \propto -\beta \, (\log{\nu} - \log{\nu_{\rm t}})^2 + \log[e^{-(\nu/\nu_{c}^{\rm lobe})}],
\label{lobe_model.eqn}
\end{equation}
where $\beta$ is a parameter representing the bending of the parabola, $\nu_{t}$ is the frequency at which the optical depth of the synchrotron-emitting plasma reaches unity and, $\nu_{c}^{\rm lobe}$ is the frequency corresponding to the cut-off energy of the lobe plasma energy distribution. The free parameters and constraints for this component are $\beta$, $\nu_{t}>1\,{\rm MHz}$ and, $30<\nu_{c}^{\rm lobe}<3000\,{\rm GHz}$.

A broken power-law is used to represent synchrotron emission from the jet, $F^{\rm jet}_{\nu}$, viewed through a dust screen, as follows \citep[see e.g.,][]{polletta_etal_00},
\begin{equation}
F^{\rm jet}_{\nu}(\nu) \propto e^{-\tau(\nu)} \left ( \frac{\nu_{0}}{\nu} \right )^{-\alpha_{1}} \left ( 1 - e^{-(\nu_{0} / \nu )^{\alpha_{1}-\alpha_{2}}} \right )\,e^{-(\nu/\nu_{c}^{\rm jet})},
\label{jet_model.eqn}
\end{equation}
where, $\tau$ is the optical depth of the dust component (discussed below), $\alpha_{1}$ and $\alpha_{2}$ are the optically thick and optically thin spectral indices respectively, $\nu_{0}$ is the frequency at which the broken power-law peaks and, $\nu_{c}^{\rm lobe}$ is the frequency corresponding to the cut-off energy of the jet plasma energy distribution. In the fits, the optically thick spectral index is fixed at $\alpha_{1}=2.5$ as expected for a homogenous source, and the free parameters and constraints are $\tau$, $3<\nu_{0}<300\,{\rm GHz}$, $6 \times 10^{13}<\nu_{c}^{\rm jet}<60 \times 10^{13}\,{\rm Hz}$, and, $\alpha_{2} \geq -1.2$.

The warm dust component is modeled as an optically thin spherical shell of graphite
and silicate grains \cite[properties from][]{draine_lee_84} surrounding a source
with an AGN accretion disk spectral energy distribution \citep[taken from] []{schartmann_etal_05}. Equilibrium
temperatures are calculated for all grains as a function of their size and
composition. The temperature of a dust component is not a well defined
quantity since the component consists of a distribution of grains at different
equilibrium temperatures. We define the ``temperature'' of the distribution to
be the composition-averaged temperature of the luminosity-dominating
grain size. Grains are distributed in size with the power-law distribution of \cite{mathis_etal_77}.
The upper grain-size cutoff is 0.25\microns\ and the lower size limit
is determined by the minimum grain size that has not sublimated at the
temperature of the distribution.

For the purposes of producing a simple method to characterize spectra, we make
the assumption that all dust emission in the IRS spectral range
originates from grains that are directly heated by UV photons from the
illuminating source. Since the extinction in the ultraviolet is more than 100
times the extinction in the mid-infrared, the distance over which all UV
photons are absorbed by the dust is much less than an optical depth of unity
in the mid-infrared. We may therefore assume that all infrared emission within the
spherical shell is optically thin. We also allow for the possibility that the
dust component is partially obscured behind a screen of cooler dust (defined as
dust not emitting significantly in the mid-infrared). Therefore, the total
model for the emergent dust emission is
\begin{equation}
  F^{\rm dust}_\nu(\lambda) \propto E_{\nu}(\lambda)\, e^{-\tau^{\rm scr}(\lambda)},
\end{equation}
where $E_{\nu}$ is the emissivity of the dust distribution and $\tau^{\rm scr}$ is
the screen optical depth.

As an alternative to the screen extinction model, we also use a ``mixed'' extinction scenario, where warm dust emitting in the mid-infrared is
mixed with cooler dust from the shaded regions of clouds. This model may be more appropriate for a ``clumpy'' torus. In this case, the dust emission model is
\begin{equation}
  F^{\rm dust}_{\nu}(\lambda) \propto E_{\nu}(\lambda) \left ( \frac{1 - e^{-\tau^{\rm mix}(\lambda)}} {\tau^{\rm mix}(\lambda)} \right ),
\end{equation}
where $\tau^{\rm mix}$ is the ``mixed'' optical depth. 

The screen and mixed models represent limiting cases where the warm and cool dust is either completely separate or completely mixed, and should therefore capture the range of possible behaviours. We perform the fits for each object with both dust models to find the screen and mixed optical depths.

Our model continuum includes a silicate emission feature at 9.7\microns . The optical depths that we determine are therefore greater then would be determined if we had simply used an interpolated smooth continuum. In principle, the extinction values we derive from the
fits give a better measure of the true extinction to the source, since they
incorporate this more realistic continuum. However, a full radiative transfer
model may provide yet another optical depth value, since the obscuring
dust will re-radiate the absorbed power, partially filling in the silicate
absorption feature.

Given the poor MIPS detection statistics at 70\microns\ and particularly 160\microns , we typically lack photometric constraints in the far-infrared. However, in order to fit to the far-infrared photometry where available, we also included in the fits a cooler dust component, with the temperature allowed to vary in the range ${\rm T}=$~30--80\,K. For those objects without IRS spectra (3C\,200, 3C\,220.3, 3C\,272, 3C\,289, and 3C\,427.1), we fitted two dust components at fixed temperatures ${\rm T}=80$ and 185\,K to the available infrared photometric data. (For the galaxy 3C\,220.3, which had the only 160\microns\ detection, we allowed the temperature of the cooler component to be a free parameter). 

\begin{table*}
\tablenum{5}
  \vbox to220mm{\vfil Landscape table to go here
  \caption{}
  \vfil}
  \label{fitting_params.tbl}
\end{table*}

For all of the objects in the primary sample with IRS spectra, we performed fits to the SED using the following combinations of components:
\begin{itemize}\addtolength{\itemsep}{-0.5\baselineskip}
\item warm dust + lobe synchrotron;
\item warm dust + lobe synchrotron + jet synchrotron;
\item warm dust + lobe synchrotron + cool dust;
\item warm dust + cool dust + lobe synchrotron + jet synchrotron, 
\end{itemize}
{\noindent with the free parameters varying simultaneously. In general, the combination of components which resulted in the best reduced chi-squared was selected as the best fit. However, an $(n+1)$-component model is chosen in preference to an $n$-component model only if the inclusion of the additional component results in a confidence value for the F-test statistic of 99\% or better. Figure~\ref{fits} shows the best fits for the quasars and galaxies in the primary sample. Figure~\ref{alts} shows how the IRS spectra help to constrain the jet synchrotron contribution by comparing the fits with a jet component replacing cool dust in three quasars. Notes on the fits for some individual objects are presented in Appendix~B. Table~\ref{fitting_params.tbl} gives the parameters of the fits as well as the estimated dust luminosity, $L_{\rm dust}$, and the ratios of thermal to total luminosity at 15 and 30\microns . The latter constitute the corrections which are applied to the characteristic luminosities in order to remove the non-thermal contribution. The estimated dust luminosity, $L_{\rm dust}$, is the integrated luminosity of the model dust components and so does not include any contribution from the jet or lobe components. Since we do not model hot dust in this work and since we lack constraints in the far-infrared, this estimate should be considered a lower limit.}

Our spectral fitting methodology is subject to the usual problems of fitting simple models of complex phenomena to sparse data. The following caveats are worth pointing out explicitly. The spectral energy distributions of the objects in our sample are sparsely sampled and we lack constraints in the sub-millimeter region in many cases. The data used in the fits were not obtained simultaneously and variations in core flux density can reach $\approx 10$\%. We assume that a single power-law adequately describes the optically thin jet synchrotron emission; however, the actual synchrotron spectrum is likely to be a complex superposition of emission from discrete compact regions. We generally lack constraints in the far-infrared, so we cannot constrain the existence of colder dust. Moreover, the errors on the estimated optical depths and non-thermal contributions are likely to be dominated by model-dependent uncertainties that are difficult to quantify.

Nevertheless, the methodology described above introduces no biases into the fitting of components for quasars with respect to galaxies and, as discussed \S\,\ref{nontherm} below, it is only the quasars in the sample which are estimated to have a large non-thermal contribution. As a consistency check, we also apply an orientation-based model for the beamed component of emission \citep{hoekstra_etal_97} to our data in \S\,\ref{nontherm} and find that this also predicts significant non-thermal contributions for the quasars in our sample, suggesting that we are not systematically introducing jet components where none would be expected. Moreover, all of the model components that we use have been well established in detailed studies of individual objects, and our fitting models are chosen conservatively.

\section{DISCUSSION\label{discussion}}

We detect powerful emission in the mid-infrared for our sample of radio galaxies and quasars. All objects in our sample except one were detected at 24\microns\ using MIPS, with luminosities $L_{24\mu{\rm m}} > 10^{22.4}$\,W\,Hz\mo\,sr\mo . In the following section we discuss the AGN physical conditions derived from the emission line measurements, the origin of the measured infrared emission, and the comparative luminosity of quasars and radio galaxies.

\begin{figure}[!t]
\epsscale{1.0}
\plotone{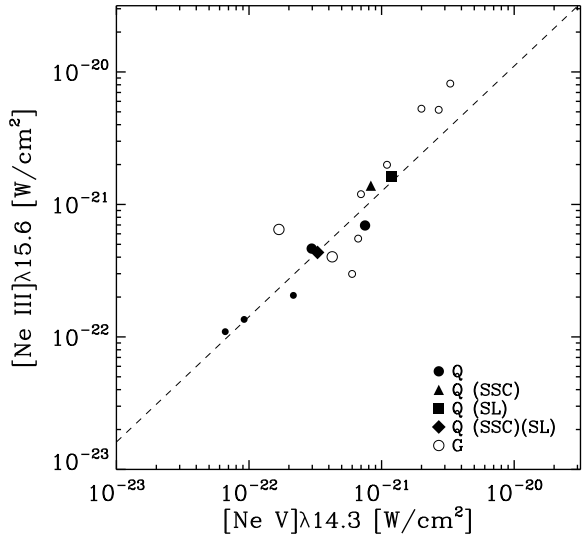}
\caption{Observed flux density for [Ne\,{\sc v}]$\lambda14.3\,\mu{\rm m}$ vs.\ [Ne\,{\sc iii}]$\lambda15.5\,\mu{\rm m}$ for the primary sample as well as other 3CRR sources from \cite{haas_etal_05} and \cite{ogle_etal_06} (plotted using smaller symbols). Only objects with detections in both lines are plotted. The dashed line (of slope 0.9) is a fit to the data for the quasars only. Symbols are as in Figure~\ref{L178vsz}. \label{neon_lines}}
\end{figure}

\begin{figure}[!t]
\epsscale{1.0}
\plotone{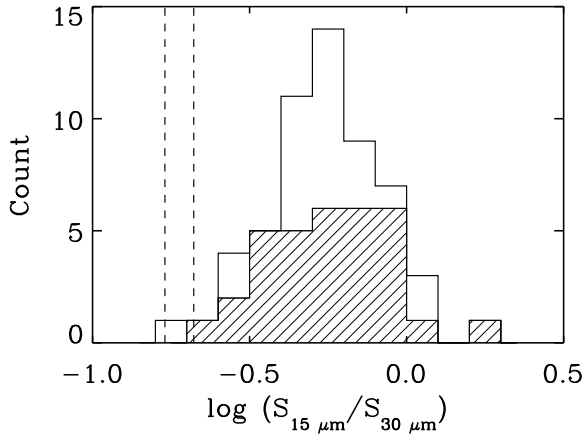}
\caption{This figure shows the distribution of the thermal 15 to 30\microns\ flux density ratio for the primary and supplementary samples together (unfilled histogram) as well the primary sample alone (cross-hatched histogram). The vertical dashed lines indicate the approximate division ($\approx$ 0.17--0.21) between AGN and starbursts found for the 15/30\microns\ ratio by \cite{brandl_etal_06}.
\label{mid_far_ir}}
\end{figure}

\subsection{AGN Physical Conditions\label{phys_cond}}

The high-excitation (97.1\,eV) [Ne\,{\sc v}]$\lambda14.3\,\mu{\rm m}$ line was detected in five quasars (see Table~4); it was also detected in two radio galaxies, 3C\,330 and 3C\,184, directly confirming the presence of a highly luminous ionizing radiation field in these sources. We can use the neon emission line ratios to probe the physical conditions in the photoionized region. The [Ne\,{\sc v}]$\lambda14.3\,\mu{\rm m}$/[Ne\,{\sc iii}]$\lambda15.5\,\mu{\rm m}$ ratio responds to the ionization parameter and the [Ne\,{\sc ii}]$\lambda12.8\,\mu{\rm m}$/[Ne\,{\sc iii}]$\lambda15.5\,\mu{\rm m}$ ratio is sensitive to the hardness of the photoionizing spectrum. Comparing to the {\sc Cloudy} model grid of Voit~(1992), the neon line ratios are consistent with values of the ionization parameter, $\log{U}$, in the range $-2$ to $-2.5$ and, assuming a power-law photoionizing spectrum $\propto \nu^{\alpha}$, power-law indices, $\alpha$, in the range $-1$ to $-1.5$. 

Figure~\ref{neon_lines} plots the [Ne\,{\sc v}]$\lambda14.3\,\mu{\rm m}$ vs.\ [Ne\,{\sc iii}]$\lambda15.5\,\mu{\rm m}$ line flux densities. The ratio of these lines varies strongly with ionization parameter \citep[$\propto U^{2}$,][]{voit_92}. The strong correlation between these line flux densities shown in Figure~\ref{neon_lines} demonstrates that there is little variation in ionization parameter amongst the 3C objects in which they are both detected. 

\subsection{Star-formation Contribution\label{starformation}}

The observed infrared emission from AGN can have contributions from thermal and non-thermal processes. As well as emission from warm dust surrounding the nucleus, which peaks in the mid-infrared, dust at lower temperatures heated by star-formation can produce emission that peaks in the far-infrared. 
\begin{figure}[!t]
\epsscale{1.0}
\plotone{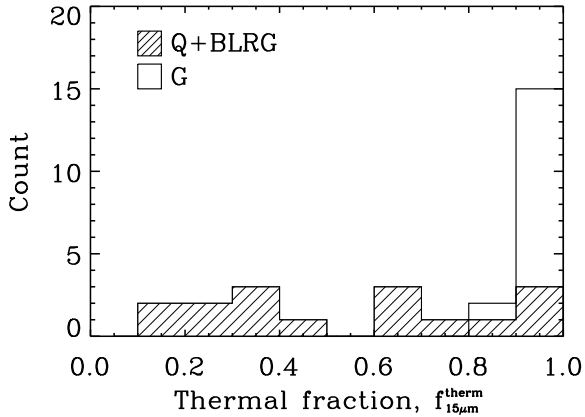}
\caption{The thermal fraction, $f^{\rm therm}_{15\mu{\rm m}} = L^{\rm therm}_{15\mu{\rm m}} / L^{\rm total}_{15\mu{\rm m}}$ for galaxies (unfilled histogram) and quasars (cross-hatched histogram). The BLRG 3C\,22 is included with the quasars.
\label{frac_hist}}
\end{figure}
\cite{sturm_etal_02} demonstrated the use of the [Ne\,{\sc v}]$\lambda14.3\,\mu{\rm m}$ to [Ne\,{\sc ii}]$\lambda12.8\,\mu{\rm m}$ ratio to determine the AGN contribution in composite AGN/starburst objects. For our sample, however, we can only constrain the 
[Ne\,{\sc v}]$\lambda14.3\,\mu{\rm m}$ to [Ne\,{\sc ii}]$\lambda12.8\,\mu{\rm m}$ ratio in a small number of cases. Recently, \cite{brandl_etal_06} have shown that the 15/30\microns\ flux density ratio is also a very good AGN/starburst discriminator. As shown in Figure~\ref{mid_far_ir}, almost all objects in the primary sample exceed the threshold 15/30\microns\ flux density ratio ($\approx 0.2$) at which the AGN contribution dominates \citep{brandl_etal_06}. We therefore conclude that star-formation does not contribute significantly to the measured infrared emission for the majority of objects in the primary sample.

\begin{figure*}
\centering
\begin{tabular}{cc}
\tablewidth{0pt}

\psfig{file=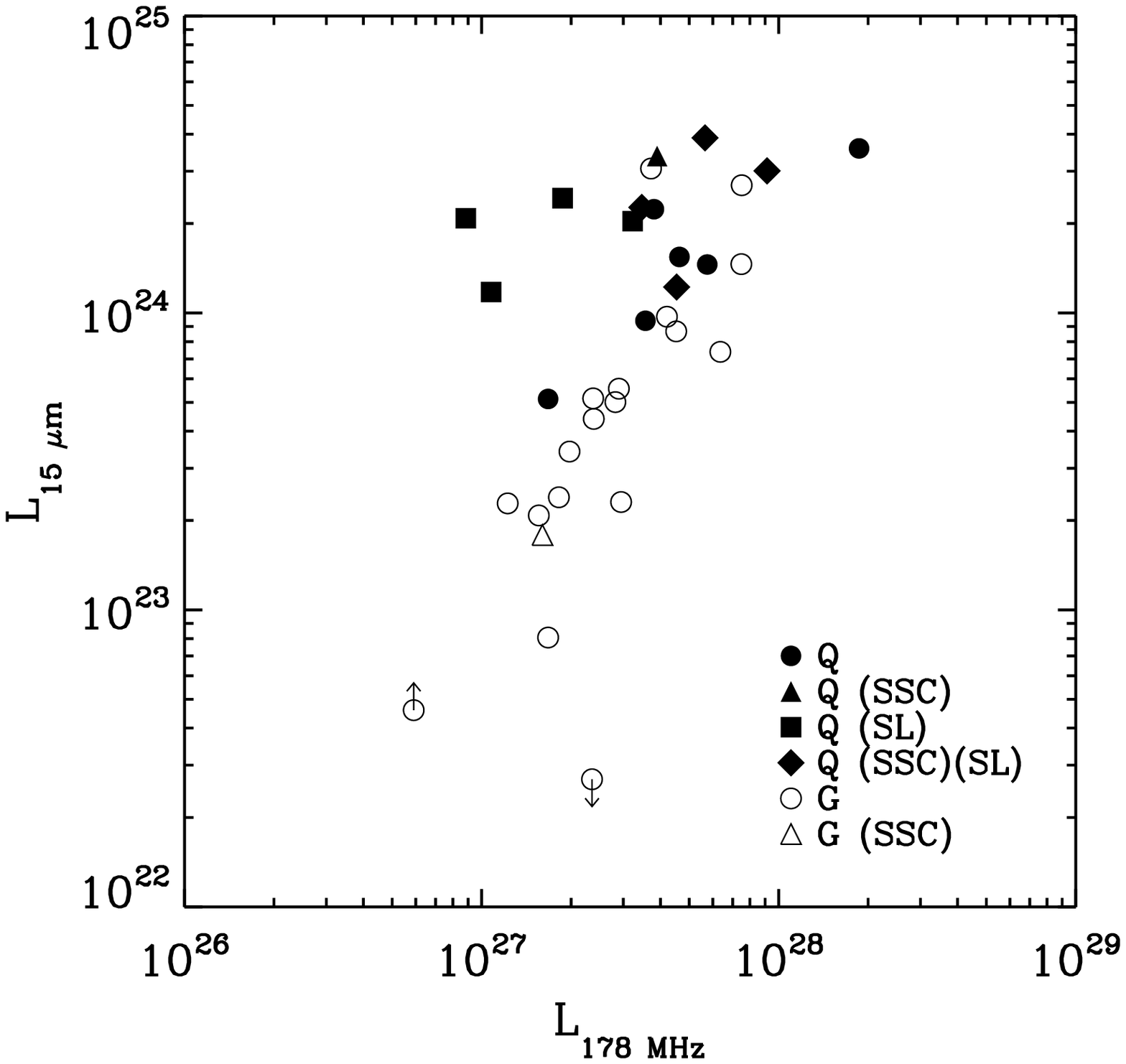,scale=0.45} & 
\psfig{file=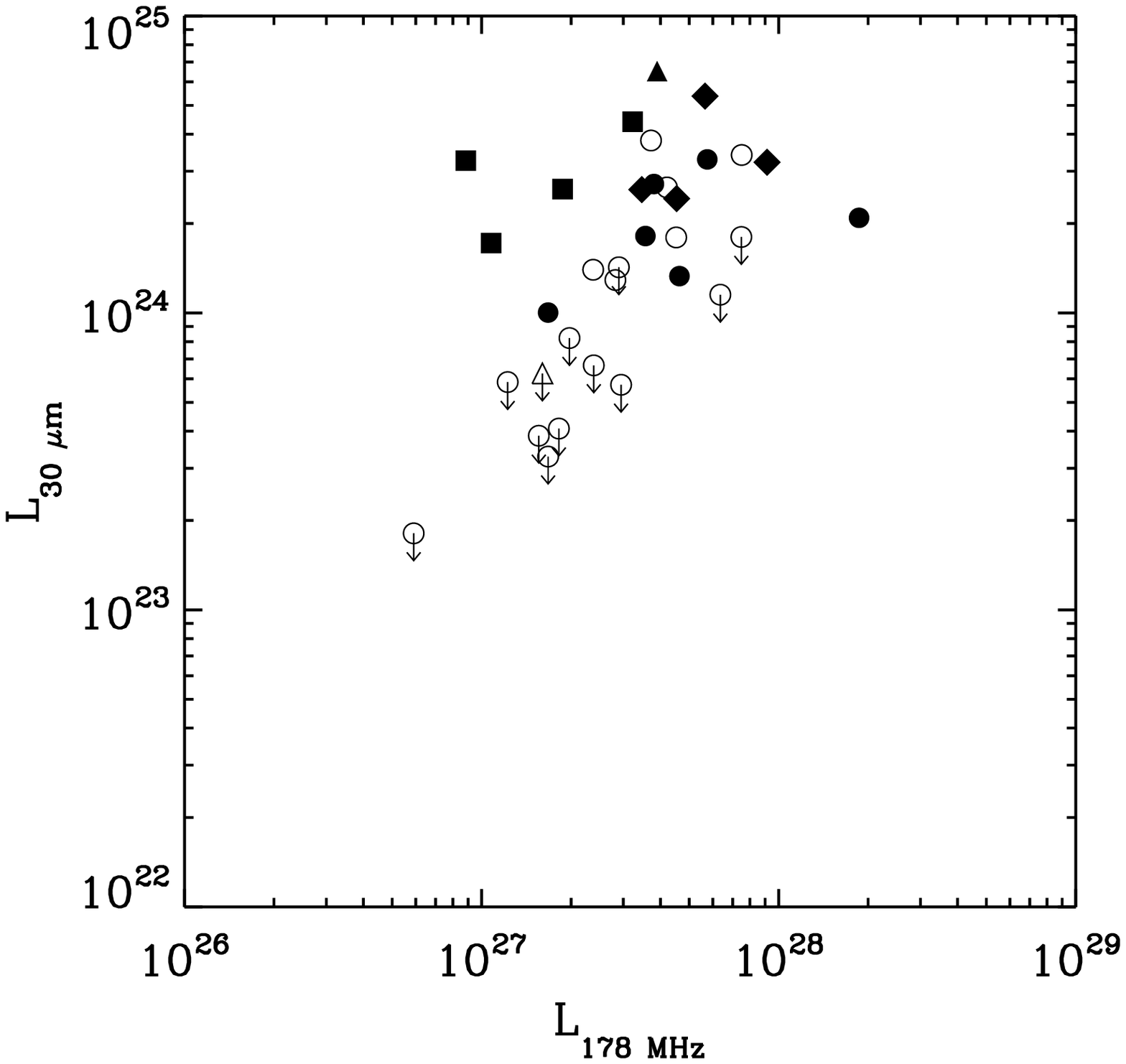,scale=0.45}\\

\psfig{file=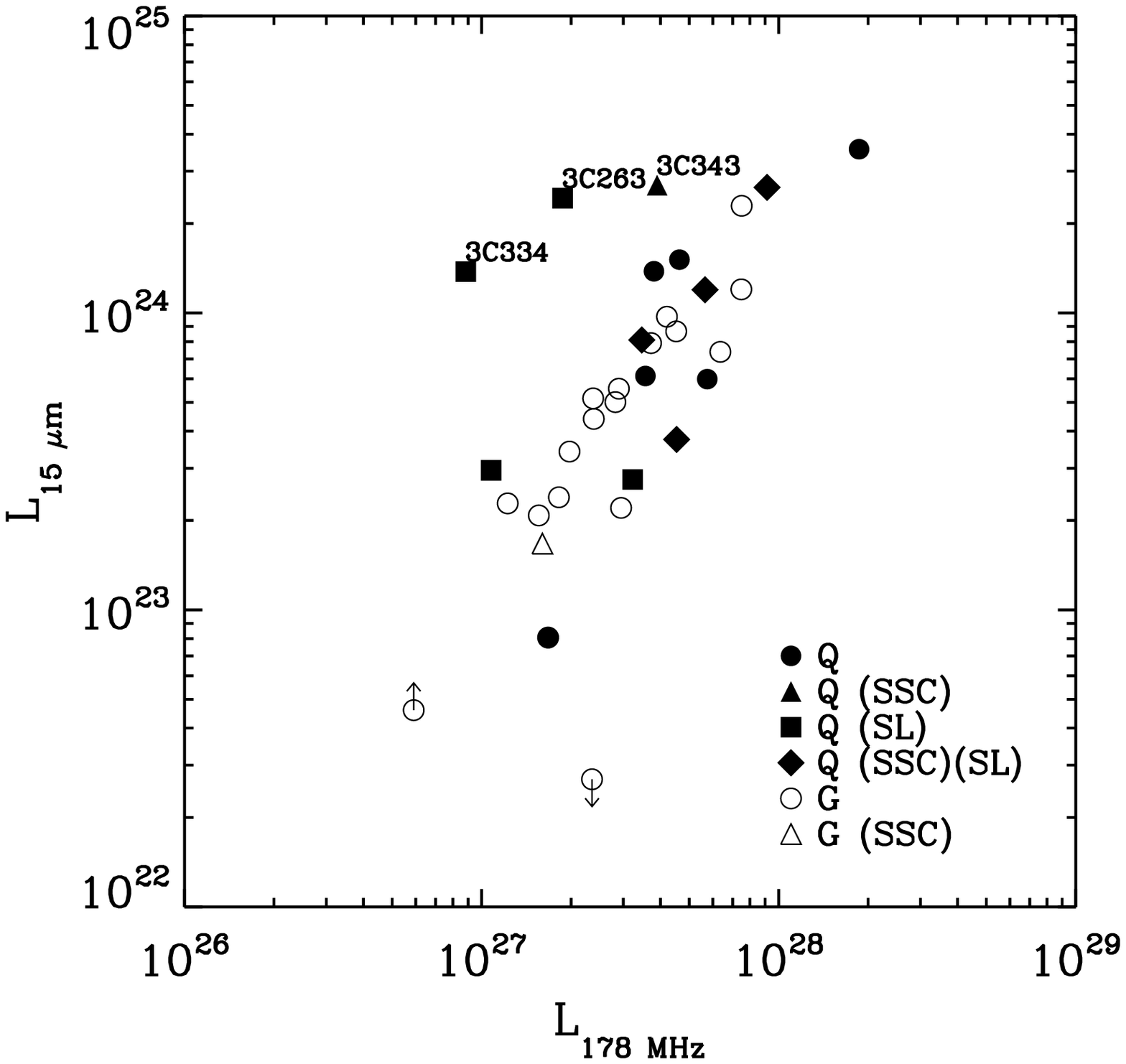,scale=0.45} & 
\psfig{file=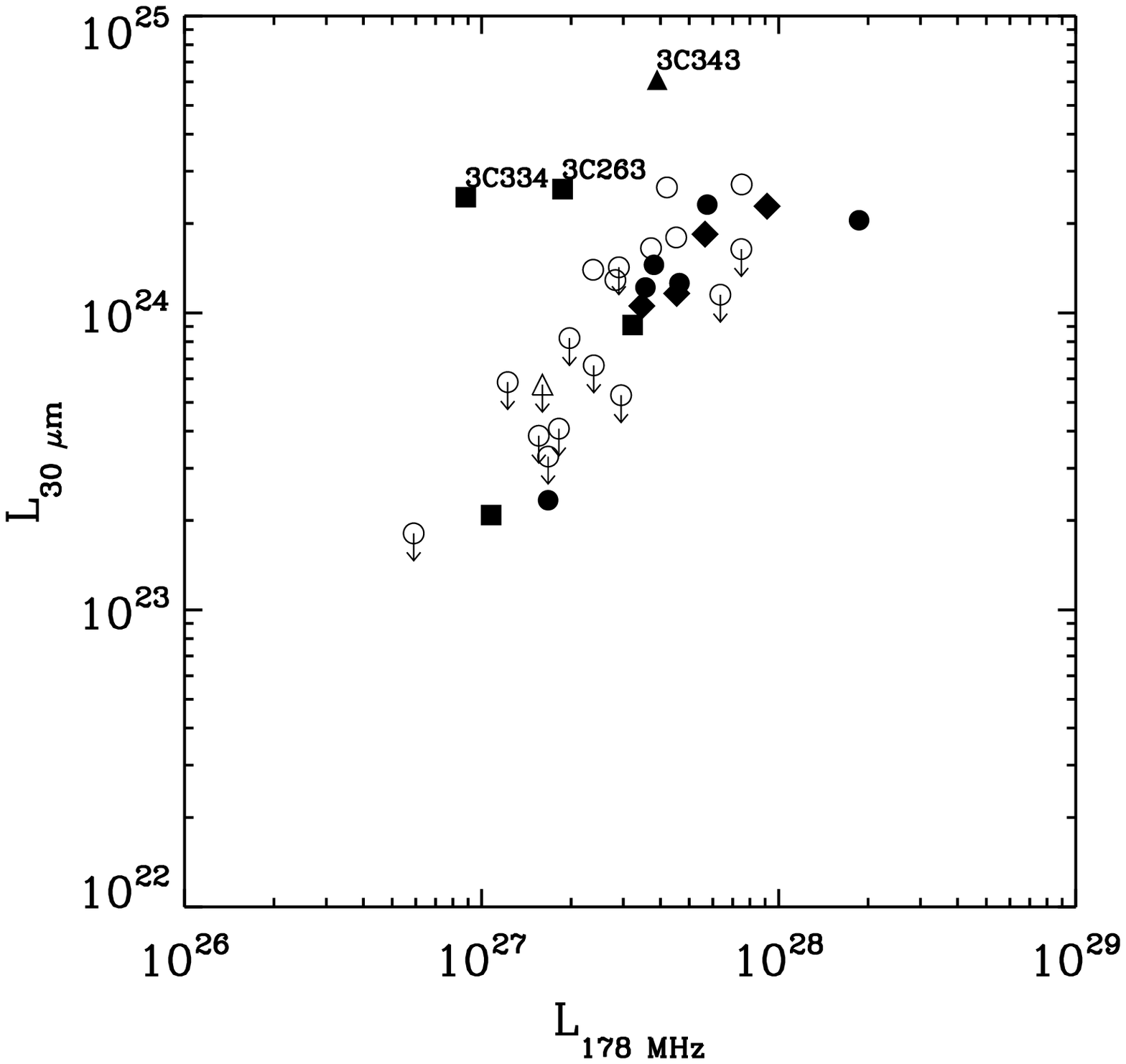,scale=0.45}\\

\end{tabular}
\caption{Luminosity (W\,Hz\mo\,sr\mo ) at rest wavelengths 15\microns\ (left) and 30\microns\ (right) vs.\ luminosity at rest-frame 178\,MHz for objects in the primary sample, before (top) and after (bottom) the correction for non-thermal emission has been applied. Symbols are as in Figure~\ref{L178vsz}. Objects with high thermal infrared luminosity for their radio-lobe power are labeled.
\label{Lirvs178}}
\end{figure*}


\begin{figure*}
\centering
\begin{tabular}{cc}
\tablewidth{0pt}

\psfig{file=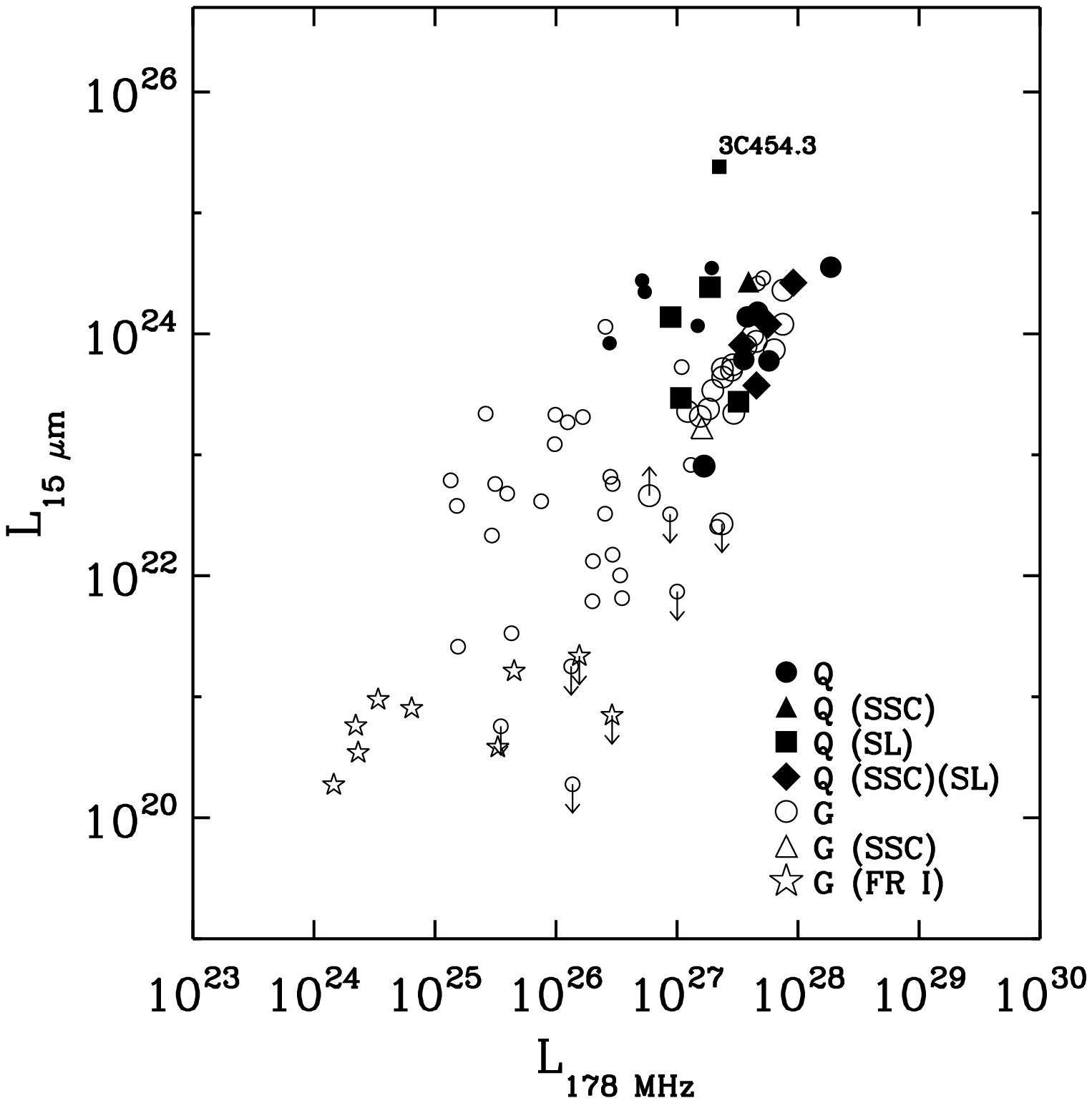,scale=0.5} & 
\psfig{file=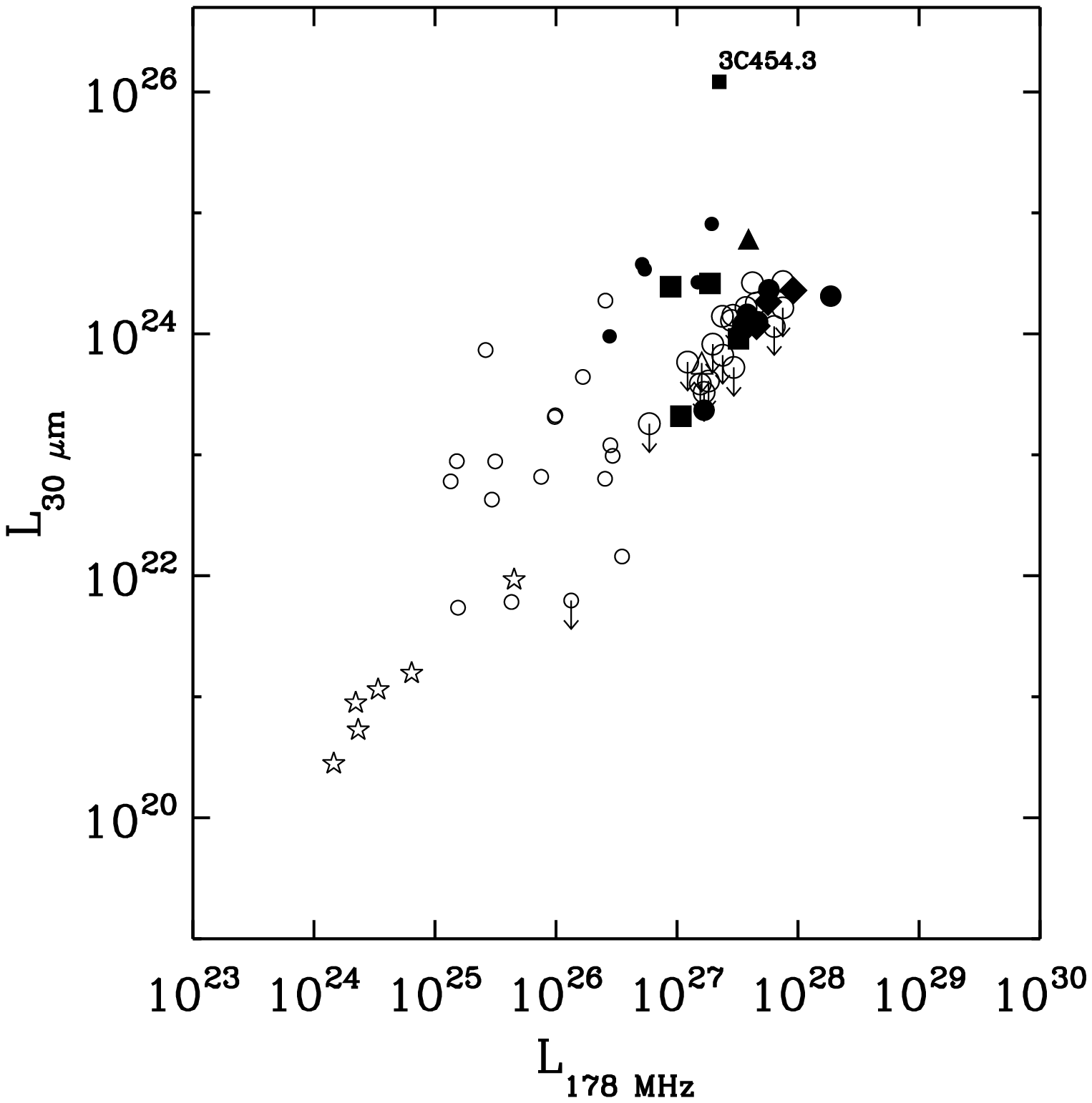,scale=0.5}\\

\end{tabular}
\caption{Luminosity (W\,Hz\mo\,sr\mo\ ) at rest wavelengths 15\microns\ (left) and 30\microns\ (right) vs.\ luminosity at rest-frame 178\,MHz for objects in the primary and supplementary samples. Symbols are as in Figure~\ref{L178vsz}. FR\,I galaxies from the supplementary sample are plotted as open stars. Data for objects from the supplementary sample are plotted using smaller symbols. The infrared luminosity of objects in the primary sample has been corrected for non-thermal emission, while that from objects in the supplementary sample has not.
\label{Lirvs178all_fr}}
\end{figure*}

\begin{deluxetable}{llccc}
\tabletypesize{\scriptsize}
\tablecaption{$R_{\rm dr}$ statistics for quasars and galaxies.}
\label{avg_lums}
\tablewidth{0pt}
\tablenum{6}
\tablehead{
 & & \colhead{Quasars} & \colhead{Galaxies} & \colhead{Q/G\tablenotemark{a}}
}
\startdata
$\langle \log R_{\rm dr}(15\mu{\rm m})\rangle$\dotfill & total		& $1.78 \pm 0.07$		& $1.21 \pm 0.06$            & $3.7^{+0.9}_{-0.7}$\phn   \\
$\langle \log R_{\rm dr}(15\mu{\rm m})\rangle$\dotfill	& thermal	& $1.44 \pm 0.10$   		& $1.19 \pm 0.05$	       & $1.8^{+0.5}_{-0.4}$\phn  \\ 
$\langle \log R_{\rm dr}(15\mu{\rm m})\rangle$\dotfill & screen		& $1.66 \pm 0.14$		& $1.66 \pm 0.11$            & $1.0^{+0.5}_{-0.3}$\phn   \\
$\langle \log R_{\rm dr}(15\mu{\rm m})\rangle$\dotfill	& mixed		& $2.00 \pm 0.25$   		& $2.74 \pm 0.36$	       & $0.2^{+0.3}_{-0.1}$\phn  \\ 
\hline\\
${\rm Med}(\log R_{\rm dr}(15\mu{\rm m}))$\dotfill	& total		& $1.85 \pm 0.07$	    	& $1.29 \pm 0.06$	       & $3.6^{+0.9}_{-0.7}$\phn  \\
${\rm Med}(\log R_{\rm dr}(15\mu{\rm m}))$\dotfill     & thermal	& $1.42 \pm 0.10$		& $1.21 \pm 0.05$	       & $1.6^{+0.5}_{-0.4}$\phn \\
${\rm Med}(\log R_{\rm dr}(15\mu{\rm m}))$\dotfill	& screen	& $1.67 \pm 0.14$	    	& $1.58 \pm 0.11$	       & $1.2^{+0.6}_{-0.4}$\phn  \\
${\rm Med}(\log R_{\rm dr}(15\mu{\rm m}))$\dotfill     & mixed		& $1.72 \pm 0.26$		& $2.41 \pm 0.37$	       & $0.2^{+0.4}_{-0.1}$\phn \\
\hline\\
$\langle \log R_{\rm dr}(30\mu{\rm m})\rangle$\dotfill & total		& $1.63 \pm 0.08$		& $1.06 \pm 0.04$\tablenotemark{b}            & $3.7^{+0.8}_{-0.7}$\phn  \\
$\langle \log R_{\rm dr}(30\mu{\rm m})\rangle$\dotfill	& thermal	& $1.33 \pm 0.09$  		& $1.06 \pm 0.04$\tablenotemark{b}  	       & $1.9^{+0.5}_{-0.4}$\phn \\
$\langle \log R_{\rm dr}(30\mu{\rm m})\rangle$\dotfill & screen		& $1.49 \pm 0.11$		& $1.15 \pm 0.10$\tablenotemark{b}            & $2.2^{+0.9}_{-0.6}$\phn  \\
$\langle \log R_{\rm dr}(30\mu{\rm m})\rangle$\dotfill	& mixed		& $1.74 \pm 0.16$  		& $1.52 \pm 0.22$\tablenotemark{b}  	       & $1.7^{+1.4}_{-0.8}$\phn \\
\hline\\
${\rm Med}(\log R_{\rm dr}(30\mu{\rm m}))$\dotfill	& total		& $1.62 \pm 0.08$		& $1.19 \pm 0.04$\tablenotemark{c} 	       & $2.7^{+0.6}_{-0.5}$\phn \\
${\rm Med}(\log R_{\rm dr}(30\mu{\rm m}))$\dotfill     & thermal	& $1.26 \pm 0.10$  		& $1.20 \pm 0.04$\tablenotemark{c} 	       & $1.2^{+0.3}_{-0.3}$\phn \\
${\rm Med}(\log R_{\rm dr}(30\mu{\rm m}))$\dotfill	& screen	& $1.37 \pm 0.11$  		& $1.50 \pm 0.06$\tablenotemark{c} 	       & $0.7^{+0.2}_{-0.2}$\phn \\
${\rm Med}(\log R_{\rm dr}(30\mu{\rm m}))$\dotfill     & mixed		& $1.50 \pm 0.18$  		& $1.95 \pm 0.22$\tablenotemark{c} 	       & $0.4^{+0.3}_{-0.2}$\phn \\
\enddata
\tablecomments{The table gives the mean and median of $\log R_{\rm dr} = \log (\nu L_{\nu}(\rm IR)/ \nu L_{\nu}(\rm{178\,MHz}))$, at the characteristic rest wavelengths of 15 and 30\microns , for quasars and galaxies in the sample with estimates of the infrared luminosity and dust optical depth. Also tabulated is the ratio of the mean or median $R_{\rm dr}$ for quasars to that for galaxies. In each case, the statistic is computed for the `total' $R_{\rm dr}$ (i.e.\ as derived from the data), the `thermal' $R_{\rm dr}$ (i.e.\ corrected for non-thermal contributions), and finally corrected for both non-thermal contributions and dust extinction, using the `screen' and `mixed' optical depths. The distribution of $R_{\rm dr}$ is adequately described as log-normal. We therefore estimate the mean of $R_{\rm dr}$ as $10^{\langle \log{R_{\rm dr}}\rangle }$, and the resulting standard error is asymmetric. Where the data include upper limits, the statistics are computed using the Kaplan-Meier (KM) estimate from the \textsc{Asurv} statistical package \citep{isobe_etal_86}. The 15\microns\ lower luminosity limit for 3C\,200 is treated as a detection. The BLRG 3C\,22 is included with the quasars. The number of quasars and galaxies included in the analysis (number of detections in parentheses) at 15\microns\ is 16~(16) and 13~(13); at 30\microns\ it is 16~(16) and 13~(2).}
\tablenotetext{a}{This column gives the ratio of the estimated mean and median $R_{\rm dr}$ for the quasars to that of the galaxies.}
\tablenotetext{b}{This estimate is biased since a censored datum was changed to a detection for the KM computation. Since the KM method does not provide an estimate of the error on the median, we estimate the error by treating all measurements as detections.}
\tablenotetext{c}{The KM estimate of the median falls outside the range of observed values due to the heavy censoring of data. Instead, we estimate the median and standard error by treating the upper limits as detections}
\end{deluxetable}

\subsection{Non-thermal Contribution\label{nontherm}}

Synchrotron emission from optically thin radio lobes has flux-density spectral indices $\alpha \approx -0.7$, so the contribution to the infrared is typically well over an order of magnitude below the observed flux density. However, Doppler-boosted synchrotron emission from dense, compact regions is a potentially serious contaminant of the infrared emission. The relative contribution of thermal and non-thermal processes to the infrared emission of AGN has been the subject of many studies. \cite{sanders_etal_89} were among the first to advance thermal dust emission as the dominant infrared emission process in quasars, based on a study of Palomar-Green (PG)  quasars. However, claims that quasars are brighter sources of infrared emission than radio galaxies \citep[e.g.,][]{heckman_etal_92, heckman_etal_94,hes_etal_95} have provided evidence for a beamed component in the infrared emission of FR\,II sources. Based on a spectral decomposition for three 3C quasars (3C\,47, 3C\,207 and 3C\,334) with {\em IRAS\/} 60\microns\ detections, \cite{vanbemmel_etal_98} found evidence for a non-thermal contribution to the IRAS flux density in one object (3C\,207).  An orientation-based model described by \cite{hoekstra_etal_97} was consistent with a significant non-thermal contribution to the {\em IRAS\/} 60\microns\ emission of FR\,II sources with higher core dominance.

The {\em Spitzer\/} data presented here have provided further constraints on the role of non-thermal processes in the infrared emission of radio galaxies and quasars. In \S\,\ref{sed_section}, we fitted the SEDs of the objects in our sample with broadband spectral components representing synchrotron emission from the radio lobes and jet as well as thermal emission from circumnuclear dust (see Fig.~\ref{cartoon}). In this way, the non-thermal contribution to the 15 and 30\microns\ luminosity was estimated, and the corrections applied to the data are given in Table~\ref{fitting_params.tbl} as the thermal fraction, $f^{\rm therm} = L^{\rm therm} / L^{\rm total}$ at 15 and 30\microns. Figure~\ref{frac_hist} shows a histogram of the thermal fractions at 15\microns\ for the quasars and galaxies in the primary sample. For the galaxies, the thermal contribution is always $>80$\%; for the quasars, the thermal contribution is in the range 10--100\%. All objects, quasars and galaxies, were fitted in a consistent manner; however it is only the quasars in the sample which were estimated to have a non-thermal contribution $>20$\%.

Since the errors on the estimated thermal fractions are likely dominated by model-dependent uncertainties, as a consistency check we also used the orientation-based model of \cite{hoekstra_etal_97} to predict the non-thermal contribution as a function of core dominance. Following \cite{hoekstra_etal_97}, we fit the ratio of rest-frame 30\microns\ to 178 MHz flux density for the primary sample with an equation of the form, 

\begin{equation}
\frac{F_{\rm 30\,\mu m}^{\rm total}}{F_{\rm 178\,MHz}} = B\,D_{\rm 30\,\mu m}\,(R+C_{\rm 30\,\mu m}),
\label{hoekstra_model1.eqn}
\end{equation}

{\noindent where, $F_{\rm 30\,\mu m}$ and $F_{\rm 178\,MHz}$ are the rest-frame 30\microns\ and 178\,MHz flux densities, $B$ is the ratio of the rest-frame extended emission at 5\,GHz and 178\,MHz, $D_{\rm 30 \mu m}$ is the ratio of the beamed emission at 30\microns\ and 5\,GHz, $C_{\rm 30 \mu m}=F_{\rm 30 \mu m}^{\rm iso}/(F_{\rm 5\,GHz}^{\rm ext}\,D_{\rm 30 \mu m})$ is the ratio of the isotropic 30\microns\ emission to the product of the extended emission at 5\,GHz and $D_{\rm 30\,\mu m}$, and $R$ is the core dominance parameter. In this model, $D$ is a measure of the strength of the beamed component of emission and $C$ is a measure of the value of $R$ at which the core emission starts to dominate over the extended emission. In order to help constrain the high core-dominance region, we also included the blazars 3C\,454.3 from the supplementary sample and 3C\,345 based on {\em IRAS\/} detections in the literature. Despite the many upper limits, we perform this fit at the characteristic wavelength of 30\microns , rather than 15\microns , in order to minimise the effect of any correlation of optical depth with core dominance. Fitting Equation~\ref{hoekstra_model1.eqn} to the data (Fig.~\ref{hoekstra_model}, left), we obtain a value for $C_{\rm 30\,\mu m}$ of 0.05, from which the non-thermal fraction as a function of core dominance can be estimated as,

\begin{equation}
f^{\rm non-therm}_{\rm 30\,\mu m} = \frac{R}{R + C_{\rm 30 \mu m}}.
\label{hoekstra_model2.eqn}
\end{equation}

Figure~\ref{hoekstra_model} (right) shows the non-thermal fractions estimated from the spectral fitting and the predictions of the \cite{hoekstra_etal_97} model are plotted as the solid curve. The dashed curve shows the predictions based on the 30\microns\ flux density corrected for the estimated screen optical depth ($C_{\rm 30\,\mu m}=0.09$). The model predicts a significant non-thermal contribution for the objects in the primary sample with higher core dominance. Although there is considerable scatter, reasonable agreement can be seen between the model predictions and the results of the spectral fitting. The broad-lined radio galaxy, 3C\,22, has a significantly higher non-thermal estimate from the spectral fitting than predicted by the model. However, this object also has a $F_{\rm 30\,\mu m}/F_{\rm 178\,MHz}$ ratio more similar to the quasars at higher core dominance. The quasar 3C\,380 has an estimated non-thermal fraction of around 0.1 from the spectral fits although it has the highest core dominance of the primary sample. However, this object is an outlier with a low $F_{\rm 30\,\mu m}/F_{\rm 178\,MHz}$ ratio. We note that the fitted value for $C_{\rm 30\,\mu m}$ of 0.05--0.09 is much less than that for  $C_{\rm 60\,\mu m}$ of 2.6 found by \cite{hoekstra_etal_97}, indicating that the onset of core-dominated emission at 30\microns\ in the primary sample occurs at larger inclinations from the radio axis than found by \cite{hoekstra_etal_97} at 60\microns\ for their sample of {\em IRAS\/}-detected 3C and 4C objects.

\begin{figure*}
\centering
\begin{tabular}{cc}
\tablewidth{0pt}

\psfig{file=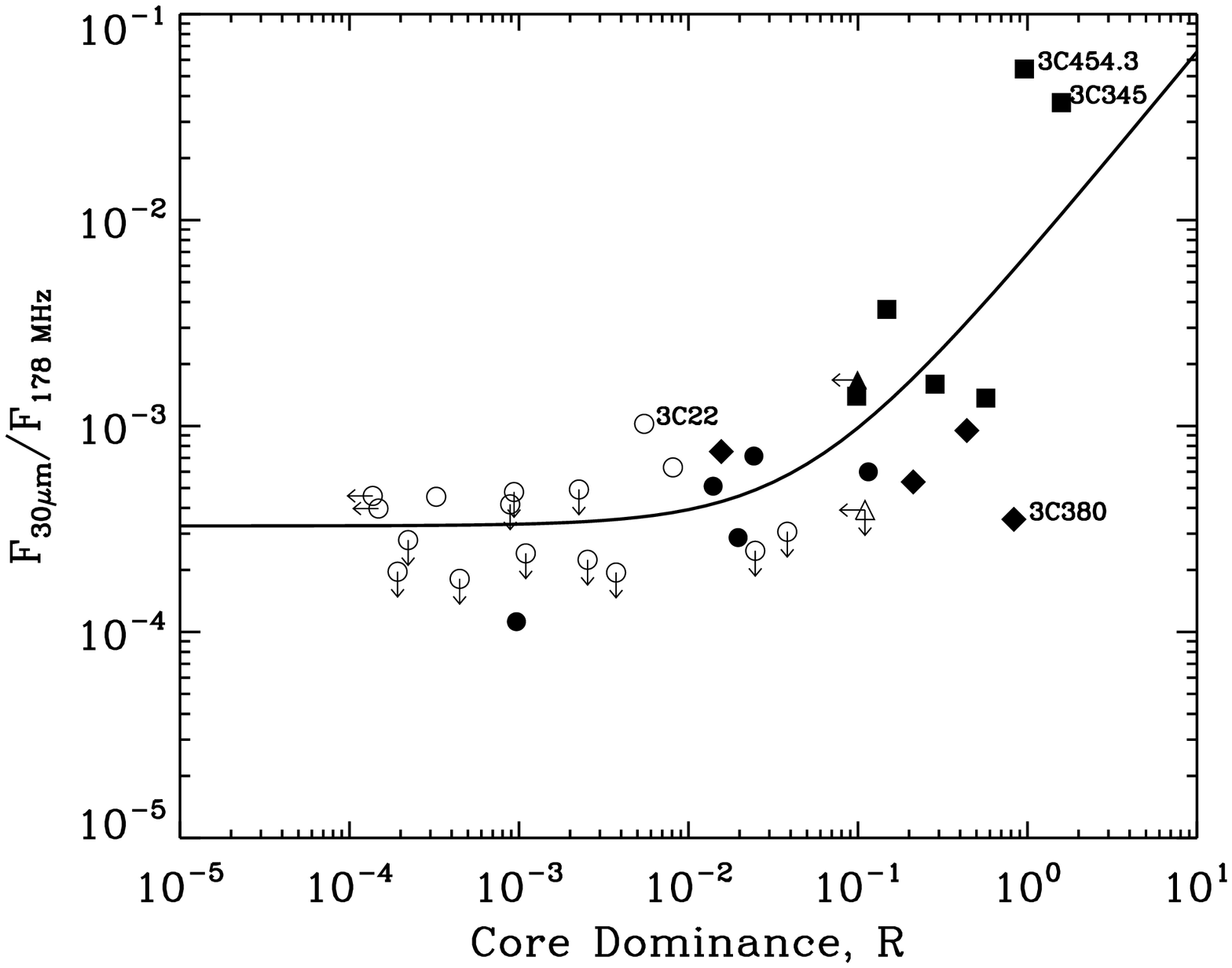,scale=0.4} & 
\psfig{file=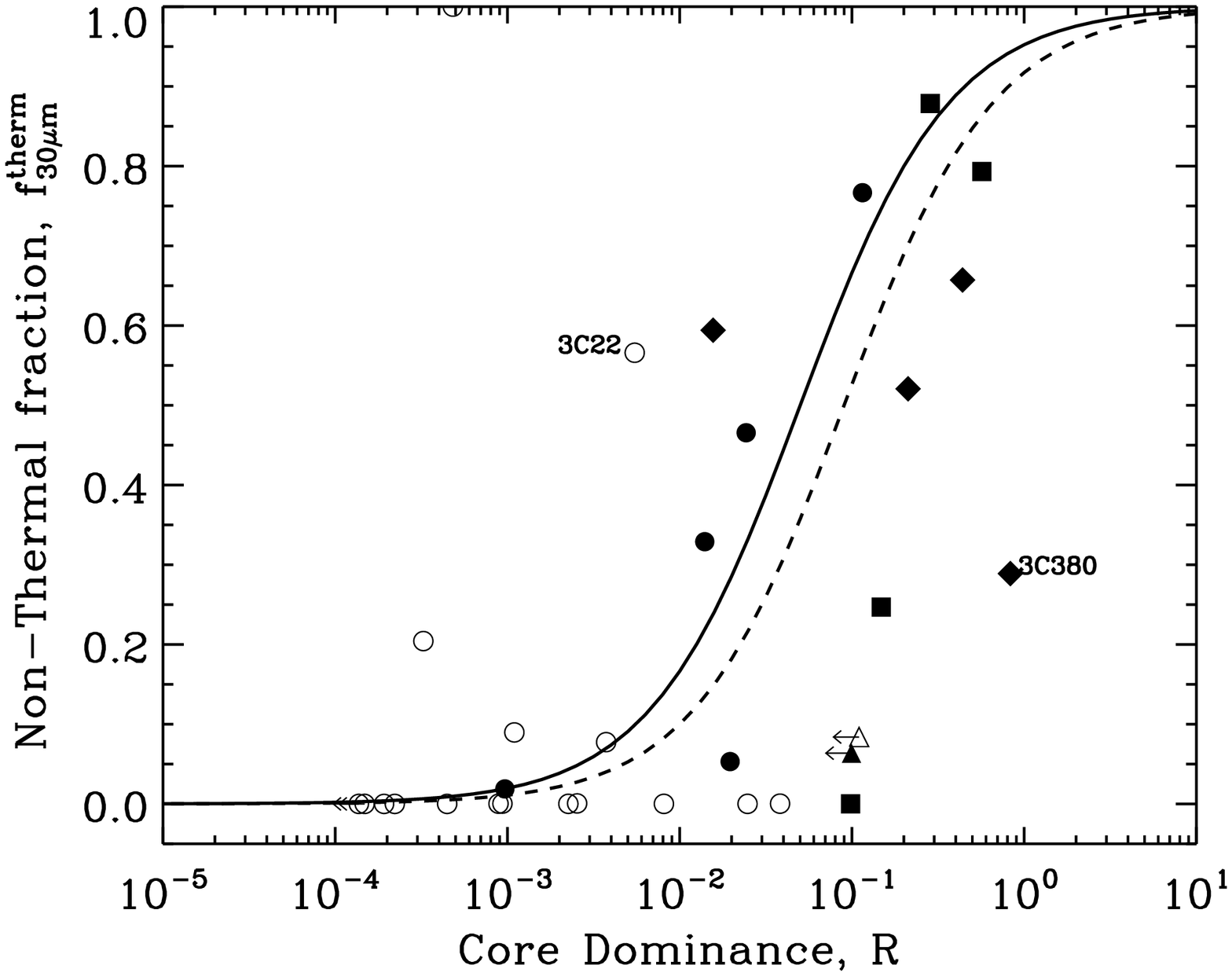,scale=0.4}\\

\end{tabular}
\caption{The orientation-based model (solid curve) of \cite{hoekstra_etal_97} is fitted to the ratio of the rest-frame 30\microns\ to 178 MHz flux density (left). Based on the fit to the data, the model can be used to predict the non-thermal fraction as a function of orientation. We plot the non-thermal fraction for each source in the primary sample estimated from the spectral fitting (right) and also the model prediction (solid curve). The dashed curve shows the predictions based on the 30\microns\ flux density corrected for the estimated screen optical depth. The model predicts significant non-thermal contributions for the more core dominated objects in the primary sample. Although there is considerable scatter, reasonable agreement can be seen between the model predictions and the results of the spectral fitting. Symbols are as in Figure~\ref{L178vsz}.
\label{hoekstra_model}}
\end{figure*}

Figure~\ref{Lirvs178} shows the 15 and 30\microns\ luminosity vs.\ 178\,MHz luminosity for objects in our sample, before and after the correction for non-thermal emission is applied. Figure~\ref{Lirvs178all_fr} shows the non-thermal corrected luminosities, but also includes objects from the supplementary sample. We were unable to fit dust models for 3C\,427.1 (labeled in Figs.~\ref{Lirvs178} and \ref{Lirvs178all_fr}) since no IRS or MIPS detections were obtained for this object. However, we have assumed a thermal fraction of $1.0$ since it is a lobe-dominated galaxy. Comparing the 15\microns\ luminosity of the objects in the primary sample before and after (Fig.~\ref{Lirvs178}) the non-thermal correction has been applied, we can see that the effect of the correction is to reduce the luminosity of most quasars to a level comparable with that of the galaxies. Note that any apparent correlations in the luminosity-luminosity plots presented here may be due to a mutual dependence on a third variable (e.g., redshift).


\begin{figure*}
\centering
\begin{tabular}{cc}
\tablewidth{0pt}

\psfig{file=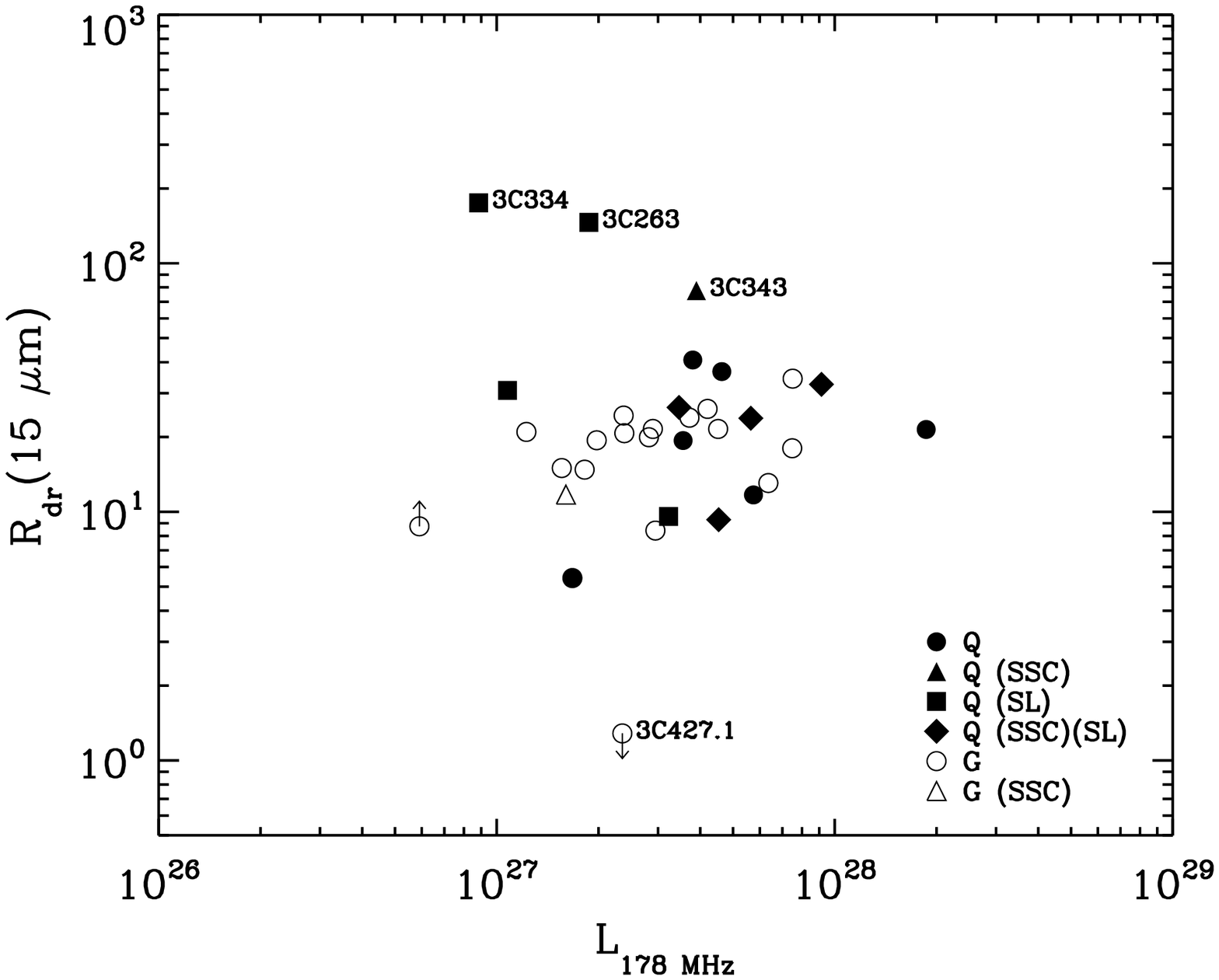,scale=0.4} & 
\psfig{file=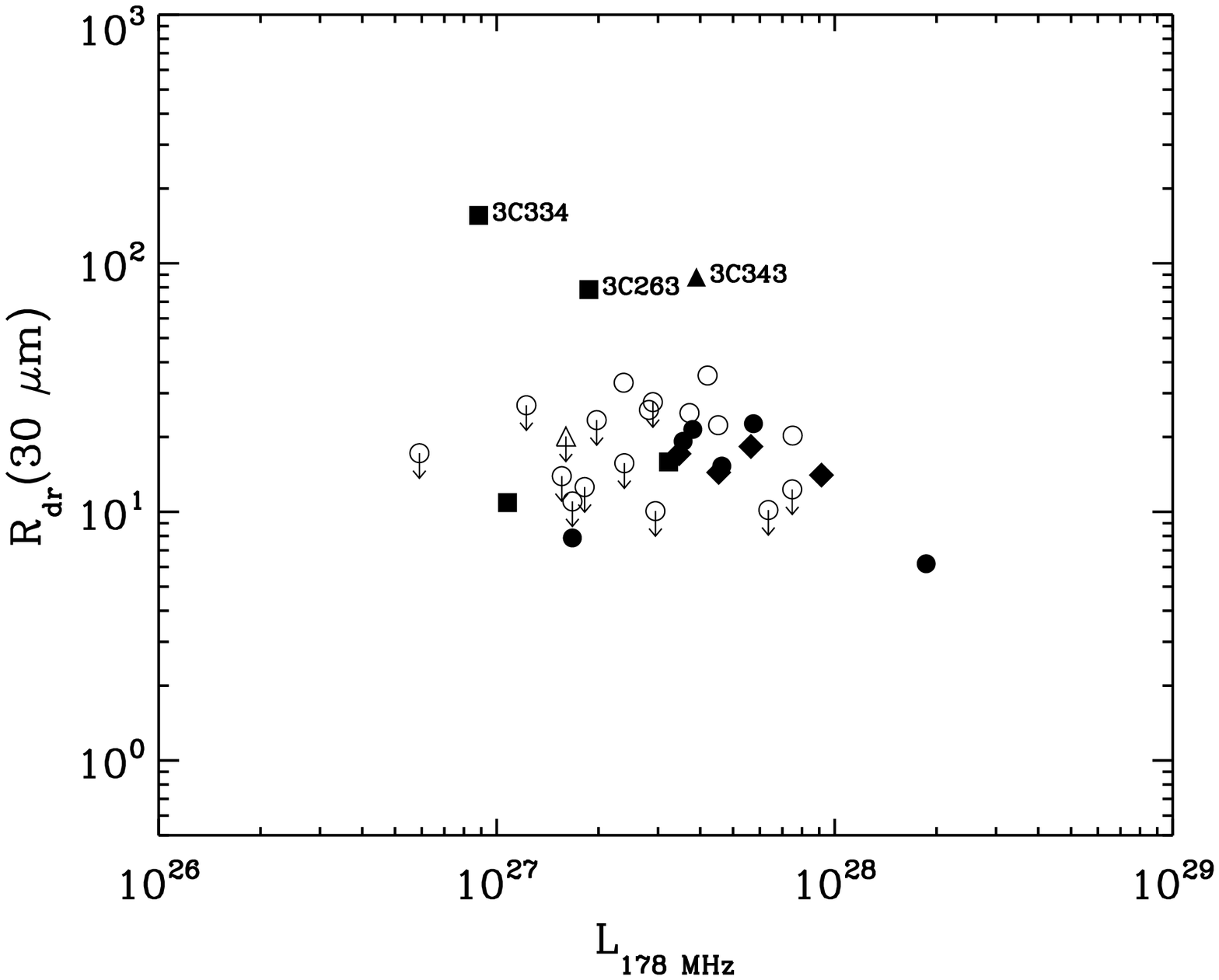,scale=0.4}\\

\end{tabular}

\caption{$R_{\rm dr}(15\micron )$ (left) and $R_{\rm dr}(30\micron )$ (right) vs.\ 178\,MHz luminosity (W\,Hz\mo\,sr\mo) for objects in the primary sample, corrected for non-thermal emission. Symbols are as in Figure~\ref{L178vsz}. Objects with high $R_{\rm dr}$ are labeled.\label{rdrIR}}
\end{figure*}


\begin{figure*}
\centering
\begin{tabular}{cc}
\tablewidth{0pt}

\psfig{file=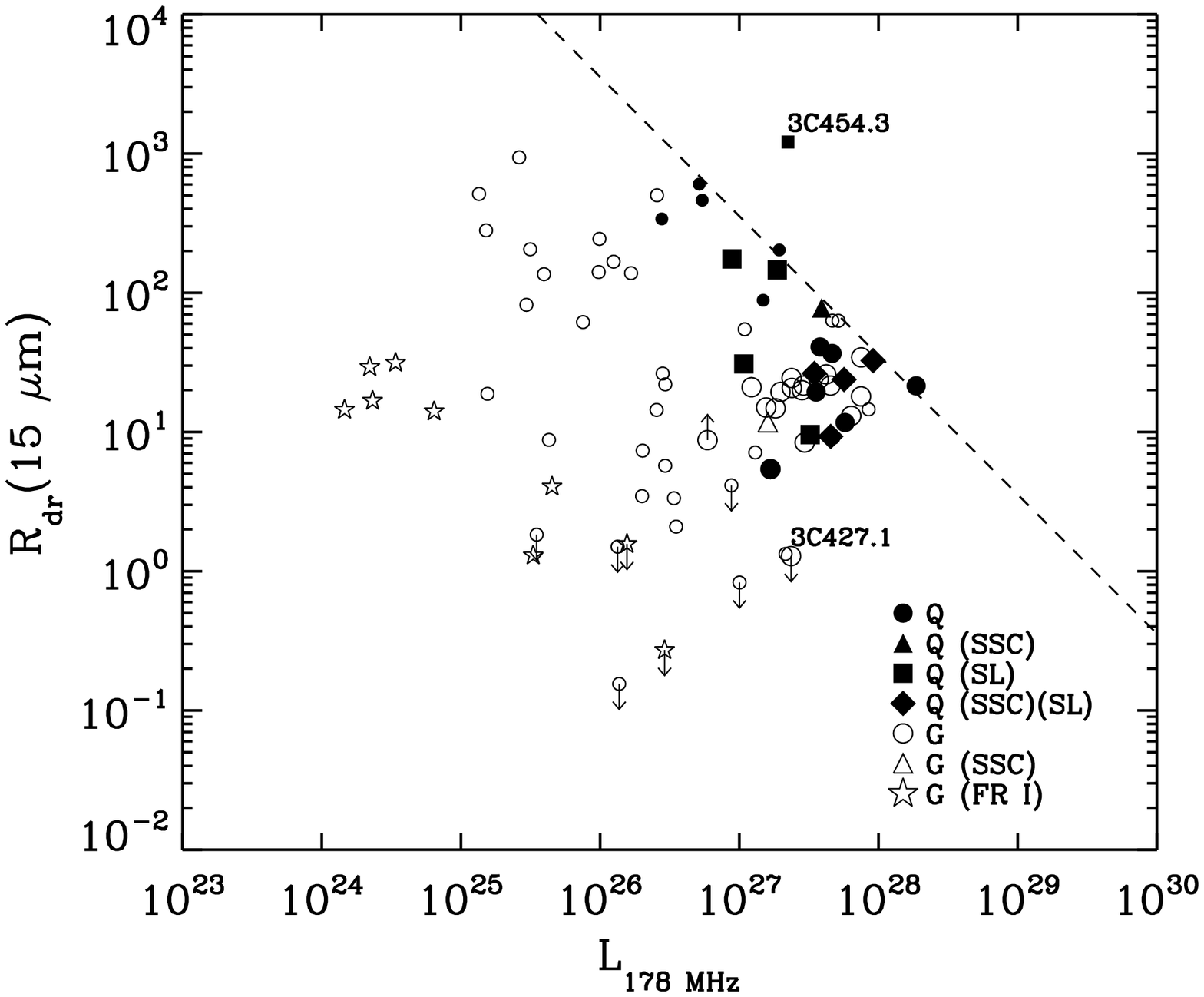,scale=0.4} & 
\psfig{file=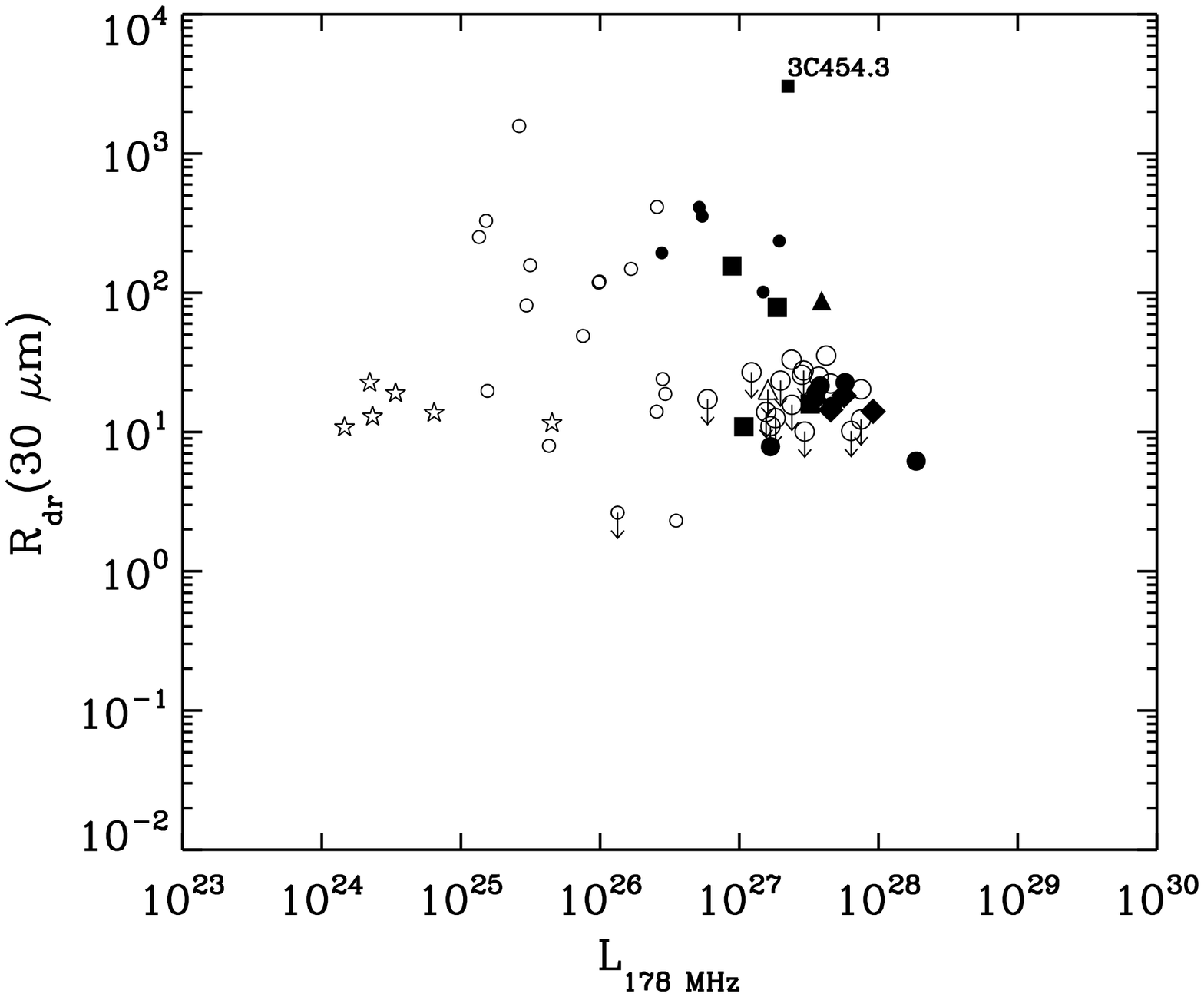,scale=0.4}\\

\end{tabular}
\caption{$R_{\rm dr}(15\micron )$ (left) and $R_{\rm dr}(30\micron )$ (right) vs.\ rest-frame 178\,MHz
luminosity (W\,Hz\mo\,sr\mo) for objects in the primary and supplementary samples. Objects in the primary sample have been corrected for non-thermal emission. Symbols are as in Figure~\ref{L178vsz}. Data for objects from the supplementary sample are plotted using smaller symbols. An upper envelope, defined by setting $R_{\rm dr}(15\,\mu{\rm m})= K\times(10^{24.5}/L_{\rm 178\,MHz})$, where $K=3\times10^{8}/(15 \times 178)$, is indicated by the dashed line. The blazar, 3C\,454.3, lies above this upper envelope.
\label{rdrIRall}}
\end{figure*}

Considering the primary and supplementary samples together, we also note that there appears to be an upper envelope to the 15\microns\ luminosity of $\approx 10^{24.5}$~W\,Hz\mo\,sr\mo\ (see Fig.~\ref{Lirvs178all_fr}). Of the objects in the primary and supplementary samples, only the blazar, 3C\,454.3 has a luminosity greater than $10^{25}$~W\,Hz\mo\,sr\mo . We will discuss this upper envelope further in the next section.

\subsection{Comparative Luminosity of 3CRR Radio Galaxies and Quasars\label{comparative}}

In order to compare the infrared luminosities of quasars and radio galaxies we must first account for the variation of central engine power amongst objects in our sample. To do this, we normalise the infrared emission by the isotropic low-frequency radio luminosity \citep[see e.g.,][]{heckman_etal_92}.  Figure~\ref{rdrIR} shows $R_{\rm dr} = \nu L_{\nu}(\rm IR)/ \nu L_{\nu}(\rm{178\,MHz})$ for our sample at 15 and 30\microns . Figure~\ref{rdrIRall} is the same, but includes objects from the supplementary sample. Table~6 shows the mean and median $\log R_{\rm dr}$ for quasars and galaxies in the primary sample, as well as the ratio between the mean or median $R_{\rm dr}$ for quasars to that for galaxies. For the purposes of this comparison, the BLRG 3C\,22 has been included with the quasars. Only objects with characteristic luminosity and optical depth estimates are included in the calculation of these statistics. Due to the low detection rate for galaxies at 70\microns\ using MIPS, we base our conclusions in this section on the 15\microns\ statistics.

The mean $R_{\rm dr}$ for quasars is $\approx 4$ times higher than for galaxies at 15\microns . Subtracting the non-thermal emission brings this down to $\approx 2$. The fitted optical depths allow us to go further and ask whether obscuration by dust  can account for this residual anisotropy of the thermal emission between quasars and galaxies. Using the fitted screen optical depths to correct for dust extinction, we find that the thermal emission from quasars and galaxies is on average equal at 15\microns . The mixed optical depths would imply intrinsically brighter galaxies than quasars. Since there is no evidence that galaxies are brighter than quasars in the far infrared, where the torus is expected to be optically thin, we conclude that the actual distribution of warm and cool dust is better approximated by the screen model. Therefore, around half of the difference in mid-infrared brightness between quasars and galaxies can be attributed to Doppler-boosted synchrotron emission present in some quasars, but not in galaxies. The other half is consistent with dust absorption in the galaxies but not in the quasars. 

\cite{haas_etal_04} find no difference in the distribution of $R_{\rm dr}$ for quasars and galaxies for their sample of 3CR radio galaxies and quasars; however, we believe that this is not inconsistent with the results presented here for the following reasons. \cite{haas_etal_04} exclude from their sub-sample those objects which do not exhibit a prominent thermal bump, thereby excluding objects with significant non-thermal contributions. Also, their mean $R_{\rm dr}$ is computed at rest-frame 70\microns\ where the effect of optical depth is much less than at 15\microns .

Figure~\ref{tauvsr} plots the screen, $\tau^{\rm scr}_{9.7 \mu {\rm m} }$, and mixed, $\tau^{\rm mix}_{9.7 \mu {\rm m} }$, optical depths against the core dominance parameter, $R = F_{\rm core}/F_{\rm extended}$ (see Table~7). Our dust model (see \S\,\ref{sed_section}) does not, a priori, impose any asymmetry in the distribution of dust surrounding the nucleus.  From Figure~\ref{tauvsr} there is evidence of an anti-correlation between optical depth and core dominance, from which we can infer an equatorial distribution of dust. The median screen (mixed) optical depth for objects with ${\rm R} > 10^{-2}$ is 0.4 (1.0) and with ${\rm R} \leq 10^{-2}$ is 1.1 (3.0). This is in line with the predictions of orientation-based unification of 3CRR quasars and radio galaxies.

At lower radio luminosities than those of the primary sample, we can see from Figure~\ref{rdrIRall} that there are quasars and galaxies of similar radio luminosity but separated by over three orders of magnitude in $R_{\rm dr}$, suggesting that differences may exist between these objects which cannot be explained by orientation alone \citep[see e.g.,][]{shi_etal_05, ogle_etal_06}. The lack of similar objects at the greater radio luminosities spanned by the primary sample may be due to selection effects, since the most radio-luminous objects in the flux-limited 3C sample are farther away. There is, however, one notable exception: the galaxy 3C\,427.1, which was not detected using MIPS or IRS. Since the MIPS 24\microns\ measurement for this object is at $\approx 15$\microns\ in the rest-frame, we obtain $\log R_{\rm dr}(15\mu{\rm m}) \ltsim 0$ (see Fig.~\ref{rdrIRall}, left), which is over two orders of magnitude below quasars of similar radio luminosity. This galaxy may therefore be different to the quasars and other galaxies in the primary sample, perhaps with an intrinsically weak source of optical/UV compared to other powerful radio sources.

We note that there is an apparent upper envelope in $R_{\rm dr}(15\,\mu{\rm m})$, due to the previously noted upper envelope of $L_{\rm{15\mu m}} \ltsim 10^{24.5}$~W\,Hz\mo\,sr\mo . This is shown by the dashed line in Figure~\ref{rdrIRall} (left), which is defined as $R_{\rm dr}(15\,\mu{\rm m})= K\times(10^{24.5}/L_{\rm 178\,MHz})$, where $K=3\times10^{8}/(15 \times 178)$. The blazar 3C\,454.3 lies above this envelope. \cite{haas_etal_04} point out that a decline of $R_{\rm dr}$ with radio luminosity is expected from the ``receding-torus'' model of \cite{lawrence_91}, yet find no evidence for such a decline in their sample of 3CR objects and so suggest a modification to the original proposal in which the torus scale height remains constant with luminosity. Taking all quasars in the primary and supplementary samples (except the blazar 3C\,454.3) and computing the mean (non-thermal corrected) $\log R_{\rm dr}(15\,\mu{\rm m})$ in two bins defined by their median radio luminosity ($\approx 10^{27.5}$~W\,Hz\mo\,sr\mo ), we find that $R_{\rm dr}$ for quasars is around $4^{+3}_{-2}$ times greater in the lower radio luminosity bin. For galaxies, $R_{\rm dr}$ is approximately equal in both bins. The observed decline in $R_{\rm dr}$ for quasars is of low significance; however if the observed upper envelope in 15\microns\ luminosity is shown to apply at higher radio-lobe luminosities than those of the {\em Spitzer\/} sample, then this would provide further support for the receding torus model as originally formulated by \cite{lawrence_91}.

\begin{figure*}
\centering
\begin{tabular}{cc}
\tablewidth{0pt}

\psfig{file=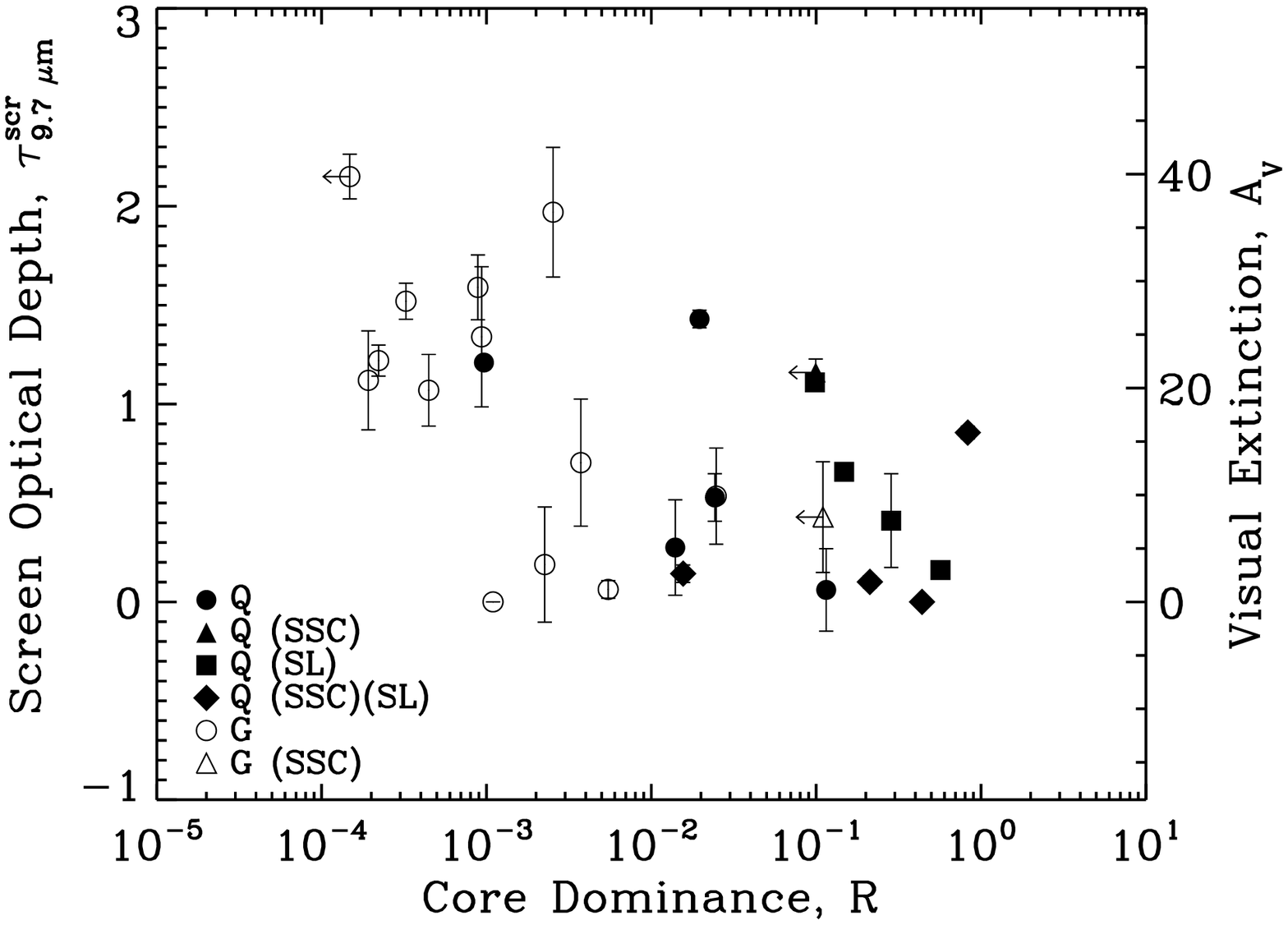,scale=0.4} & 
\psfig{file=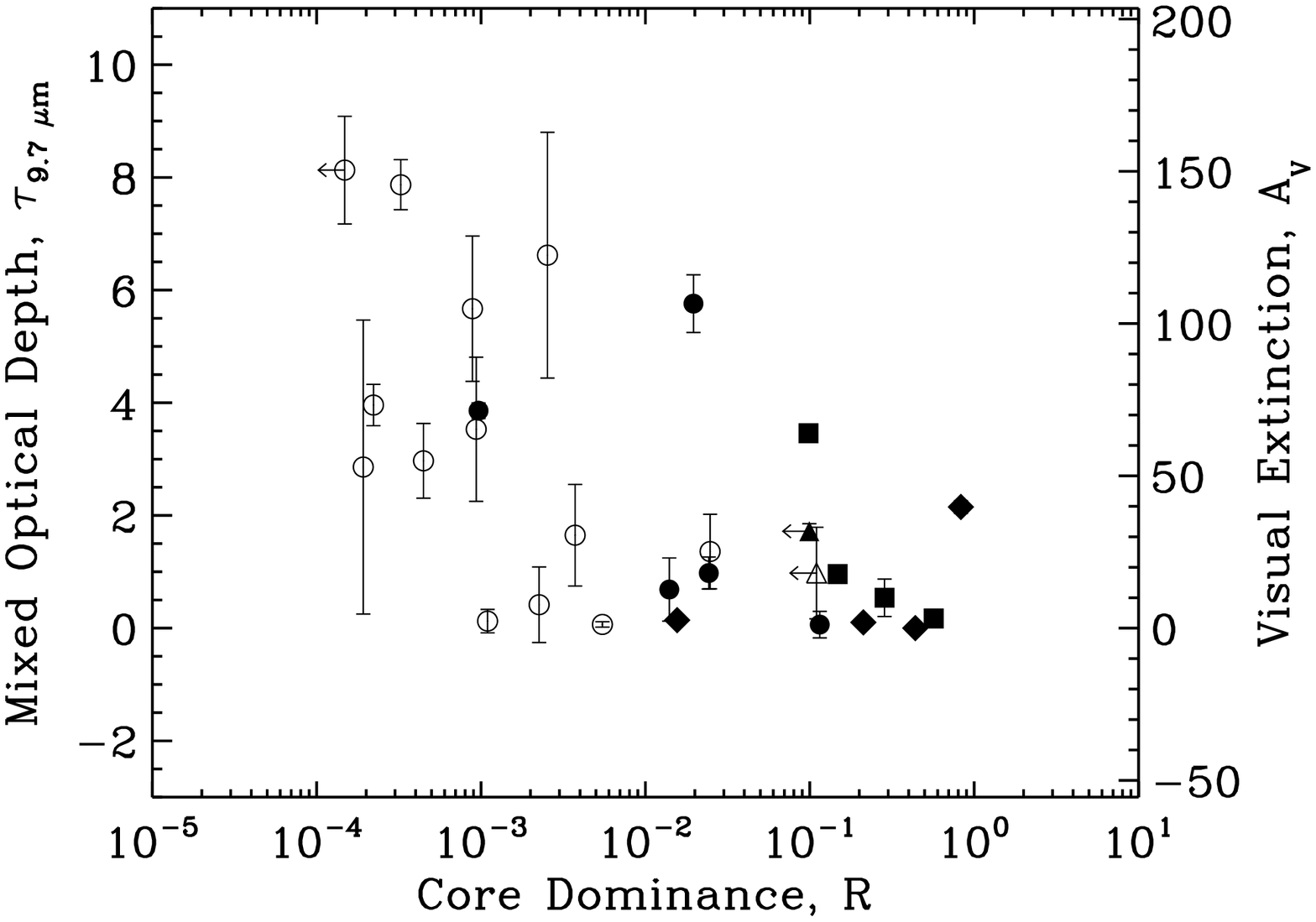,scale=0.4}\\

\end{tabular}
\caption{The 9.7\microns\ screen (left) and mixed (right) optical depths estimated from from the fits to the data versus the core dominance parameter, $R = F_{\rm core}/F_{\rm extended}$. The right-hand $y$-axis on each plot indicates the visual extinction, based on $A_{\rm V}/\tau_{\rm 9.7\mu m} =  18.5$, typical for the local diffuse ISM \citep{draine_03}. Symbols are as in Figure~\ref{L178vsz}. The error bars indicate the formal uncertainties in the fits. From the apparent anti-correlation of optical depth with core dominance, we infer an equatorial distribution of cool dust consistent with the ``dusty torus'' of orientation-based unification schemes.
\label{tauvsr}}
\end{figure*}


\section{CONCLUSIONS}

We have observed a sample of extremely powerful radio sources with the IRS and MIPS instruments aboard {\em Spitzer}. The sample was selected at low frequency, and is orientation independent. Our aim was (i) to determine the mid-infrared properties of an orientation-unbiased sample of the most powerful radio sources, (ii) to assess the relative contribution of the AGN, star-formation and Doppler-boosted jet synchrotron and (iii) to study the relative importance of physical differences and orientation effects in powerful radio sources.

Our main results can be summarised as follows.

- We have detected extremely luminous emission in the mid-infrared for our sample of radio galaxies and quasars. All but one object observed in our sample with MIPS were detected at 24\microns , with specific luminosities, $L_{\nu} > 10^{22.4}$\,W\,Hz\mo\,sr\mo .

- The MIPS and, in particular, IRS measurements provide powerful constraints on the spectral energy distributions of the sources in our sample. They have allowed us to fit continuum synchrotron and dust components representing the dominant emission processes and thereby estimate the non-thermal contribution to the mid-infrared luminosity. We find that non-thermal processes can contribute a significant proportion (up to 90\%) of the measured infrared emission in some quasars.

- Warm, optically thin dust components with temperatures in the approximate range 100--300 K,  obscured by cooler, optically thick dust, provide very good fits to the IRS spectra (when combined with appropriate synchrotron components from the lobes and jet). The optical depths resulting from the fits are typically $\tau^{\rm scr}_{9.7 \mu {\rm m} } \lesssim 2$ ($A_{\rm V} \lesssim 40$) when using a dust screen model and $\tau^{\rm mix}_{9.7 \mu {\rm m} } \lesssim 8$ ($A_{\rm V} \lesssim 150$) when considering a mixed extinction scenario.

- For the most powerful radio sources ($L_{\rm 178\,MHz} \gtsim 10^{26.5}$\,W\,Hz\mo\,sr\mo ), the mid-infrared luminosity appears to have an upper envelope of around $L_{\rm{15\mu m}}=10^{24.5}$\,W\,Hz\mo\,sr\mo . If this envelope applies at even higher radio-lobe luminosities than those of the {\em Spitzer\/} sample, this may provide support for the ``receding torus'' model of \cite{lawrence_91}.

- Normalising by the isotropic radio luminosity (to account for variations in central engine power), the quasars are $\approx 4$~times brighter than the galaxies in our sample at 15\microns . Half of this difference can be attributed to the presence of non-thermal emission in the quasars but not in the galaxies. The other half is consistent with absorption by a dust screen in the radio galaxies but not the quasars.

- We find an anti-correlation between the estimated optical depth and core dominance. The median screen optical depth for objects with core-dominance factor ${\rm R} > 10^{-2}$ is
$\approx 0.4$; for objects with ${\rm R} \leq 10^{-2}$, it is $\approx 1.1$. This is consistent with the picture of the MIR-emitting dust in FR\,II galaxies being observed through an equatorial distribution of cooler dust, while for the quasars, the MIR-emitting dust is seen directly as the torus is viewed face-on.

Previous investigators have shown that quasars are brighter than galaxies in the mid-infrared, in agreement with the results presented above. It is also clear from previous work that beamed synchrotron emission and dust extinction modulate the mid-infrared emission of quasars and radio galaxies to some degree. In the work presented here, however, we have quantified these effects for an orientation-unbiased sample of powerful radio sources and shown that once they are taken into account, quasars and radio galaxies are on average equally bright in the mid-infrared.

\acknowledgments

We thank the anonymous referee for comments and suggestions which have improved the overall presentation of this paper. We also thank Mark Birkinshaw, Bernhard Brandl, Bernhard Schulz, Nick Seymour, Michael Werner and Diana Worrall for helpful comments. This work is based on observations made with the Spitzer Space Telescope, which is operated by the Jet Propulsion Laboratory, California Institute of Technology under a contract with NASA. This research has made use of the NASA/IPAC Extragalactic Database (NED) which is operated by the Jet Propulsion Laboratory, California Institute of Technology, under contract with NASA. 

\appendix
\clearpage
\section{CHARACTERISTIC LUMINOSITIES}

\begin{table*}
\tablenum{7}
  \vbox to220mm{\vfil Landscape table to go here
  \caption{}
  \vfil}
  \label{char_lums}
\end{table*}

\clearpage

\section{FITS TO THE SPECTRAL ENERGY DISTRIBUTIONS}
\begin{figure*}[!ht]
\figurenum{13}
\centering
\begin{tabular}{cc}
\tablewidth{0pt}
\psfig{file=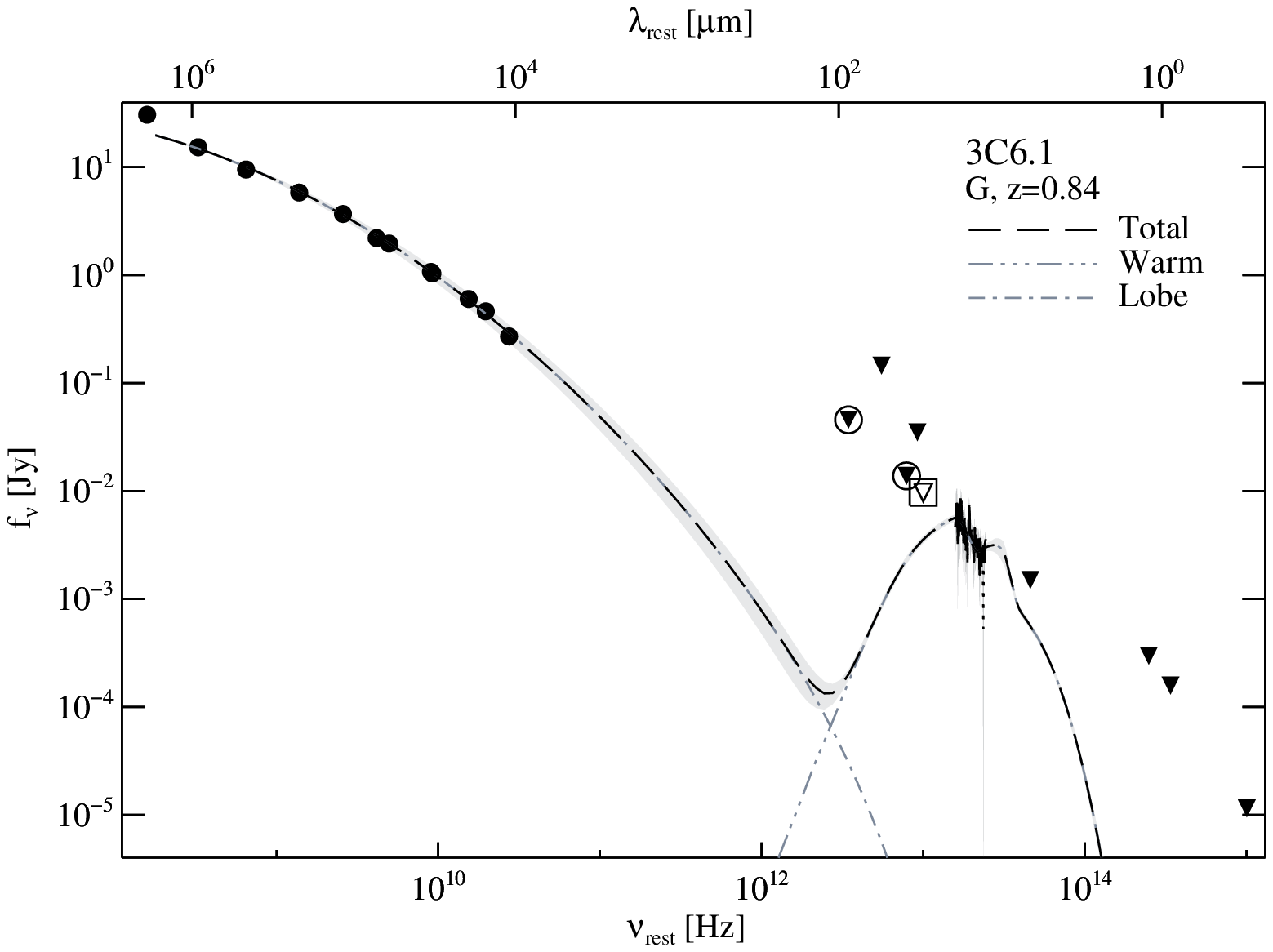,angle=0,width=8cm,height=6.0cm,clip=} & 
\psfig{file=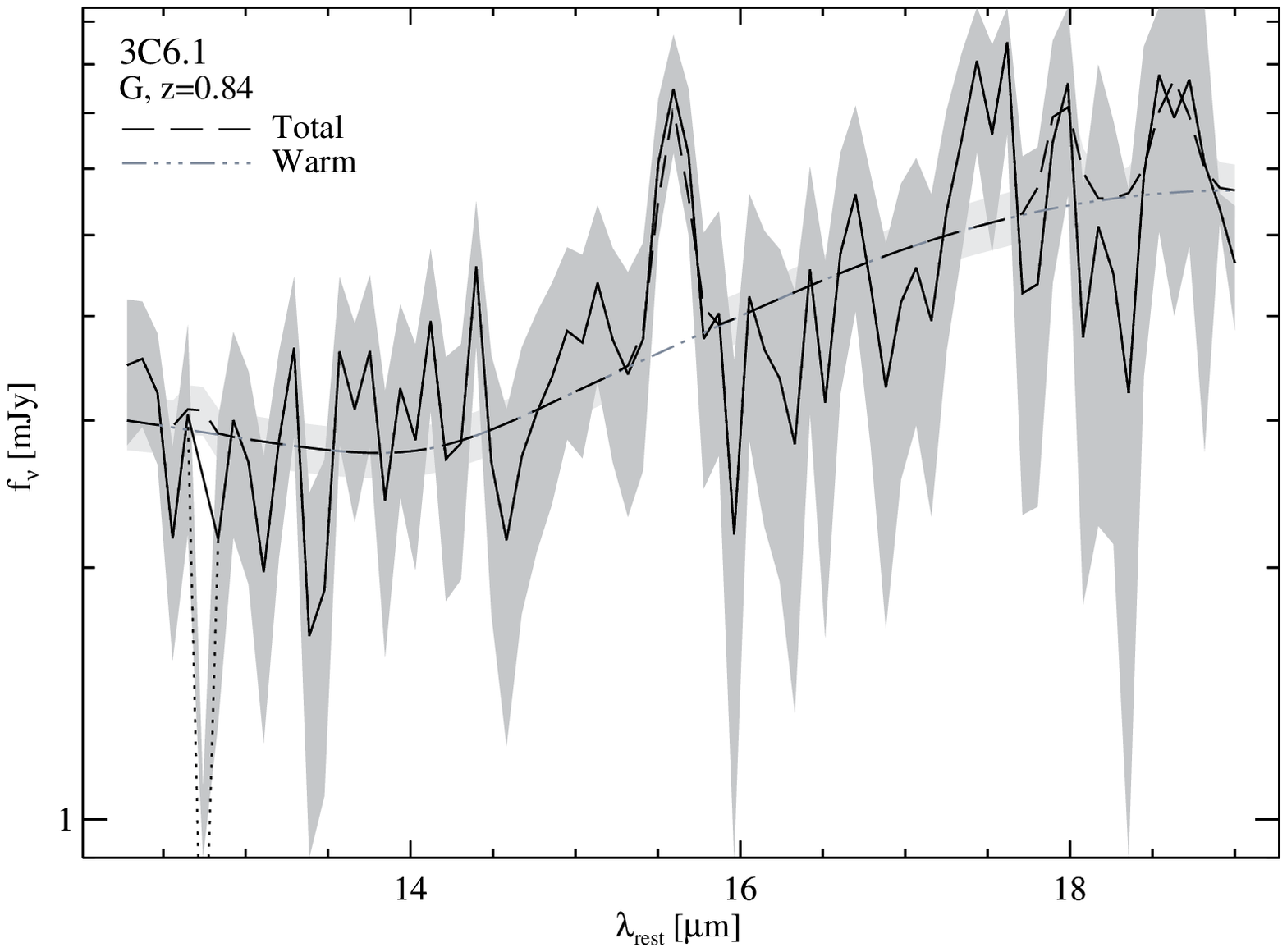,angle=0,width=8cm,height=5.5cm,clip=}\\

\psfig{file=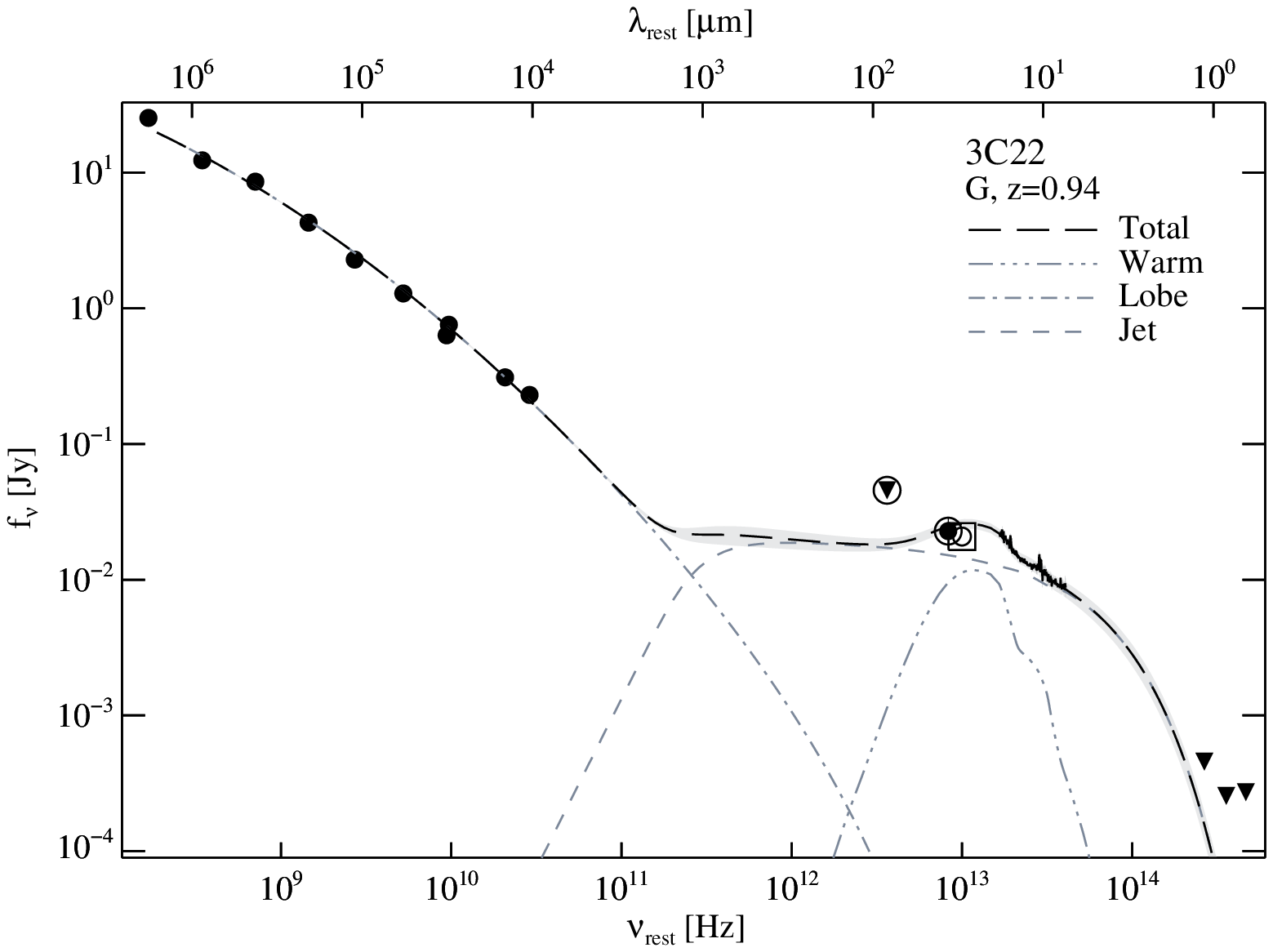,angle=0,width=8cm,height=6.0cm,clip=} & 
\psfig{file=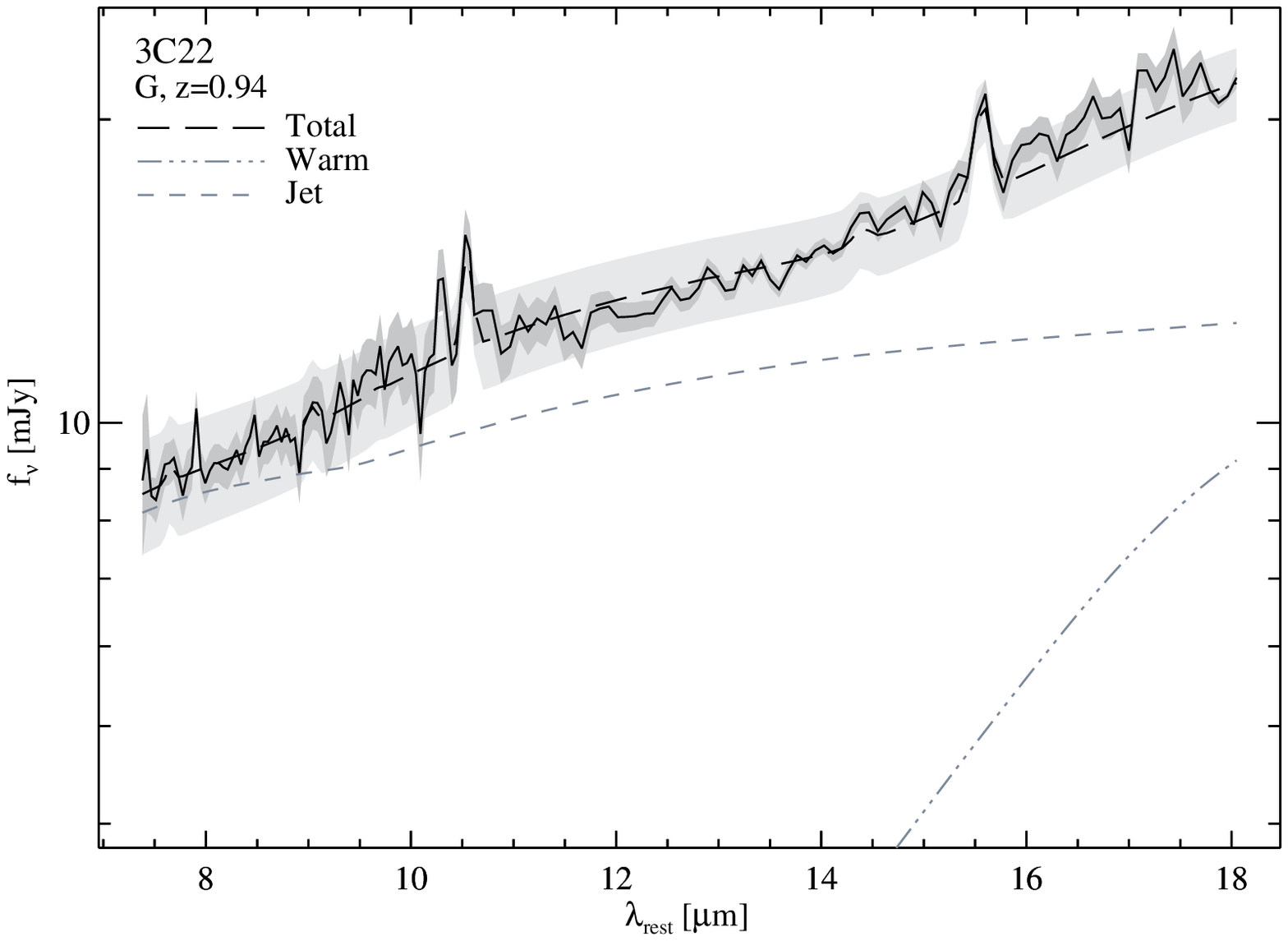,angle=0,width=8cm,height=5.5cm,clip=}\\

\end{tabular}
\caption{Fits to the IRS spectra and photometric data (left). The fit to the IRS data (where available) is also shown in detail (right). Filled circles and triangles represent detections and upper limits respectively. Symbols within circles represent the MIPS measurements. For clarity, the MIPS 24\microns\ measurement is excluded from the plot unless no IRS spectrum was obtained. Data from the literature which appear discrepant with {\em Spitzer\/} data are plotted as open symbols and excluded from the fits. The characteristic 30\microns\ flux density is plotted as an open circle within a square. The dark-shaded error contours represent the uncertainty in the IRS spectrum obtained by differencing the two nod positions. The light-shaded contours represent the formal $1\sigma$ uncertainty in the fitted models (note that the fit uncertainties are greatest in regions where the dust and synchrotron components overlap). IRS data with signal-to-noise ratio $<1$ are connected with dotted lines and excluded from the fit.  
\label{fits}}
\end{figure*}

\begin{figure*}
\figurenum{13}
\centering
\begin{tabular}{cc}
\tablewidth{0pt}

\psfig{file=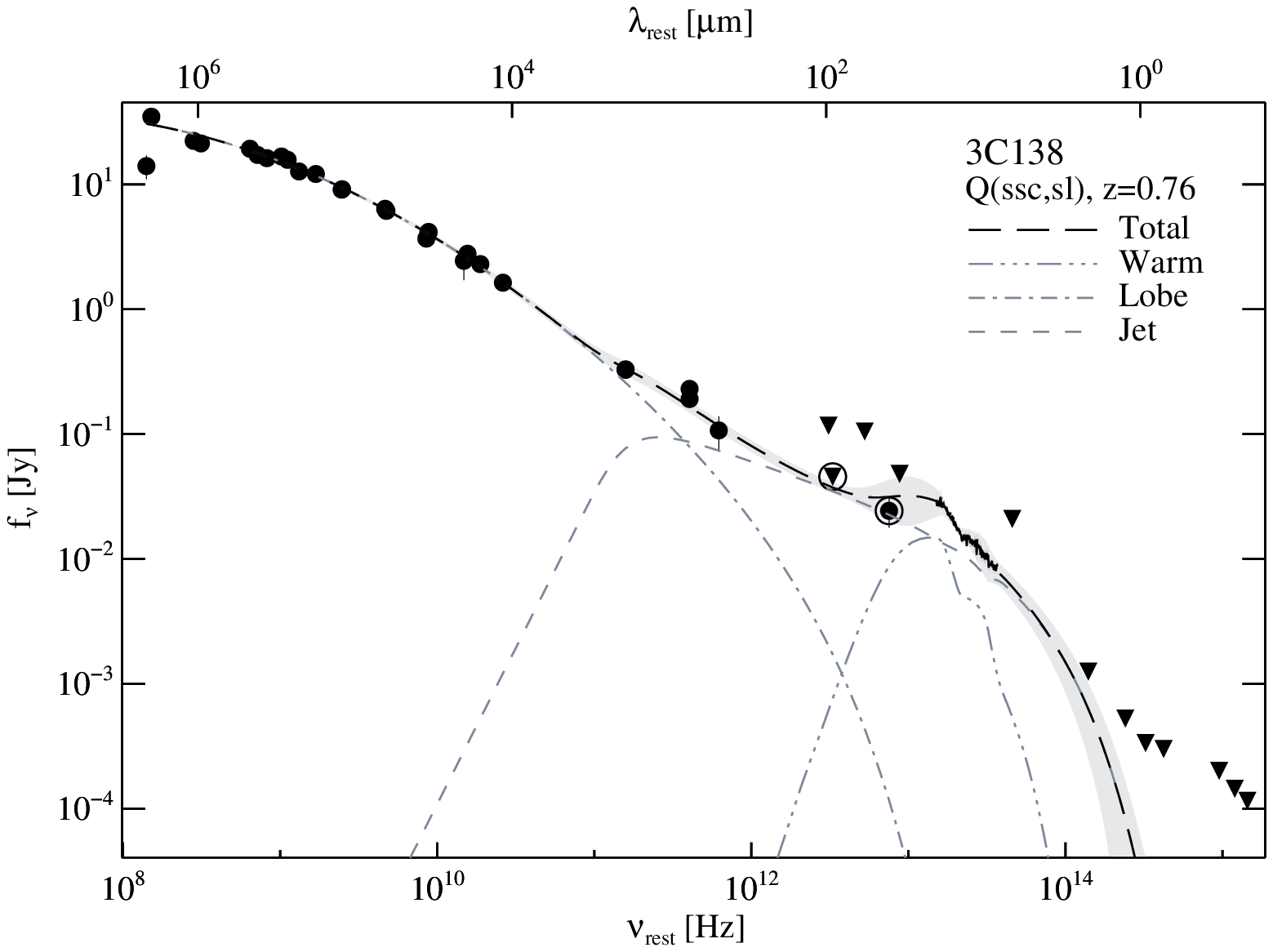,angle=0,width=8cm,height=6.0cm,clip=} & 
\psfig{file=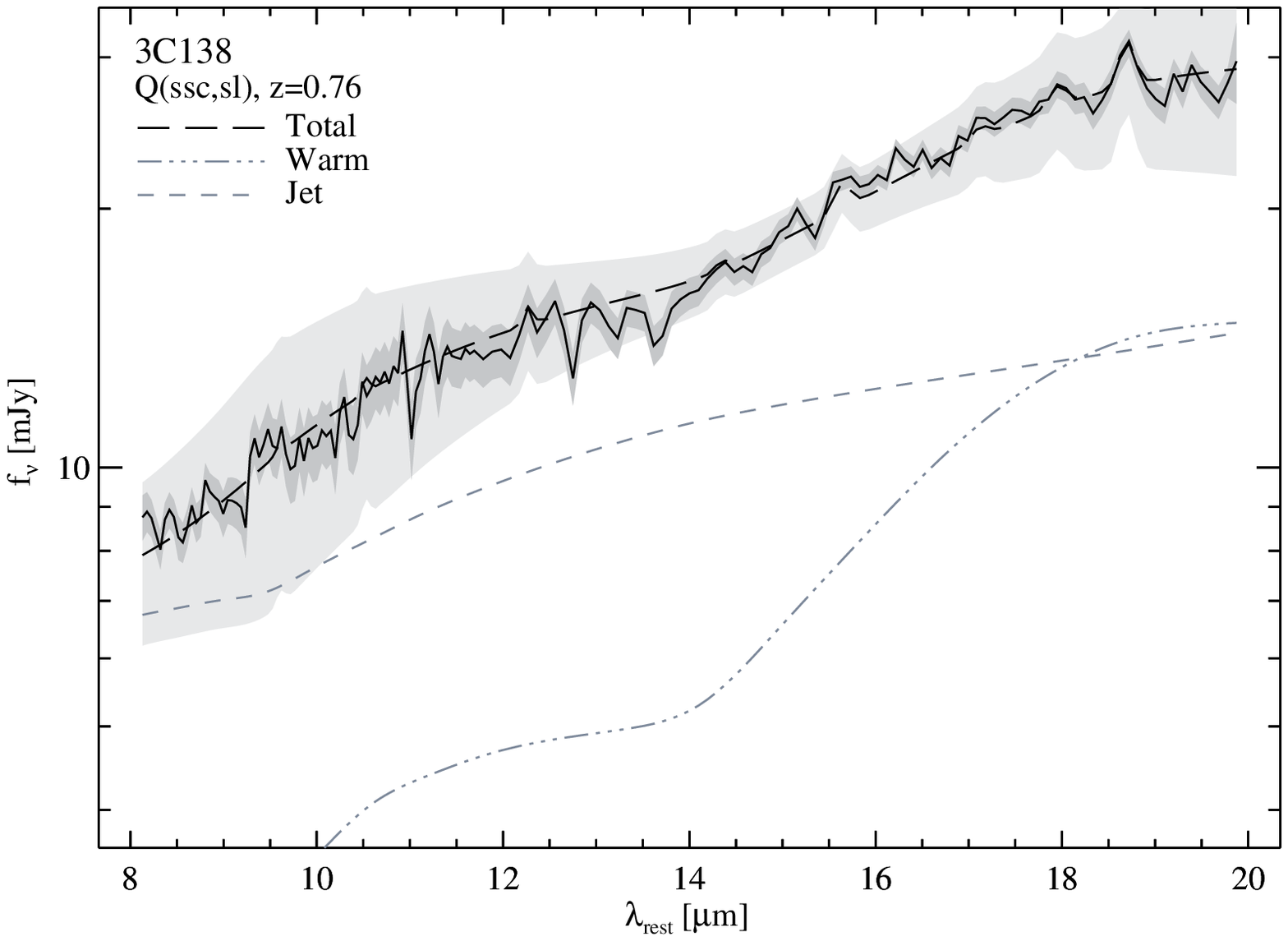,angle=0,width=8cm,height=5.5cm,clip=}\\

\psfig{file=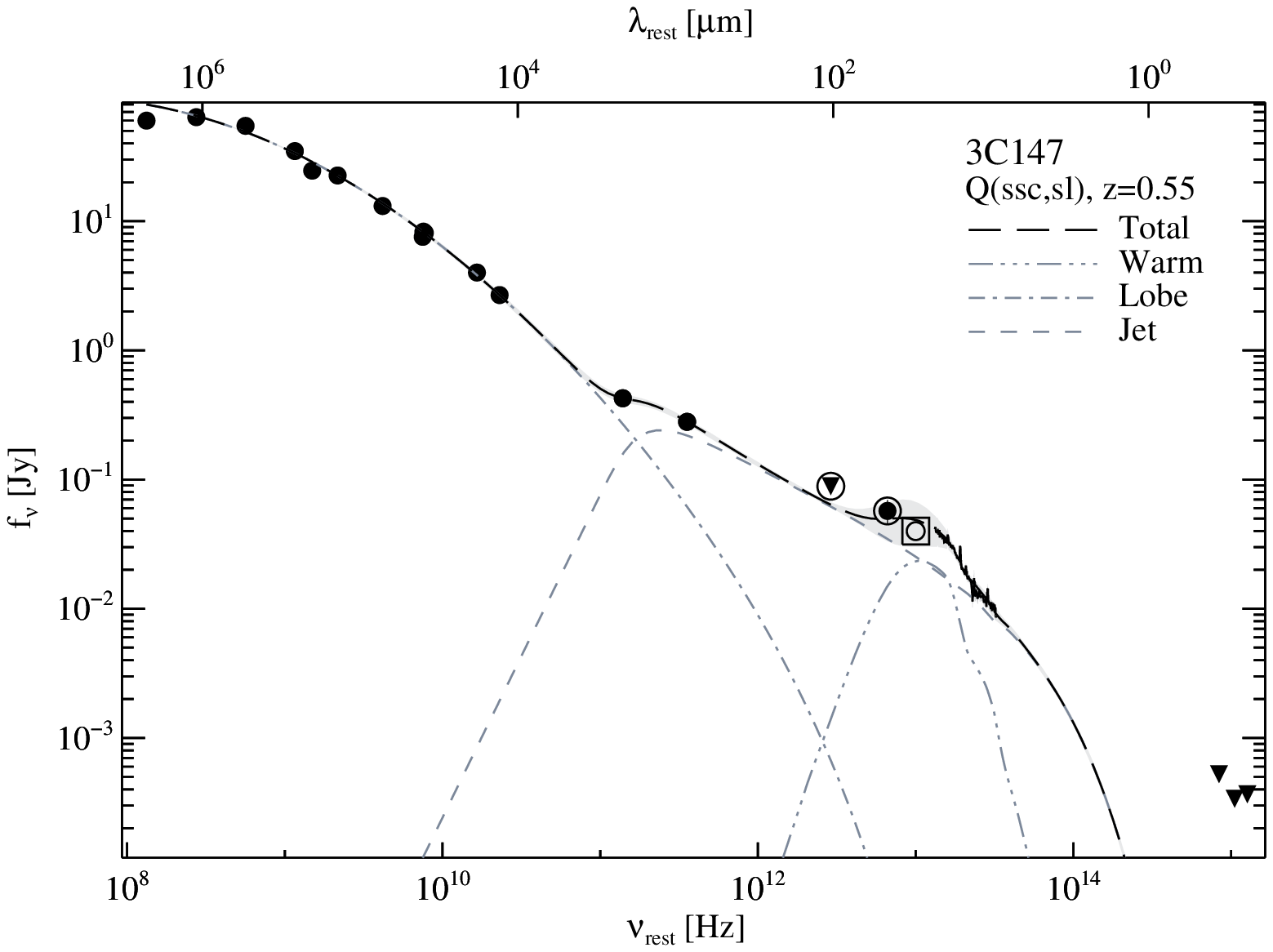,angle=0,width=8cm,height=6.0cm,clip=} & 
\psfig{file=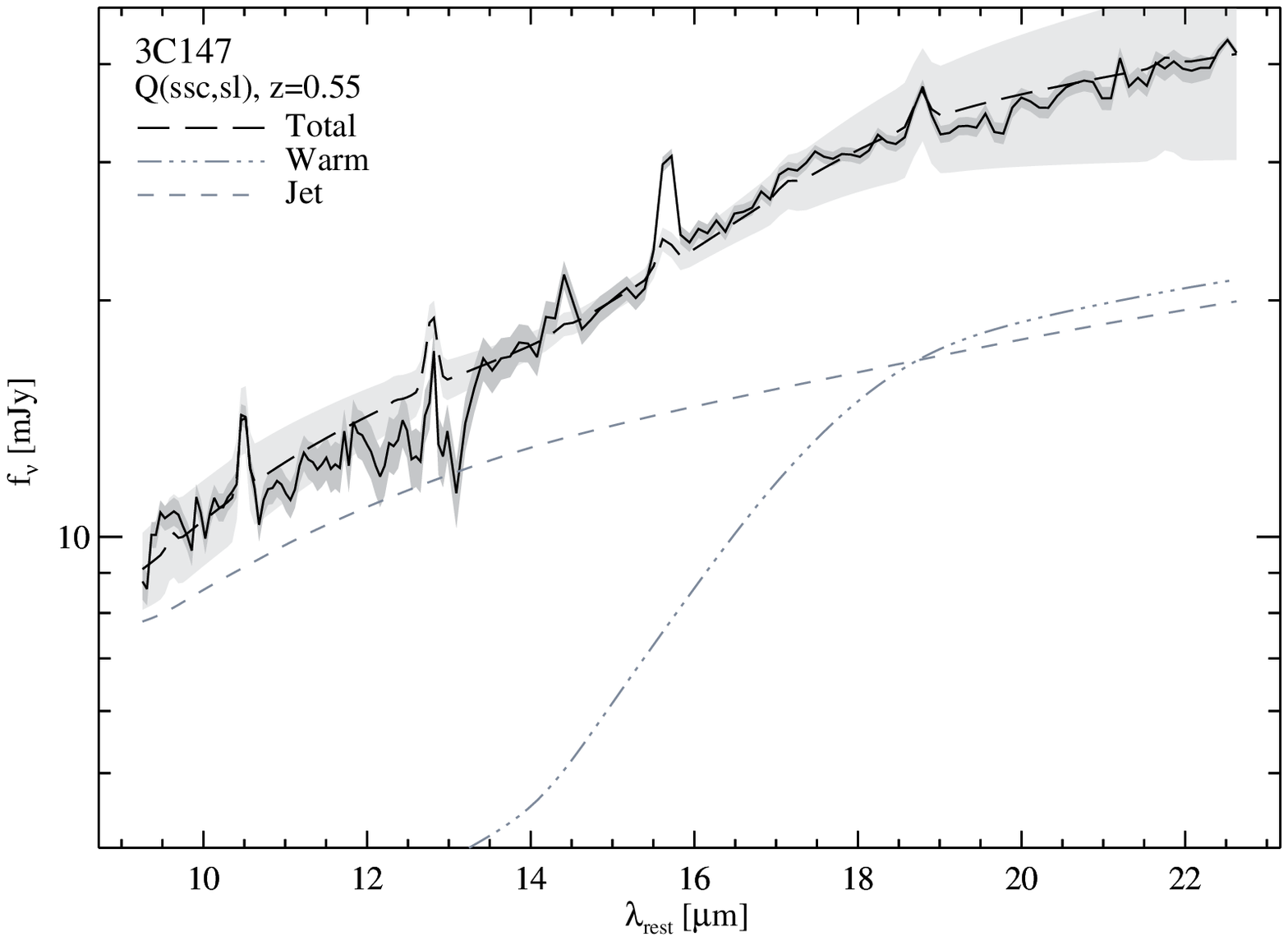,angle=0,width=8cm,height=5.5cm,clip=}\\

\psfig{file=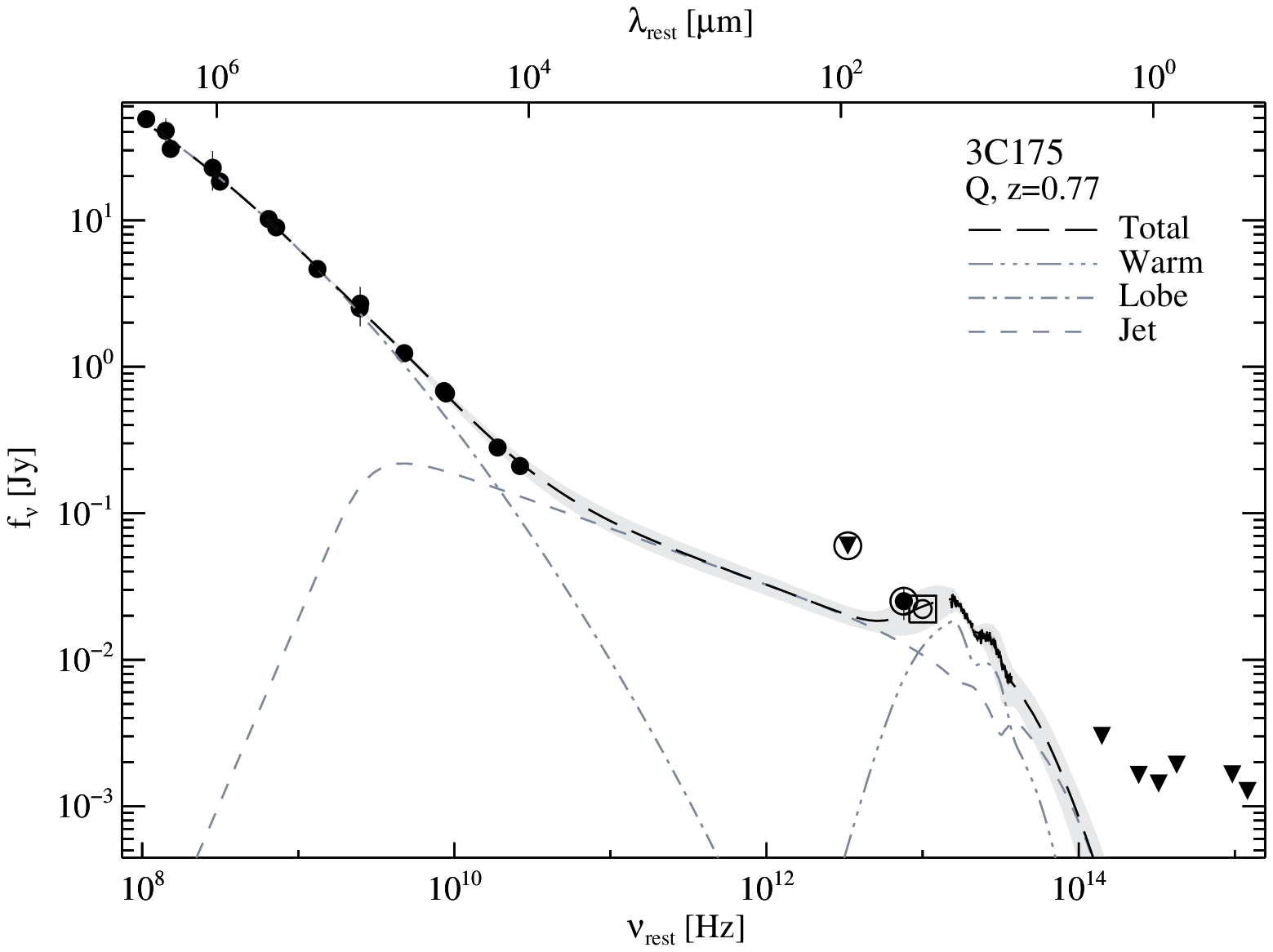,angle=0,width=8cm,height=6.0cm,clip=} & 
\psfig{file=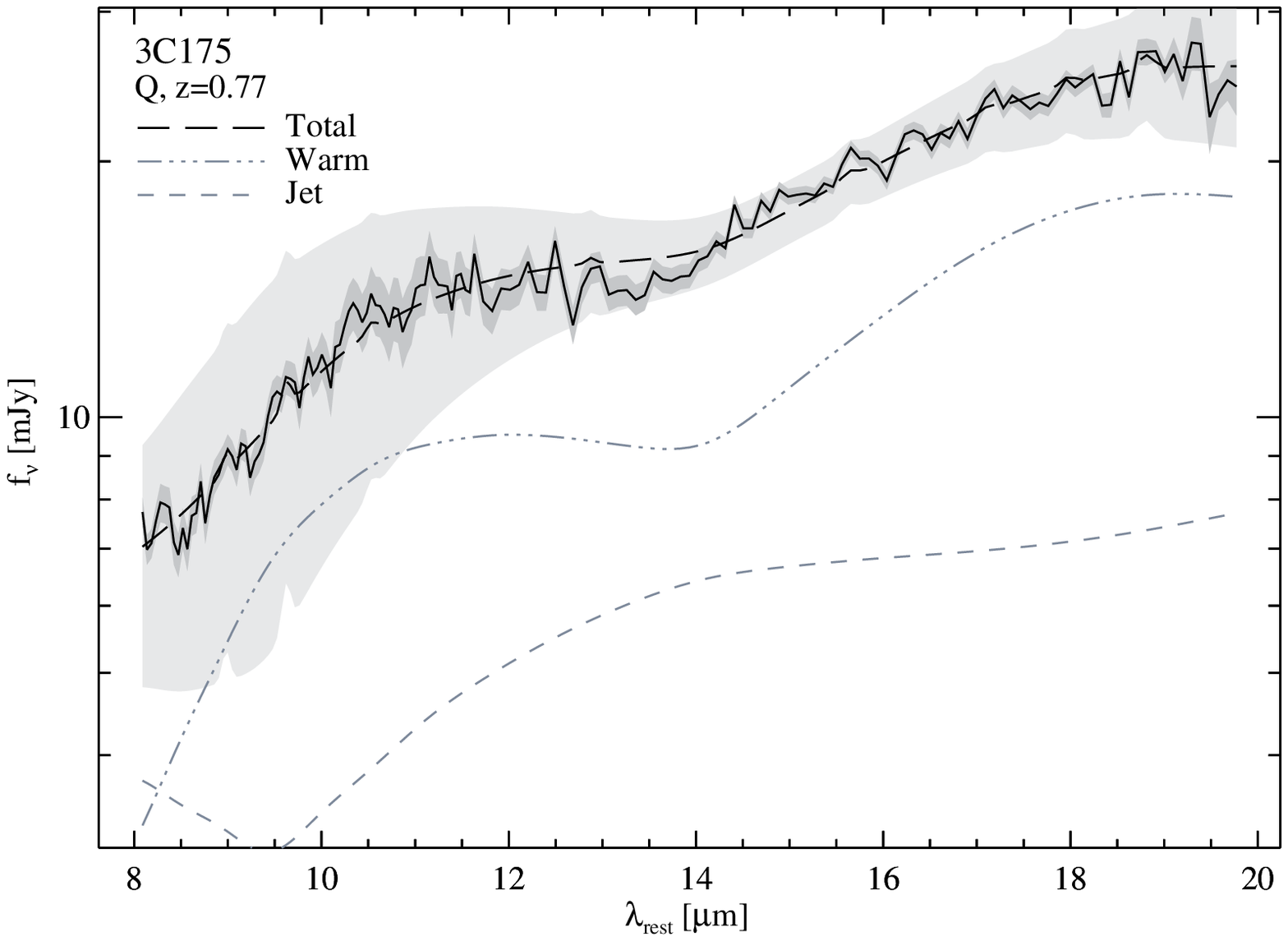,angle=0,width=8cm,height=5.5cm,clip=}\\

\end{tabular}
\caption{{\em Continued}}
\end{figure*}

\begin{figure*}
\figurenum{13}
\centering
\begin{tabular}{cc}
\tablewidth{0pt}

\psfig{file=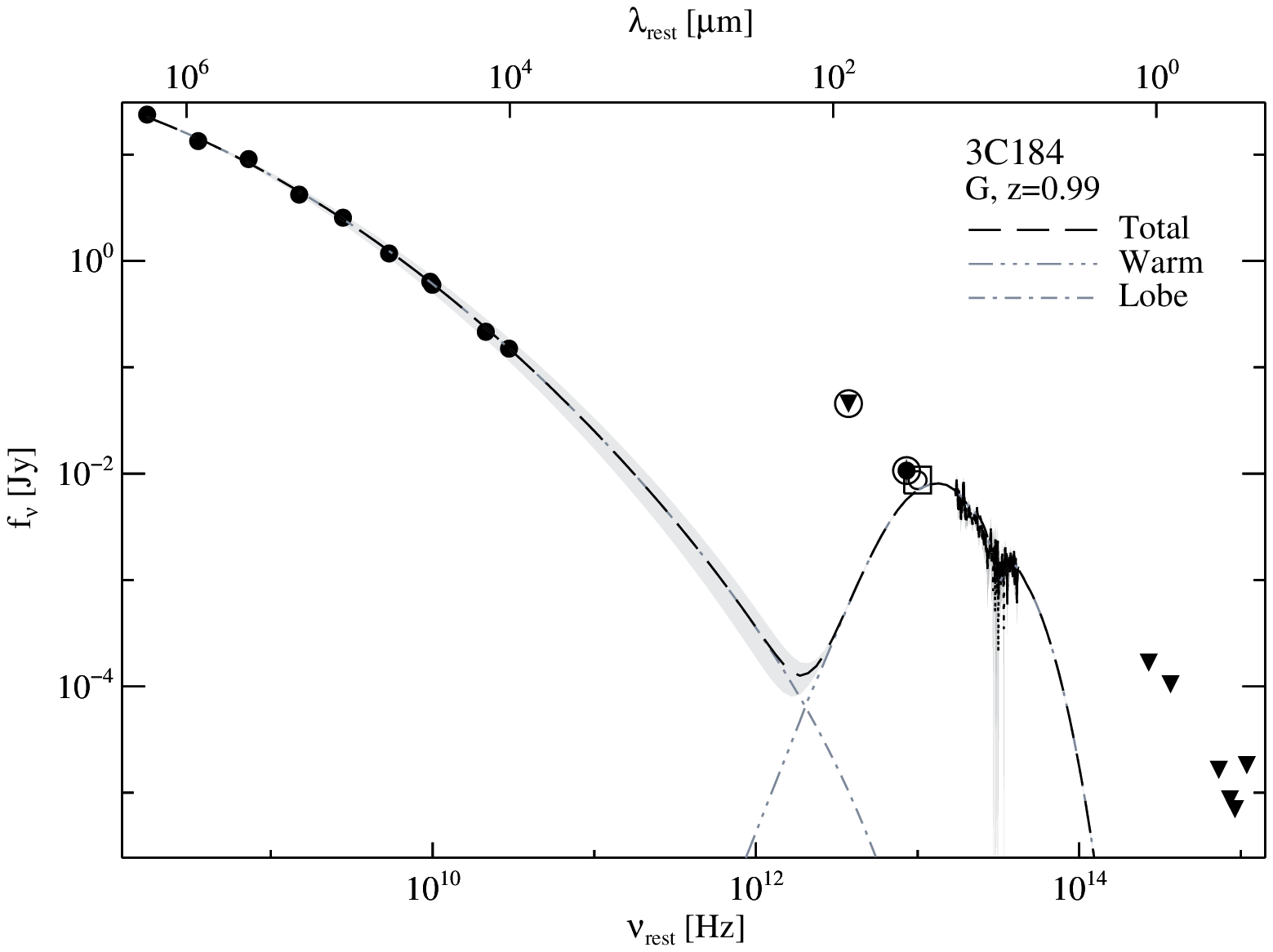,angle=0,width=8cm,height=6.0cm,clip=} & 
\psfig{file=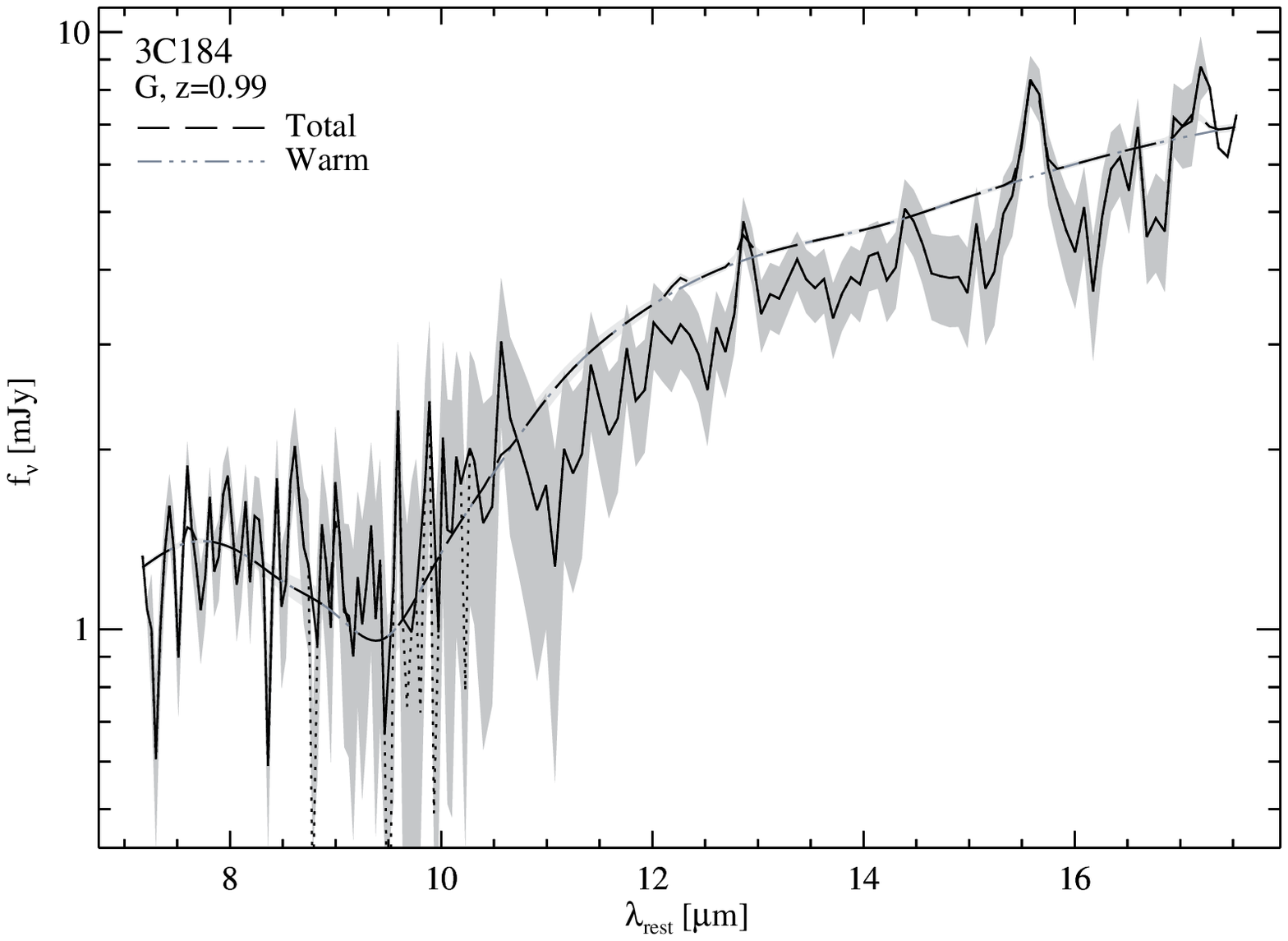,angle=0,width=8cm,height=5.5cm,clip=}\\

\psfig{file=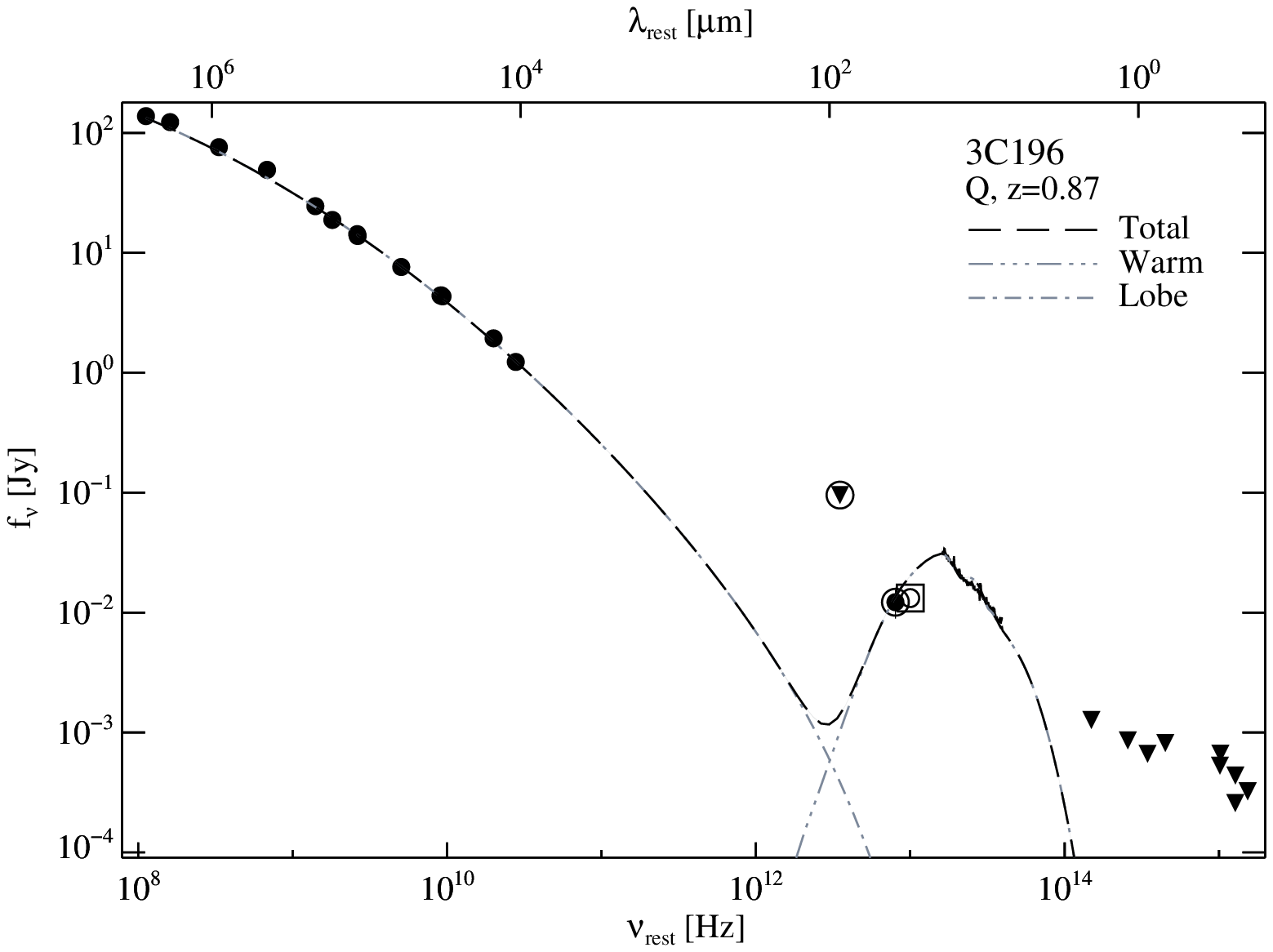,angle=0,width=8cm,height=6.0cm,clip=} & 
\psfig{file=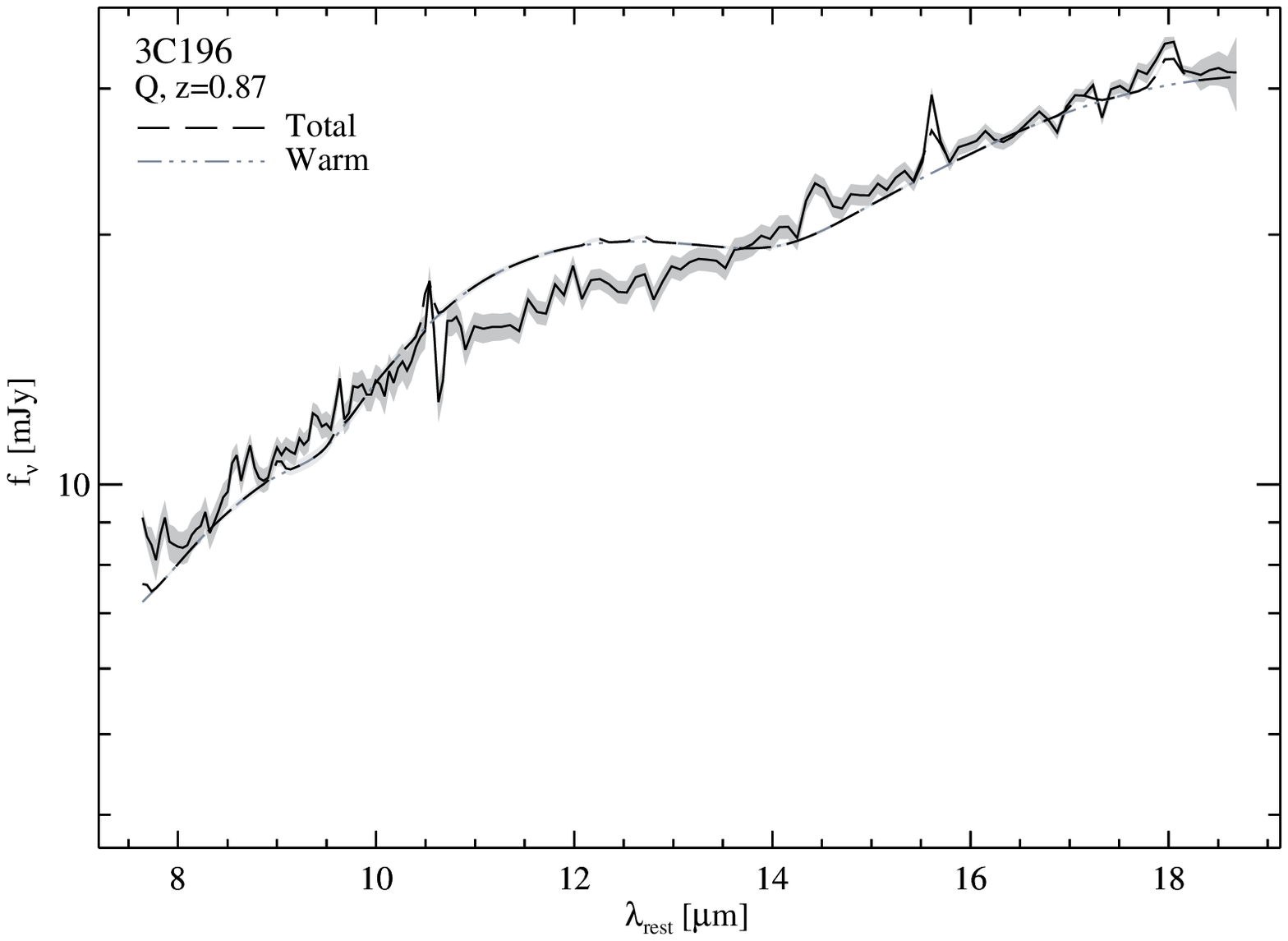,angle=0,width=8cm,height=5.5cm,clip=}\\

\psfig{file=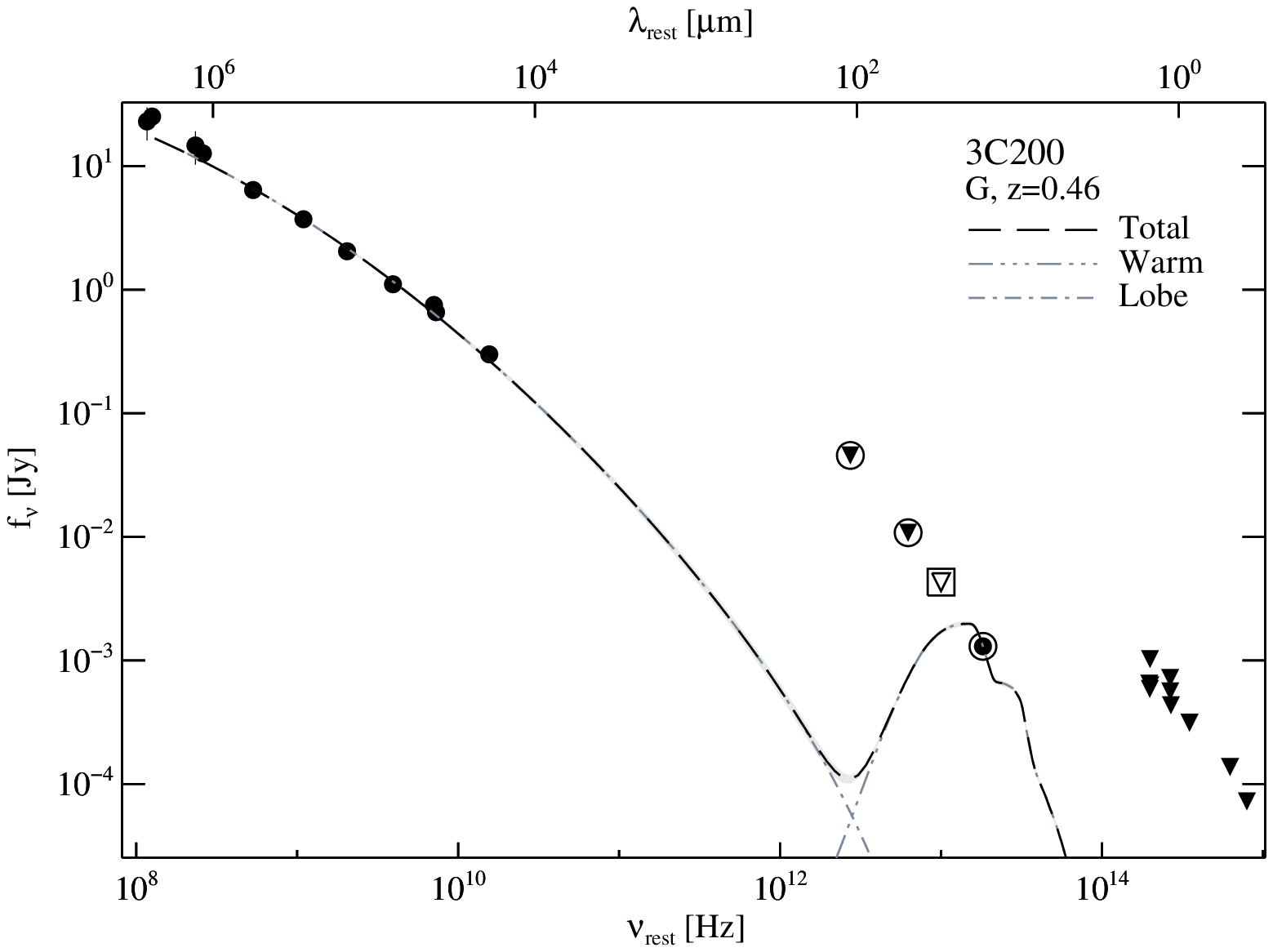,angle=0,width=8cm,height=6.0cm,clip=} & 
\psfig{file=f13p.eps,angle=0,width=8cm,height=5.5cm,clip=}\\

\end{tabular}
\caption{{\em Continued}}
\end{figure*}

\begin{figure*}
\figurenum{13}
\centering
\begin{tabular}{cc}
\tablewidth{0pt}

\psfig{file=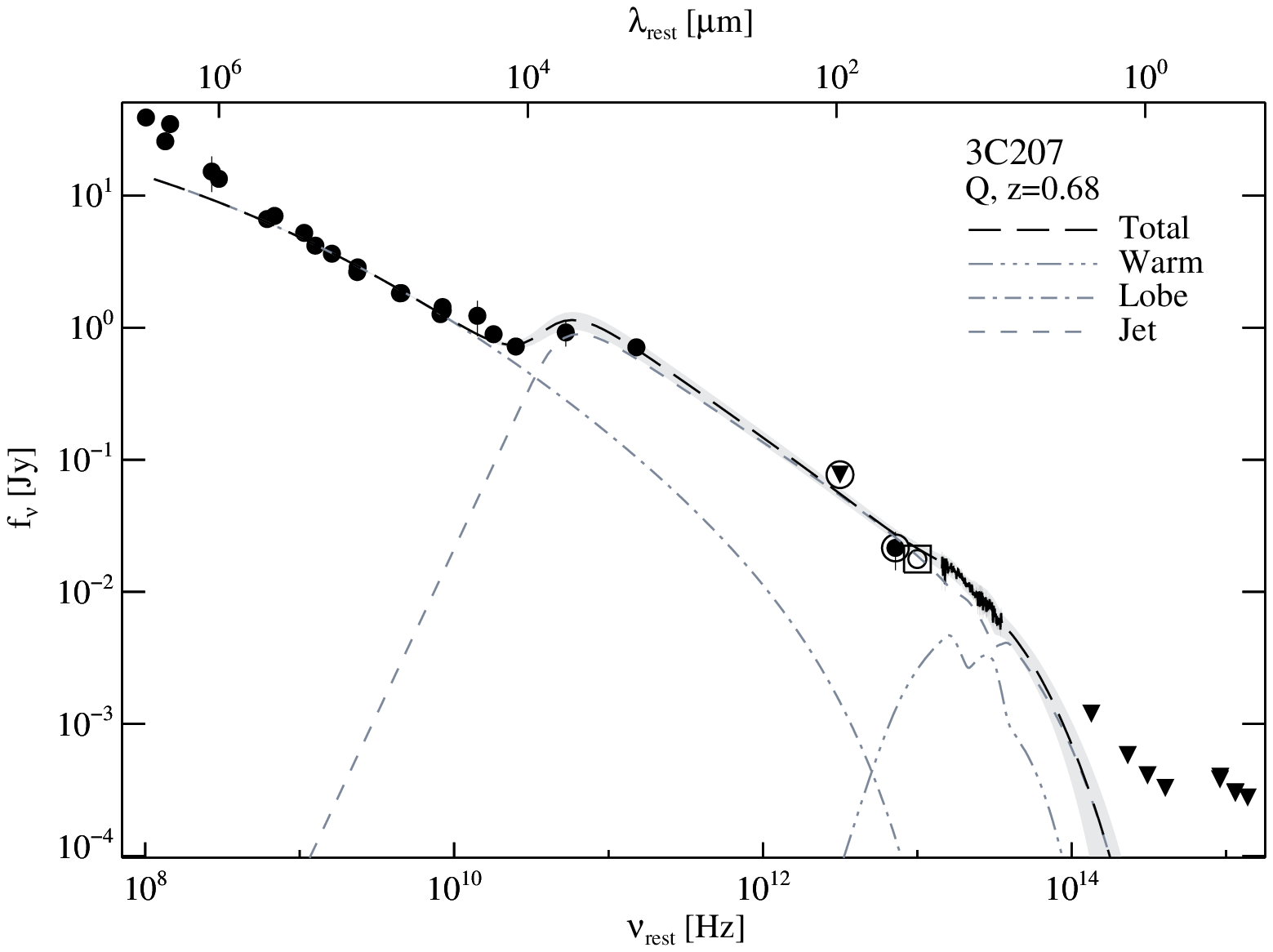,angle=0,width=8cm,height=6.0cm,clip=} & 
\psfig{file=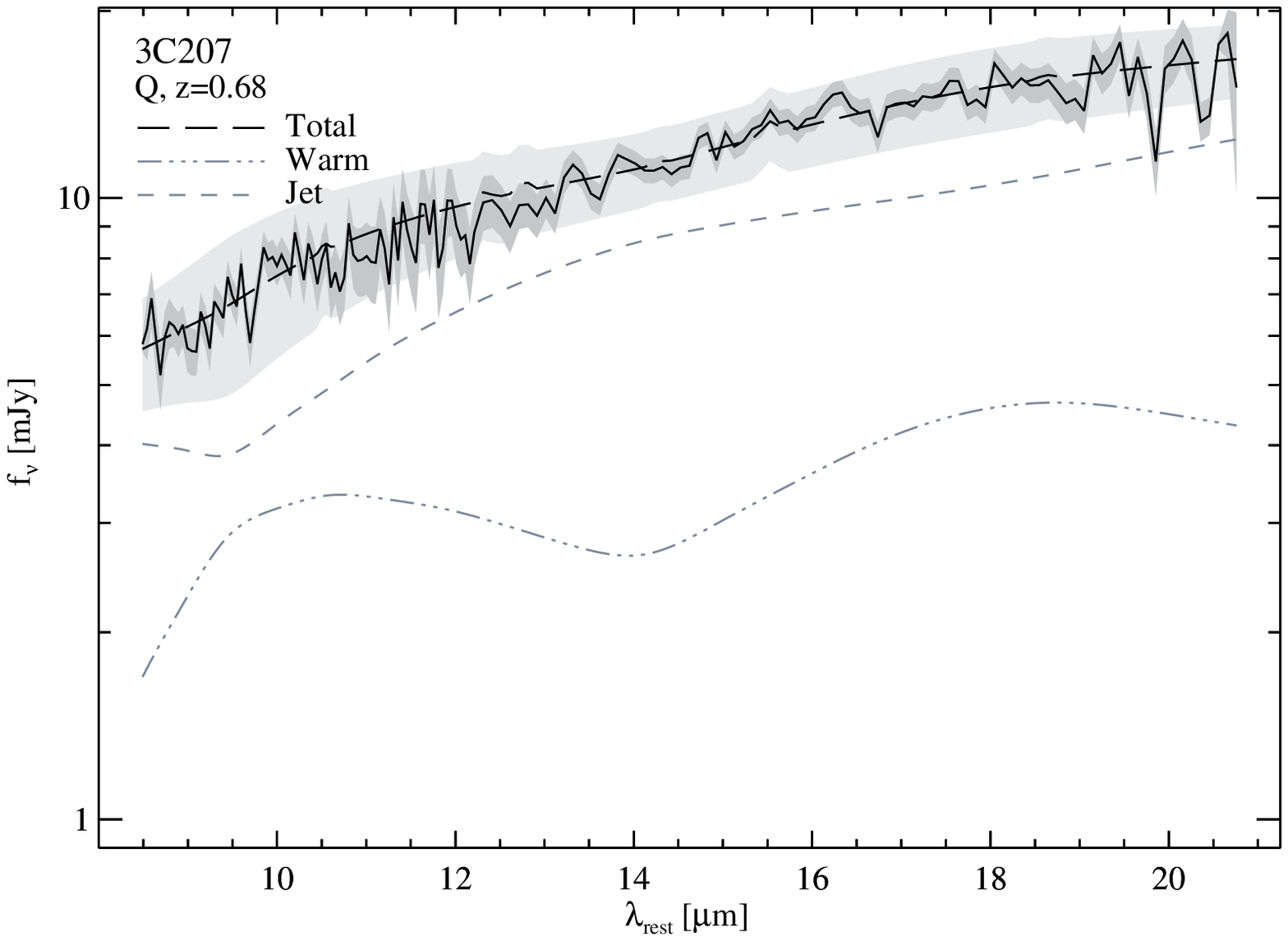,angle=0,width=8cm,height=5.5cm,clip=}\\

\psfig{file=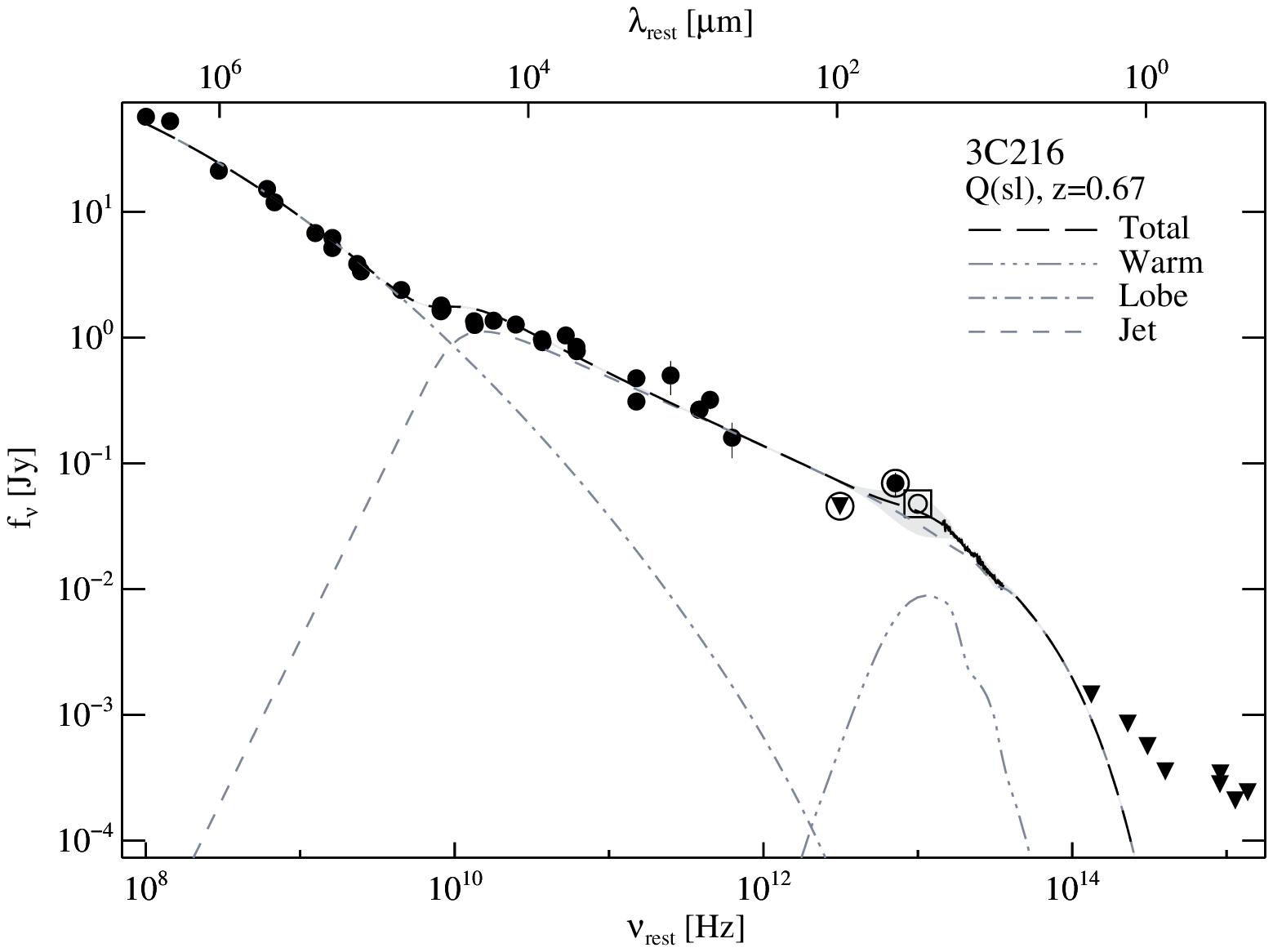,angle=0,width=8cm,height=6.0cm,clip=} & 
\psfig{file=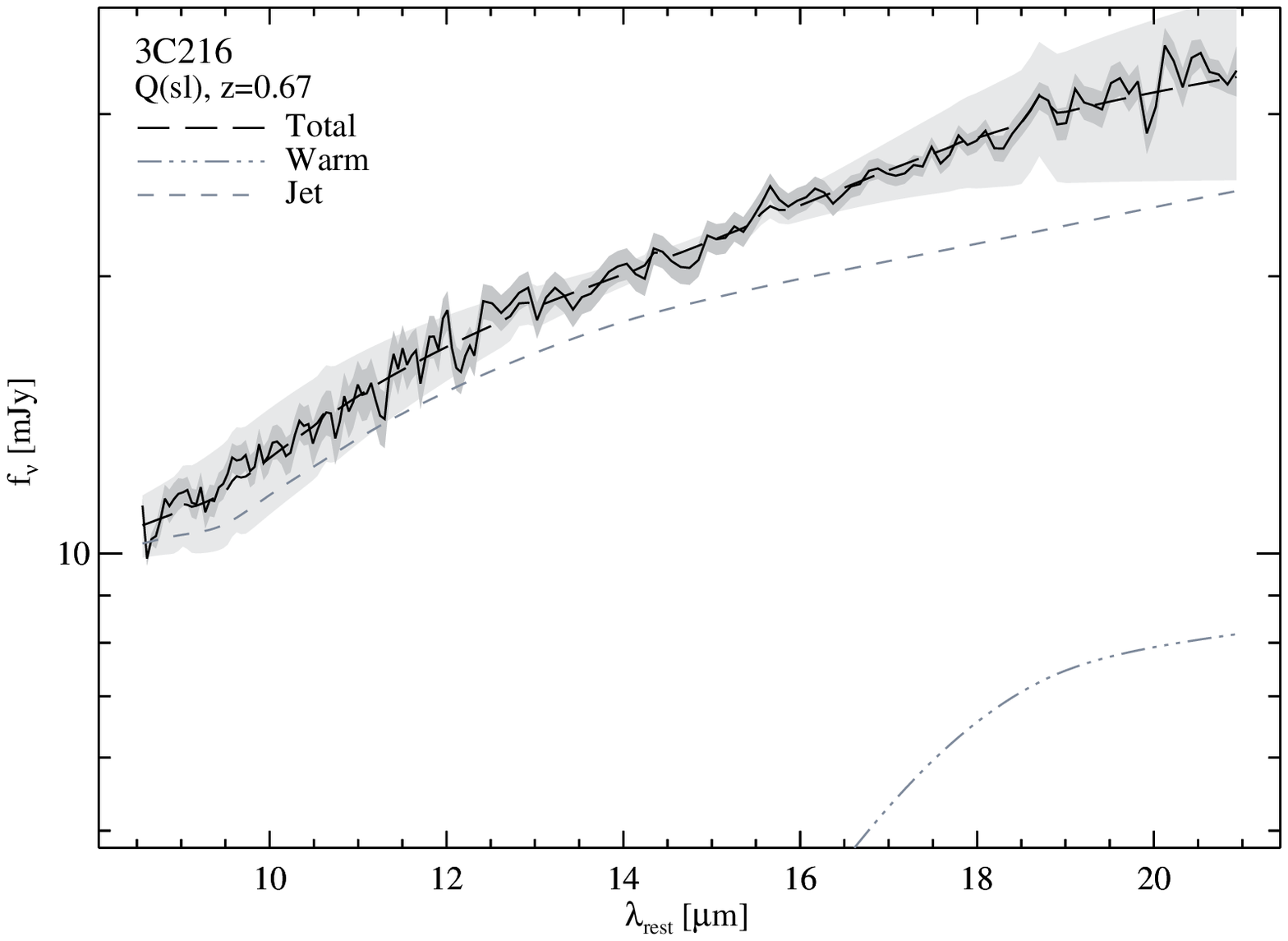,angle=0,width=8cm,height=5.5cm,clip=}\\

\psfig{file=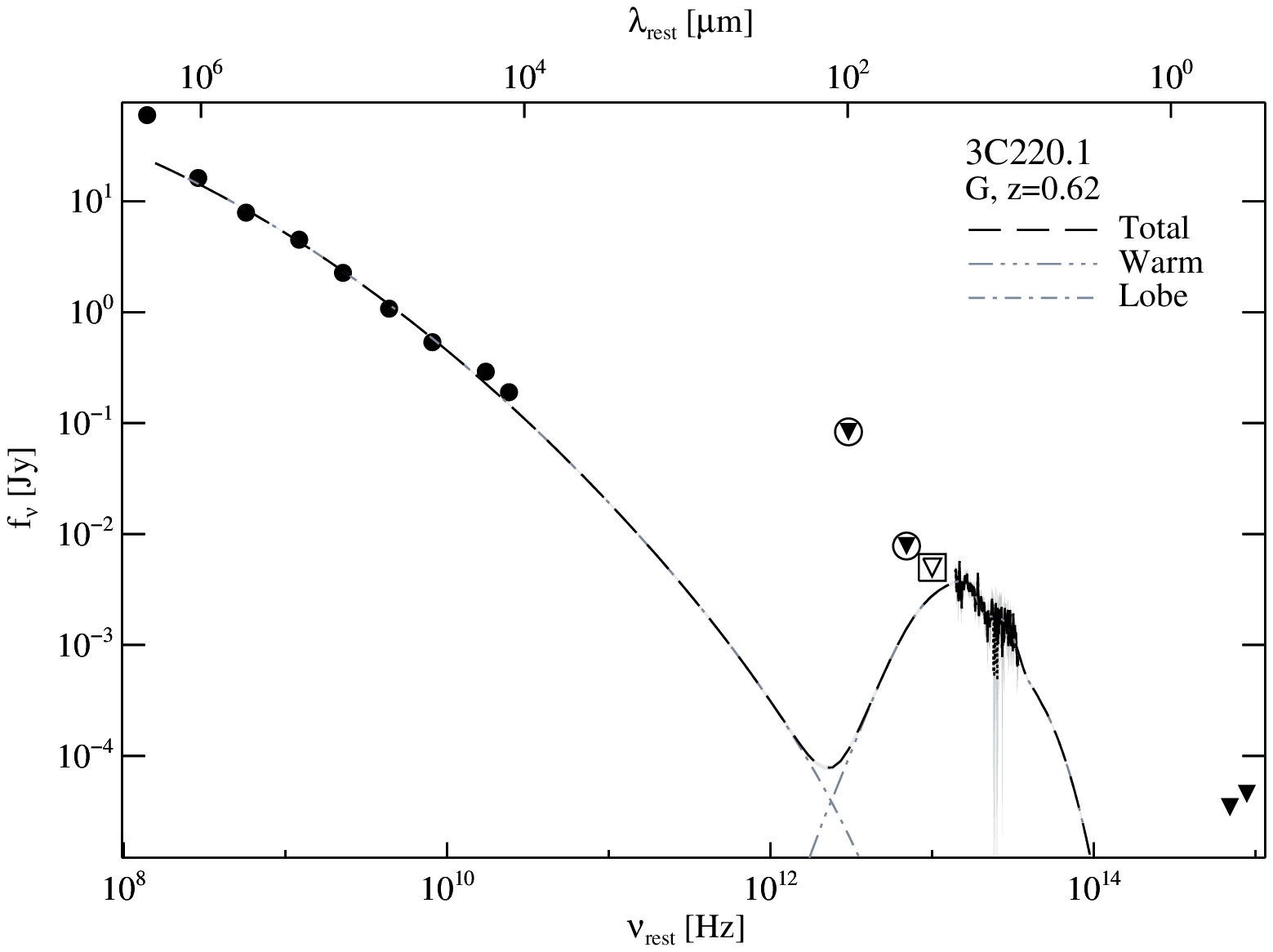,angle=0,width=8cm,height=6.0cm,clip=} & 
\psfig{file=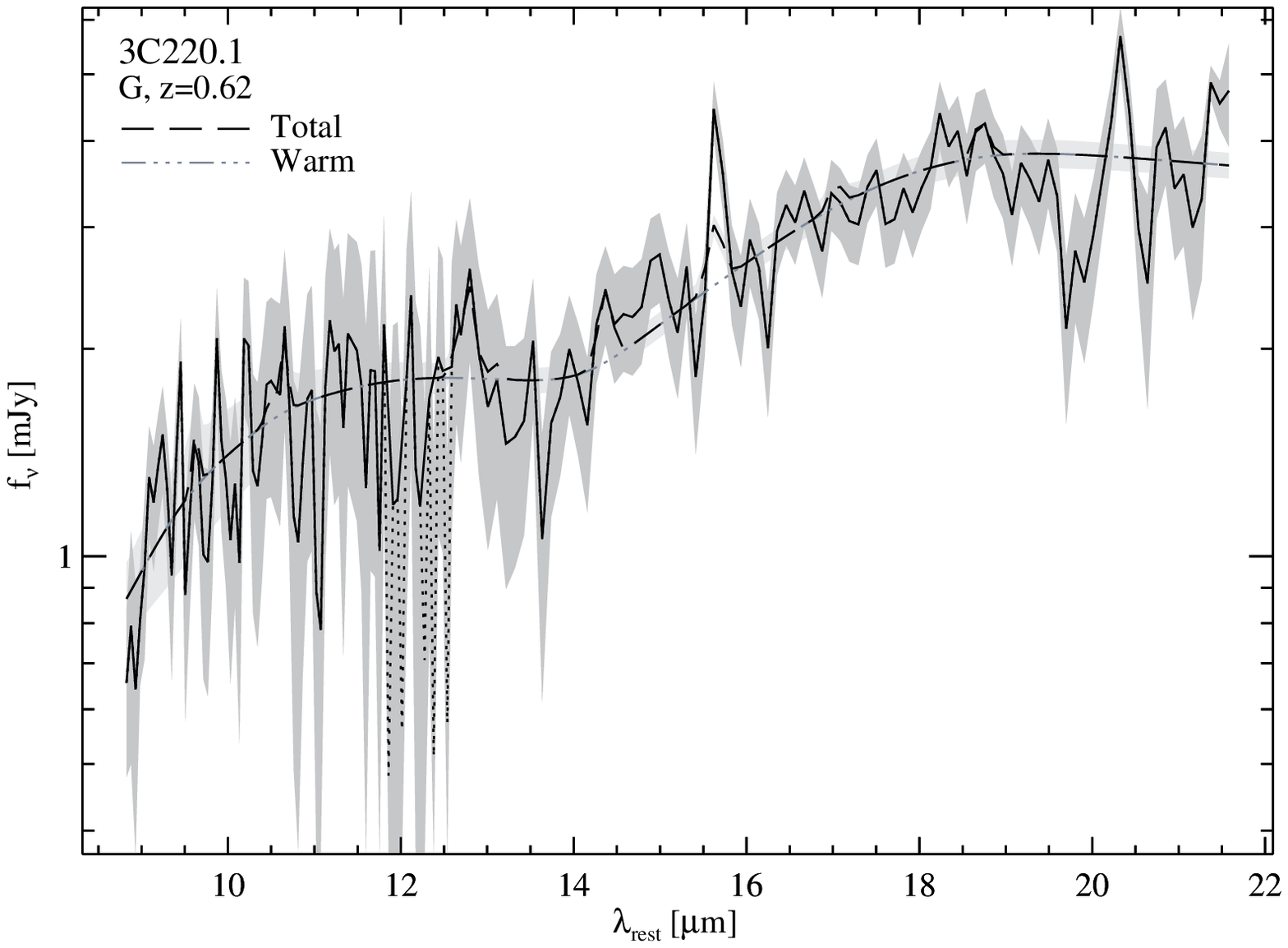,angle=0,width=8cm,height=5.5cm,clip=}\\

\end{tabular}
\caption{{\em Continued}}
\end{figure*}

\begin{figure*}
\figurenum{13}
\centering
\begin{tabular}{cc}
\tablewidth{0pt}

\psfig{file=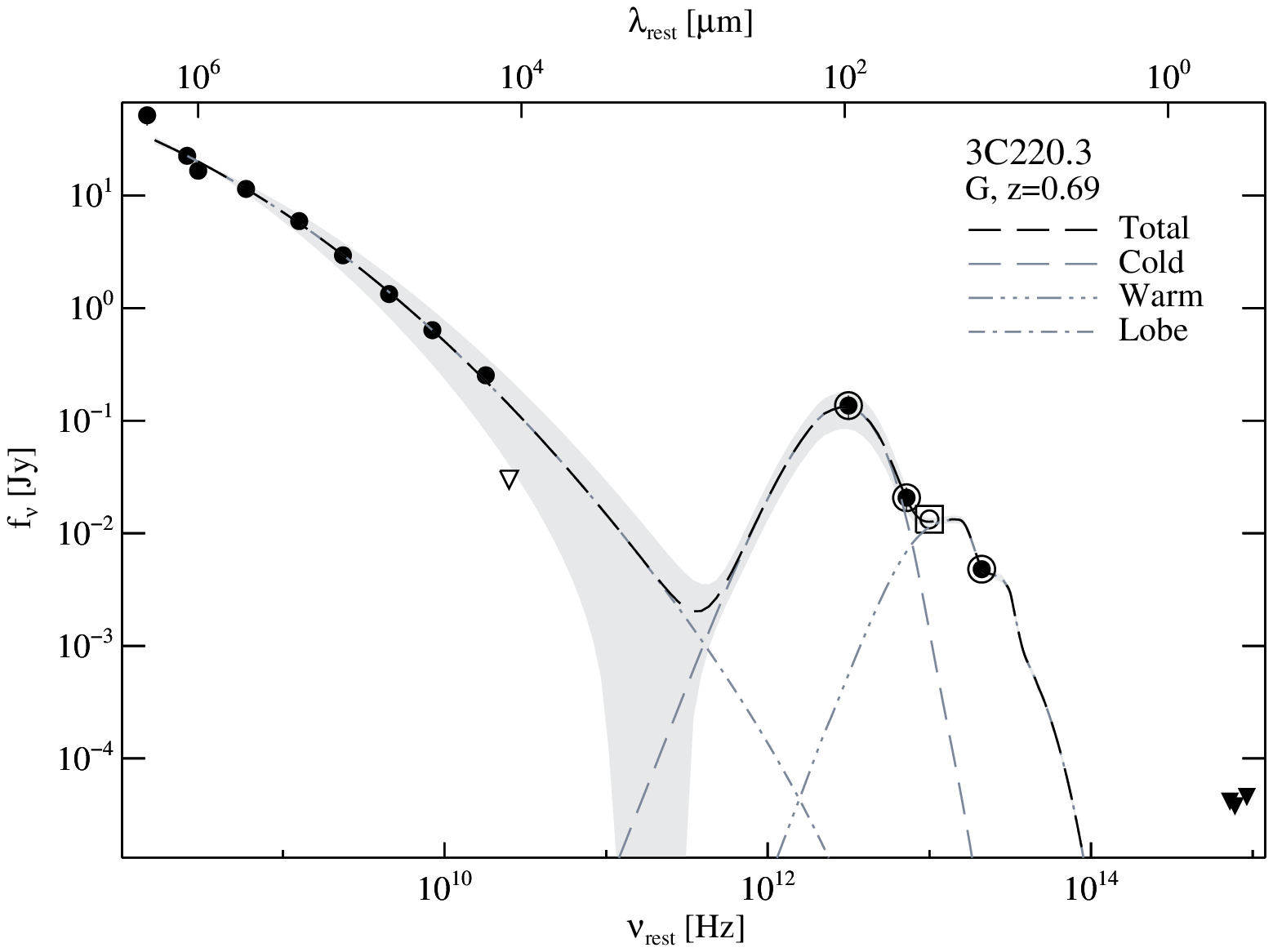,angle=0,width=8cm,height=6.0cm,clip=} & 
\psfig{file=f13x.eps,angle=0,width=8cm,height=5.5cm,clip=}\\

\psfig{file=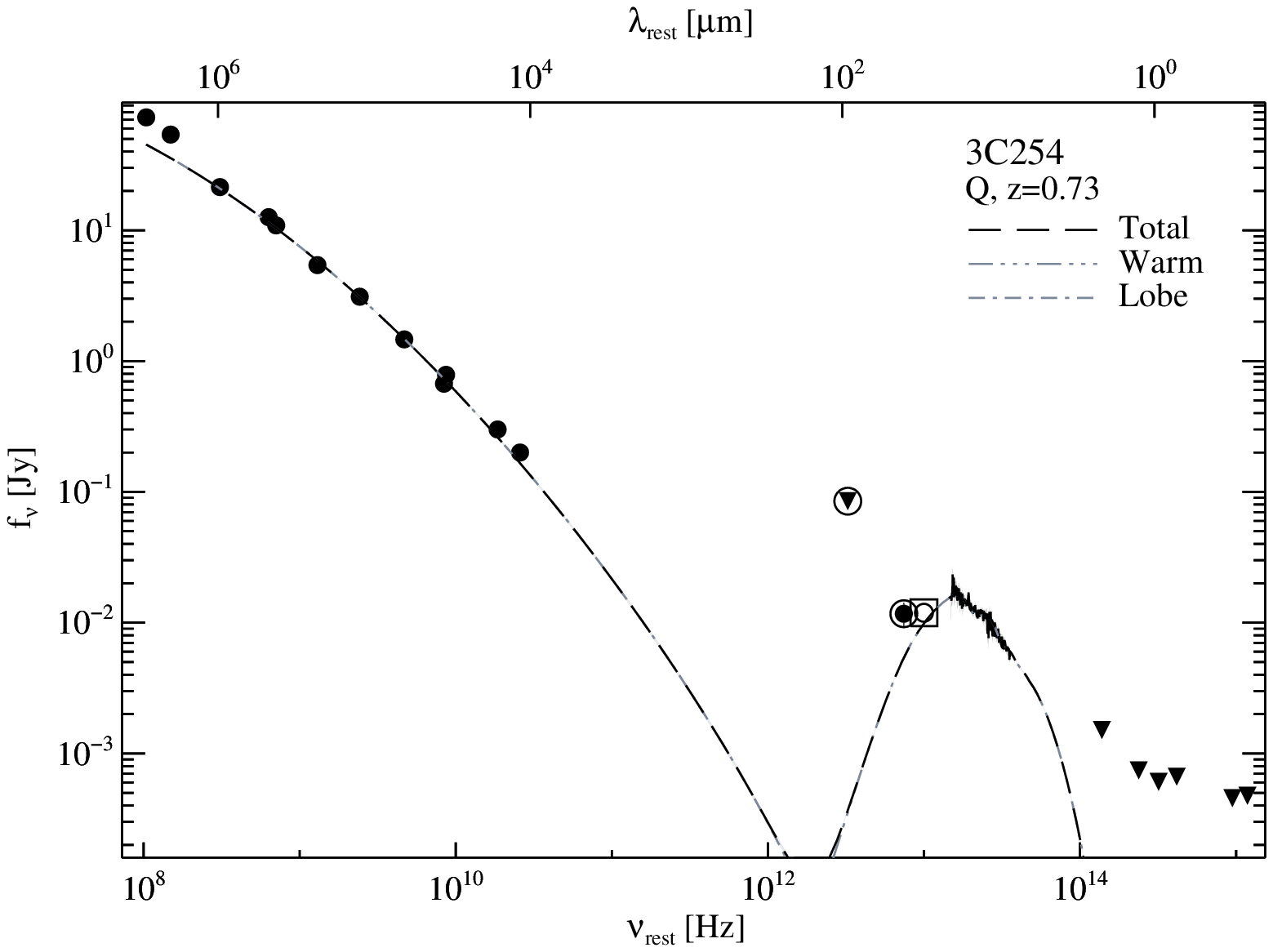,angle=0,width=8cm,height=6.0cm,clip=} & 
\psfig{file=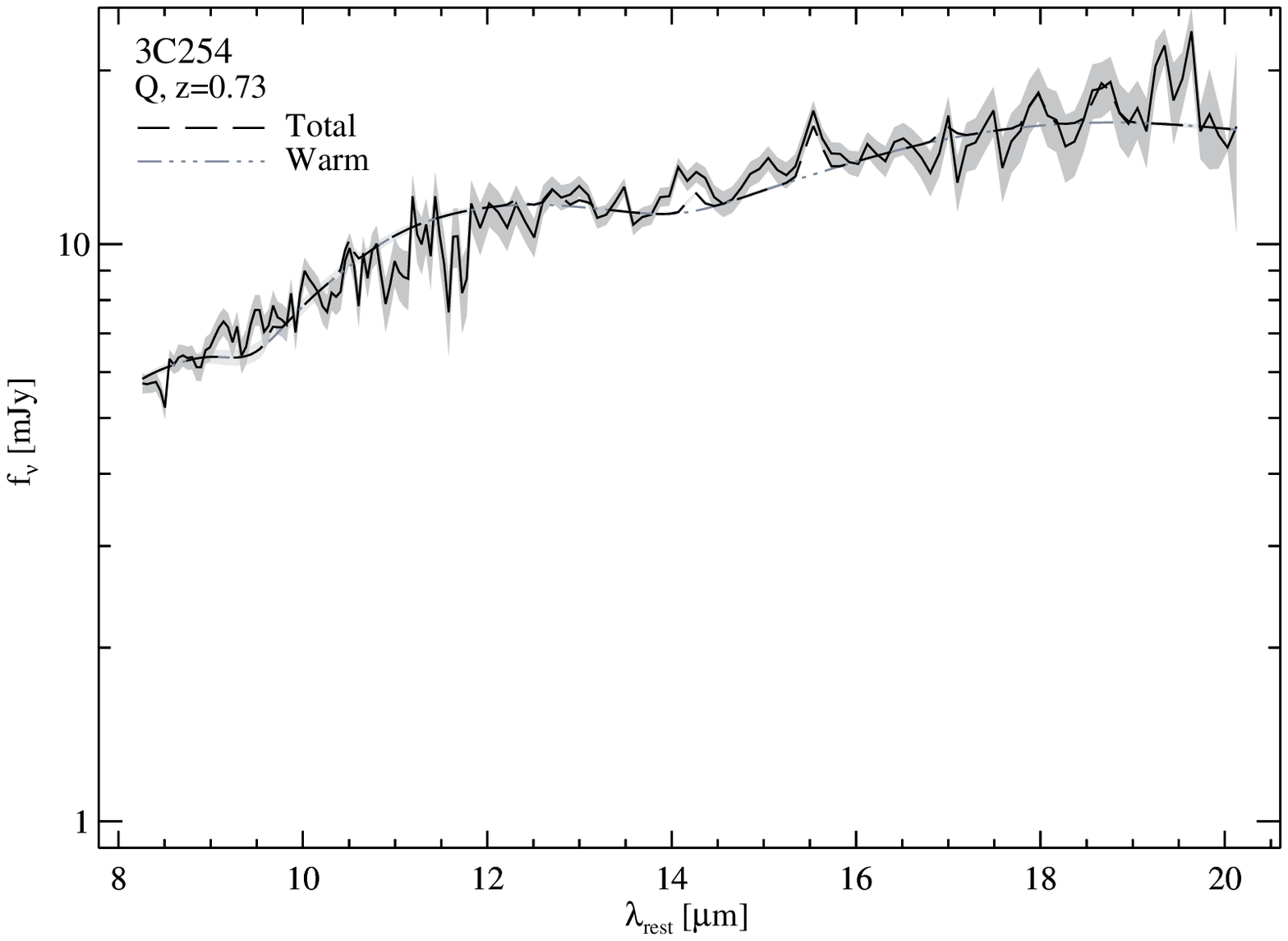,angle=0,width=8cm,height=5.5cm,clip=}\\

\psfig{file=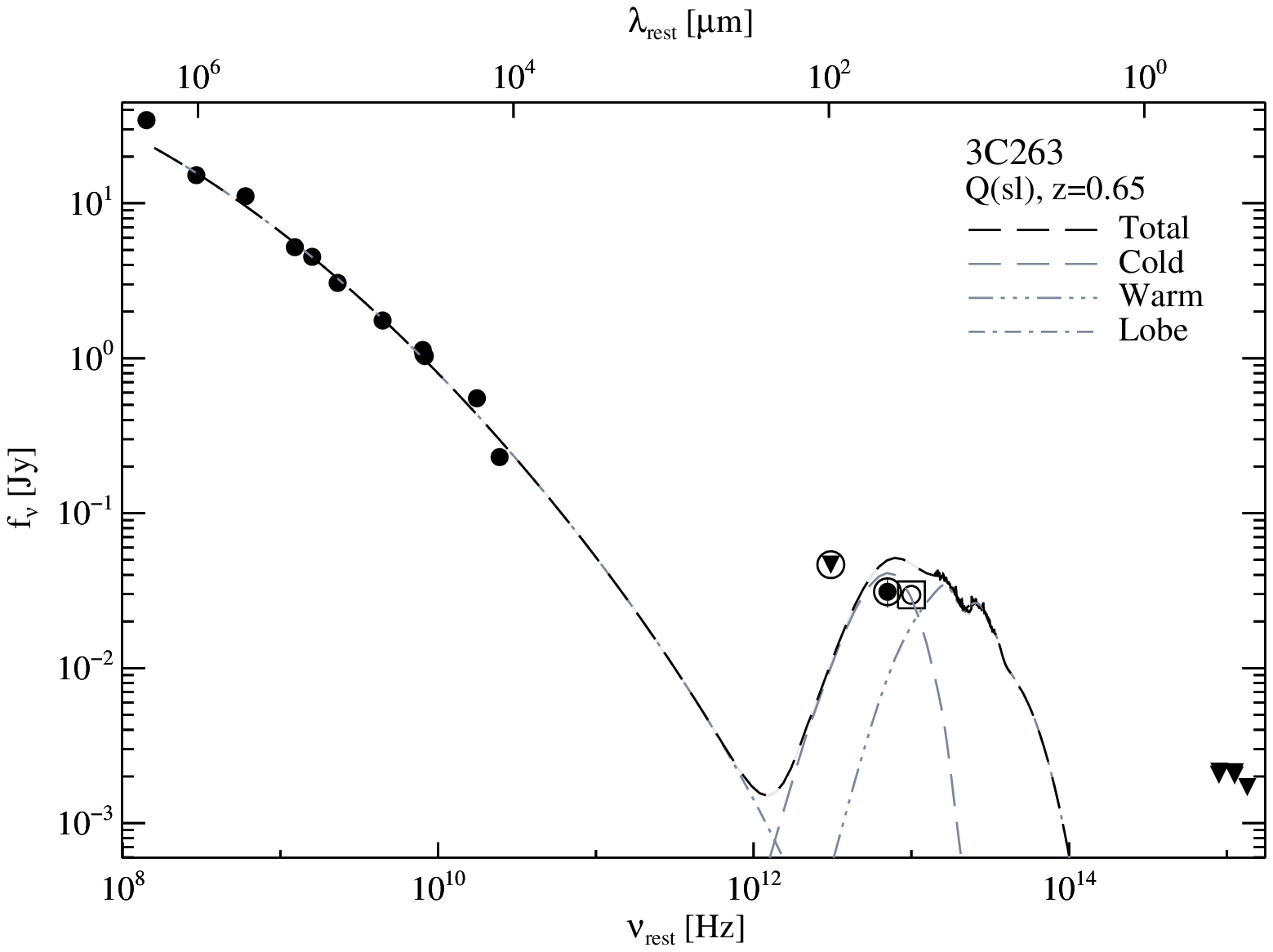,angle=0,width=8cm,height=6.0cm,clip=} & 
\psfig{file=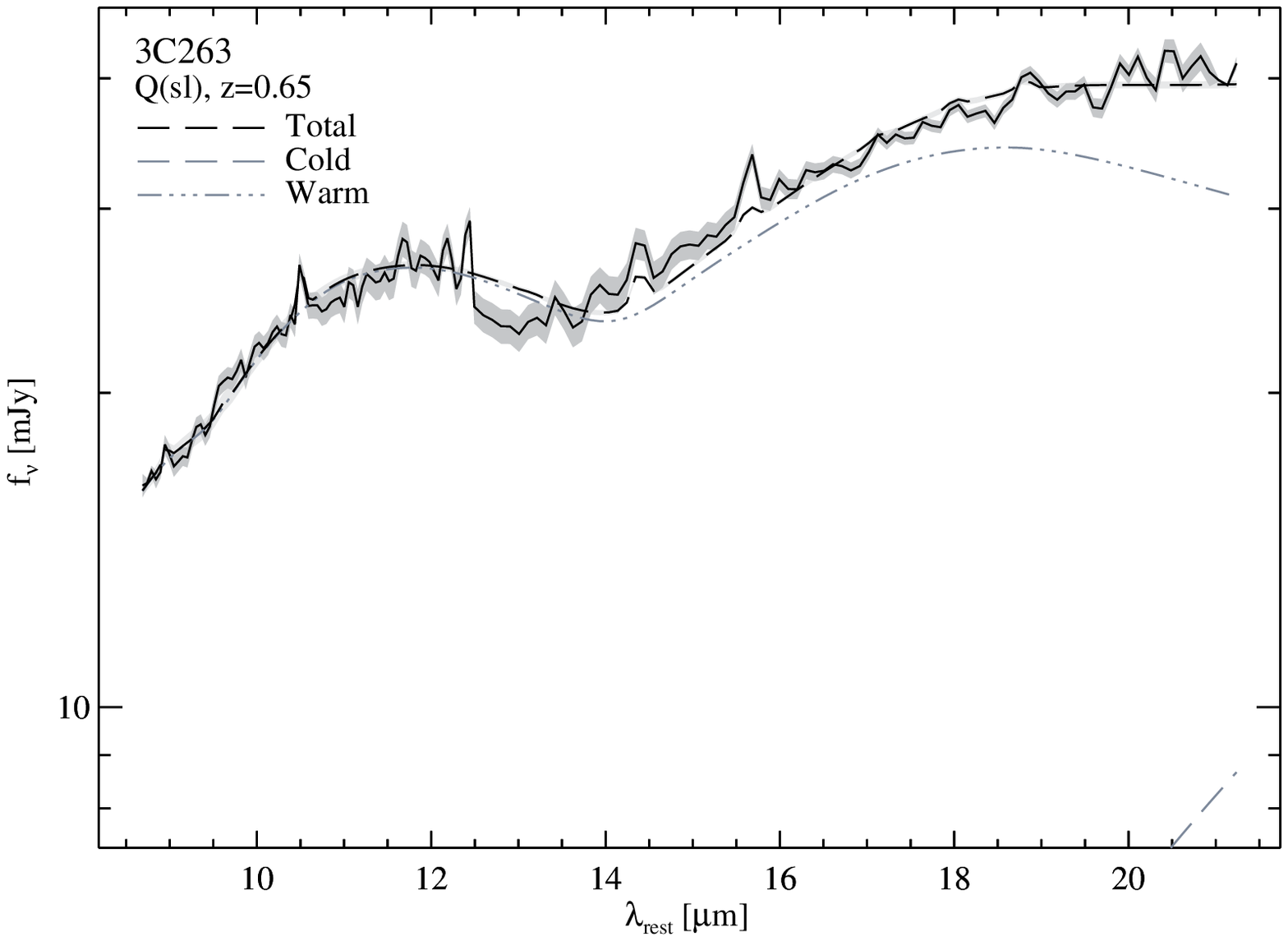,angle=0,width=8cm,height=5.5cm,clip=}\\

\end{tabular}
\caption{{\em Continued}}
\end{figure*}

\begin{figure*}
\figurenum{13}
\centering
\begin{tabular}{cc}
\tablewidth{0pt}

\psfig{file=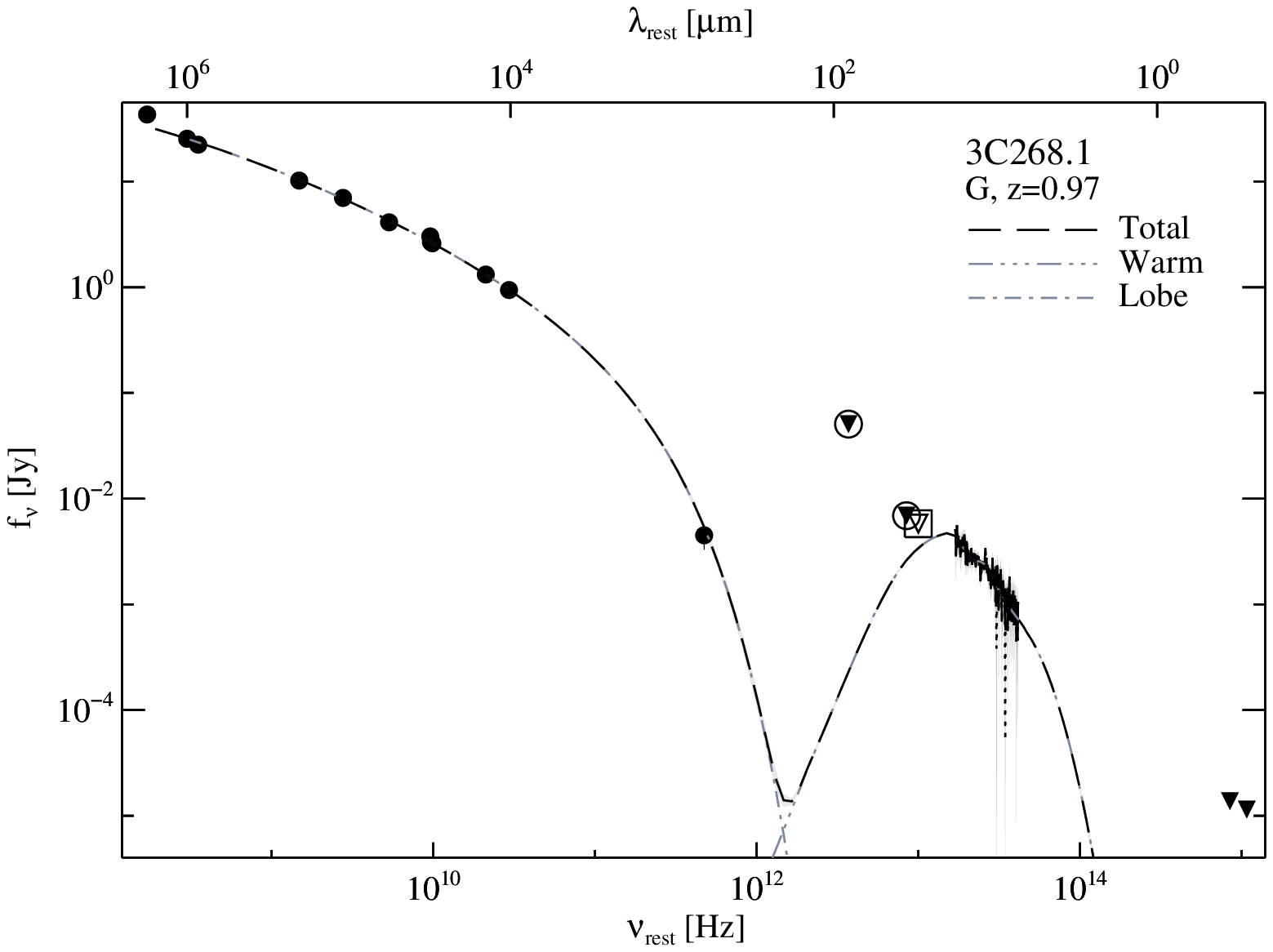,angle=0,width=8cm,height=6.0cm,clip=} & 
\psfig{file=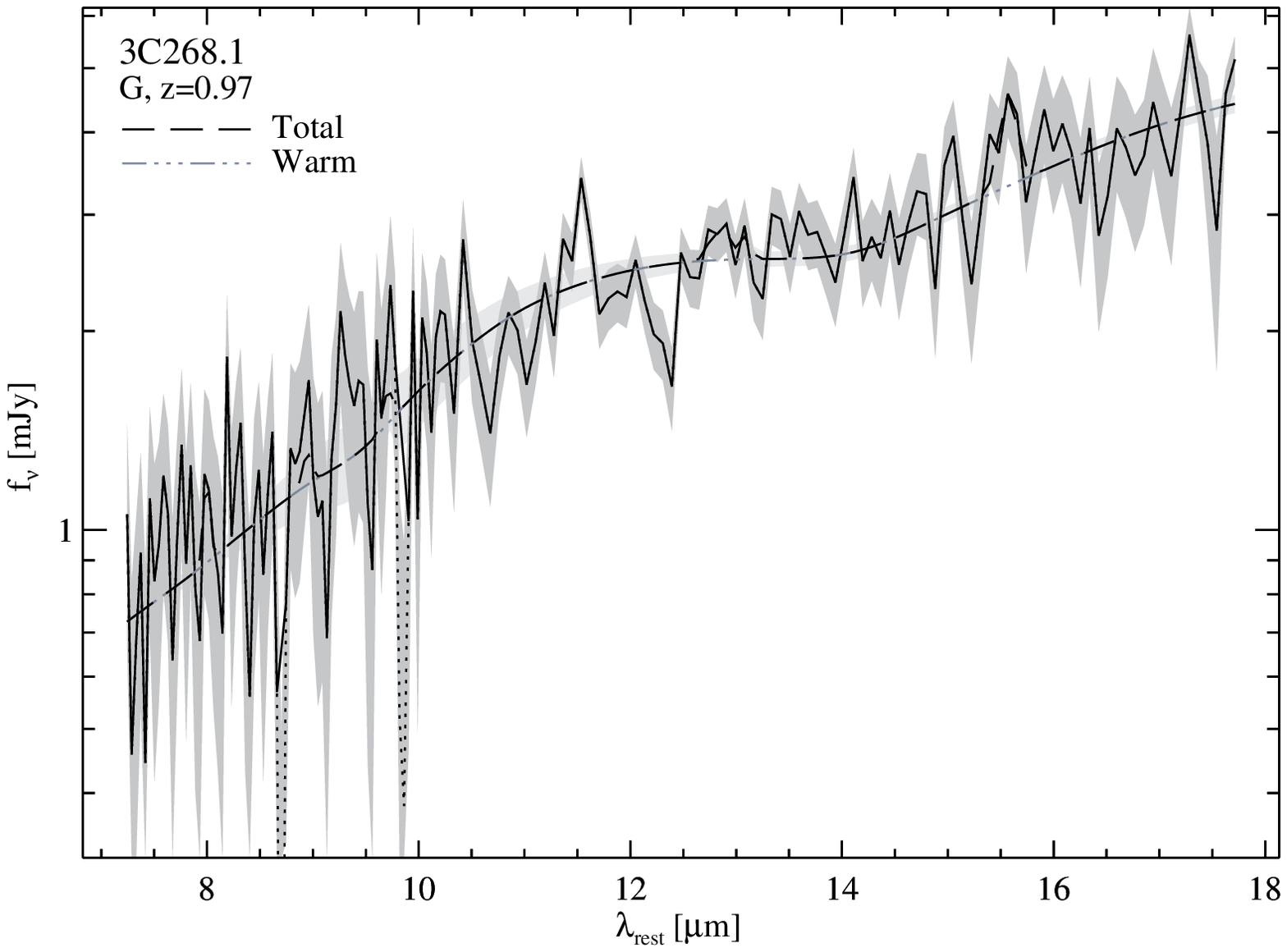,angle=0,width=8cm,height=5.5cm,clip=}\\

\psfig{file=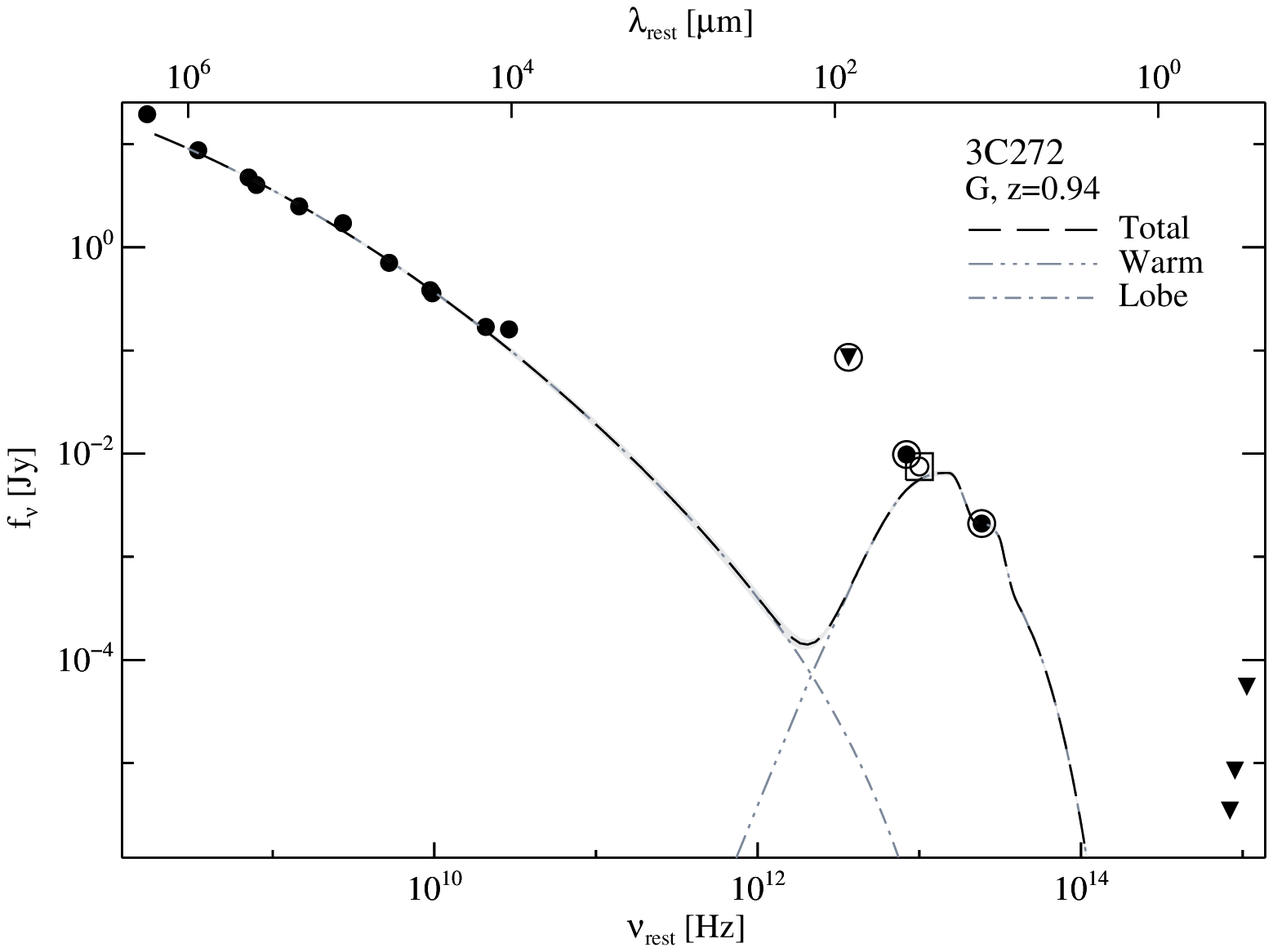,angle=0,width=8cm,height=6.0cm,clip=} & 
\psfig{file=f13af.eps,angle=0,width=8cm,height=5.5cm,clip=}\\

\psfig{file=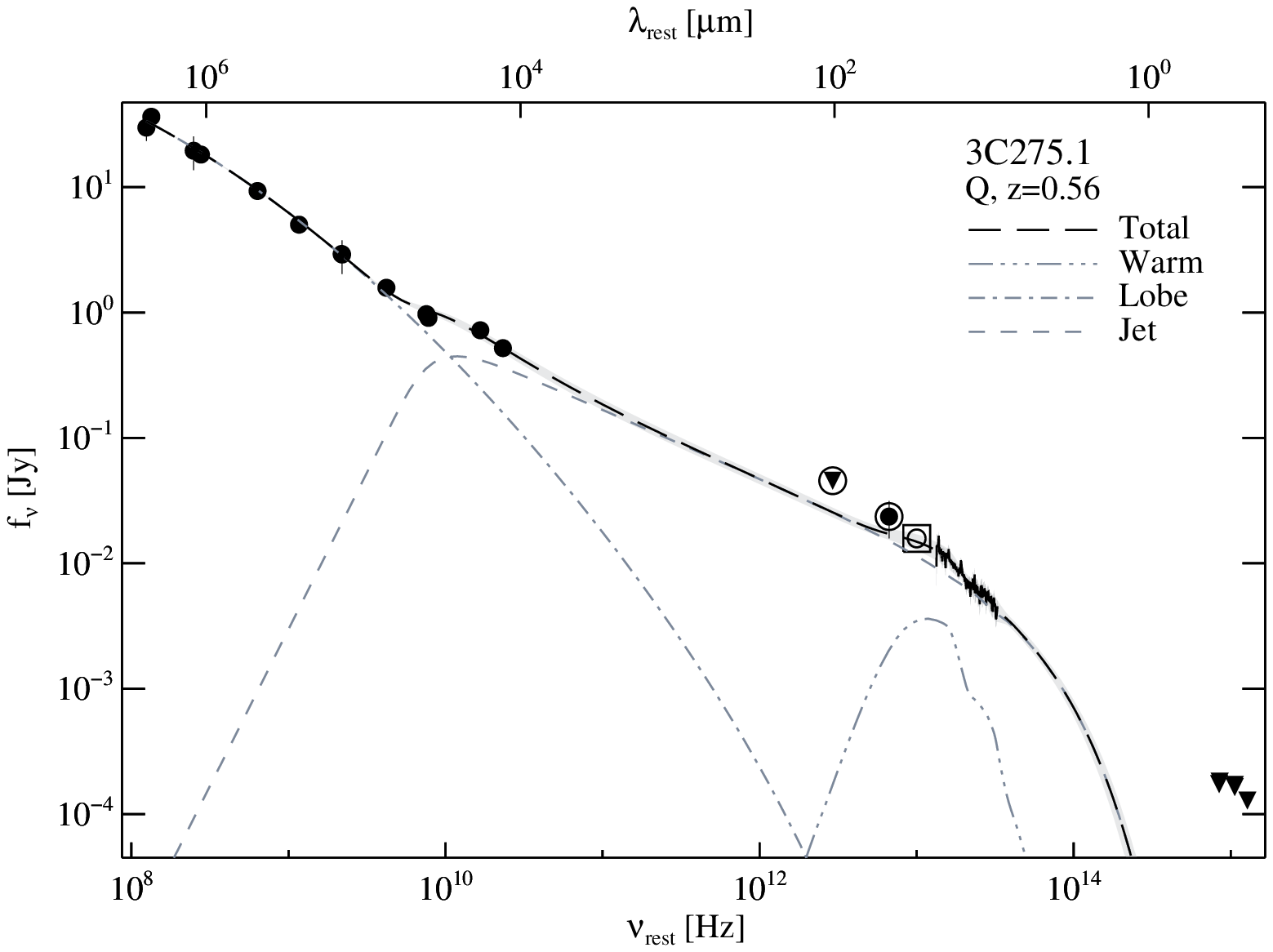,angle=0,width=8cm,height=6.0cm,clip=} & 
\psfig{file=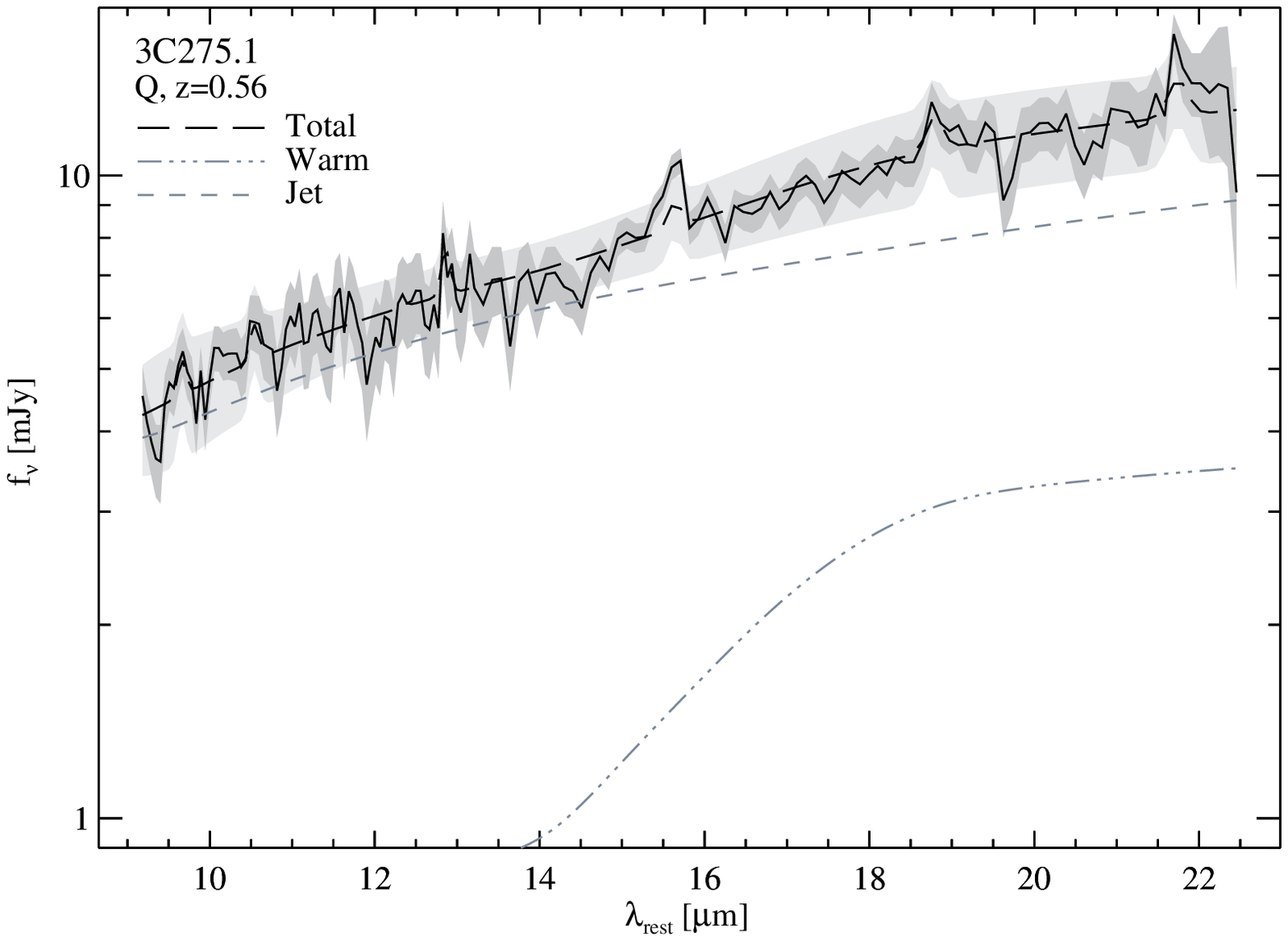,angle=0,width=8cm,height=5.5cm,clip=}\\

\end{tabular}
\caption{{\em Continued}}
\end{figure*}

\begin{figure*}
\figurenum{13}
\centering
\begin{tabular}{cc}
\tablewidth{0pt}

\psfig{file=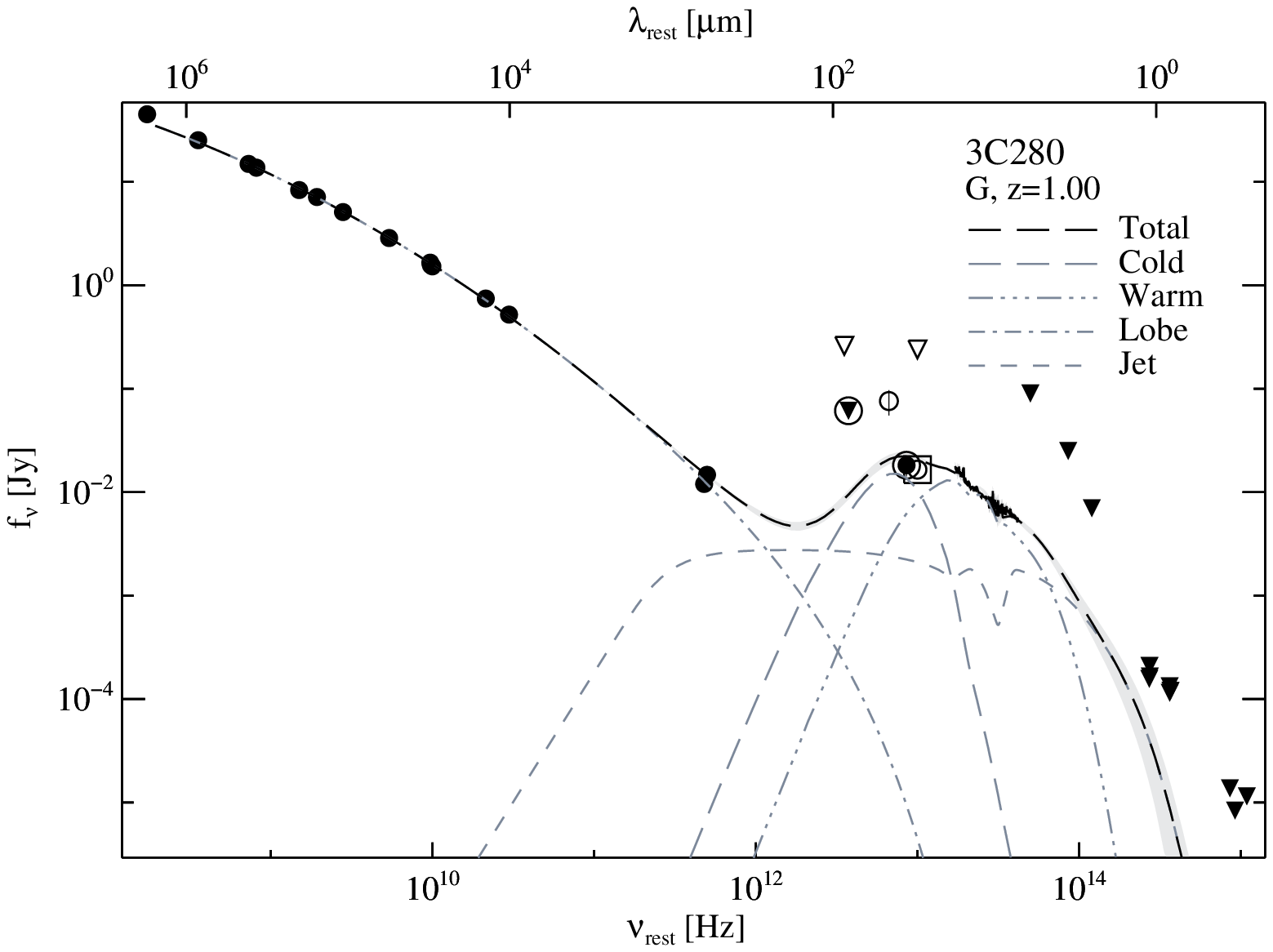,angle=0,width=8cm,height=6.0cm,clip=} & 
\psfig{file=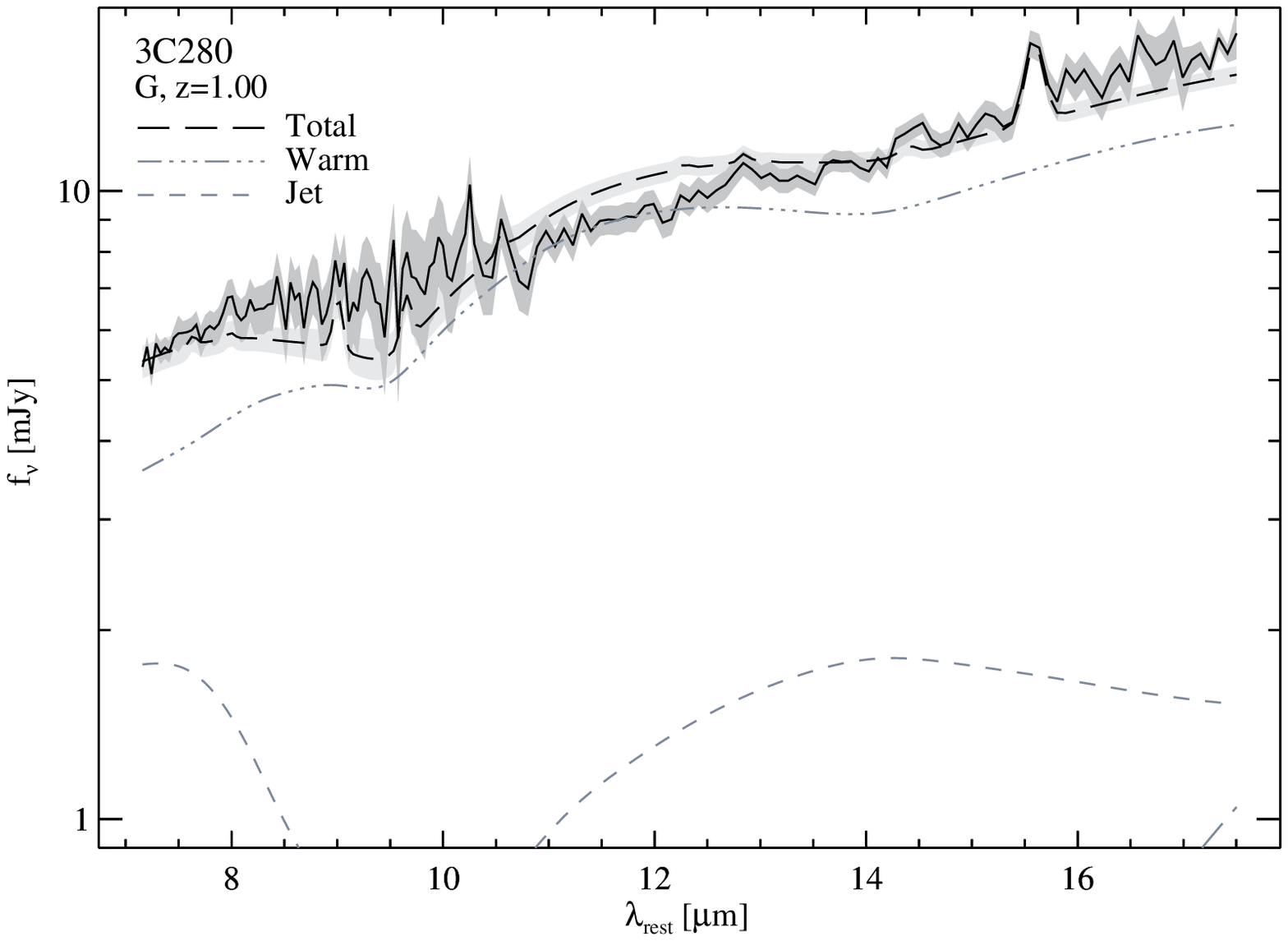,angle=0,width=8cm,height=5.5cm,clip=}\\

\psfig{file=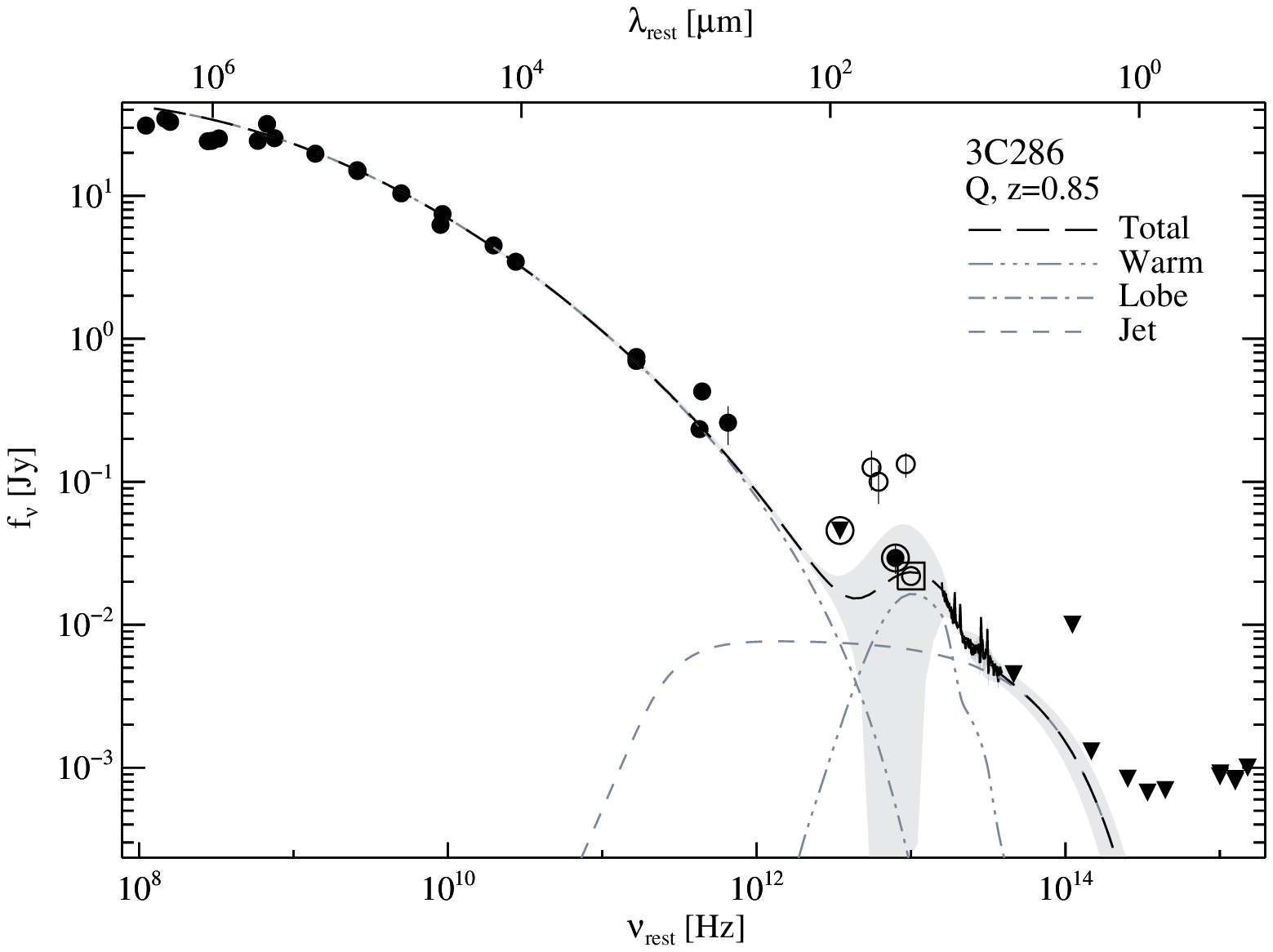,angle=0,width=8cm,height=6.0cm,clip=} & 
\psfig{file=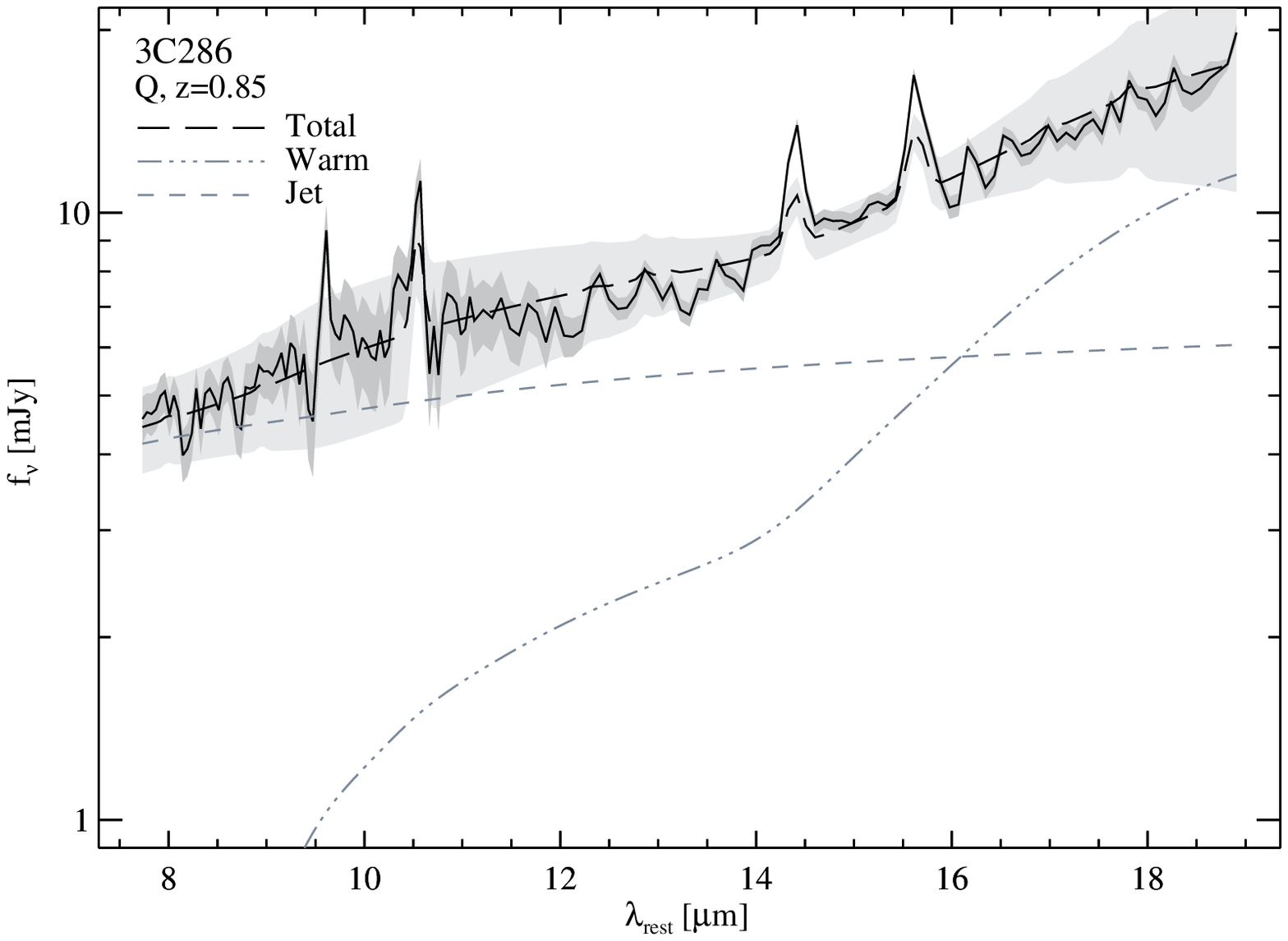,angle=0,width=8cm,height=5.5cm,clip=}\\

\psfig{file=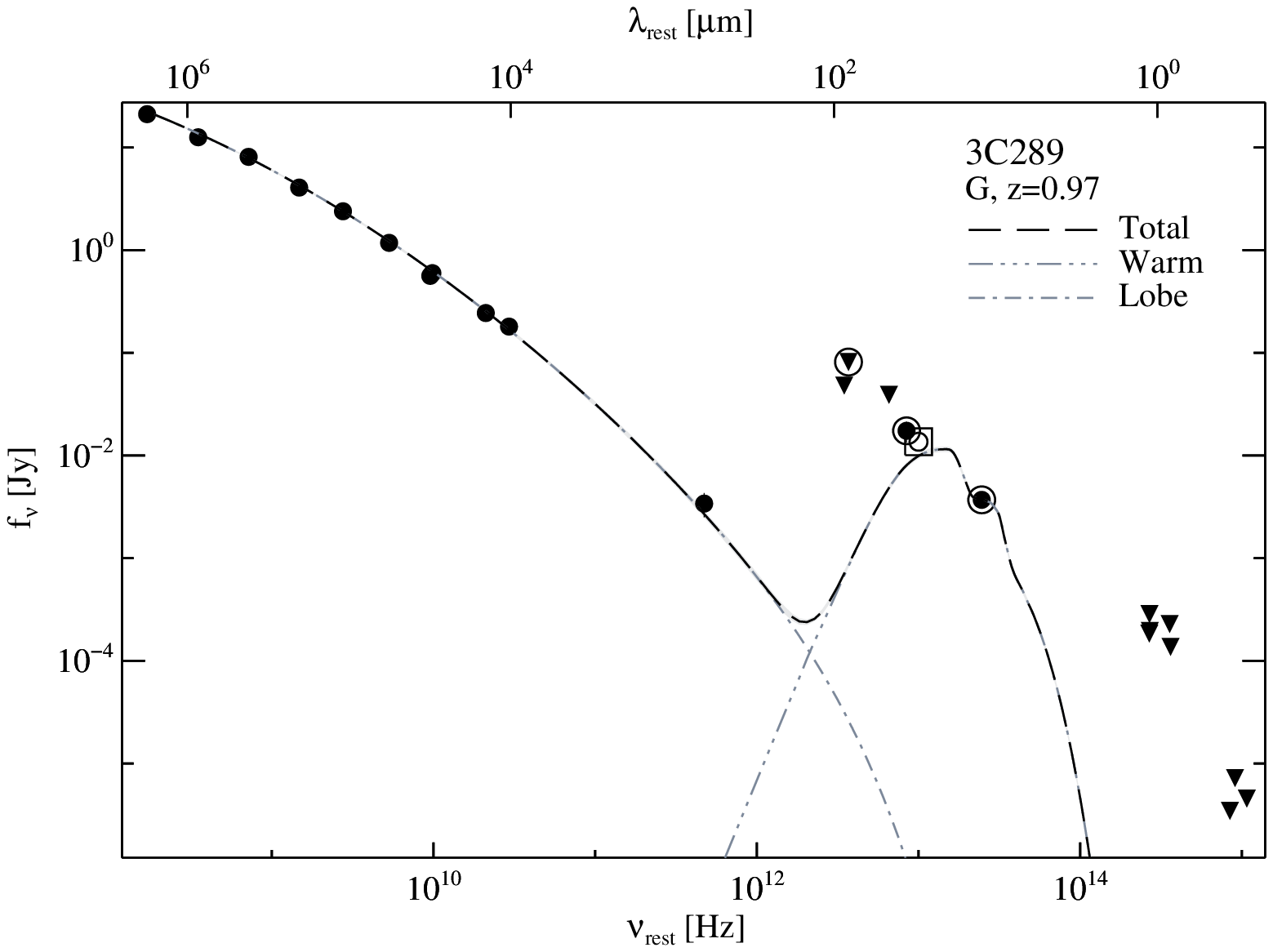,angle=0,width=8cm,height=6.0cm,clip=} & 
\psfig{file=f13an.eps,angle=0,width=8cm,height=5.5cm,clip=}\\

\end{tabular}
\caption{{\em Continued}}
\end{figure*}

\begin{figure*}
\figurenum{13}
\centering
\begin{tabular}{cc}
\tablewidth{0pt}

\psfig{file=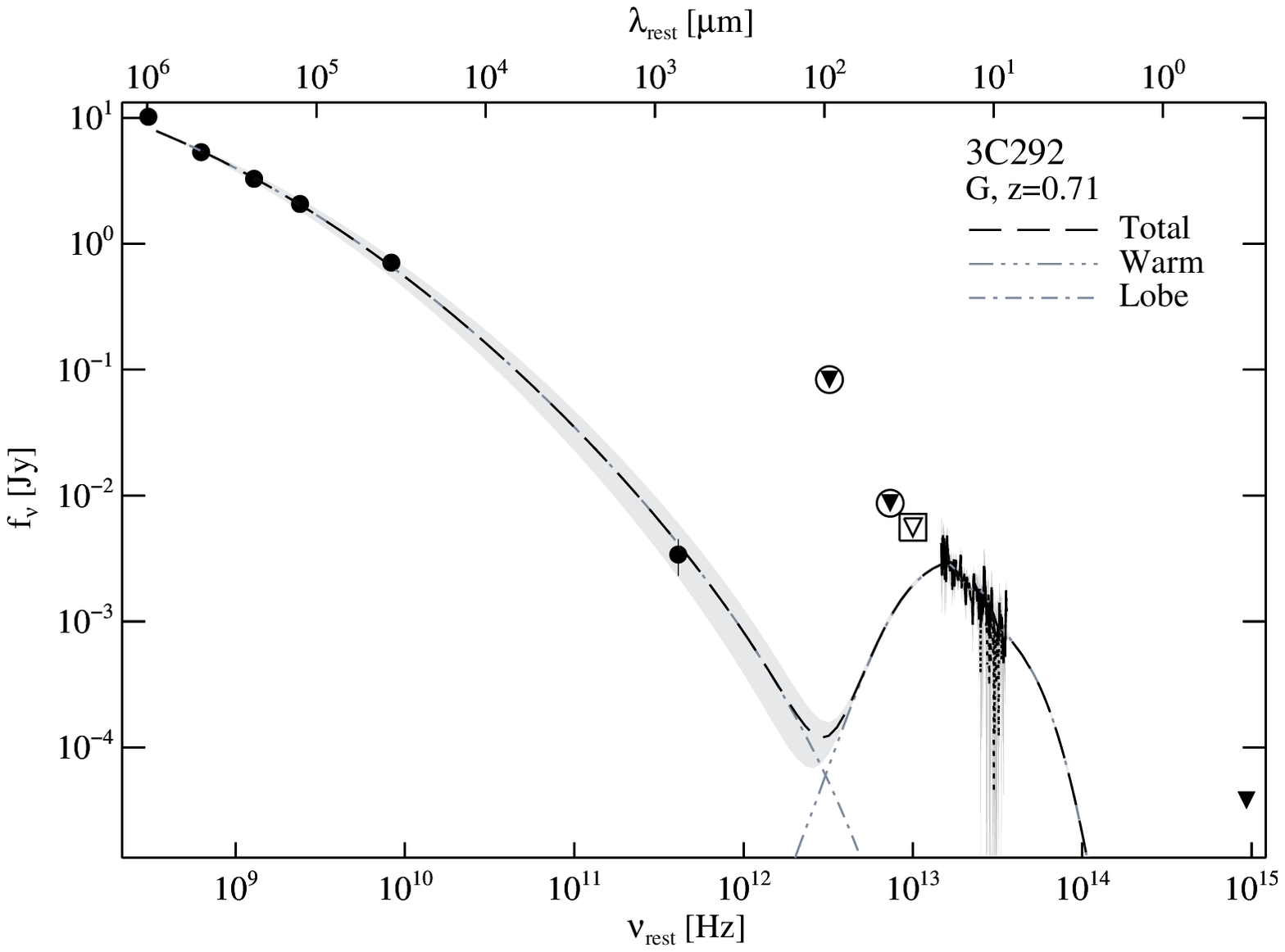,angle=0,width=8cm,height=6.0cm,clip=} & 
\psfig{file=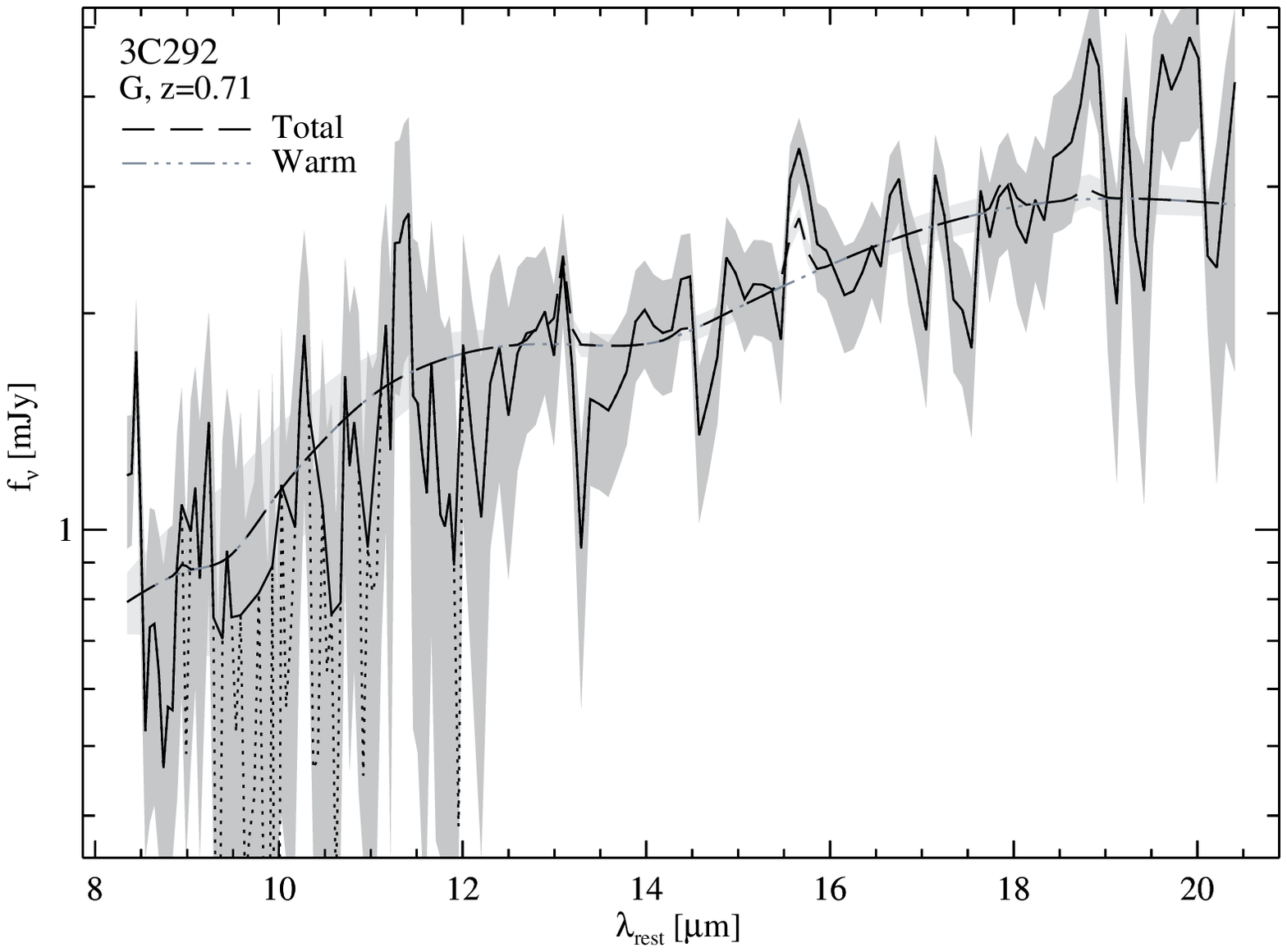,angle=0,width=8cm,height=5.5cm,clip=}\\

\psfig{file=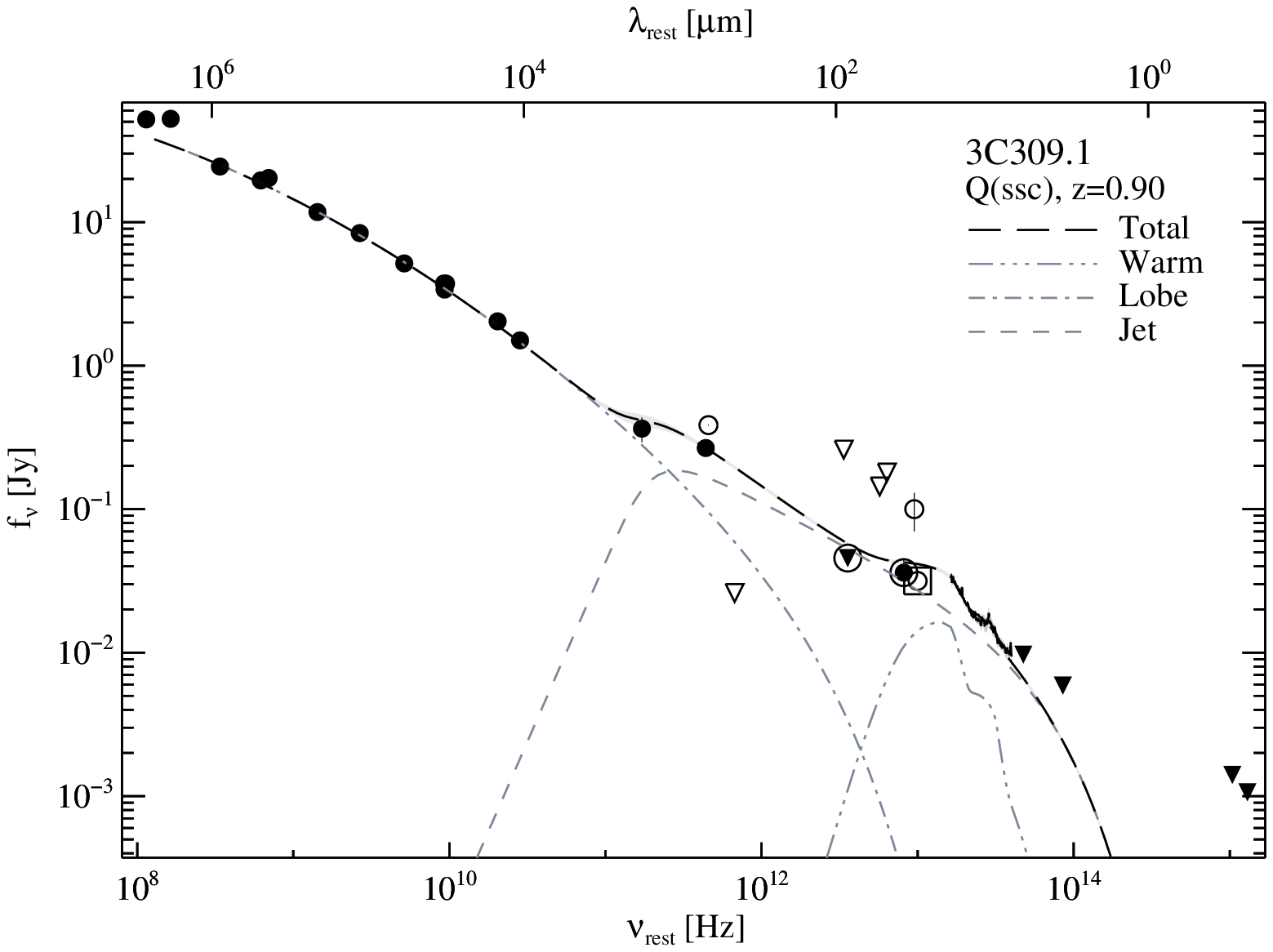,angle=0,width=8cm,height=6.0cm,clip=} & 
\psfig{file=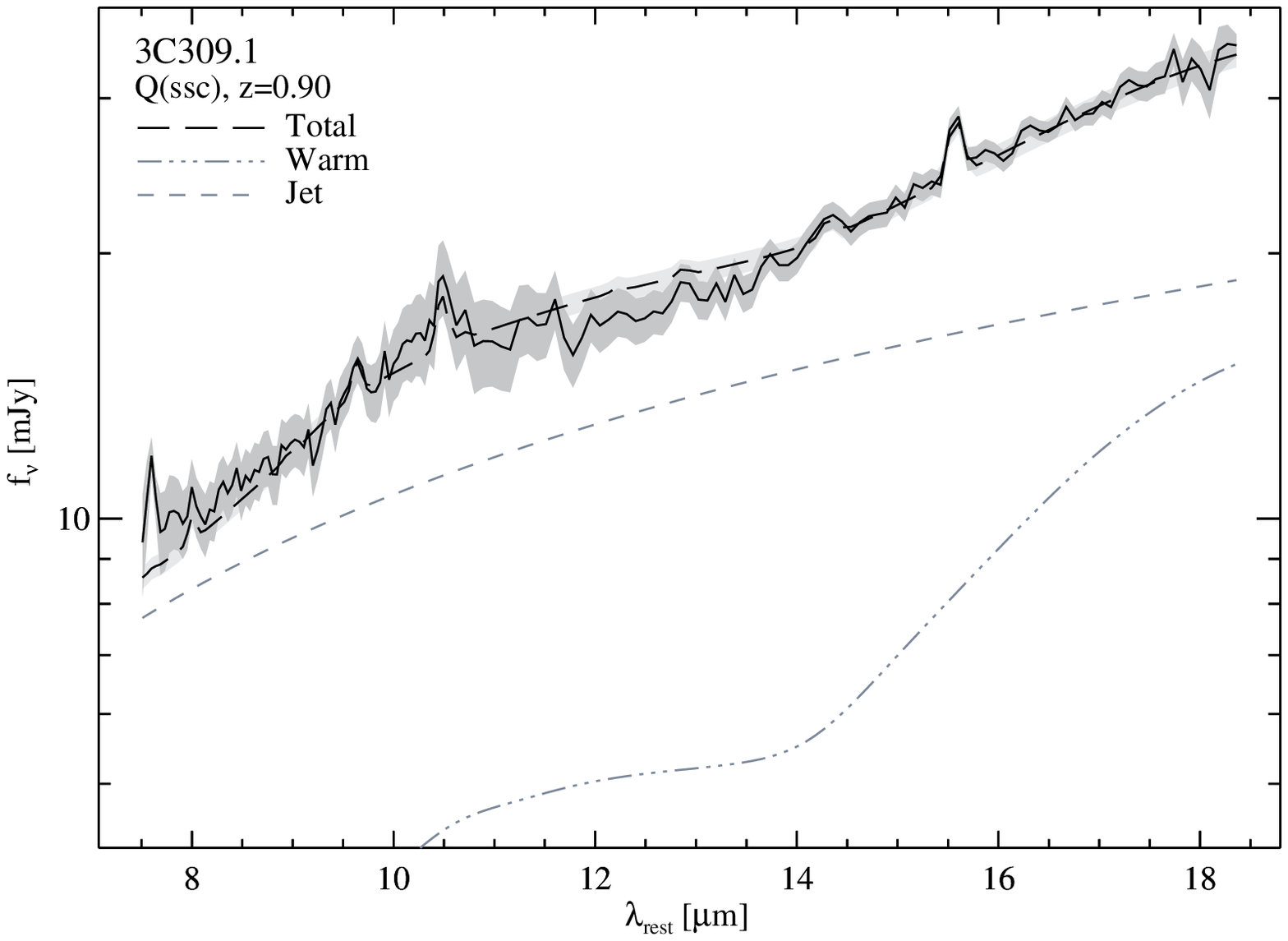,angle=0,width=8cm,height=5.5cm,clip=}\\

\psfig{file=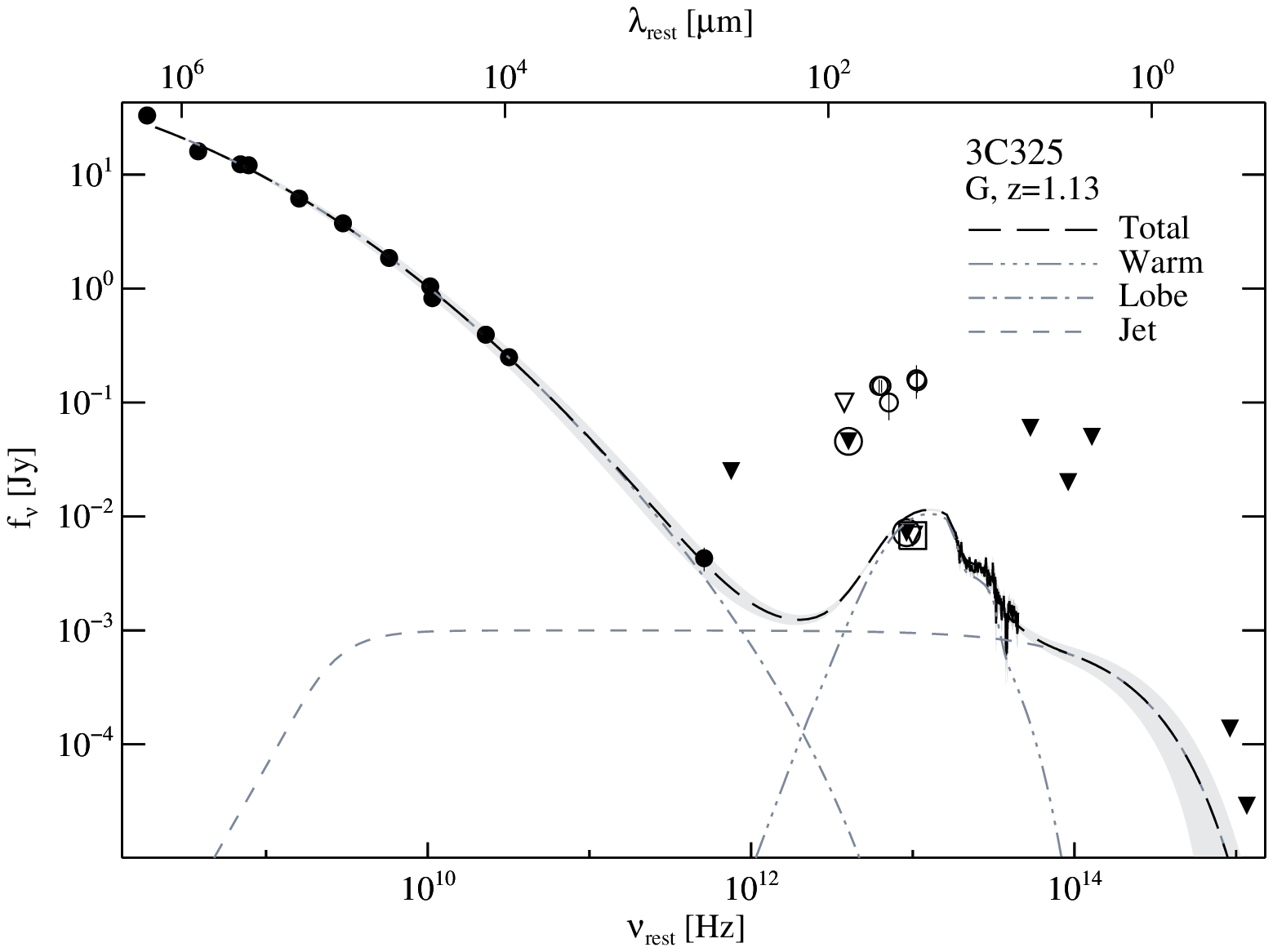,angle=0,width=8cm,height=6.0cm,clip=} & 
\psfig{file=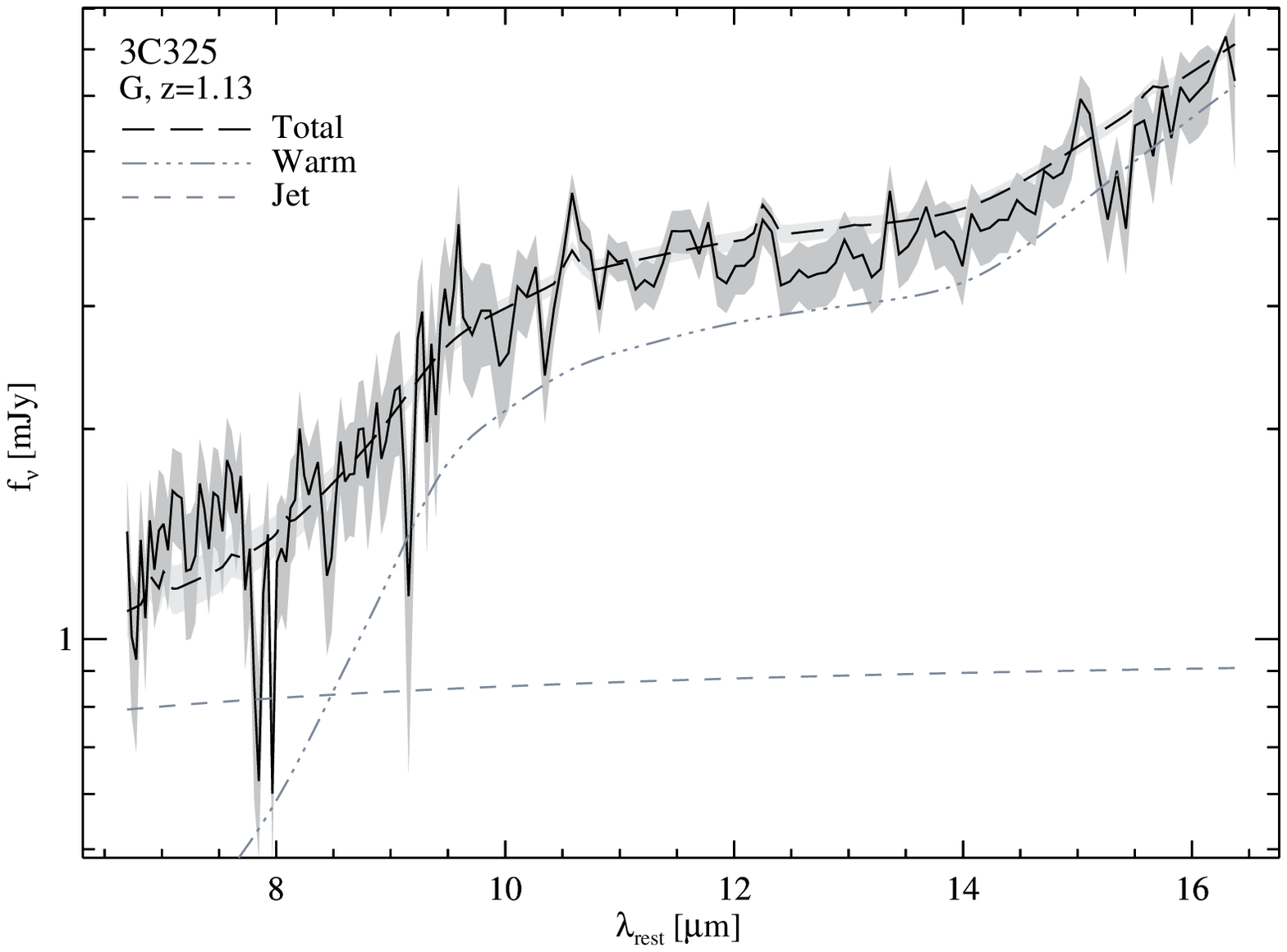,angle=0,width=8cm,height=5.5cm,clip=}\\

\end{tabular}
\caption{{\em Continued}}
\end{figure*}

\begin{figure*}
\figurenum{13}
\centering
\begin{tabular}{cc}
\tablewidth{0pt}

\psfig{file=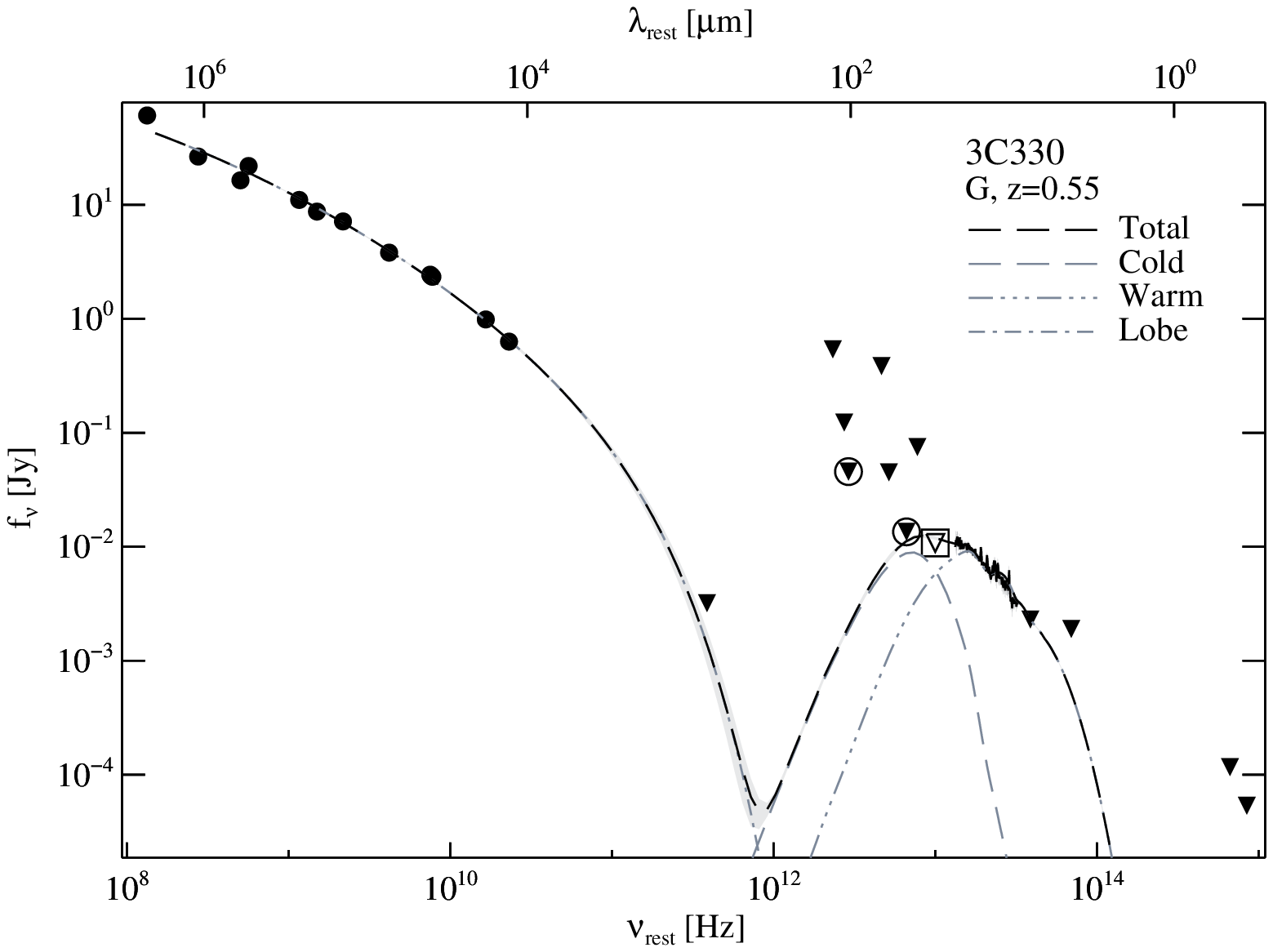,angle=0,width=8cm,height=6.0cm,clip=} & 
\psfig{file=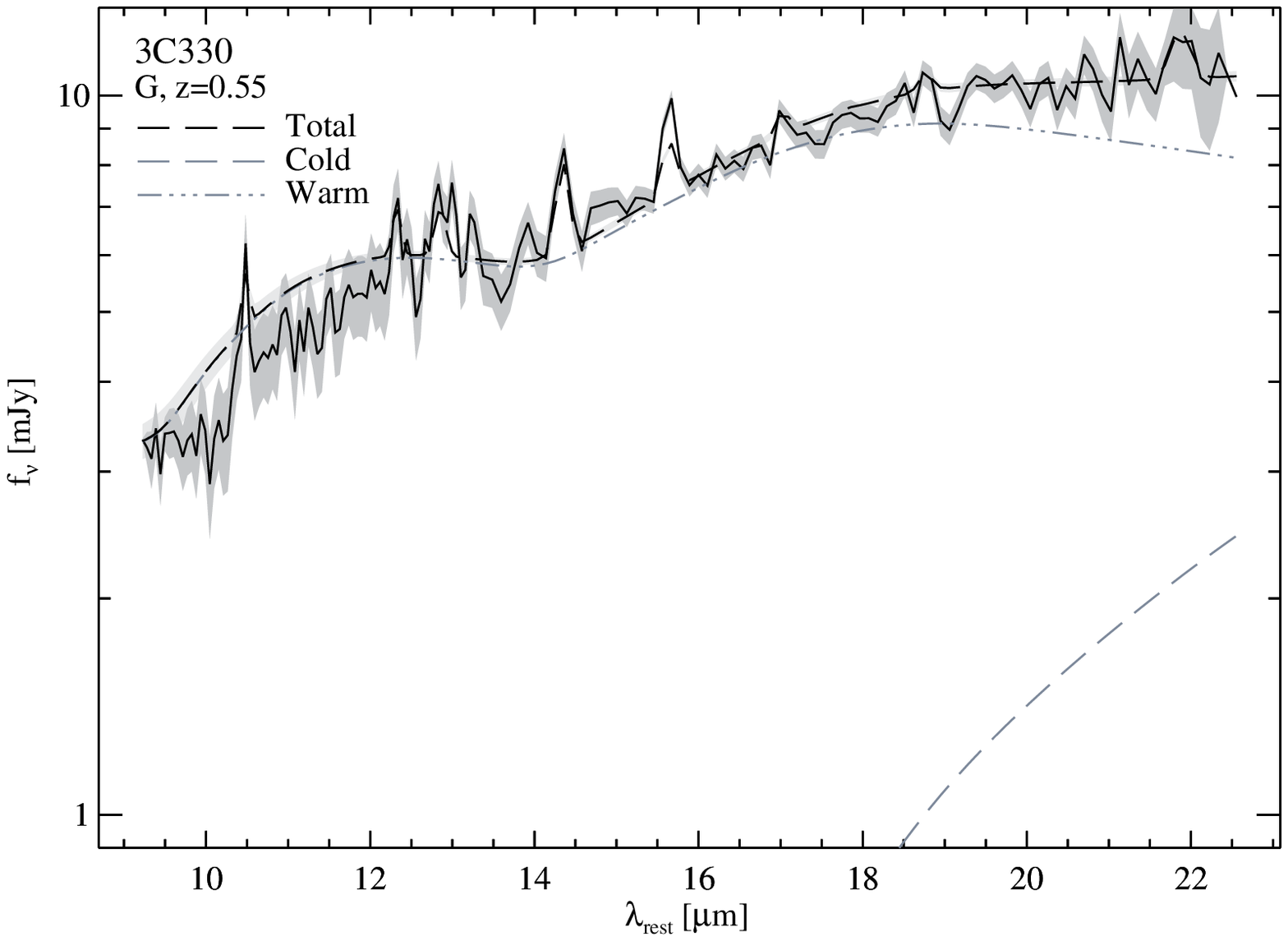,angle=0,width=8cm,height=5.5cm,clip=}\\

\psfig{file=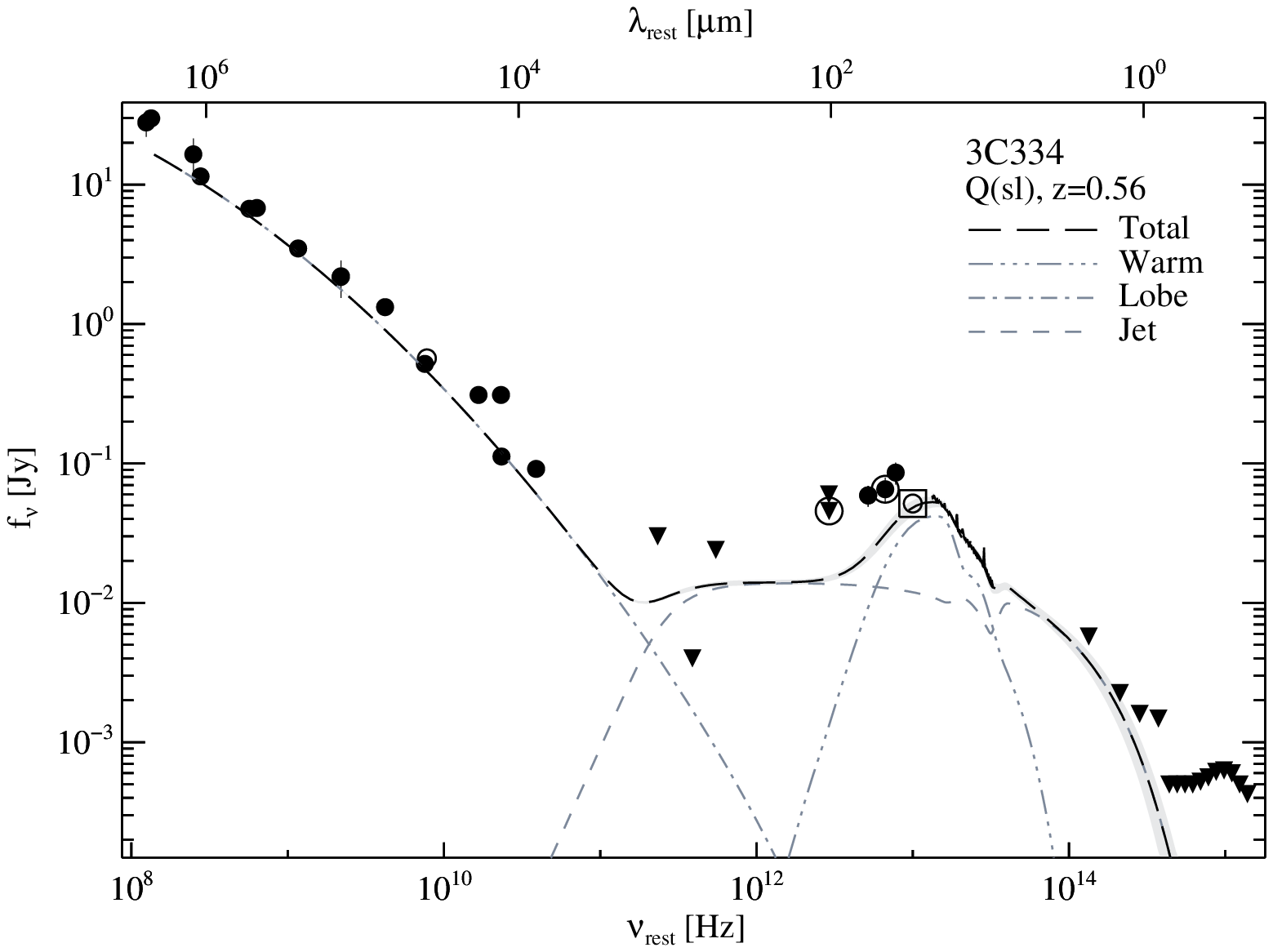,angle=0,width=8cm,height=6.0cm,clip=} & 
\psfig{file=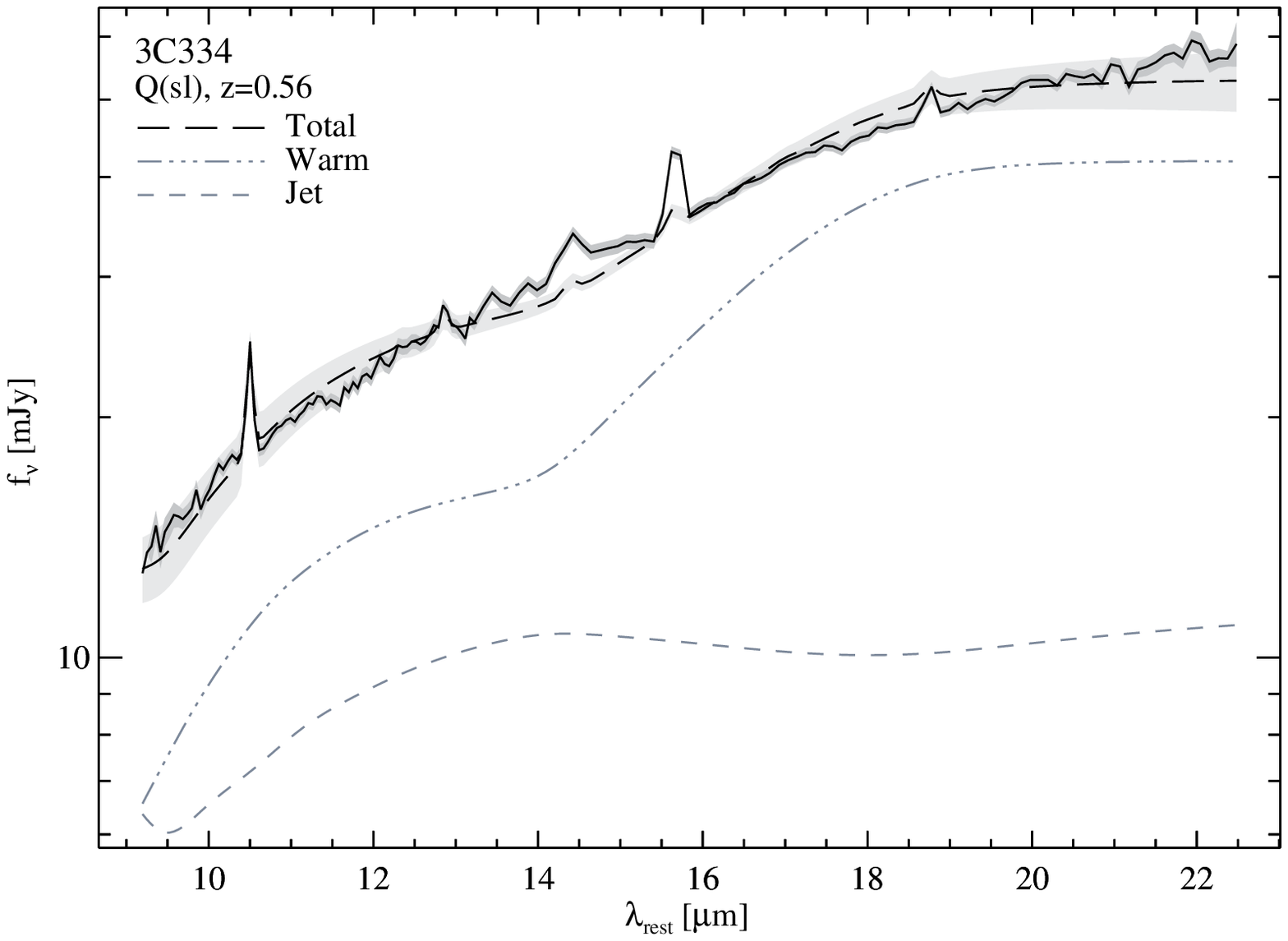,angle=0,width=8cm,height=5.5cm,clip=}\\

\psfig{file=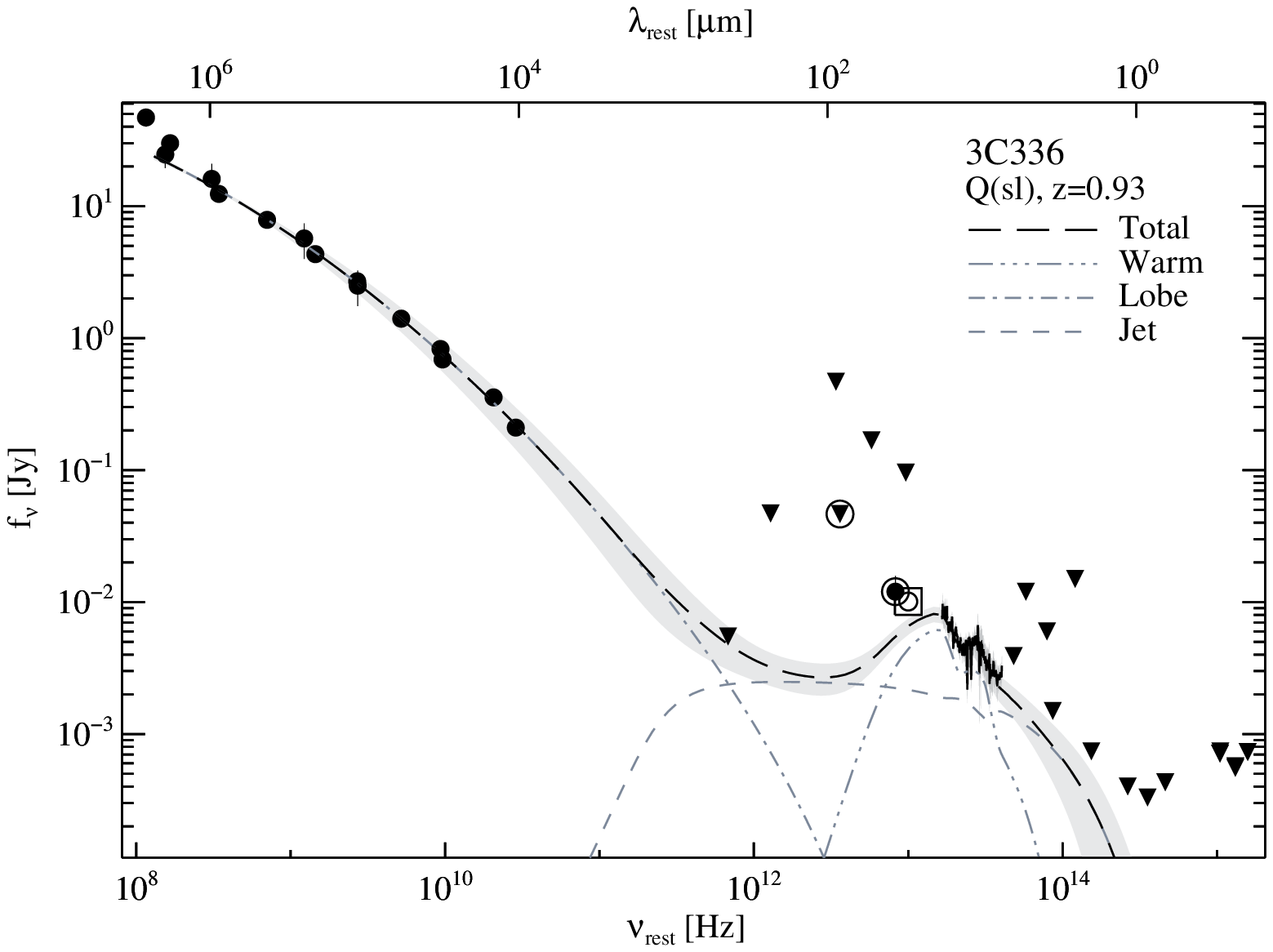,angle=0,width=8cm,height=6.0cm,clip=} & 
\psfig{file=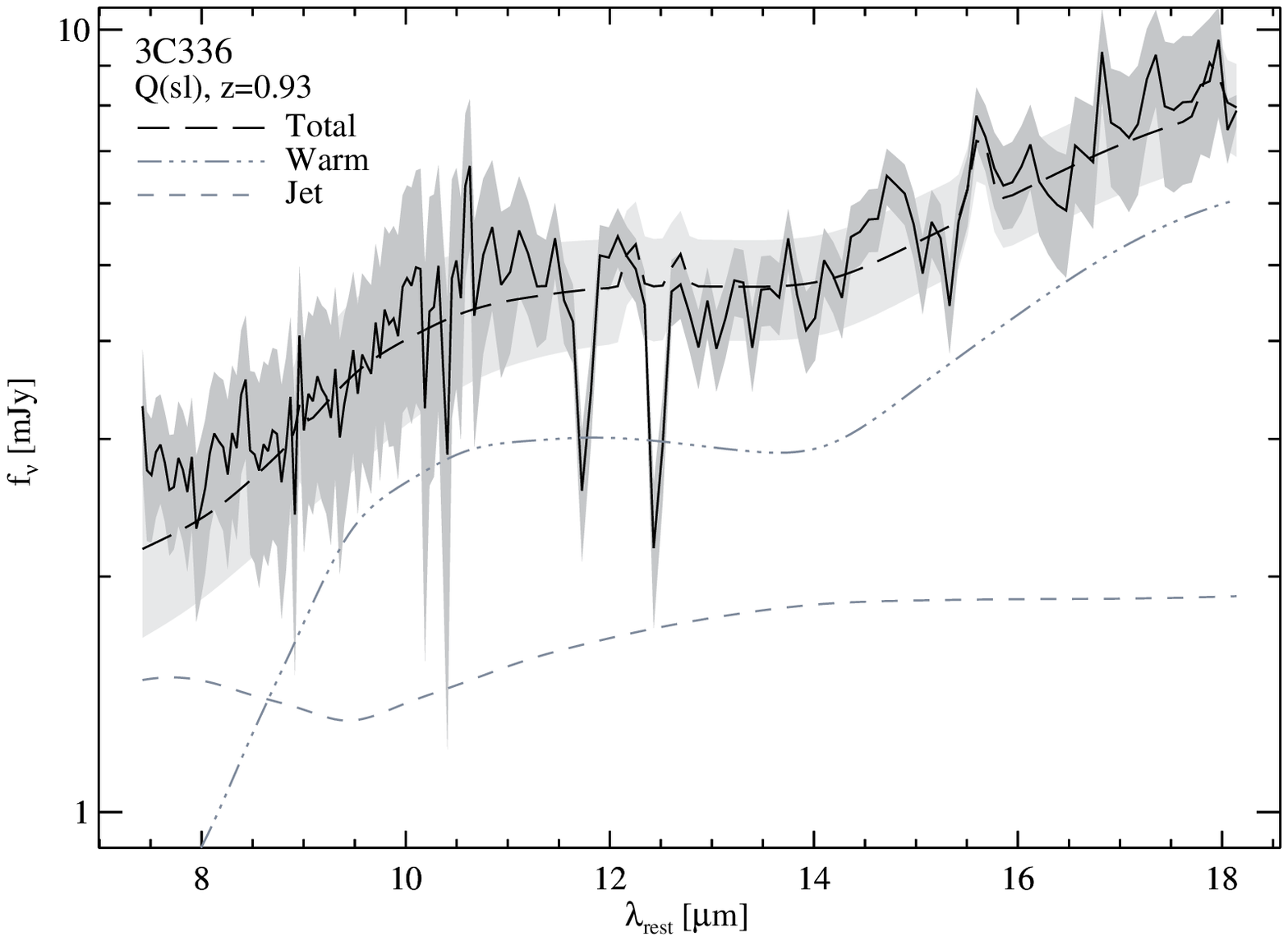,angle=0,width=8cm,height=5.5cm,clip=}\\

\end{tabular}
\caption{{\em Continued}}
\end{figure*}

\begin{figure*}
\figurenum{13}
\centering
\begin{tabular}{cc}
\tablewidth{0pt}

\psfig{file=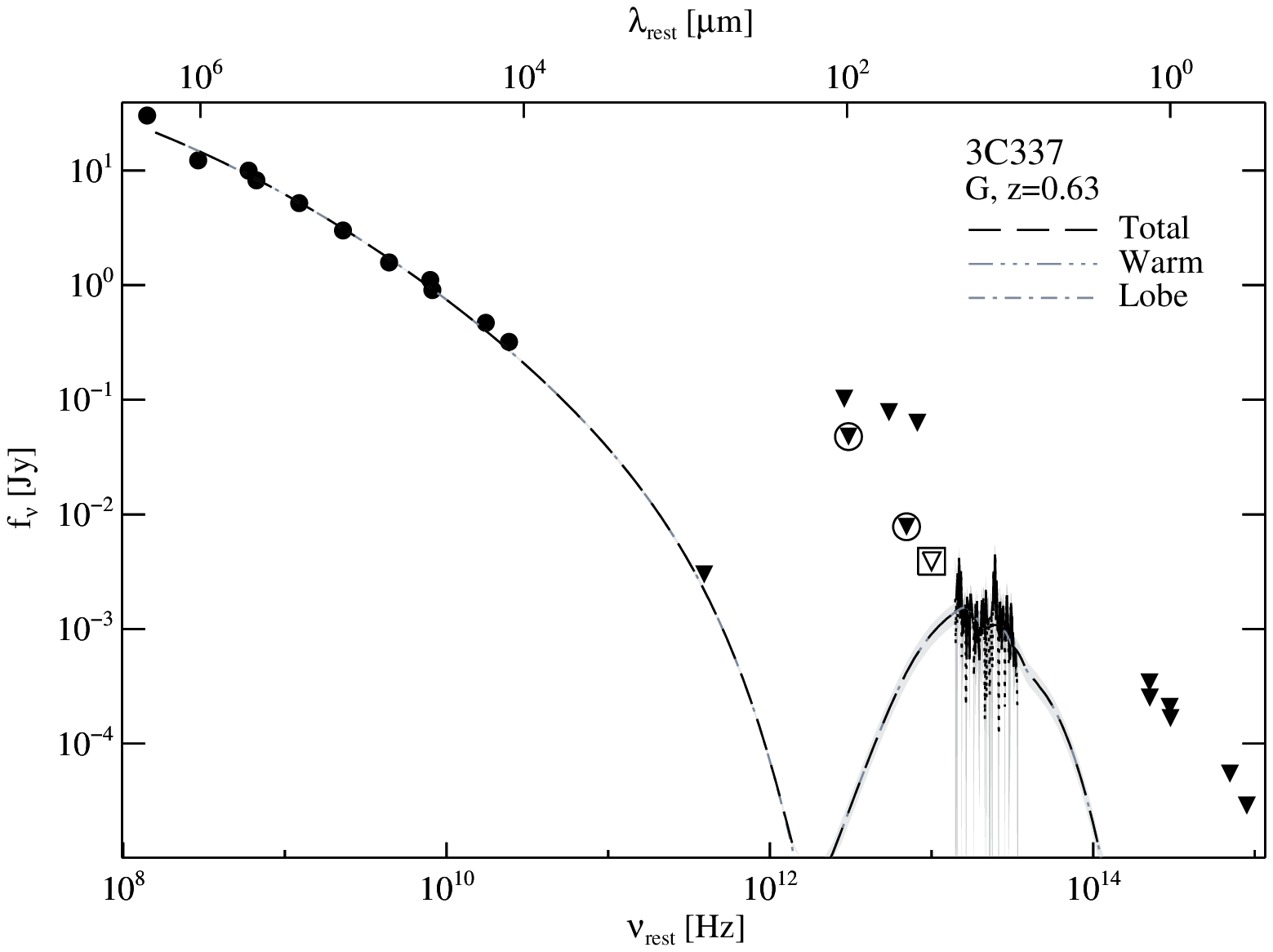,angle=0,width=8cm,height=6.0cm,clip=} & 
\psfig{file=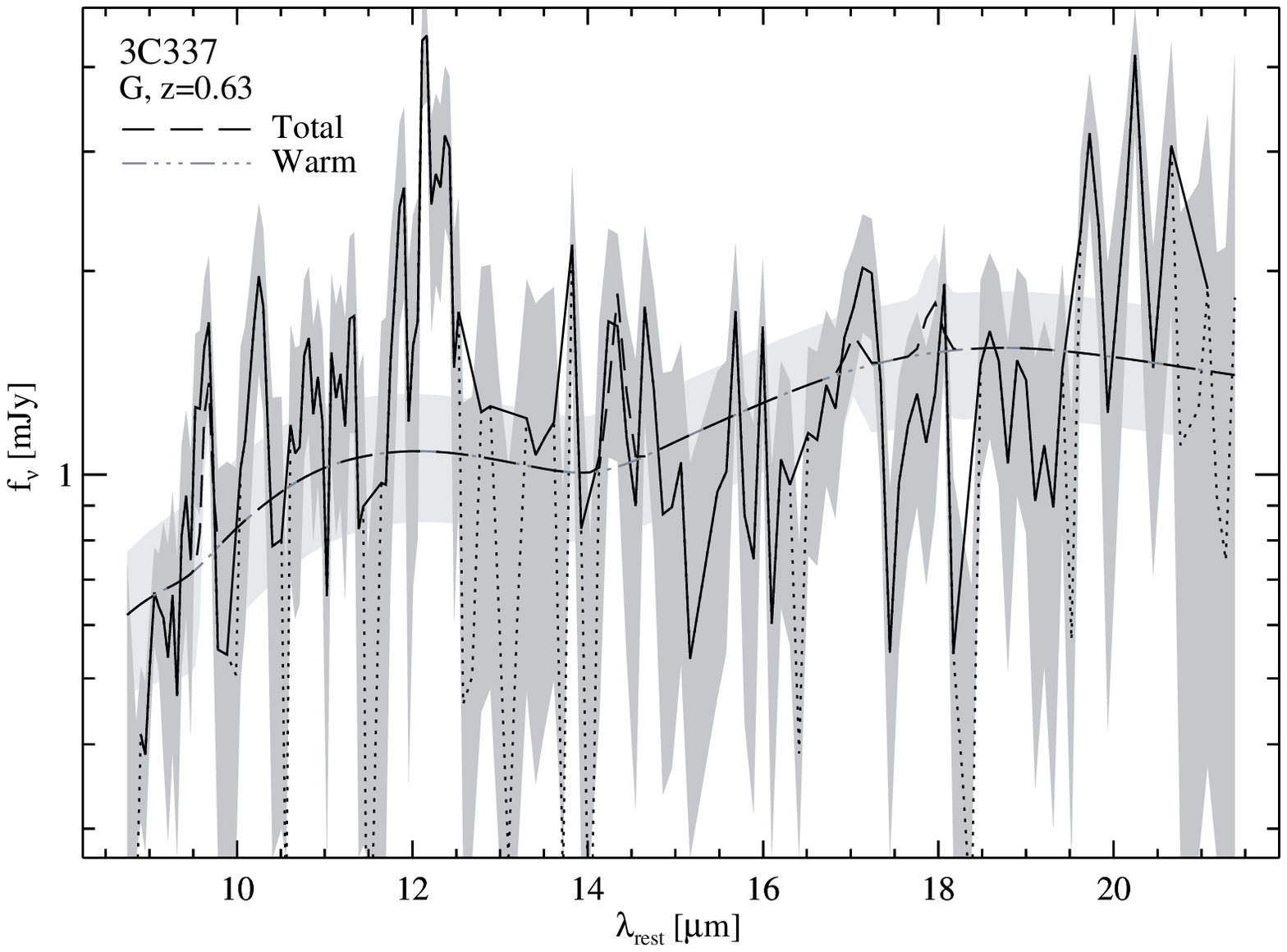,angle=0,width=8cm,height=5.5cm,clip=}\\

\psfig{file=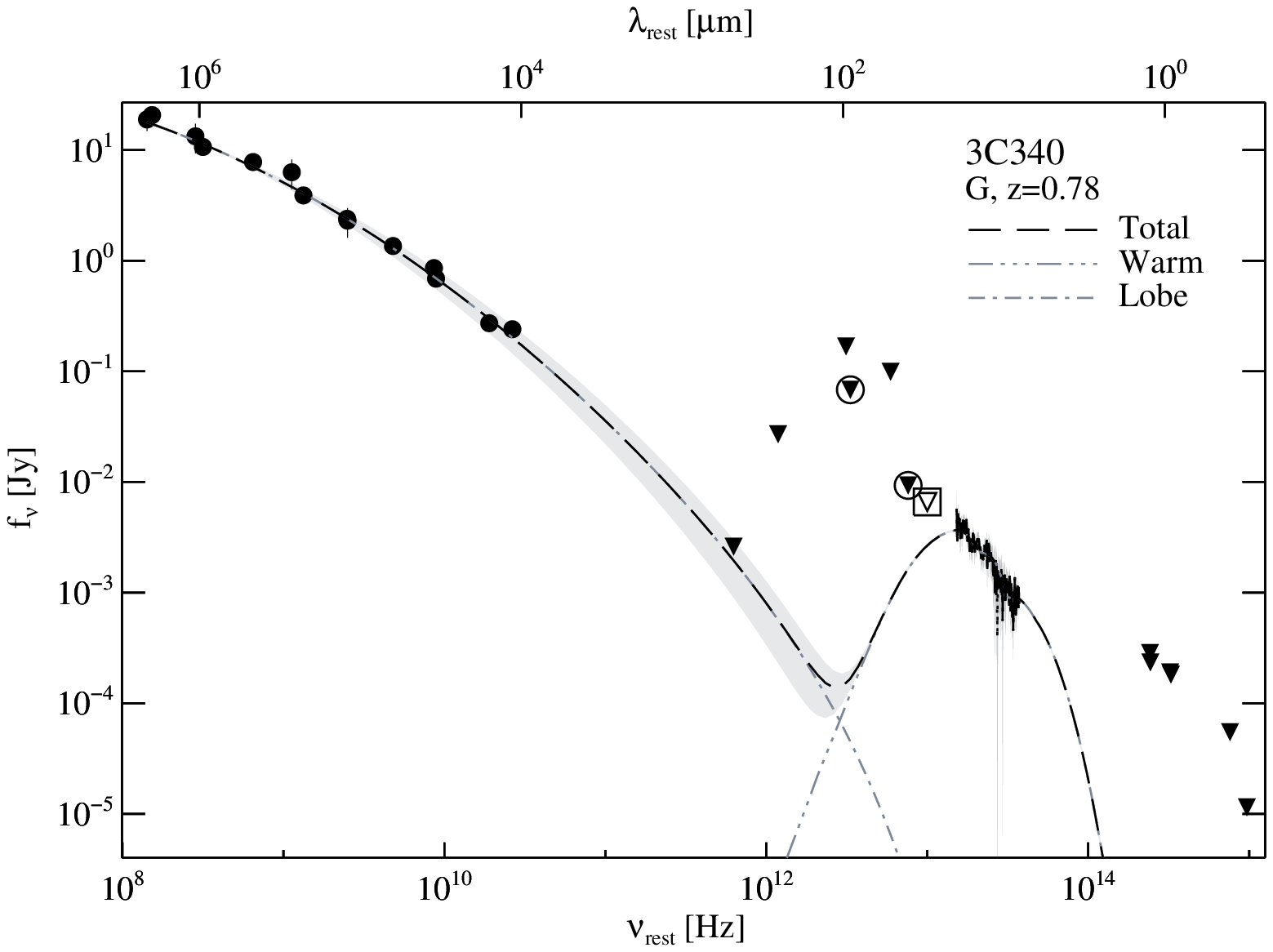,angle=0,width=8cm,height=6.0cm,clip=} & 
\psfig{file=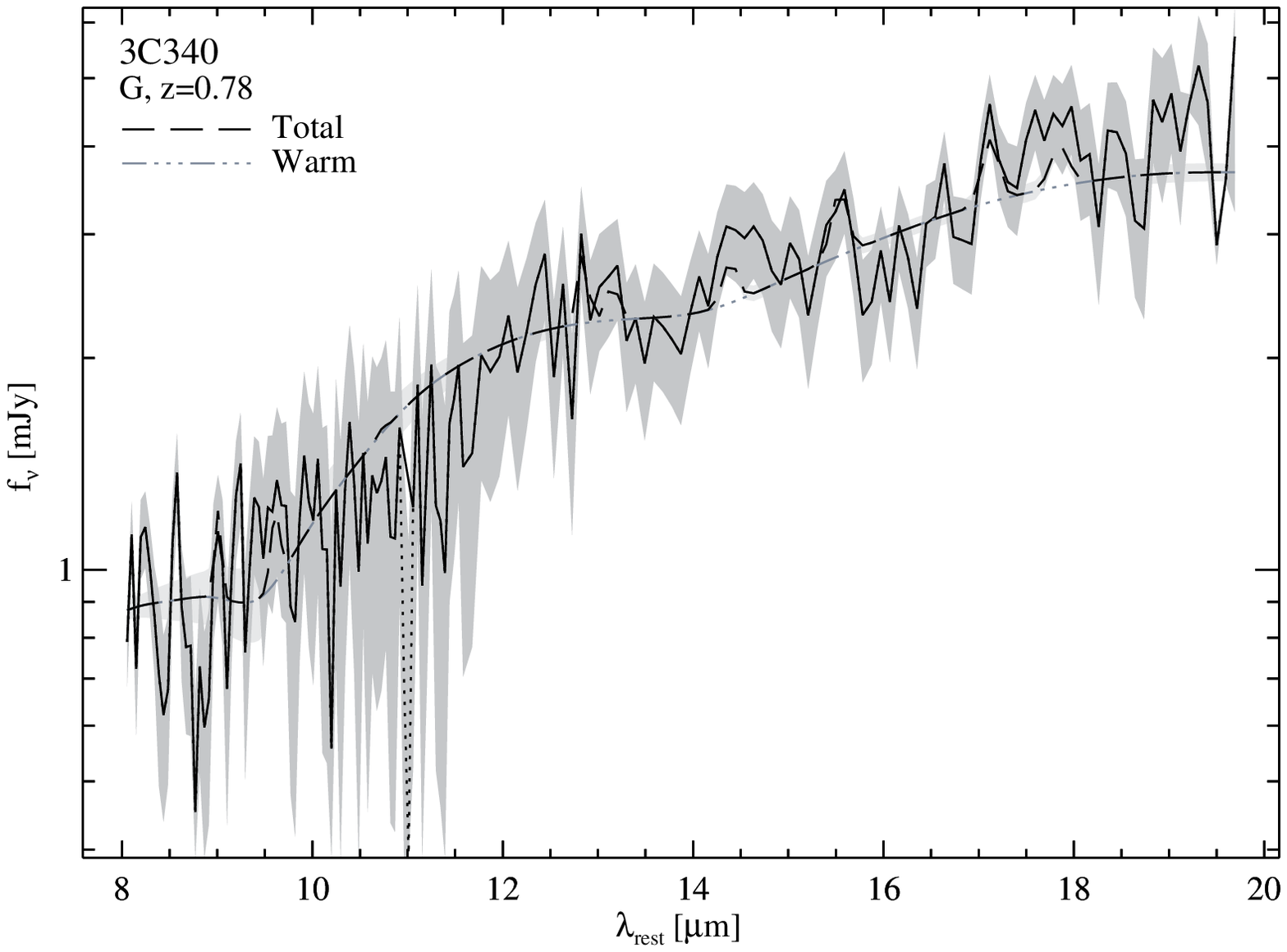,angle=0,width=8cm,height=5.5cm,clip=}\\

\psfig{file=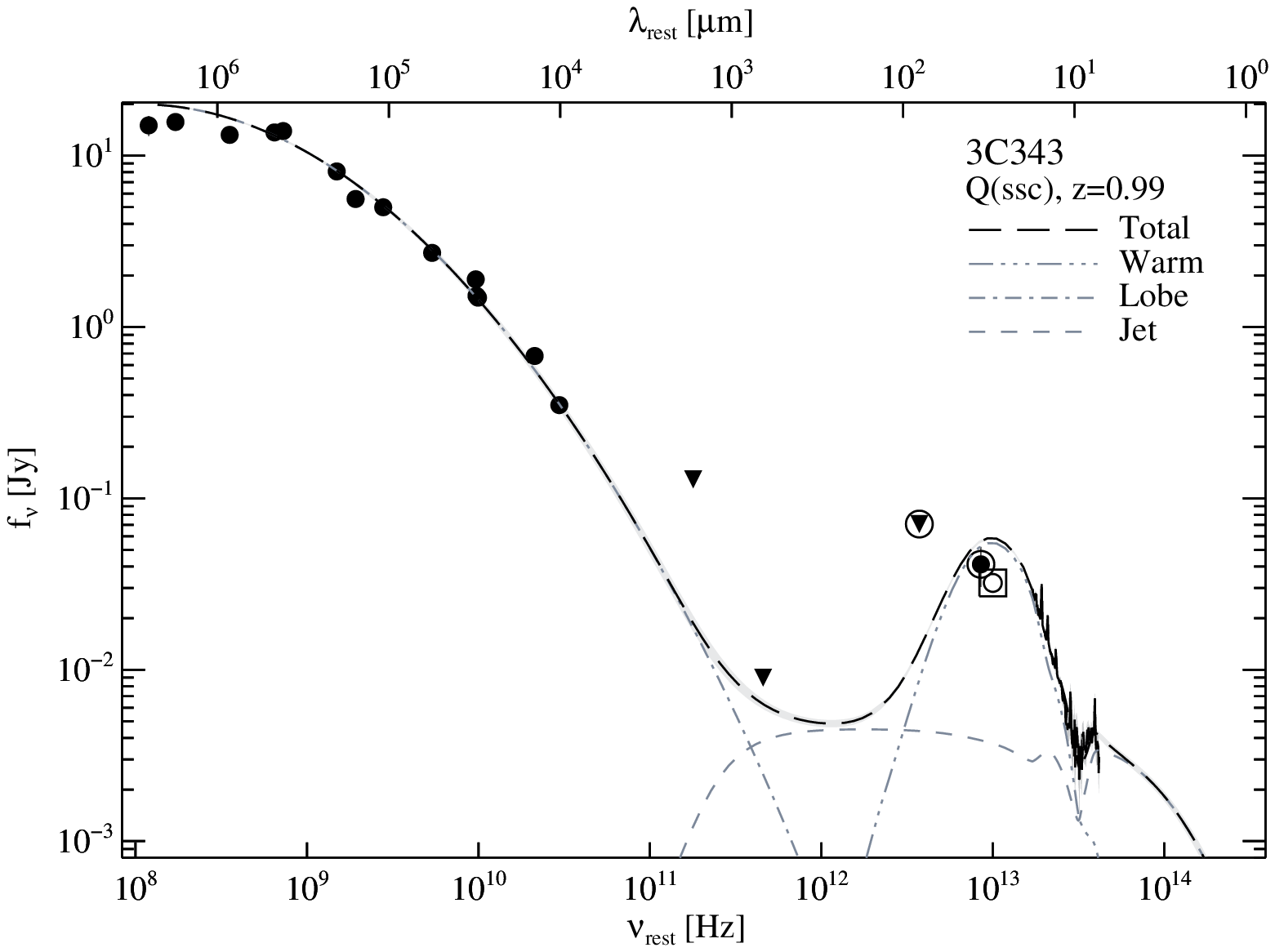,angle=0,width=8cm,height=6.0cm,clip=} & 
\psfig{file=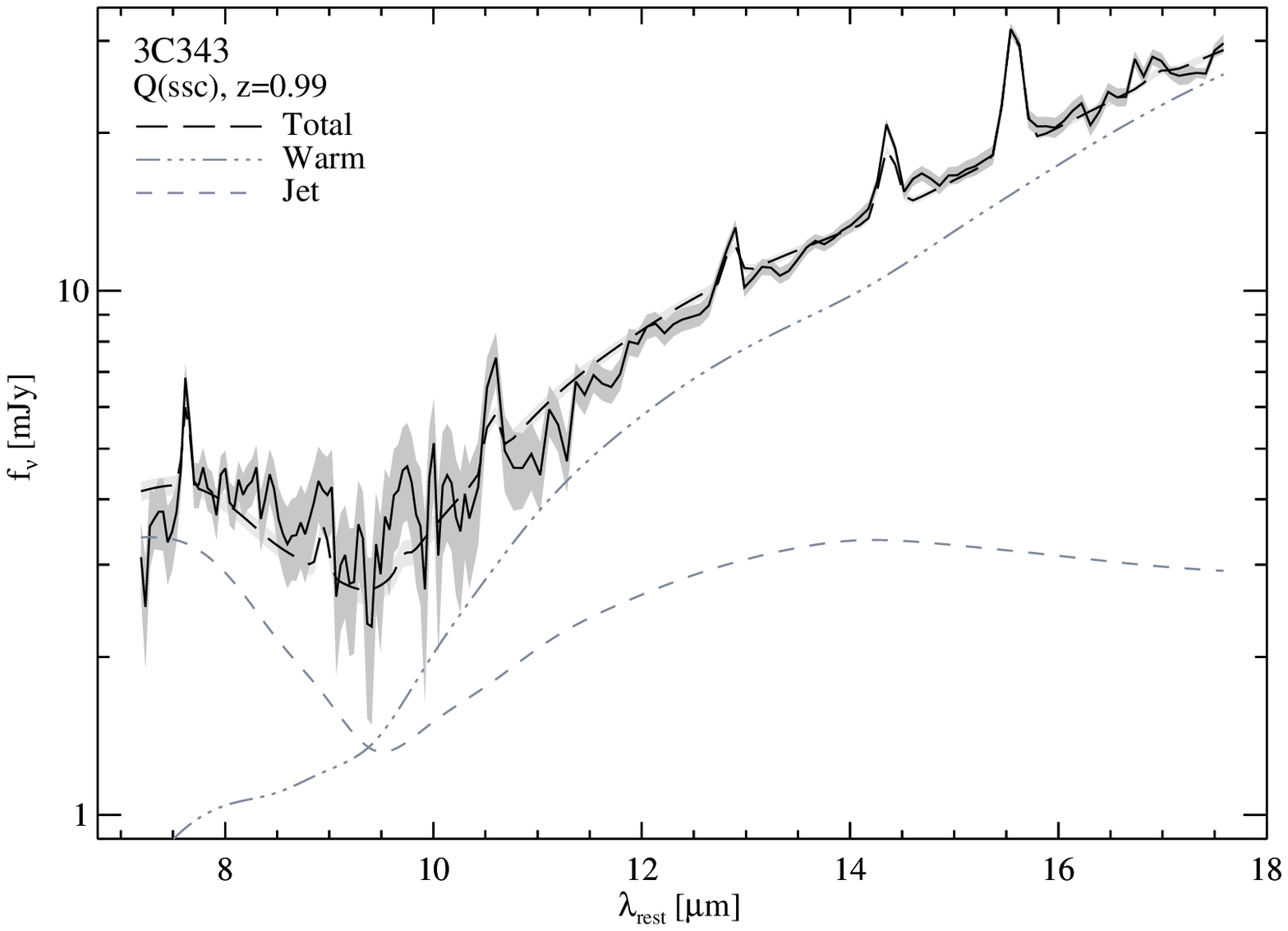,angle=0,width=8cm,height=5.5cm,clip=}\\

\end{tabular}
\caption{{\em Continued}}
\end{figure*}
\clearpage

\begin{figure*}[!ht]
\figurenum{13}
\centering
\begin{tabular}{cc}
\tablewidth{0pt}

\psfig{file=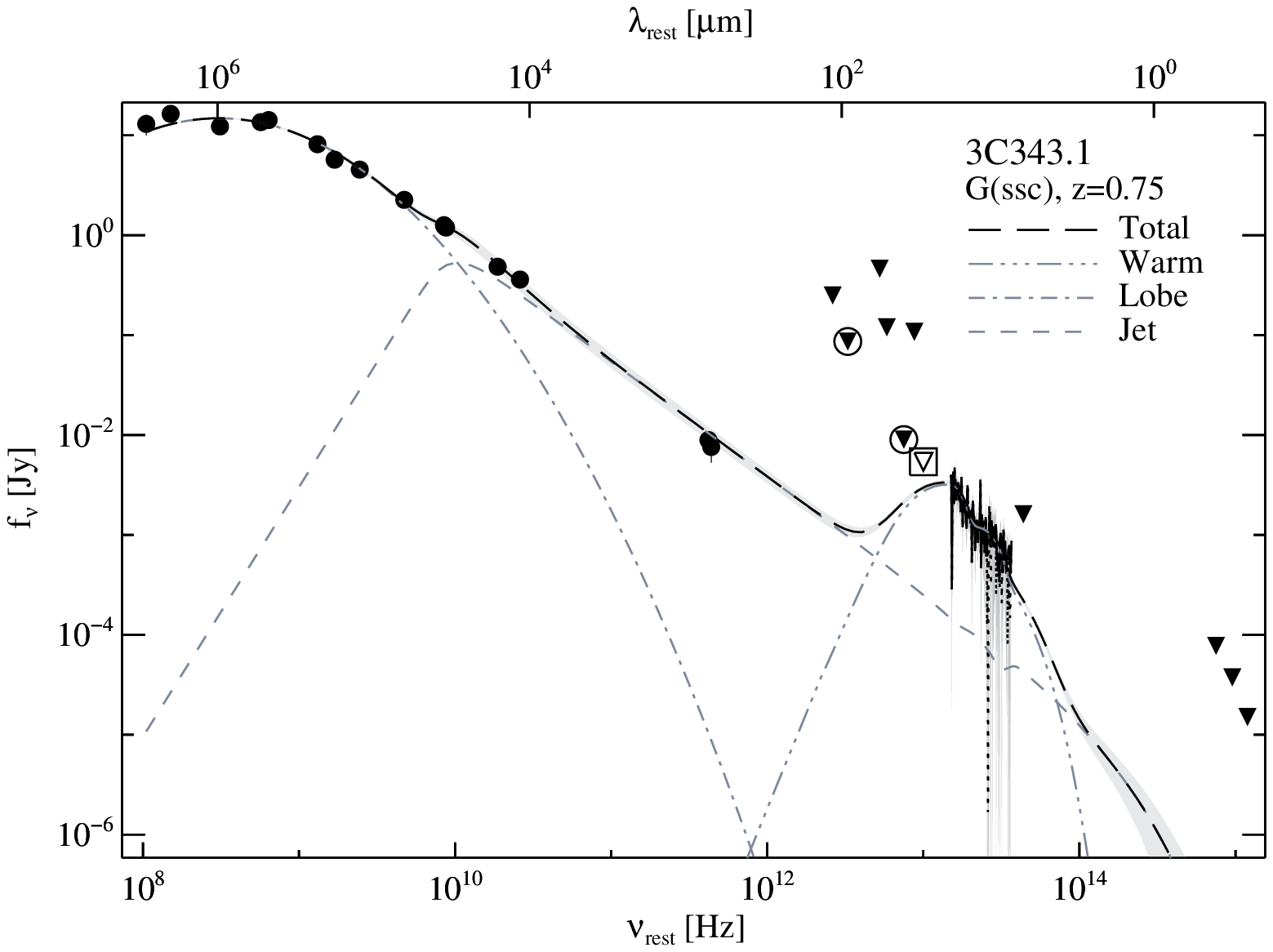,angle=0,width=8cm,height=6.0cm,clip=} & 
\psfig{file=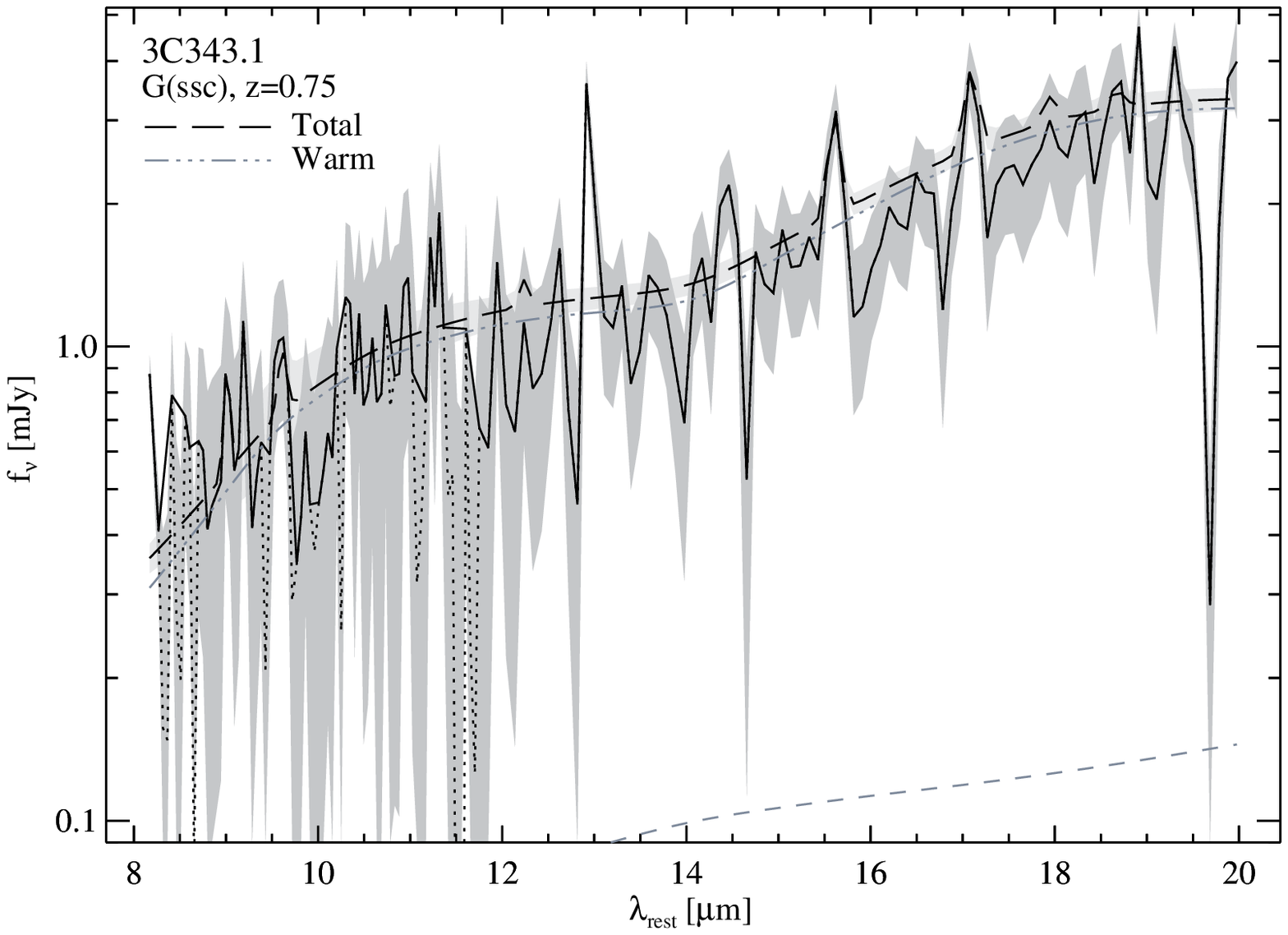,angle=0,width=8cm,height=5.5cm,clip=}\\

\psfig{file=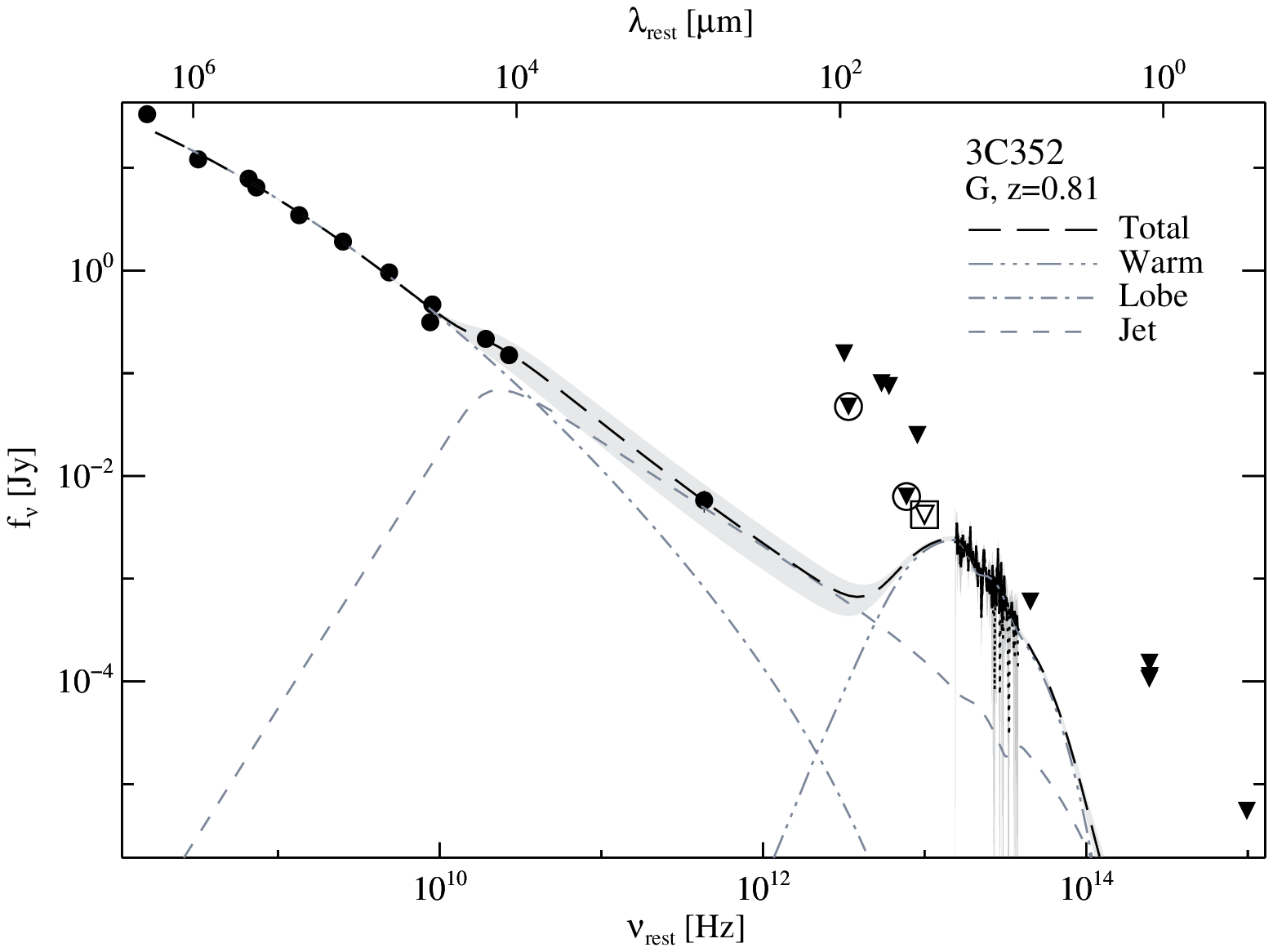,angle=0,width=8cm,height=6.0cm,clip=} & 
\psfig{file=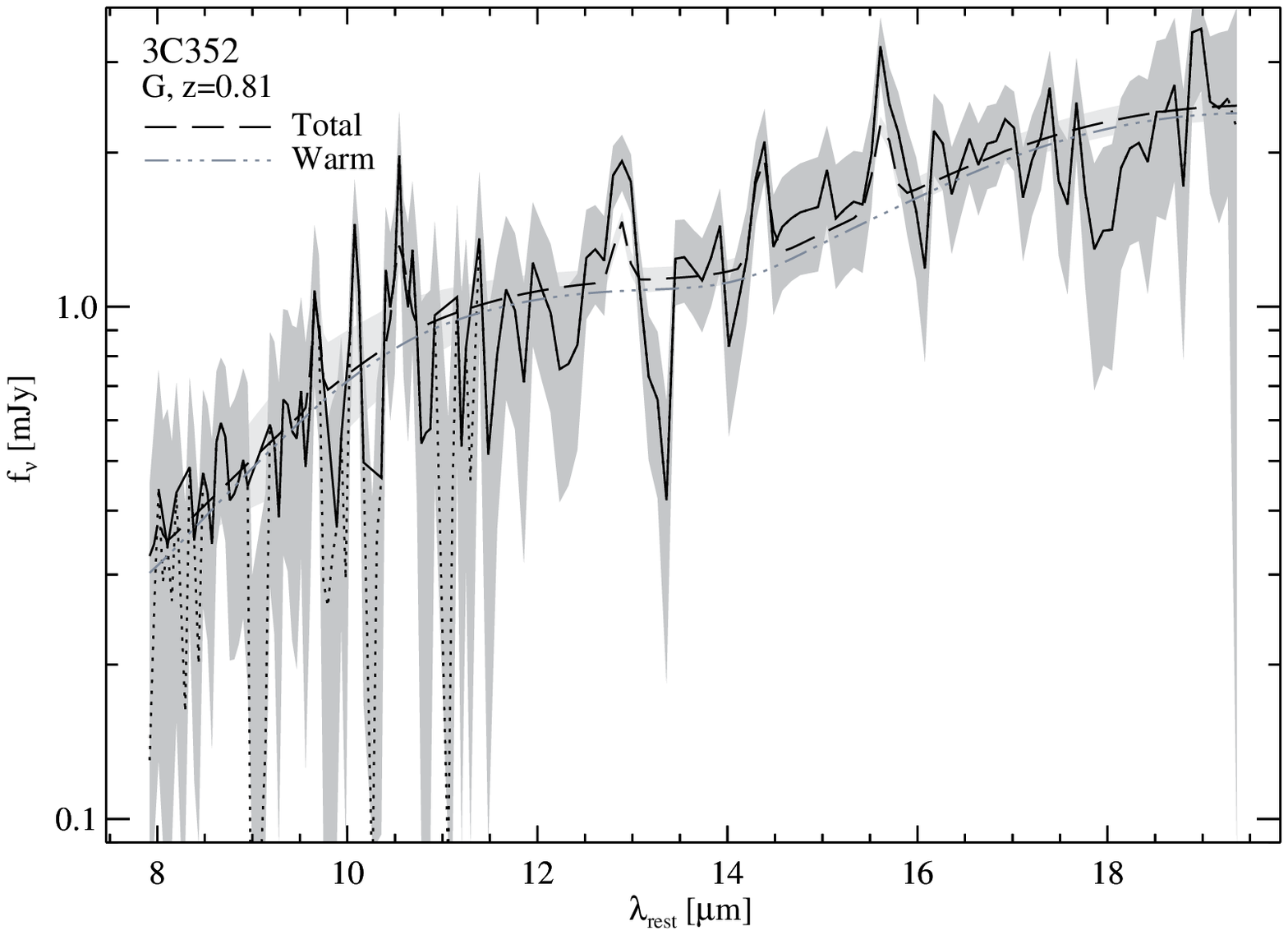,angle=0,width=8cm,height=5.5cm,clip=}\\

\psfig{file=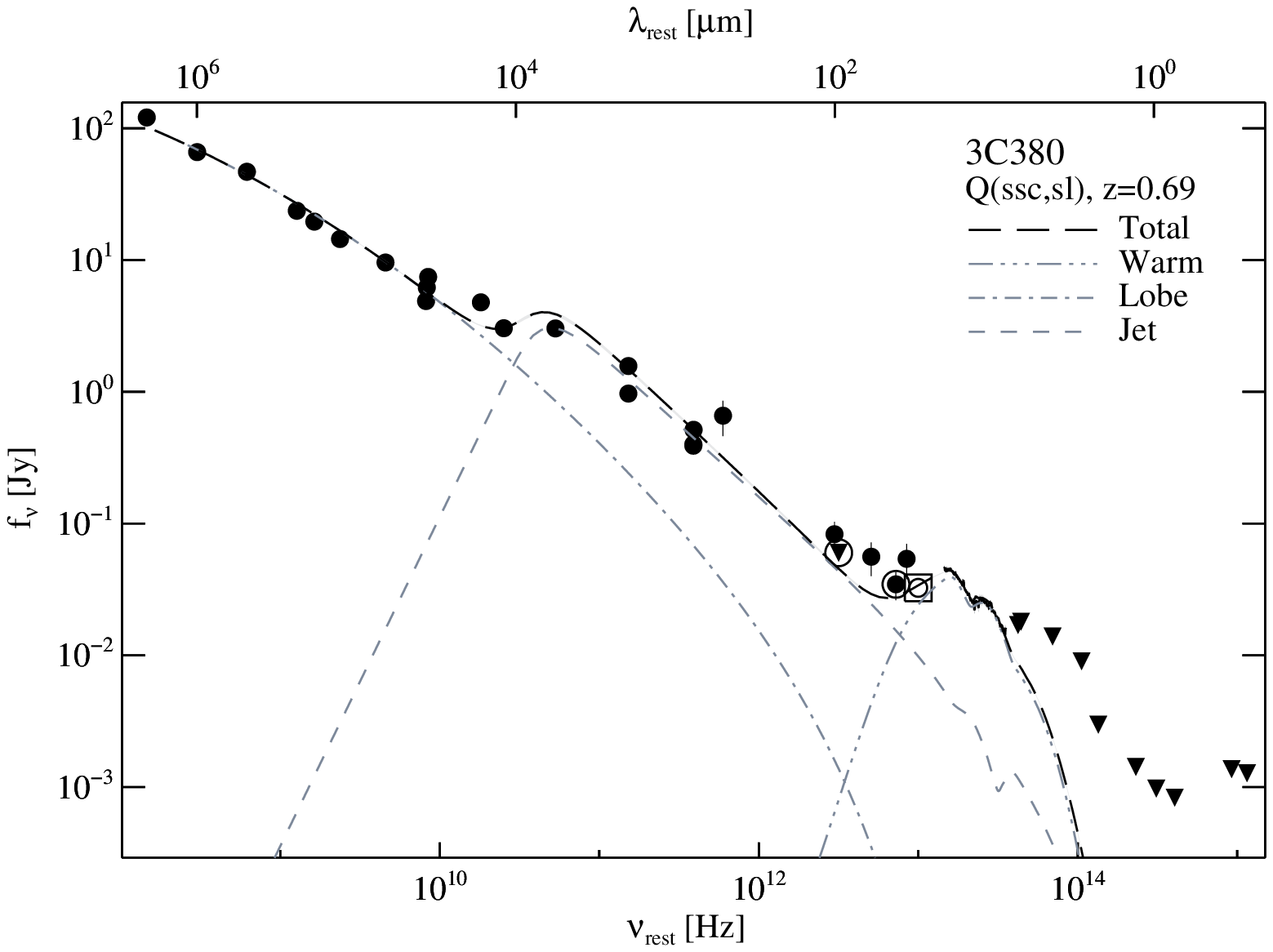,angle=0,width=8cm,height=6.0cm,clip=} & 
\psfig{file=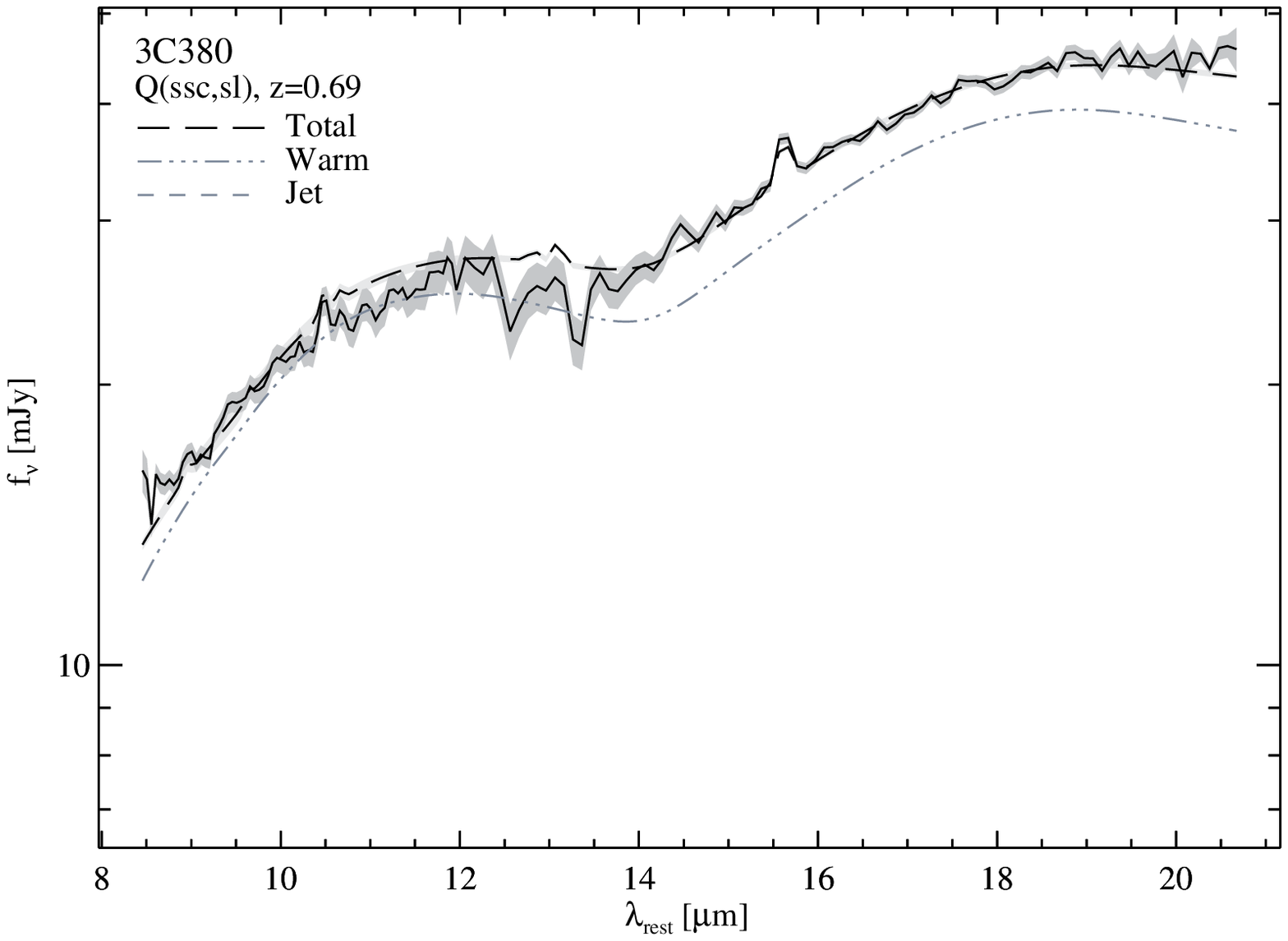,angle=0,width=8cm,height=5.5cm,clip=}\\

\end{tabular}
\caption{{\em Continued}}
\end{figure*}

\begin{figure*}
\figurenum{13}
\centering
\begin{tabular}{cc}
\tablewidth{0pt}

\psfig{file=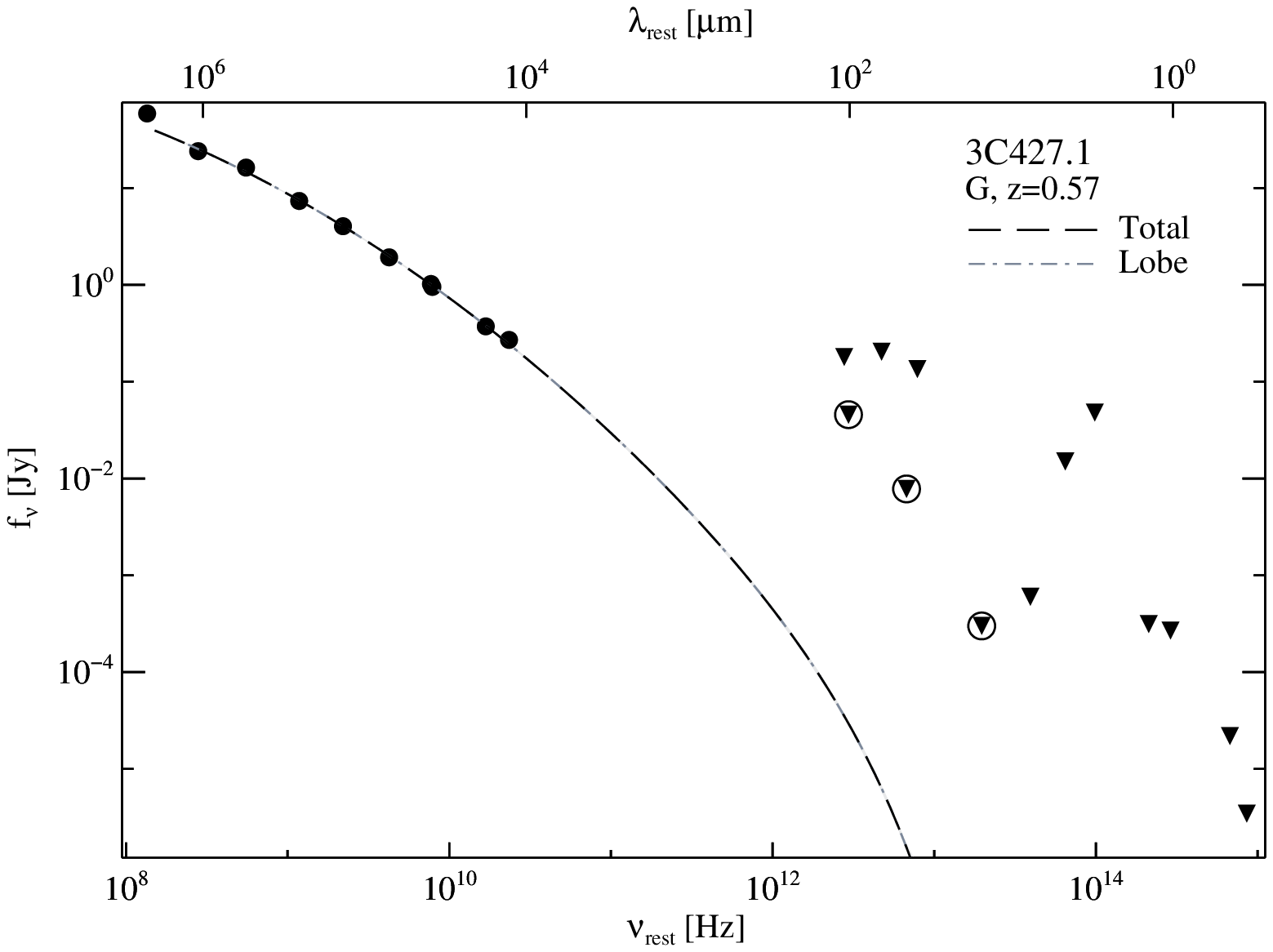,angle=0,width=8cm,height=6.0cm,clip=} & 
\psfig{file=f13bn.eps,angle=0,width=8cm,height=5.5cm,clip=}\\

\psfig{file=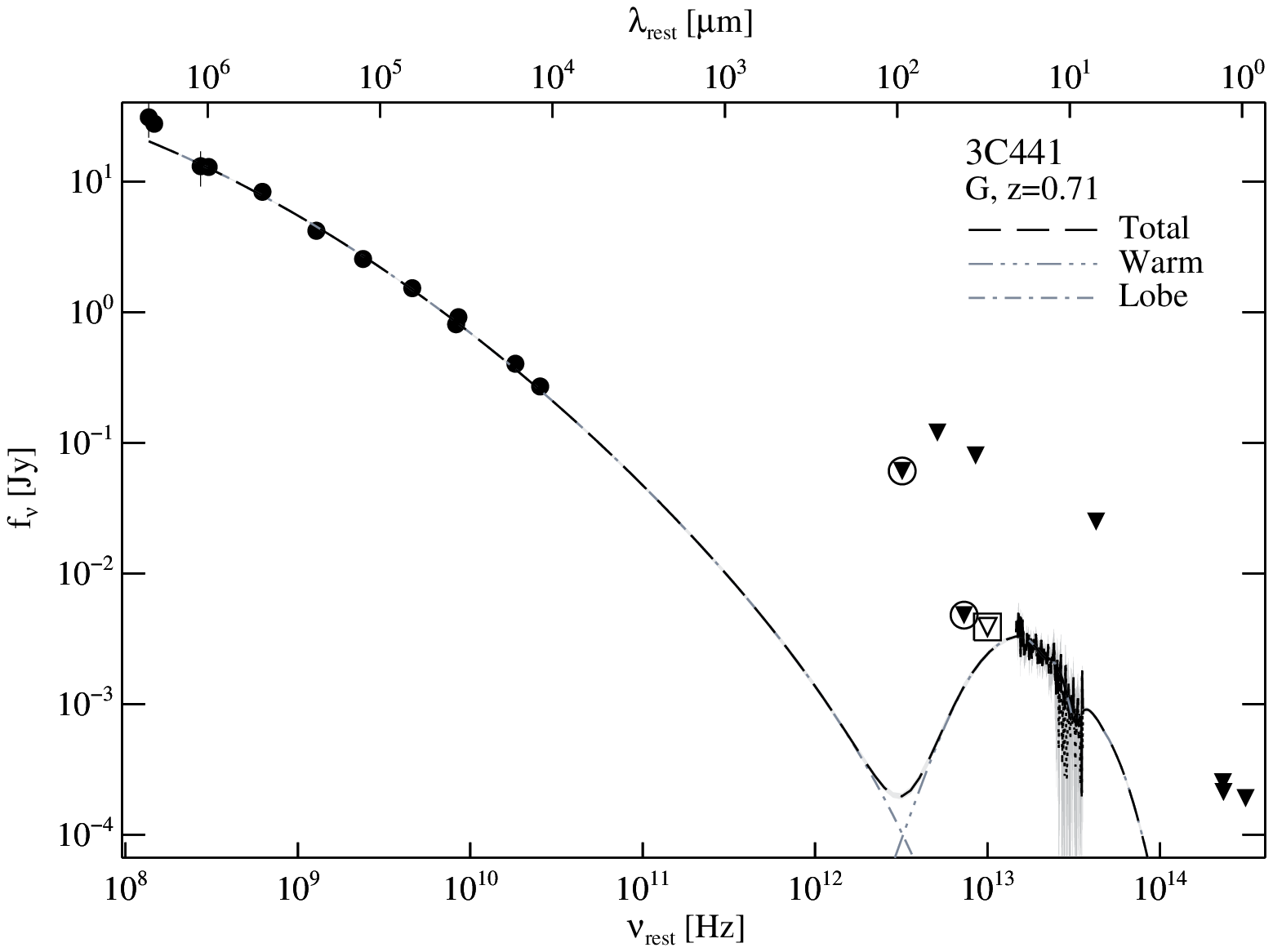,angle=0,width=8cm,height=6.0cm,clip=} & 
\psfig{file=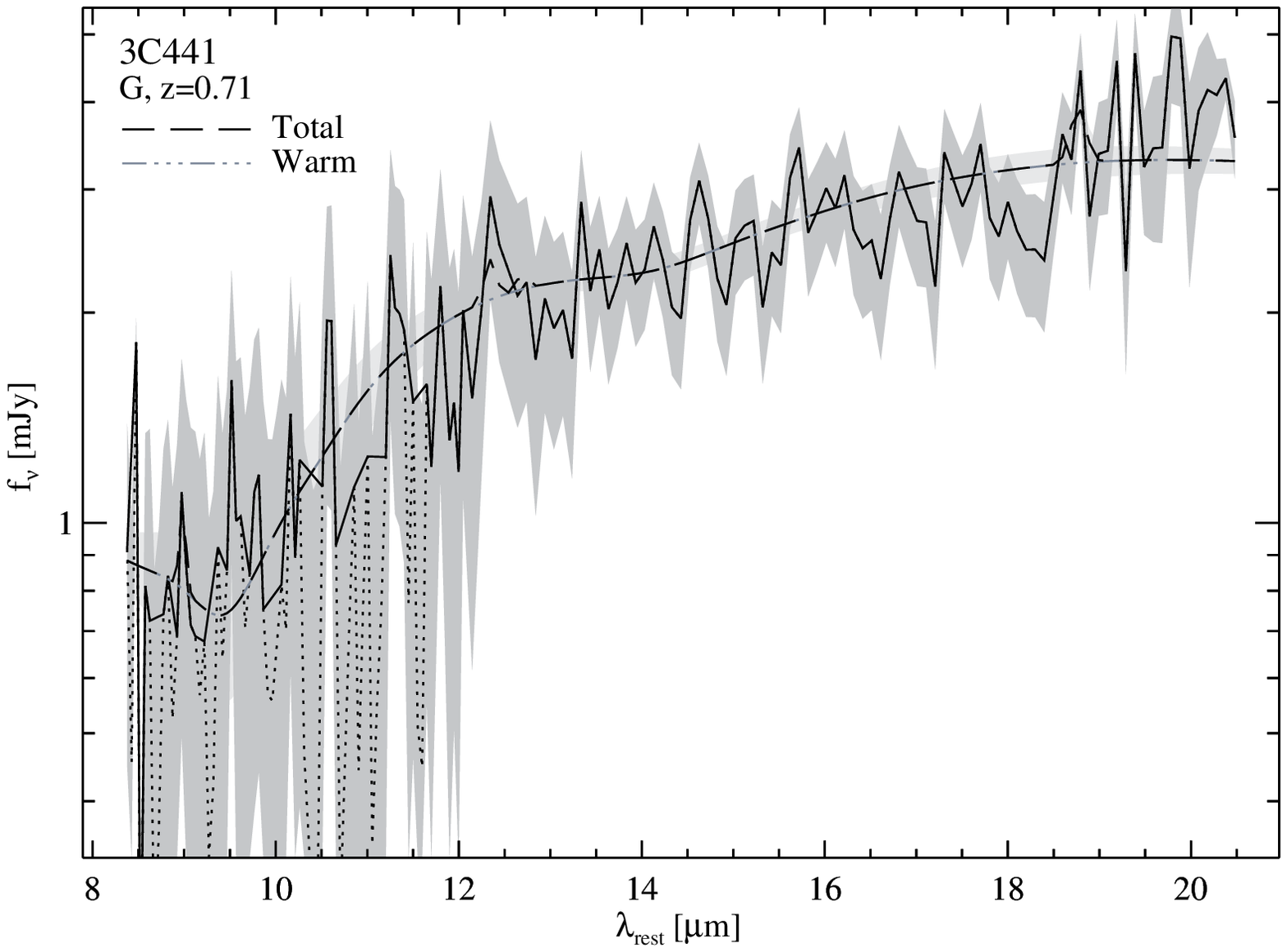,angle=0,width=8cm,height=5.5cm,clip=}\\

\end{tabular}
\caption{{\em Continued}}
\end{figure*}

\begin{figure*}
\figurenum{14}
\centering
\begin{tabular}{cc}
\tablewidth{0pt}

\psfig{file=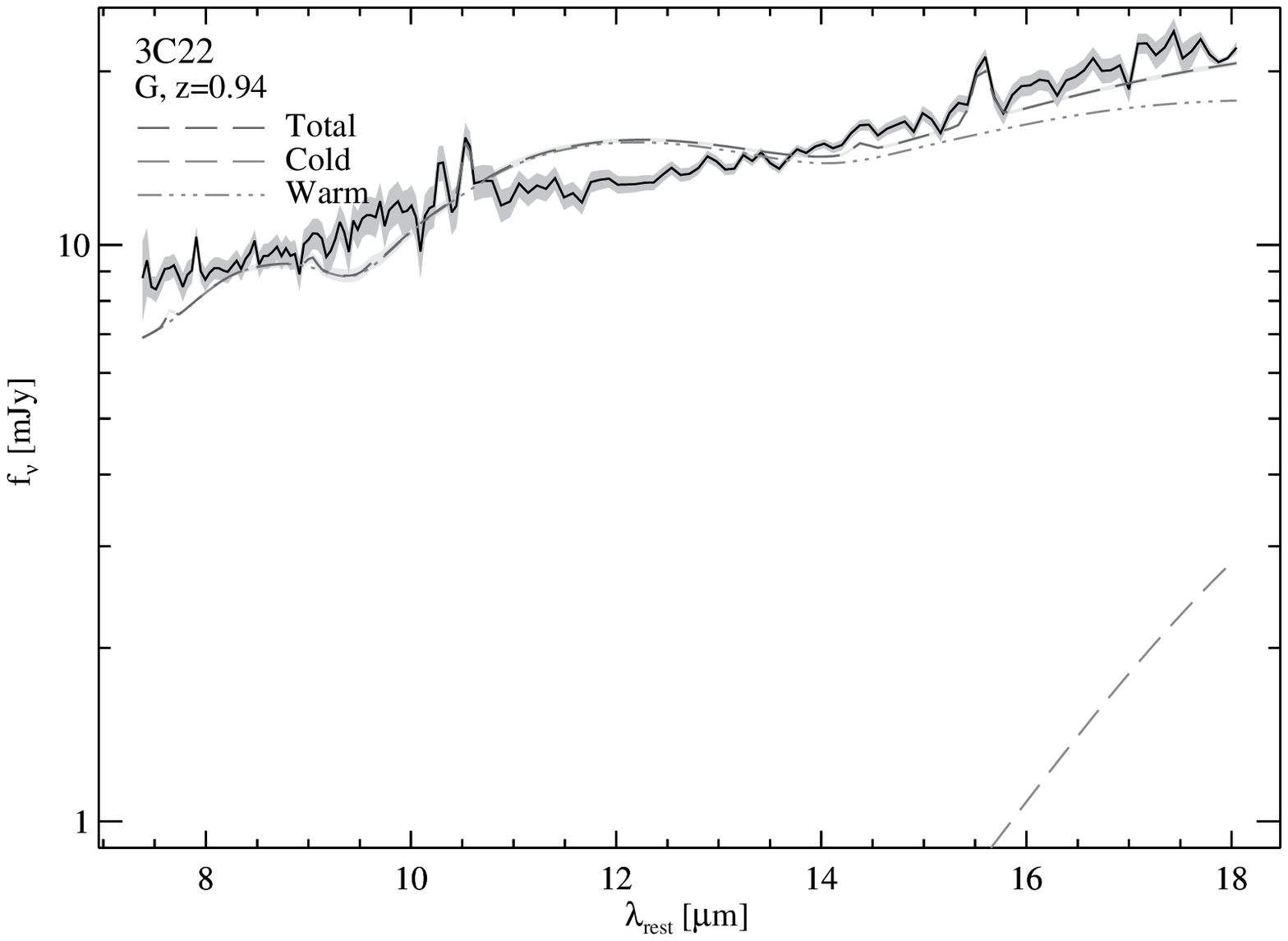,angle=0,width=8cm,height=6.0cm,clip=} & 
\psfig{file=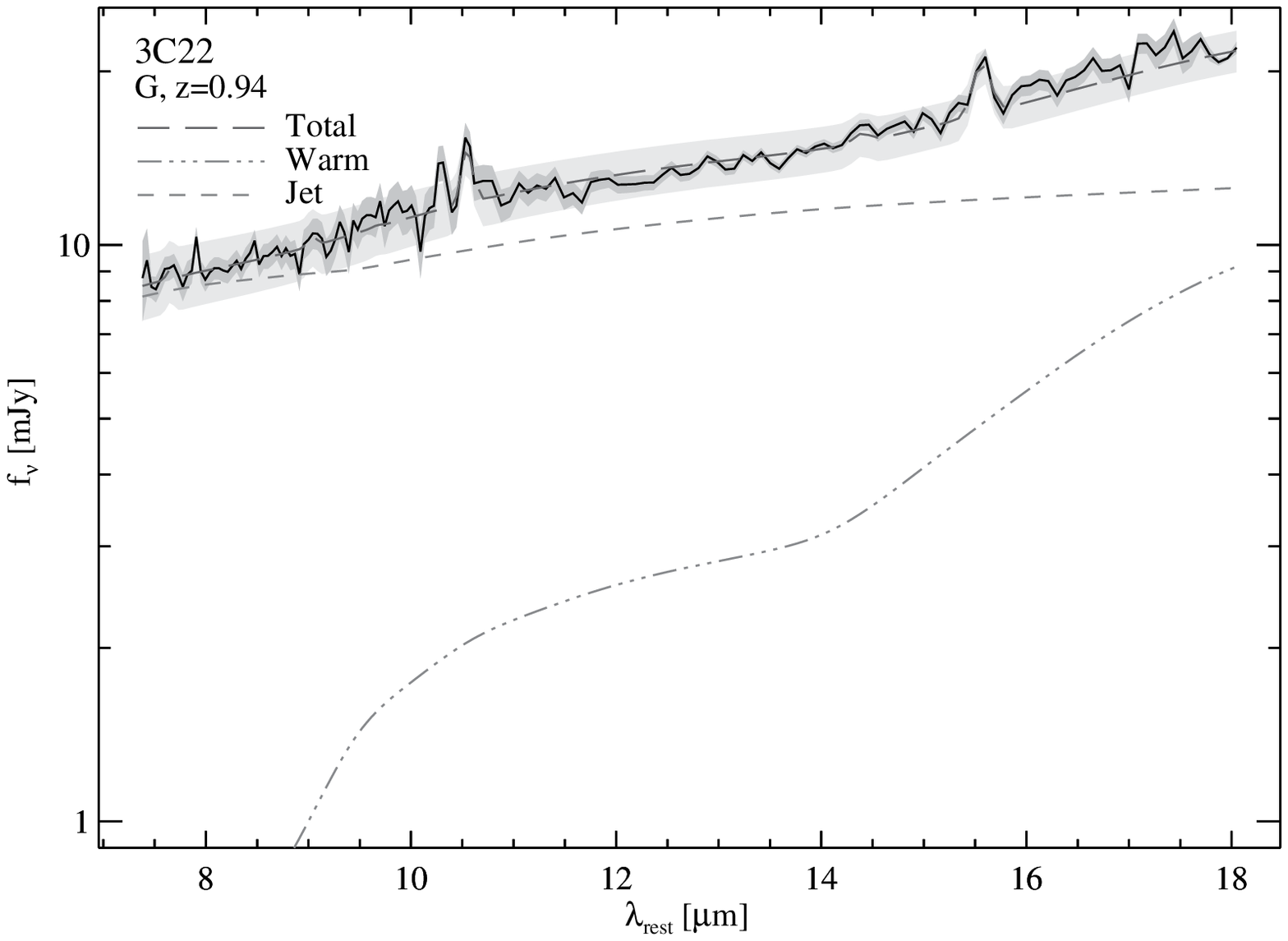,angle=0,width=8cm,height=6.0cm,clip=}\\

\psfig{file=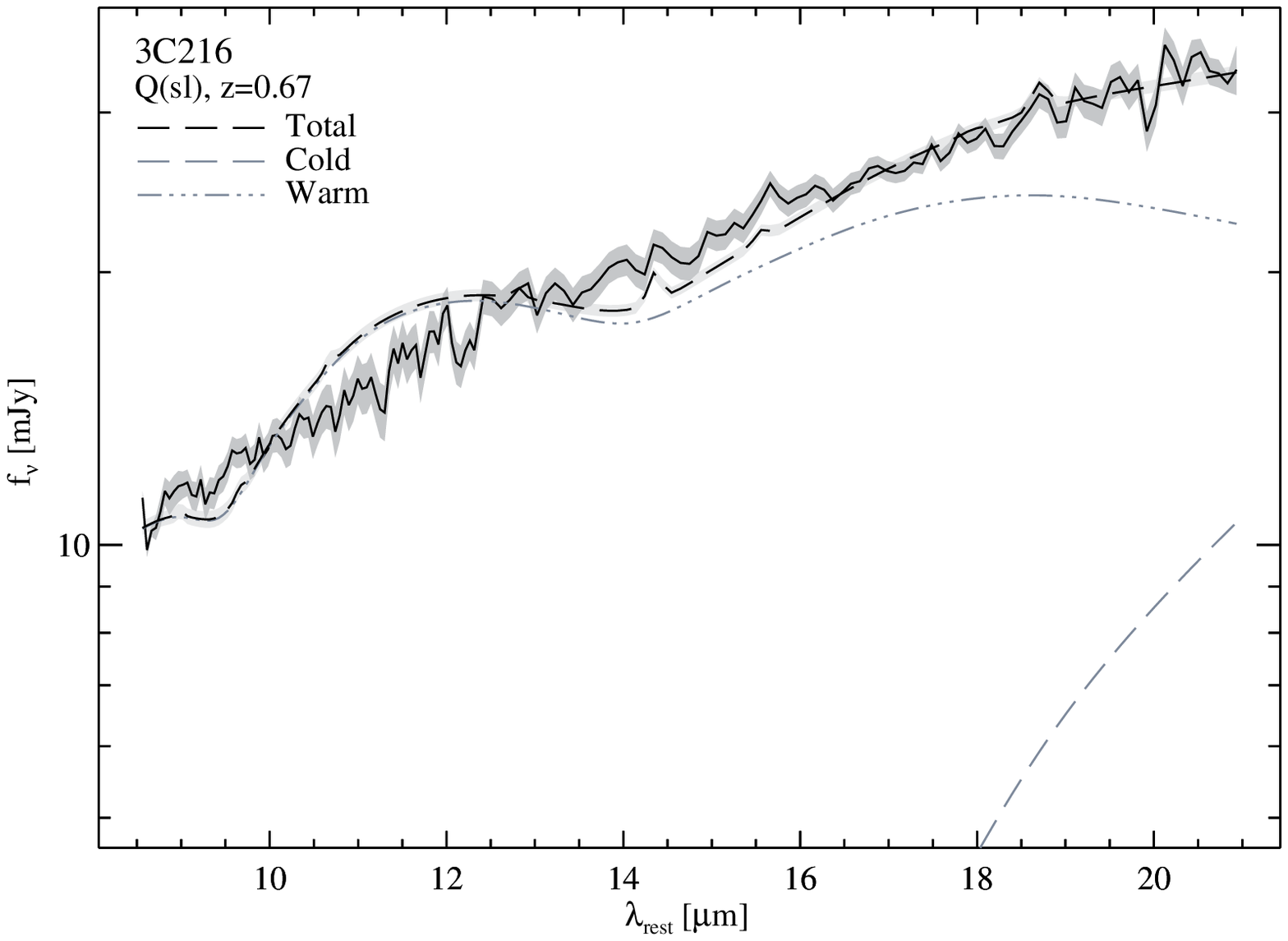,angle=0,width=8cm,height=6.0cm,clip=} & 
\psfig{file=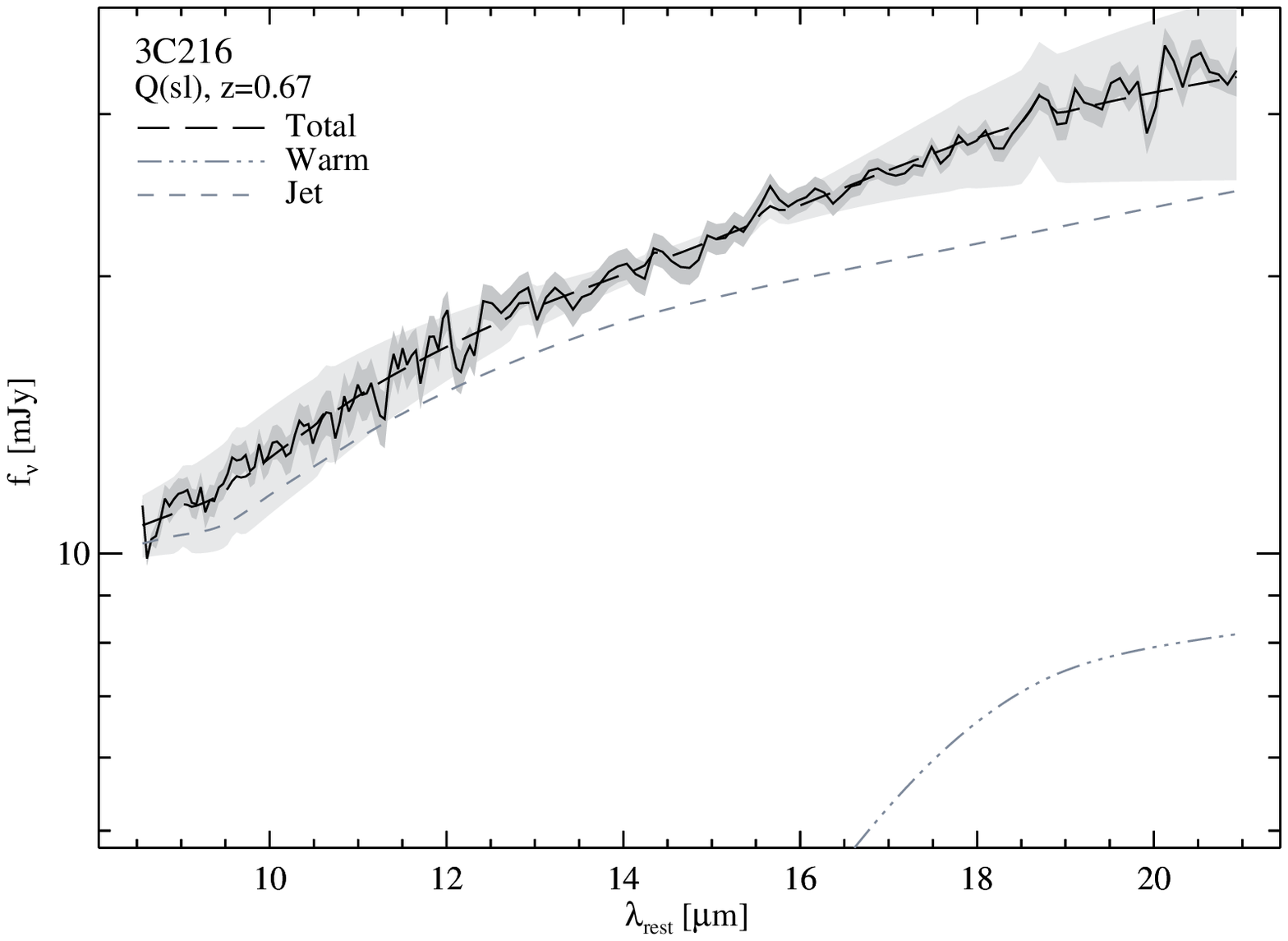,angle=0,width=8cm,height=6.0cm,clip=}\\

\psfig{file=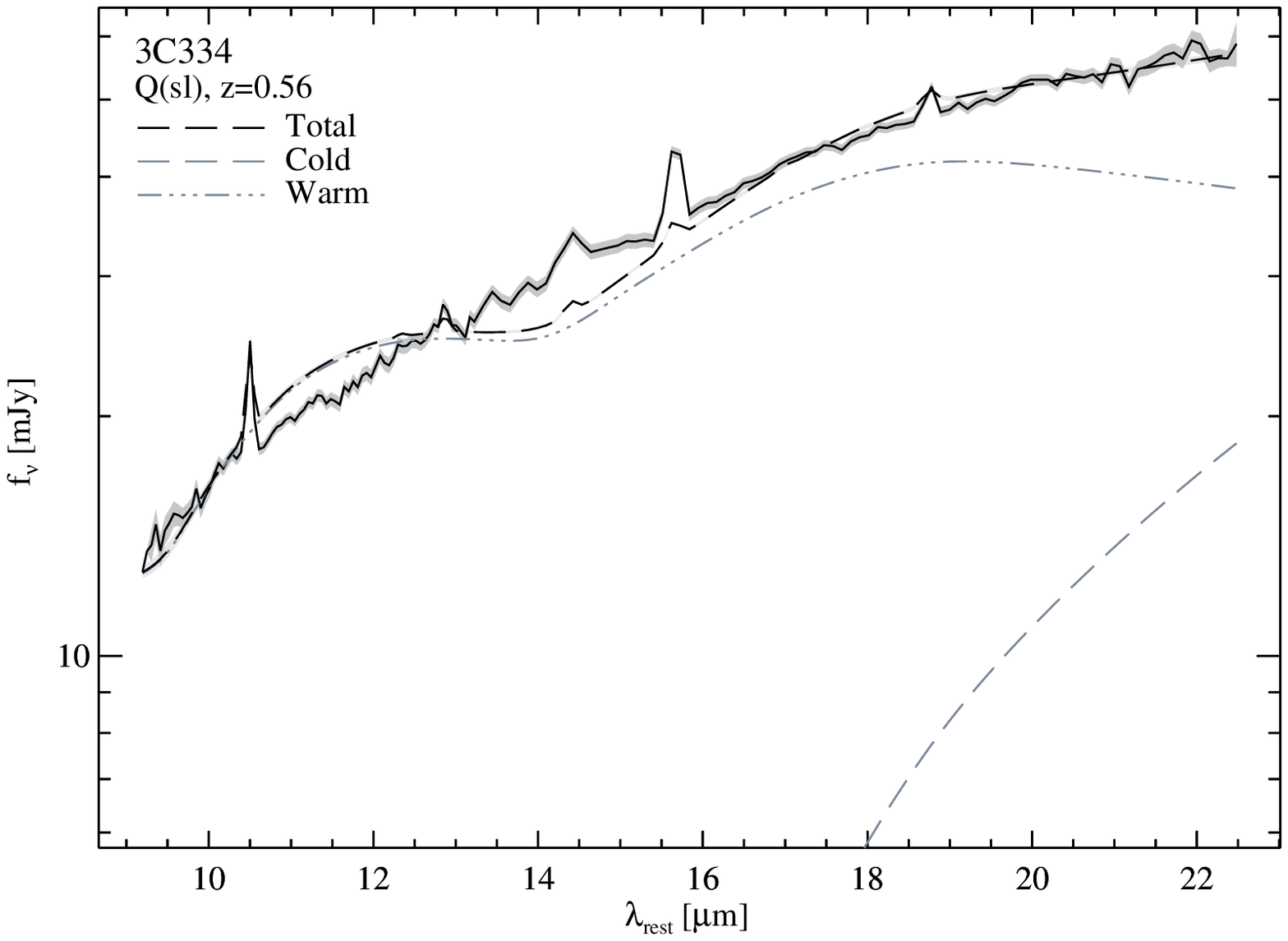,angle=0,width=8cm,height=6.0cm,clip=} & 
\psfig{file=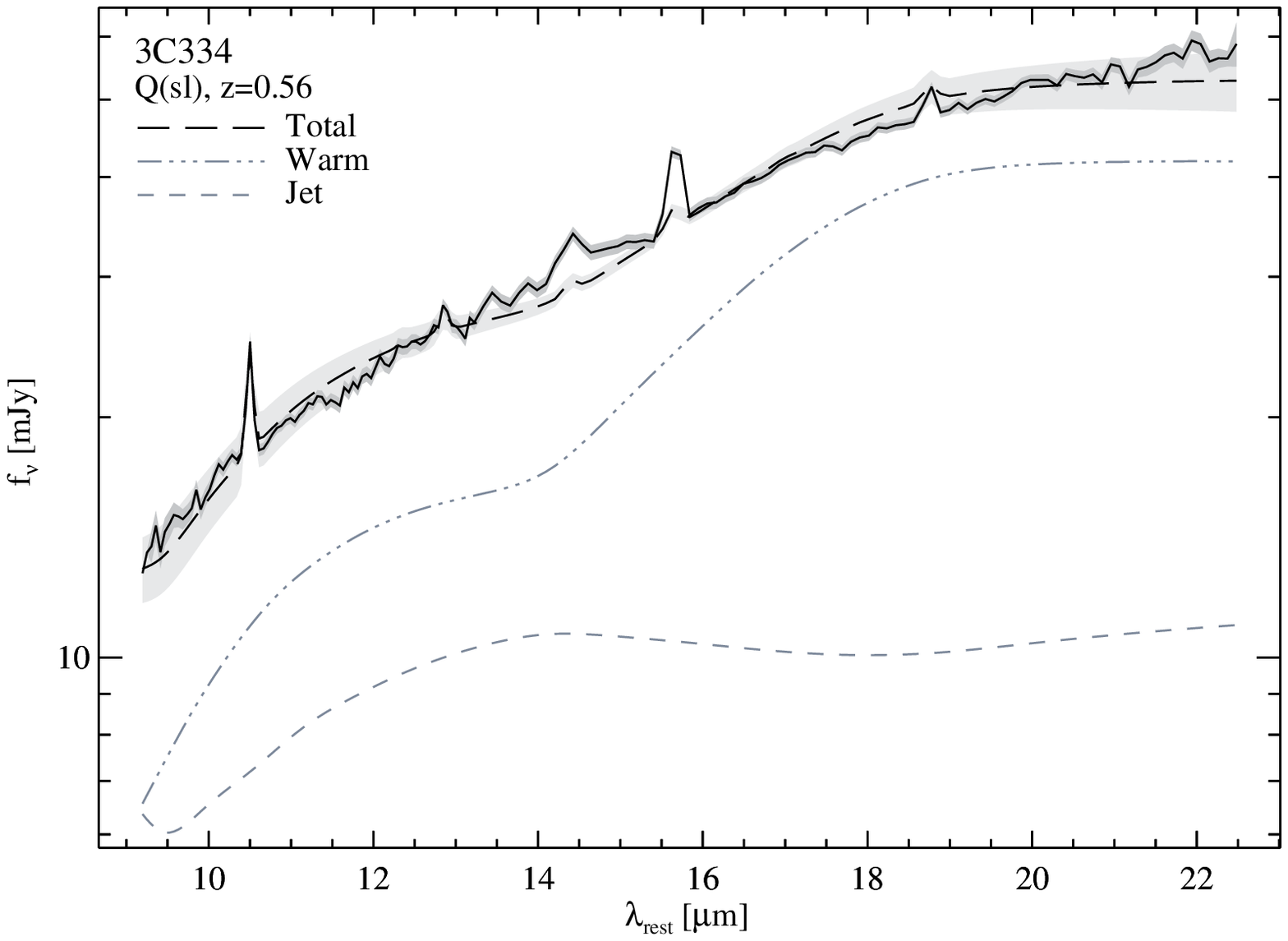,angle=0,width=8cm,height=6.0cm,clip=}\\

\end{tabular}
\caption{The fit to the IRS spectrum for three quasars with warm dust + lobe synchrotron + cool dust (left) and warm dust  + lobe synchrotron + jet synchrotron (right). The importance of the constraints from the IRS spectrum on the fits is clear.
\label{alts}}
\end{figure*}

\clearpage

\section{NOTES ON THE FITS FOR SOME INDIVIDUAL OBJECTS}

\noindent\textbf{3C\,6.1}: IRS SL2 was not detected for this source and SL1 alone provides poor constraints on the optical depth from the fits.

\noindent\textbf{3C\,22}: The best fitting combination of components includes a jet functional form, since this component greatly improves the fit to the IRS spectrum. The next-best fit involved two dust components and a lobe synchrotron component, as shown in Figure~\ref{alts}, but the fit to the IRS spectrum is much poorer. Since this is a broad-line galaxy, it is not unreasonable to expect some contribution from the jet, however sub-millimeter measurements of this source are required to settle the issue.

\noindent\textbf{3C\,200}: No IRS spectrum was available for this galaxy due to an IRS peak-up failure. Instead, a warm dust component with the temperature fixed to ${\rm T} = 185$\,K was fitted to the MIPS 24\microns\ detection.

\noindent\textbf{3C\,207}: This object has been examined in detail by \cite{vanbemmel_etal_98} who performed a spectral decomposition based on simultaneous cm and mm data and estimate that substantial non-thermal contamination of the {\it IRAS\/} 60\microns\ flux density is possible. The orientation-based model of \cite{hoekstra_etal_97} yields an estimate for the non-thermal contribution of around 15\% at 60\microns . We estimate a non-thermal contribution of approximately 90\% at rest-frame 30\microns . 

\noindent\textbf{3C\,216}: This is a quasar with known superluminal motion and beamed jet emission is clearly present in the radio and millimetric data. A combination of three components best fits the data: a dust component with ${\rm T} = 158$\,K, low-frequency synchrotron emission from the lobes, and Doppler-boosted synchrotron emission from the jet. We estimate that the jet contributes around 90\% of the flux density at 15\microns .

\noindent\textbf{3C\,220.3}: This galaxy provided the only MIPS detection at 160\microns , as well as detections at 24 and 70\microns . The measured 160\microns\ flux density was very high, at 136\,mJy. No IRS spectrum was available for this galaxy due to an IRS peak-up failure. Instead, a warm dust component with fixed temperature ${\rm T} = 185$\,K and a cool dust component whose temperature was a free parameter were included in the fits. The resulting best-fitting cool component has a temperature ${\rm T} = 31$\,K and comprises $\approx 69$\% of the total fitted dust luminosity, a proportion comparable to that seen in starburst galaxies \citep[see e.g.,][]{marshall_etal_06}. We omitted a mm upper limit from the fit in order to achieve a good fit to the radio data using the lobe synchrotron functional form.

\noindent\textbf{3C\,272}: No IRS spectrum was available for this galaxy due to an IRS peak-up failure. Instead, a warm dust component with the temperature fixed to ${\rm T} = 185$\,K was fitted to the MIPS 24 and 70\microns\ detections.

\noindent\textbf{3C\,280}: We have not included an apparently discrepant {\it ISO\/} detection from \cite{andreani_etal_02} in the fit.

\noindent\textbf{3C\,286}: We have not included apparently discrepant {\it ISO\/} detections from \cite{meisenheimer_etal_01} and \cite{andreani_etal_02} in the fit.

\noindent\textbf{3C\,289}: No IRS spectrum was available for this galaxy due to an IRS peak-up failure. Instead, a warm dust component with the temperature fixed to ${\rm T} = 185$\,K was fitted to the MIPS 24 and 70\microns\ detections.

\noindent\textbf{3C\,309.1}: The best fit for this SSC quasar has synchrotron components from the radio lobes and jet, as well as a dust component with $T = 181$\,K. However, there is a discrepant {\it ISO\/} detection at 60\microns\ \citep{andreani_etal_02} which we have excluded from the fit. We were unable to achieve a good fit to the IRS data without also excluding the upper limit at 850\microns\ \citep{haas_etal_04}.

\noindent\textbf{3C\,325}: The {\it ISO\/} 60 and 90\microns\ detections from \cite{andreani_etal_02} appear inconsistent with MIPS upper limits and so they have been excluded from the fit.

\noindent\textbf{3C\,334}: As well as lobe synchrotron and warm dust, the best fit includes a jet functional form which exceeds a sub-millimeter upper limit. Although some far-infrared data are not well fit by the model, a cool dust component was not also included in the best fit since the addition of this component resulted in a confidence level for the F-test statistic of just 40\%. The inclusion of a jet in the best fit is largely due to the IRS data (see Fig.~\ref{alts}) and we estimate the non-thermal contribution to be approximately 20\% at rest-frame 30\microns . This object has also been examined in detail by \cite{vanbemmel_etal_98} who performed a spectral decomposition based on simultaneous cm and mm data and estimate a negligible non-thermal contribution to the  {\it IRAS\/} 60\microns\ flux density. In addition, the orientation-based model of \cite{hoekstra_etal_97} yields a non-thermal contribution of much less than 15\% at 60\microns . However, as noted by \cite{vanbemmel_etal_98}, the existence of a variable sub-millimeter component, not accounted for in their analysis, is likely in this superluminal object.

\noindent\textbf{3C\,427.1}: No constraints were available in the infrared since this galaxy was not detected with MIPS or IRS.

\end{document}


\special{papersize=11in,8.5in}

\begin{deluxetable}{llcccccccc}
\centering
\rotate
\tabletypesize{\tiny}
\tablecaption{Characteristic luminosities for the primary 3CRR Sample.}
\label{char_lums.tbl}
\tablewidth{0pt}
\tablenum{7}
\tablehead{
\colhead{} & \colhead{} & \colhead{$F_{\rm178/(1+z)\,MHz}$} & \colhead{$F_{\rm 5/(1+z)\,GHz}$} & \colhead{$F^{\rm core}_{\rm 5\,GHz}$} & \colhead{$F_{15(1+z)\mu{\rm m}}$} &\colhead{$F_{30(1+z)\mu{\rm m}}$} & \colhead{$\log L_{\rm 178\,MHz}$} & \colhead{$\log L_{15\mu{\rm m}}$} & \colhead{$\log L_{30\mu{\rm m}}$}\\
\colhead{Name} & \colhead{Type} & \colhead{[Jy]} & \colhead{[Jy]} & \colhead{[mJy]} & \colhead{[mJy]} & \colhead{[mJy]} & \colhead{[W\,Hz$^{-1}$\,sr$^{-1}$]} & \colhead{[W\,Hz$^{-1}$\,sr$^{-1}$]} & \colhead{[W\,Hz$^{-1}$\,sr$^{-1}$]}
}
\startdata
3C\,138  & Q(SSC)(SL) 	&	28.9  &	\phn6.0  &	\phm{$<$}\phn\phn94.0  	   & \phm{$<$}18.9  	&	\phm{$<$}21.7      	   &	27.5 &	\phm{$<$}24.4  &	\phm{$<$}24.4 \\
3C\,147  & Q(SSC)(SL) 	&	74.5  &	11.8 	 &	\phm{$<$}2500.0		   & \phm{$<$}20.1  	&	\phm{$<$}39.9  	   &	27.7 &	\phm{$<$}24.1  &	\phm{$<$}24.4   \\
3C\,175  & Q      	&	31.0  &	\phn1.0  &	\phm{$<$}\phn\phn23.5  	   & \phm{$<$}18.2  	&	\phm{$<$}22.2  	   &	27.6 &	\phm{$<$}24.3  &	\phm{$<$}24.4   \\
3C\,196  & Q      	&	118.0 &	\phn7.3  &	\phm{$<$}\phn\phn\phn7.0   & \phm{$<$}22.7  	&	\phm{$<$}13.2  	   &	28.3 &	\phm{$<$}24.6  &	\phm{$<$}24.3   \\
3C\,207  & Q      	&	11.1  &	\phn1.8  &	\phm{$<$}\phn510.0 	   & \phm{$<$}12.1  	&	\phm{$<$}17.7  	   &	27.0 &	\phm{$<$}24.1  &	\phm{$<$}24.2   \\
3C\,216  & Q(SL)  	&	34.9  &	\phn1.9  &	\phm{$<$}1050.0		   & \phm{$<$}22.0  	&	\phm{$<$}47.7  	   &	27.5 &	\phm{$<$}24.3  &	\phm{$<$}24.6   \\
3C\,254  & Q      	&	41.4  &	\phn1.0  &	\phm{$<$}\phn\phn19.0  	   & \phm{$<$}13.8  	&	\phm{$<$}11.9  	   &	27.7 &	\phm{$<$}24.2  &	\phm{$<$}24.1   \\
3C\,263  & Q(SL)  	&	21.3  &	\phn1.6  &	\phm{$<$}\phn157.0 	   & \phm{$<$}27.7  	&	\phm{$<$}29.7  	   &	27.3 &	\phm{$<$}24.4  &	\phm{$<$}24.4   \\
3C\,275.1& Q      	&	26.3  &	\phn1.1  &	\phm{$<$}\phn130.0 	   & \phm{$<$}\phn8.1   &	\phm{$<$}15.8  	   &	27.2 &	\phm{$<$}23.7  &	\phm{$<$}24.0   \\
3C\,286  & Q      	&	38.3  &	10.6 	 &	\phn\phn\ldots			   & \phm{$<$}\phn9.7   &	\phm{$<$}21.9  	   &	27.8 &	\phm{$<$}24.2  &	\phm{$<$}24.5   \\
3C\,309.1& Q(SSC) 	&	33.2  &	\phn5.4  &	\phm{$<$}2350.0		   & \phm{$<$}22.8  	&	\phm{$<$}31.6  	   &	27.8 &	\phm{$<$}24.6  &	\phm{$<$}24.7   \\
3C\,334  & Q(SL)  	&	14.0  &	\phn0.8  &	\phm{$<$}\phn111.0 	   & \phm{$<$}33.0  	&	\phm{$<$}51.6  	   &	26.9 &	\phm{$<$}24.3  &	\phm{$<$}24.5   \\
3C\,336  & Q(SL)  	&	19.9  &	\phn1.5  &	\phm{$<$}\phn\phn20.4  	   & \phm{$<$}\phn5.3   &	\phm{$<$}10.1  	   &	27.6 &	\phm{$<$}24.0  &	\phm{$<$}24.3   \\
3C\,343  & Q(SSC) 	&	19.2  &	\phn3.0  &	\phn$<$300.0		   & \phm{$<$}16.6  	&	\phm{$<$}32.1  	   &	27.6 &	\phm{$<$}24.5  &	\phm{$<$}24.8   \\
3C\,380  & Q(SSC)(SL) 	&	92.3 &	\phn9.0  &	\phm{$<$}7447.0		   & \phm{$<$}30.4  	&	\phm{$<$}32.5  	   &	28.0 &	\phm{$<$}24.5  &	\phm{$<$}24.5   \\
3C\,6.1  & G		&	19.7  &	\phn2.0  &	\phm{$<$}\phn\phn\phn4.4   & \phm{$<$}\phn3.8   &	\phn$<$9.7	   & 	27.5 &	\phm{$<$}23.7  &	$<$24.2  \\
3C\,22   & G      	&	20.3  &	\phn1.4  &	\phm{$<$}\phn\phn\phn7.8   & \phm{$<$}16.7  	&	\phm{$<$}20.8  	   &	27.6 &	\phm{$<$}24.5  &	\phm{$<$}24.6   \\
3C\,184  & G      	&	22.0  &	\phn1.4  &	\phn\phn\phn$<$0.2  	   & \phm{$<$}\phn4.2   &	\phm{$<$}\phn8.7   &	27.7 &	\phm{$<$}23.9  &	\phm{$<$}24.3   \\
3C\,200  & G      	&	13.9  &	\phn0.9  &	\phm{$<$}\phn\phn35.1  	   & \phn$>$1.1  	&	\phn$<$4.3  	   &	26.8 &	$>$22.7        &	$<$23.3  \\
3C\,220.1& G      	&	20.3  &	\phn1.0  &	\phm{$<$}\phn\phn25.0  	   & \phm{$<$}\phn2.7   &	\phn$<$5.0  	   &	27.2 &	\phm{$<$}23.3  &	$<$23.6  \\
3C\,220.3& G      	&	29.0  &	\phn1.2  &	\phn\phn\phn$<$0.2 	   & \phm{$<$}\phn5.2   &	\phm{$<$}13.3  	   &	27.5 &	\phm{$<$}23.7  &	\phm{$<$}24.1   \\
3C\,268.1& G      	&	32.3  &	\phn4.5  &	\phm{$<$}\phn\phn\phn2.0   & \phm{$<$}\phn3.8   &	\phn$<$5.8  	   &	27.8 &	\phm{$<$}23.9  &	$<$24.1  \\
3C\,272  & G      	&	12.8  &	\phn0.8  &	\phn\phn\ldots			   & \phm{$<$}\phn2.8   &	\phm{$<$}\phn7.5   &	27.4 &	\phm{$<$}23.7  &\phm{$<$}24.1   \\
3C\,280  & G      	&	36.4  &	\phn3.1  &	\phm{$<$}\phn\phn\phn1.0   & \phm{$<$}13.1  	&	\phm{$<$}16.5  	   &	27.9 &	\phm{$<$}24.4  &	\phm{$<$}24.5   \\
3C\,289  & G      	&	21.6  &	\phn1.3  &	\phm{$<$}\phn\phn10.6  	   & \phm{$<$}\phn5.0   &	\phm{$<$}13.6  	   &	27.6 &	\phm{$<$}24.0  &	\phm{$<$}24.4   \\
3C\,292  & G      	&	11.7  &	\phn1.1  &	\phm{$<$}\phn\phn\phn1.0   & \phm{$<$}\phn2.2   &	\phn$<$5.6  	   &	27.1 &	\phm{$<$}23.4  &	$<$23.8  \\
3C\,325  & G      	&	28.1  &	\phn2.2  &	\phm{$<$}\phn\phn\phn2.4   & \phm{$<$}\phn5.5   &	\phn$<$6.7     	   &	27.9 &	\phm{$<$}24.2  &	$<$24.3  \\
3C\,330  & G      	&	38.6  &	\phn3.3  &	\phm{$<$}\phn\phn\phn0.7   & \phm{$<$}\phn7.1   &	$<$10.8    	   &	27.4 &	\phm{$<$}23.6  &	$<$23.8  \\
3C\,337  & G      	&	20.1  &	\phn1.5  &	\phm{$<$}\phn\phn\phn0.3   & \phm{$<$}\phn1.0   &	\phn$<$3.9  	   &	27.2 &	\phm{$<$}22.9  &	$<$23.5  \\
3C\,340  & G      	&	15.8  &	\phn1.2  &	\phm{$<$}\phn\phn\phn1.1   & \phm{$<$}\phn2.7   &	\phn$<$6.6  	   &	27.3 &	\phm{$<$}23.5  &	$<$23.9  \\
3C\,343.1& G(SSC) 	&	13.7  &	\phn1.8  &	\phn$<$200.0		   & \phm{$<$}\phn1.5   &	\phn$<$5.4  	   &	27.2 &	\phm{$<$}23.3  &	$<$23.8  \\
3C\,352  & G      	&	21.8  &	\phn0.9  &	\phm{$<$}\phn\phn\phn3.2   & \phm{$<$}\phn1.7   &	\phn$<$4.2  	   &	27.5 &	\phm{$<$}23.4  &	$<$23.8  \\
3C\,427.1& G      	&	35.0  &	\phn1.7  &	\phm{$<$}\phn\phn\phn0.8   & \phn$<$0.4  	&	\phn\ldots      	   &	27.4 &	$<$22.4        &	\phn\ldots \\
3C\,441  & G      	&	17.5  &	\phn1.4  &	\phm{$<$}\phn\phn\phn3.5   & \phm{$<$}\phn2.3   &	\phn$<$3.9  	   &	27.3 &	\phm{$<$}23.4  &	$<$23.6  \\
\enddata
\tablecomments{The table gives the flux density in the observed frame, $F_{\nu}$, at frequencies, $\nu$,  of $178/(1+z)$\,MHz and $5/(1+z)$\,GHz, using the best fitting lobe synchrotron model fits (see Table~5) to the radio data. The observed core flux density at a frequency of 5\,GHz, $F^{\rm core}_{\rm 5\,GHz}$, is also given. Assuming a core flux density spectral index $\alpha=0$, the core-dominance parameter is calculated as $R=F^{\rm core}_{\rm 5\,GHz}/F_{\rm 5/(1+z)\,MHz}$. The observed flux density at a wavelength corresponding to 15\microns\ in the rest-frame, $F_{15(1+z)\mu{\rm m}}$, is computed from the IRS spectrum as the median flux density in a 0.2\microns\ bin centered on 15(1+z)\microns . The flux density at a wavelength corresponding to 30\microns\ in the rest-frame, $F_{30(1+z)\,\mu{\rm m}}$, is computed from the MIPS 24 and 70\microns\ flux densities as $F_{30(1+z)\,\mu{\rm m}} = F_{24\mu{\rm m}}(30(1+z)/24)^{-\alpha}$, where $\alpha$ is the 24--70\microns\ flux density spectral index ($F_{\nu}\propto \nu^{\alpha}$). Note that the 30\microns\ luminosities are derived from MIPS photometric measurements with a specific spectral response, whereas the 15\microns\ flux densities are derived from spectroscopic measurements averaged over a narrow wavelength bin. Care must therefore be taken if directly comparing these quantities. The rest-frame luminosities, $L_{\rm 178\,MHz}$, $L_{15\mu{\rm m}}$ and $L_{30\mu{\rm m}}$, are calculated per steradian as $L_{\nu_{\rm e}} = F_{\nu_{\rm e}/(1+z)} D_{\rm L}^{2}/(1+z)$, where $\nu_{\rm e}$ is the emitted frequency and $D_{\rm L}$ is the luminosity distance.}
\end{deluxetable}